\documentclass[a4paper,12pt,twoside]{report}
\usepackage[utf8]{inputenc}
\usepackage[T1]{fontenc}
\usepackage[unicode,breaklinks=true]{hyperref}
\PassOptionsToPackage{defaults=hu-min}{magyar.ldf}
\usepackage[magyar,english]{babel}
\usepackage{graphicx}
\usepackage{amsmath}
\usepackage{amssymb}
\usepackage[style=english]{csquotes}
\usepackage{indentfirst}

\DeclareMathOperator{\sech}{sech}
\DeclareMathOperator{\re}{Re}
\DeclareMathOperator{\im}{Im}

\allowdisplaybreaks[1]

\begin{document}

\pagenumbering{roman}

\begin{titlepage}
 \begin{center}

  \vspace*{\stretch{0.1}}

  \huge
  \textbf{A review on radiation of\\
  oscillons and oscillatons
  }

  \vspace{\stretch{1}}

  \large
  A review based on the English translation of a\\
  thesis for the Doctor of Sciences (D.Sc.) degree\\
  of the Hungarian Academy of Sciences

  \vspace{\stretch{1}}

  \textbf{Gyula Fodor}

  \vspace{\stretch{0.1}}

  \Large
  Wigner Research Centre for Physics,\\
  Institute for Particle and Nuclear Physics

  \vspace{\stretch{0.4}}

  \includegraphics[width=50mm]{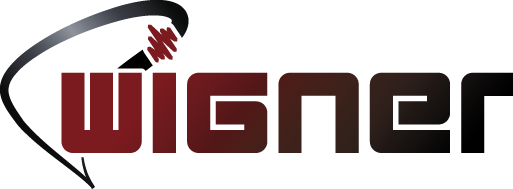}

  \vspace{\stretch{1}}

  Budapest, 2019
 \end{center}
\end{titlepage}

\newpage

\begin{abstract}
 Numerical simulations show that a massive real scalar field in a nonlinear theory can form long-lived oscillating localized states. For a self-interacting scalar on a fixed background these objects are named oscillons, while for the self-gravitating case they are called oscillatons. This extensive review is about the history and various general properties of these solutions, though mainly focusing on the small but nonzero classical scalar field radiation emitted by them. The radiation for higher amplitude states can be calculated by a spectral numerical method. For small and moderately large amplitudes an analytical approach based on complex extension, asymptotic matching and Borel summation can be used. This procedure for the calculation of the energy loss rate is explained in a detailed way in this review, starting with the simplest one-dimensional scalar oscillons at first, and reaching to $3+1$ dimensional self-gravitating oscillatons based on that experience.
\end{abstract}

\newpage

\pagenumbering{arabic}

\tableofcontents

\chapter{Introduction}

A fundamental question of physics is how the available matter can form spatially localized lumps, and what prevents the formed objects from falling apart, evaporating or radiated out, such that the matter and energy occupy uniformly all the available space. In this review we approach this problem from a theoretical perspective where we study the simplest classical field theoretical models where long-lived localized states may appear. In this way we can get insight on what properties a theory must have so that these type of structures would be allowed to exist at all. The objects corresponding to the states appearing in the simplest models can also be expected to develop in more general, complicated, physically more relevant theories. The typical size of the formed structures may range from the size of particles to galaxy clusters, depending on the parameters in the theory, such as the mass of the scalar field, for example.

Presumably, the simplest classical relativistic field theoretical model is the theory in which there is only a single real (not complex) scalar field $\phi$ on a fixed $1+1$ dimensional Minkowski spacetime background. The scalar field $\phi$ depends on the time coordinate $t$ and spatial coordinate $x$, and satisfies the differential equation
\begin{equation}
 -\frac{\partial^2\phi}{\partial t^2}+\frac{\partial^2\phi}{\partial x^2}=U'(\phi) \ ,
\end{equation}
where $U'(\phi)$ denotes the derivative of the scalar potential $U(\phi)$ with respect to $\phi$. For appropriate choice of $U(\phi)$ various localized states can already appear in this simple model. The most widely investigated cases are the $\phi^4$ potential with two symmetrically placed minimums, $U(\phi)=(\phi^2-1)^2/4$, and the sine-Gordon potential $U(\phi)=1-\cos\phi$, which has infinitely many identical minimums. In both models there are \emph{kink} solutions, which appear at the boundary of two different vacuum domains. The kink solution belongs to the class of \emph{solitons}. Solitons are nonlinear wave packets that propagate with constant speed, keep their shape during the propagation, and they show considerable stability even when colliding with each other. Solitons appear in several physical systems. Their discovery is associated to the name of John Scott Russell, a Scottish civil engineer, who in 1834 observed a stable wave packet propagating along a canal, and followed it on horseback for a few kilometers.

If we observe the solution from a system moving along with the kink, then we obtain the simplest static kink solution. The shape of this solution for the case of the sine-Gordon theory is $\phi=4\arctan(\pm\exp x)$. Although the energy density of the $1+1$ dimensional kink is essentially restricted to a bounded region, it is not a localized solution in the sense that it tends to different vacuum states in the negative and positive $x$ directions. The $3+1$ dimensional generalization of this solution is not a particle-like state localized around a point, but rather a domain wall. From Derrick's theorem we know that there are no stable static localized solutions for three or more spatial dimensions \cite{Derrick64}.

The other particle-like solution in the sine-Gordon theory is known as the \emph{sine-Gordon breather}, the form of which will be written out later in equation \eqref{eqsinegex}. There is a one-parameter family of sine-Gordon breather solutions, which can be parametrized by the frequency of the oscillation of the field. The breather is a time-periodically oscillating exponentially localized exact solution, which is stable and do not radiate energy out to infinity. For the sine-Gordon breather the scalar field tends to the same vacuum value at both the negative and positive directions, hence there is a possibility for the existence  of a $3+1$ dimensional spherically symmetric generalization.

The spatially reflected version of the kink solution is named antikink. Surprisingly, a sine-Gordon breather cannot be formed by the collision of a kink and an antikink. According to the exact solution describing the collision, the two objects go through each other without any change in their velocity. This is an exceptional property of the integrable sine-Gordon theory. In spite of this, the sine-Gordon breather may be still considered as a bound state of a kink and an antikink.

The $1+1$ dimensional sine-Gordon theory is also very special in the sense that it is the only theory involving a real scalar field where an exactly time-periodic localized breather can exist. In other models with analytic potential, including the higher dimensional sine-Gordon one, not only an exact solution is not known, but even with numerical methods it is not possible to find a finite energy localized time-periodic solution. It can also be proved by precise mathematical methods, that for the $1+1$ dimensional case one-parameter families of breathers can only exist for the sine-Gordon potential \cite{Kichenassamy91}. In this way, from the intended localized solutions not only the static but the time-periodic solutions are also excluded.

Surprisingly, in spite of all this, as numerical experience shows, a self-interacting real scalar field can still form long-lived localized oscillating states. However, these solutions always lose energy slowly by the radiation of scalar field, and hence their frequency is also changing gradually. These objects were called pulsons originally \cite{Bogolubsky-77b,bogolyubskii-77}, but later the name \emph{oscillon} became widespread in the literature \cite{gleiser-94,copeland-95}. For example, in the $1+1$ dimensional $\phi^4$ theory oscillons can easily form by the collision of a kink and an antikink, after radiating out the unnecessary energy \cite{kudryavtsev-75}.

The importance of oscillons is greatly increased by the observation that they are not exceptional states, they evolve from quite general initial data in several cases. For more than one spatial dimensions oscillons become spherically symmetric quickly by radiating out scalar field \cite{adib2002}. Oscillons can be produced rather easily by numerically following the spherically symmetric time-evolution of a general bell-shaped initial data. According to more general numerical simulations, oscillons can form from randomly chosen not spherically symmetric initial data as well \cite{Kolb94,hindmarsh2006,Amin10}. This makes it likely that they may have played a role in the early universe, evolving from the inflaton field or from some of the other scalars coupled to it. At the forming and decaying of oscillons the emission of gravitational waves may be increased, which might be observed as peaks on the spectrum of gravitational waves \cite{Zhou13,Antusch17,Antusch18a,LiuGuo18,Antusch18b,Amin18,Kitajima18,LiuGuo19,Sang19,Lozanov19}.

The nonlinearity of the theory is an essential feature for the existence of oscillons. The oscillons cannot be described by first order linear perturbation. There are small-amplitude oscillons, but the smaller the amplitude is the larger their spatial size becomes. This is necessary in order to allow the influence of the nonlinearity become effective even for small amplitude states. The shape of the oscillon is getting determined from conditions arising at higher orders in the perturbational analysis.

Oscillons have been observed to form in theories containing several kind of interacting fields, such as for example in the bosonic sector of the standard model, when the Higgs mass is twice as large as the mass of the $W^{\pm}$ boson \cite{Farhi05,Graham07a,Graham07b}. Oscillons can also form in Abelian Higgs models at the decay of sphalerons, at the collision of vortex antivortex pairs, or as a result symmetry breaking \cite{Arnold88,Rebbi96,Gleiser07,Gleiser09,Gleiser12,Achilleos13,Diakonos15}.

Although the scalar field oscillons are classical field theoretical solutions, they can also be considered as collective quantum states formed by large number of identical particles. Before the numerical discovery of oscillons, in their paper appeared in 1975, Dashen, Hasslacher and Neveu have already arrived at the study of the $1+1$ dimensional sine-Gordon breathers and $\phi^4$ oscillons, while applying the semiclassical WKB method discovered by them \cite{dashen-75}. Considering the small-amplitude oscillons as stable time-periodic solutions and quantizing the perturbations around them they have calculated the energy levels at weak coupling. The semiclassical method for the calculation of the radiation of oscillons has been applied by Hertzberg \cite{Hertzberg10}. Considering the oscillons as quantum systems and using inhomogeneous Hartree approximation and numerical methods, the time evolution and radiation of oscillons have been investigated in paper \cite{Saffin14}. The radiation of the sine-Gordon breather have been calculated by the classical-quantum correspondence method in \cite{Olle19}.

In Chapter \ref{chapteroscillon} of this review, after presenting the most important papers about oscillons, the results that we have obtained in our papers \cite{Fodor2006,Fodor2008,Fodor2009a,Fodor2009b} concerning their radiation are discussed in detail. Apart from the exceptional sine-Gordon breather, all oscillon solutions emit energy slowly by the radiation of scalar field. In certain cases this radiation can be so weak that one may have chance to detect it only by extremely precise numerical methods. For a long time it was unclear whether or not there may exist exactly periodic nonradiating solutions. By varying the parameters of a Gaussian-type initial data, in 2002 Honda and Choptuik found 125 resonance peaks, which appeared to correspond to periodic oscillons \cite{HondaChoptuik2002}. In our first paper about oscillons, we have shown that the states belonging to these peaks are actually oscillons belonging to the low amplitude unstable domain, which, even if very slightly, but necessarily radiate in a detectable way \cite{Fodor2006}.

If we compensate the energy loss of an oscillon by a same amplitude and frequency incoming wave, then we obtain an exactly time-periodic state with a small amplitude standing wave tail outside the core region. The spherically symmetric standing wave tail goes out to arbitrarily large distances, and the decrease of its amplitude is so slow that the whole energy of the system turns out to be infinite. Although the obtained solution is not localized, it still has a core region where the energy density is many orders of magnitude larger than in the tail zone. At some given frequency, for the solution with the minimal amplitude tail we have introduced the naming \emph{quasibreather} \cite{Fodor2006}. The amplitude of the standing wave tail of the quasibreather agrees with the tail-amplitude of the radiating oscillon. A big advantage of the introduction of the quasibreather is that numerically it can be much more precisely calculated than the corresponding oscillon state.

The core region of oscillons can be described quite precisely by a power series expansion in terms of an amplitude parameter
\cite{Kosevich75,dashen-75,SegurKruskal87,Buslaev1977,bogolyubskii-77c,Kichenassamy91}. This parameter is usually denoted by $\varepsilon$. The relation between the oscillon's frequency $\omega$ and the amplitude parameter can always be chosen in the form $\omega^2=1-\varepsilon^2$. The expansion worked out to higher orders has been presented in our paper \cite{Fodor2008}. The formalism also gives a result for the $\varepsilon$ dependence of the oscillon's energy, which is valid for low and moderate amplitudes. For higher amplitude states we can obtain the precise value of the energy by a spectral numerical code. The dependence of the energy on the central amplitude is important because it determines the stability of the system. If increasing the central amplitude the total energy of the system also increases, then the system is stable, otherwise it is unstable. From the analysis it follows that $3+1$ dimensional oscillons are stable only if their amplitude is above a certain limit. If their amplitude decreases below that value because of the radiation loss, then they suddenly decay. In case of $1$ and $2$ spatial dimensions all oscillons are stable under a certain amplitude. Since the energy loss rate decreases exponentially when the amplitude decreases, these lower dimensional oscillons never decay.

Although the small-amplitude expansion gives a very good representation of the core region, it is unable to describe the radiating tail, which is exponentially small in $\varepsilon$. This is closely related to the fact that this expansion is not convergent, it is an asymptotic series representation. Extending the formalism to complex values of the radial coordinate $r$, it is possible to determine the amplitude of the tail by investigating the behavior of the quantities near a singularity on the complex $r$ plane. This method was introduced by Segur and Kruskal in 1987, for the case of $1+1$ dimensional oscillons \cite{SegurKruskal87}. In our paper \cite{Fodor2009a} we have extended the procedure by a Borel summation method, and for symmetric $U(\phi)$ potentials we have determined the radiation amplitude by a purely analytic way. For two and three spatial dimensions we have generalized the procedure in our paper \cite{Fodor2009b}. In Section \ref{secsuganal} of this review we explain in a more comprehensible way this rather complicated procedure, which has been presented in a rather concise way in our papers.

In Chapter \ref{fejrelgrav} of this review we investigate localized states formed by a real scalar field interacting with gravity in the framework of general relativity. In this case, the potential $U(\phi)$ determining the self-interaction of the scalar field may have the form $U(\phi)=\frac{1}{2}m\phi^2$ corresponding to the Klein-Gordon case, since gravity already provides the necessary nonlinearity. On a fixed flat background the equations describing the Klein-Gordon scalar are linear, hence they cannot form oscillons. In the gravitational case the most studied case is just the Klein-Gordon scalar. Maybe this is the reason why the theory of localized states formed by self-gravitating scalars was born and developed in a completely separated way from that of the flat background oscillons. These configurations were discovered in 1991 by Seidel and Suen, who found apparently time-periodic localized states formed by a self-gravitating scalar field using numerical methods \cite{seidel-91}. They gave the name \emph{oscillaton} to these objects \cite{seidel-94}. In spite of the markedly similar naming, generally there is no reference to the other topic in any of the papers about one of the two subjects.

Majority of the published studies on oscillatons are restricted to the physically important $3+1$ dimensional case. As we have mentioned earlier, oscillons defined on a $3+1$ dimensional background suddenly decay after a few thousand oscillations, at most. In contrast to this, the lifetime of $3+1$ dimensional self gravitating oscillatons is infinite, their behavior is similar to the $1+1$ and $2+1$ dimensional oscillons.

Oscillatons can easily develop from Gaussian shaped initial data, and non-symmetric initial data also develops into spherically symmetric oscillatons by quick radiation of the surplus mass. The radiation is so small that for a long time it was tacitly assumed that oscillatons are time-periodic and nonradiating. This is understandable if we know that even in the most favorable case, at the maximal mass oscillaton, when the ratio of the amplitude of the radiating tail to the central amplitude is maximal, the tail amplitude is of the order $10^{-8}$, while the central amplitude is about $0.5$.

It was first pointed out by Don Page in 2003 that oscillatons must necessarily radiate, hence they cannot be exactly time periodic and localized \cite{PageDon04}. Because of the mass loss, their amplitude and frequency slowly change as time passes. In our paper \cite{fodor2010a} the method applied for oscillons earlier has been generalized for the case of oscillatons. In a subsequent paper we have also calculated the strength of the radiation by a spectral numerical method, and we have obtained results consistent with our analytical calculations \cite{grandclement2011}. Small but positive cosmological constant further increases the amplitude of the radiation, but it still remains exponentially small in terms of the cosmological constant \cite{fodor2010b}. In Chapter \ref{fejrelgrav} of this review the results published in our papers \cite{fodor2010a,grandclement2011,fodor2010b} are presented in detail.

Several physical applications of oscillatons have been proposed in the literature up to now. Oscillatons formed by scalar fields in cosmological models may be suitable for describing dark matter in galaxies \cite{Matos01,Alcubierre02,Susperregi03,Malakolkalami16}. The real scalar field necessary for the formation of oscillons or oscillatons may be most naturally provided by axions or similar weakly interacting hypothetical bosonic particles \cite{Tkachev86,Hogan88,Kolb93,Kolb94,Arvanitaki10,Marsh16,Marsh17,Krippendorf18}. If the self-interaction of the scalar field is determined by a potential $U(\phi)$ typical for axions, then in several papers the forming oscillatons are called axion stars, see e.g.~\cite{Visinelli18,Braaten18,Chavanis18}. Cosmological simulations describing the evolution of axion-like fields show that in the central parts of the forming diffuse matter structures solitonic cores are developing, which also correspond to oscillatons \cite{Schive14,Veltmaat16}. Fuzzy dark matter constituting of very low mass axion-like particles may provide an explanation of why in galaxies we cannot observe dark matter lumps which are below a certain size \cite{HuAtal00,Marsh15,Schwabe16,HuiAtal17}. Scalar dark matter may also be accumulated inside stars, and may form oscillaton-like cores there \cite{Brito15,Brito16}. Massive real vector fields under the influence of gravitation can also form localized states, the so called Proca stars, with a time-periodically oscillating metric similar to that of oscillatons \cite{Garfinkle03,Brito16}.

\chapter{Localized states on fixed bacground: oscillons} \label{chapteroscillon}

\section{Nonlinear scalar fields}

We consider a $d+1$ dimensional spacetime, the geometry of which is described by a metric tensor $g_{ab}$. The behavior of a minimally coupled real scalar field $\phi$ on this fixed background can be determined by the Lagrangian density
\begin{equation}
 \mathcal{L}_M=-\sqrt{-g}\left[\frac{1}{2}\nabla^a\phi\nabla_a\phi+U(\phi)\right] ,
 \label{eqlagrmatter}
\end{equation}
where $U(\phi)$ is the potential determining the self interaction of the scalar field, $\nabla_a$ is the derivative operator belonging to the metric $g_{ab}$, and the determinant of the metric is $g$. Variation of the action integral
\begin{equation}
 A_M=\int\mathrm{d}t\,\mathrm{d}^d x\, \mathcal{L}_M  \qquad
\end{equation}
gives the equation describing the scalar field,
\begin{equation}
 \nabla^a\nabla_a\phi=U'(\phi) \ , \label{fieldeq1}
\end{equation}
where the prime denotes the derivative of the potential with respect to $\phi$. Variation with respect the metric provides the stress-energy tensor of the scalar field,
\begin{equation}
 T_{ab}=-\frac{2}{\sqrt{-g}}\,\frac{\delta A_M}{\delta g^{ab}}
 =\nabla_a\phi\nabla_b\phi-g_{ab}\left[\frac{1}{2}\nabla^c\phi\nabla_c\phi
 +U(\phi)\right]  . \label{streneq}
\end{equation}
It can be easily checked that if \eqref{fieldeq1} holds, the stress-energy tensor is divergence-free, $\nabla^b T_{ab}=0$. The choice $U(\phi)=m^2\phi^2/2$ corresponds to the Klein-Gordon field with mass  $m$. Unless explicitly stated, we use $c=G=\hbar=1$ Planck units.

If there is a timelike Killing vector $k^a$ on the spacetime, $\nabla^{(a}k^{b)}=0$, then the vector $J^a=T^{ab}k_b$ is divergence-free, $\nabla^a J_a=0$, and hence defines conserved quantities. If the constant $t$ time surfaces have the future pointing unit length normal vector $t^a$, then the energy density is defined by $\mathcal{E}=t^a J_a$.

Localized spherically symmetric scalar states on fixed anti-de Sitter background have been studied in detail in our paper \cite{Fodor14}. Because of the effective attraction provided by the negative $\Lambda$ cosmological constant, in that case there always exist non-radiating breather solutions \cite{Avis78}. In the remaining part of the chapter we only consider scalar fields defined on a flat Minkowski background.

\section{Scalar field on Minkowski background}

For a flat Minkowski background $g_{ab}=\mathrm{diag}(-1,1,...1)$, and equation \eqref{fieldeq1} describing the evolution of the self-interacting scalar field can be written as
\begin{equation}
 -\frac{\partial^2\phi}{\partial t^2}+\Delta\phi=U'(\phi) \ , \label{fieldeq2}
\end{equation}
where for $d$ spatial dimensions the Laplacian in Cartesian coordinates is
\begin{equation}
 \Delta=\sum_{i=1}^{d}\frac{\partial^2}{\partial {x^i}^2} \ .
\end{equation}
For Minkowski background the timelike Killing vector $k^a$ and the the normal vector $t^a$ are both $k^a=t^a=\left(\frac{\partial}{\partial t}\right)^a$, and hence according to the definition $\mathcal{E}=t^a J_a$ of the energy density,
\begin{equation}
\mathcal{E} = \frac{1}{2} \left(\frac{\partial\phi}{\partial t}\right)^2
+ \frac{1}{2} \sum_{i=1}^{d}\left(\frac{\partial\phi}{\partial x^i}\right)^2
+U(\phi) \ ,
\end{equation}
and the total energy in a given moment of time is $E = \int\mathcal{E}\mathrm{d}^{d}x$.

For a spherically symmetric scalar field configuration in spherical coordinates the Laplacian is
\begin{equation}
 \Delta\phi=\frac{\partial^2\phi}{\partial r^2}
 +\frac{d-1}{r}\frac{\partial\phi}{\partial r}
 =\frac{1}{r^{d-1}}\frac{\partial}{\partial r}\left(r^{d-1}\frac{\partial\phi}{\partial r}\right)
 \ , \label{eqlapsph}
\end{equation}
where $r^2=\displaystyle\sum_{i=1}^{d}{(x^i)}^2$. The energy density is then
\begin{equation}
\mathcal{E} = \frac{1}{2} \left(\frac{\partial\phi}{\partial t}\right)^2
+ \frac{1}{2} \left(\frac{\partial\phi}{\partial r}\right)^2
+U(\phi) \ , \label{eqendensgen}
\end{equation}
and the energy inside a sphere of radius $\bar r$ is
\begin{equation}
 E(\bar r)=\frac{2\pi^{\frac{d}{2}}}{\Gamma\left(\frac{d}{2}\right)}
 \int_0^{\bar r}r^{d-1}\mathcal{E}\mathrm{d}r \ .
 \label{eqensph}
\end{equation}
Since we have chosen the normal vector $t^a$ future directed, for the validity of Stoke's theorem we need to choose the radially directed normal vector inward pointing (see appendix B.2 of Wald's book \cite{wald1984general}). Then the outward directed energy current density is
\begin{equation}
 \mathcal{S}=n^a J_a=-\frac{\partial\phi}{\partial t}\frac{\partial\phi}{\partial r}
 \label{eqsrbar} \ ,
\end{equation}
and the energy current going out through the sphere of radius $\bar r$ is
\begin{equation}
 \mathrm{S}(\bar r)=-\frac{\mathrm{d}E(\bar r)}{\mathrm{d}t}=
 \frac{2\pi^{\frac{d}{2}}}{\Gamma\left(\frac{d}{2}\right)}
 {\bar r}^{\,d-1}_{}\mathcal{S}_{r=\bar r}
 \ . \label{eqencurr}
\end{equation}

\section{The sine-Gordon breather}

This review is focused on long-lived localized states. The simplest such configurations are the well known breather solutions of the $1+1$ dimensional sine-Gordon equations. The potential describing the interaction in this case is $U(\phi)=1-\cos\phi$. Expanding around the minimum, it can be seen that this corresponds to a self-interacting scalar field with mass $m=1$,
\begin{equation}
U(\phi)=\frac{1}{2}\phi^2-\frac{1}{24}\phi^4
+\frac{1}{720}\phi^6+\mathcal{O}(\phi^8) \ . \label{eqsinegoexporig}
\end{equation}
For $d=1$ spatial dimension, the one-parameter family of \emph{sine-Gordon breather} solutions are known:
\begin{equation}
\phi(x,t)=4\arctan\left[
\frac{\varepsilon\cos(\omega t)}{\omega\cosh(\varepsilon x)}\right] \ , \label{eqsinegex}
\end{equation}
where the parameter $\varepsilon$ determines the amplitude of the state. The parameter can take any value in the $0<\varepsilon<1$ interval, and the frequency is connected to the amplitude by the relation $\omega^2=1-\varepsilon^2$.

It was anticipated for a long time that for $1+1$ dimensions the only analytic potential is the sine-Gordon potential which allows the existence of truly localized breather solutions \cite{Eleonskii1984,Eleonsky1991}. First, for a certain class of potentials \cite{Vuillermot1987}, and then for general analytic potentials it was shown that there are no other $1+1$ dimensional breathers \cite{Kichenassamy91}. From our detailed considerations described in the following sections, it also follows that for analytic potentials the family of sine-Gordon breathers is the only truly periodic localized one-parameter family of solutions which tends to the vacuum solution for the small value of the parameter. Sine-Gordon breathers only exist on $1+1$ dimensional background. We will also see that if $d>1$, then for $d+1$ dimensional flat background there is no potential for which a family of exactly periodically oscillating localized breather solutions can exist. Numerical investigations also show that the existence of isolated breathers is unlikely, even for large amplitudes \cite{Hormuzdiar99b}.

Nevertheless, time-periodic localized breather solutions do exist if the potential $U(\phi)$ is not analytic at its minimum. One such example is the so called signum-Gordon model, where $U(\phi)=|\phi|$, and hence $U'(\phi)=\mathrm{sign}\,\phi$. The breathers in this theory are generally named oscillons \cite{Arodz08,Arodz11,Swierczynski17,Klimas18}. In this case only the scalar and its first derivative is continuous, the second derivative cannot be calculated at some points, so $\phi$ is a weak solution of the field equation. The other case where breathers are known is the potential $U(\phi)=\frac{1}{2}\phi^2\left[1-\ln\left(\phi^2\right)\right]$, which is differentiable once at the minimum, and $U'(\phi)=-\phi\ln\left(\phi^2\right)$. In this case the breather solution can be searched in the product form $U(\phi)=a(t) u(\mathbf{r})$, separating the temporal and spatial dependence \cite{Maslov90,Koutvitsky05,Koutvitsky06,Bialynicki79,Koutvitsky11}.
In the following parts of this review we will assume that the potential is analytic in a neighborhood of its minimum.

The small-amplitude expansion of the sine-Gordon breather can be achieved by introducing rescaled space and time coordinates, defining $\xi=\varepsilon x$ and $\tau=\omega t$, where $\omega^2=1-\varepsilon^2$. Then the breather solutions can be expanded with respect to the  parameter $\varepsilon$ as
\begin{align}
\phi&=4\arctan\left[
\frac{\varepsilon\cos\tau}{\omega
\cosh\xi}\right] \nonumber\\
&=4\varepsilon\frac{\cos\tau}{\cosh\xi}
+\varepsilon^3\frac{\cos\tau}{\cosh\xi}\left(2-\frac{1}{\cosh^2\xi}\right)
-\varepsilon^3\frac{\cos(3\tau)}{3\cosh^3\xi}
+\mathcal{O}(\varepsilon^5) \ . \label{eqsgphiepsexp}
\end{align}
For small amplitude configurations the first term is dominant, and the amplitude is proportional to $\varepsilon$. The characteristic size of the object is determined by the $\cosh\xi$ term in the denominator, and hence the size for small amplitudes is proportional to $1/\varepsilon$. When the amplitude tends to zero, the size of the breathers become arbitrarily large. In some sense they tend to the homogeneously oscillating $\phi=a\cos t$ solution, which however, would be of infinite energy for any finite amplitude $a$. For small amplitudes the breathers oscillate according to $\cos(\omega t)$, but at higher order in the expansion $\cos(3\omega t)$, $\cos(5\omega t), \ldots$ modes also arise.

It was realized already in the seventies that a similar small-amplitude expansion can be applied for general analytic potentials \cite{Kosevich75,dashen-75,Buslaev1977,bogolyubskii-77c,Makhankov-78}, although it is not convergent. The applied expansion yields the approximation of a weakly nonlocal periodic solution in the form of asymptotic series. In spite of the lack of convergence the asymptotic expansion turns out to be very useful, since for fixed $\varepsilon$ it gives a gradually better and better approximation up to some $n_\varepsilon$ order in the expansion. However, higher than $n_\varepsilon$ order terms start to give larger and larger contributions, and finally the series diverges. The smaller $\varepsilon$ is, the higher order the expansion can be used. For every $n>0$ integer there exist a small $\varepsilon$, such that $n_\varepsilon>n$.

\section{The history of oscillons (pulsons)}

Although apart from the $1+1$ dimensional sine-Gordon breathers there are no exactly time-periodically oscillating spatially localized solutions of the flat background \eqref{fieldeq2} equations, still there exist very long-lived localized oscillating configurations. The possibility of the existence of these type of solutions for not exactly solvable $1+1$ dimensional theories was studied first by Kosevich and Kovalev \cite{Kosevich75}, and also by Dashen, Hasslacher and Neveu \cite{dashen-75} in 1975 using perturbational methods. In the same year it was shown using numerical methods by Kudryavtsev that by the collision of two $1+1$ soliton states ($\phi^4$ kinks) such long-lived localized structures can be really formed \cite{kudryavtsev-75}. These states are not completely periodic, since they lose energy by emitting scalar radiation to infinity. At the same time, their frequency and amplitude slowly changes. For the bound states formed by two solitons the name bion was introduced \cite{Makhankovbook}. The time-periodic bions are called breathers, and if the periodicity is only approximate the pulson or oscillon name is generally used.

\begin{figure}[!hbt]
\centering
\includegraphics[width=7cm]{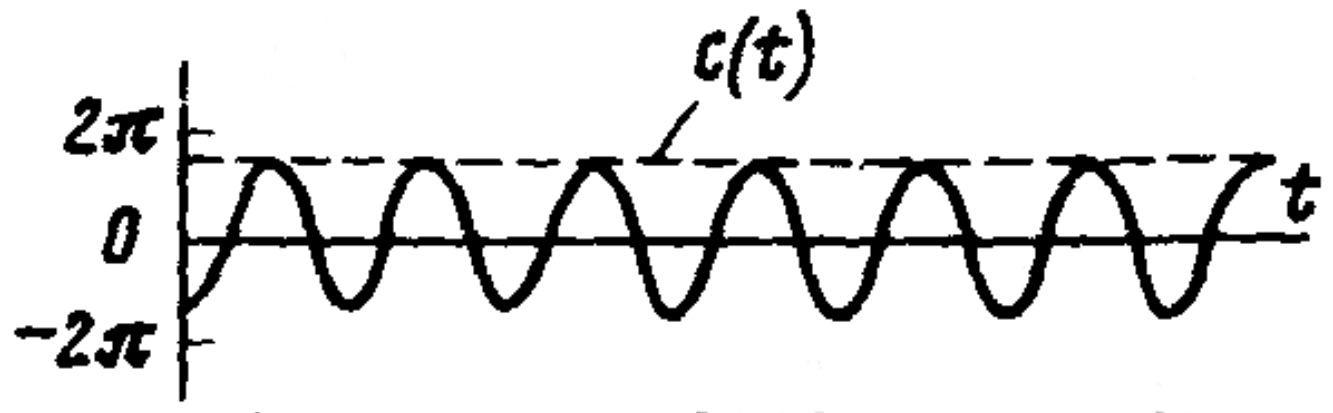}
\caption{\label{f:bogol1} The central value of the scalar field as a function of time, for $3+1$ dimension and sine-Gordon potential. The period of oscillation is $T\approx 7.4$. (Source of figure: Bogolyubskii and Makhan'kov \cite{bogolyubskii-77}.)}
\end{figure}

\begin{figure}[!hbt]
\centering
\includegraphics[width=9cm]{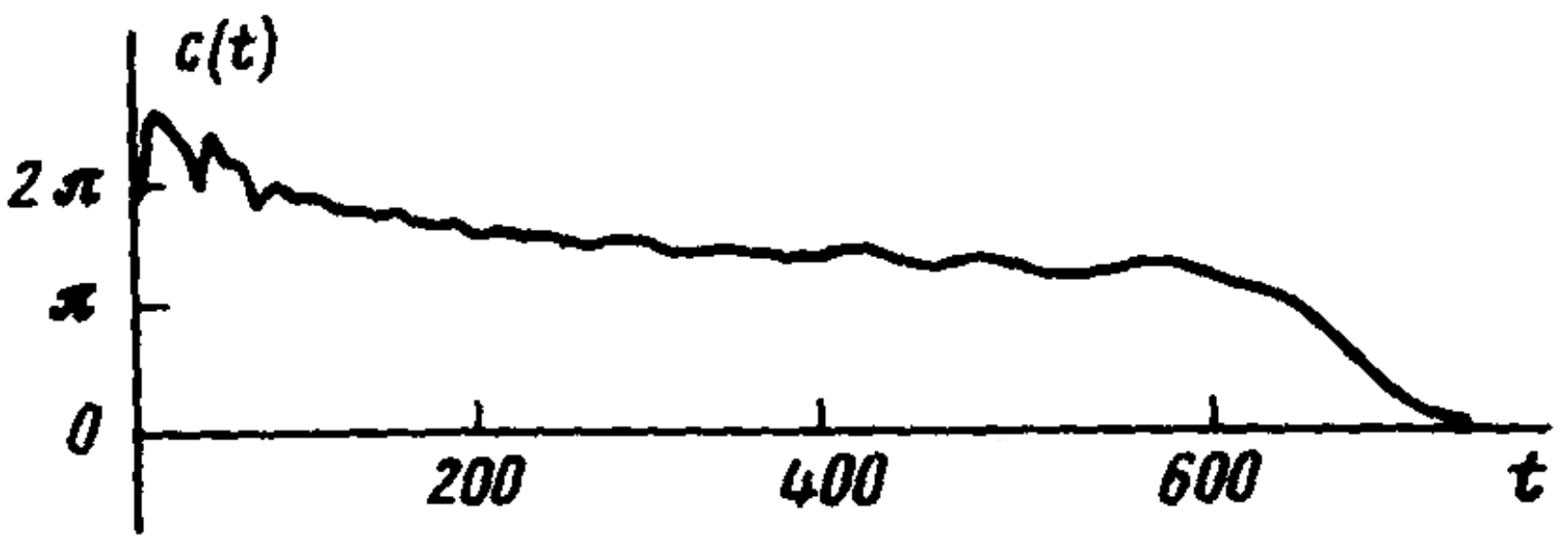}
\caption{\label{f:bogol2} The upper envelope of the oscillations of the scalar field at the center for a longer period of time, for the same system as on Fig.~\ref{f:bogol1}. (Source of figure: Bogolyubskii and Makhan'kov \cite{bogolyubskii-77}.)}
\end{figure}

Spherically symmetric long-lived oscillating states have been found first by Bogolyubskii and Makhan'kov in 1977 in Dubna, using numerical time-evolution code \cite{Bogolubsky-77b,bogolyubskii-77,Makhankov-78}. The solutions were originally called pulsons, but later the naming oscillon became widespread. As an initial condition, the shape of a $1+1$ dimensional sine-Gordon breather was used
\begin{equation}
\phi_{t=0}=4\arctan\left[\frac{\varepsilon}{\omega\cosh(\varepsilon r)}\right] 
\ , \qquad \omega^2=1-\varepsilon^2 \ ,
\end{equation}
with the choice $\varepsilon/\omega=10$, where $r$ is the distance from the center. In case of three-dimensional space, still with the $U(\phi)=1-\cos\phi$ \, sine-Gordon potential choice, this evolves as a localized oscillating matter lump, as it can be seen on Figure \ref{f:bogol1}.
However, the oscillation is only approximately periodic. As it can be seen on Fig.~\ref{f:bogol2},
taking into account a longer time period the amplitude decreases, the scalar gets radiated out to infinity, and finally the oscillon decays. In the same paper Bogolyubskii and Makhan'kov report similarly long-lived $3+1$ dimensional spherically symmetric states for the $\phi^4$ potential choice, when $U(\phi)=(\phi^2-1)^2/4$. Then the scalar field oscillates around the minimum $\phi=1$ or $\phi=-1$ of the potential. For a potential that is symmetric around its minimum the lower envelope is essentially the mirror image of the upper one, but it has a similar shape for non-symmetric potentials as well.

Spherically symmetric $n+1$ dimensional oscillons, in case of $n\geq 3$, similarly to the example shown on Fig.~\ref{f:bogol2}, after having their amplitude decreased below a certain value, suddenly decay. On the contrary, for $1+1$ and $2+1$ dimensional oscillons no such sudden decay can be observed. These oscillons keep on existing for arbitrarily long time, with less and less quickly decreasing amplitudes. One-dimensional $\phi^4$ oscillons were investigated using a more precise method by Geicke in 1983, observing many thousands of oscillations periods with extremely slowly decreasing amplitudes \cite{geicke-83}. Similarly long-lived circularly symmetric oscillons were found by Geicke in case of $2+1$ spacetime dimensions, for sine-Gordon and $\phi^4$ potentials as well \cite{geicke-84}. For smaller amplitude initial values the amplitude decrease and the frequency change can be so small that it cannot be observed by the applied numerical method, and the configuration appears to be time-periodic. Later Gleiser and Sornborger \cite{GleiserSornborger} have studied $2+1$ oscillons in more details, employing a more precise numerical method.

The interest about oscillons began to grow more quickly after 1994, following the work of Marcelo Gleiser, who studied the formation of oscillons by numerical methods, for symmetric and asymmetric double well $\phi^4$ potentials \cite{gleiser-94}. Next year, in their more detailed paper, Copeland, Gleiser and Müller started to refer to these objects as oscillons instead of pulsons, with the reasoning that the new name characterizes better the uniform almost periodic oscillation of the core.

It was already apparent from the initial investigations that apparently similar oscillons evolve from several different spherically symmetric initial data, which indicates the stability of oscillons with respect to spherically symmetric perturbations. Later investigations have also shown that oscillons are stable with respect to general perturbations, since a non-symmetric initial configuration also develops quickly into a spherically symmetric oscillon state, quickly radiating out the non-symmetric component into infinity \cite{christiansen97,Minzoni2004167,adib2002,hindmarsh2006}.

The investigation of oscillons in case of the sine-Gordon potential has been developed quite separately from the studies on $\phi^4$ oscillons \cite{christiansen81,christiansen97,piette98,alfimov2000,Minzoni2004167,Bratsos2007251}. In these sine-Gordon papers the naming oscillon was not used. Because of the near periodic nature of the states they were called breathers, and referring to the work of Bogolyubskii and Makhan'kov, the pulson name also remained.

Studying a scalar field representing axions, Kolb and Tkachev have observed the formation of localized dense pseudo-soliton configurations on an expanding cosmological background \cite{Kolb93,Kolb94}. In this context, when the self-interaction of the scalar field is represented by the axion potential, the resulting non-gravitating oscillons are also known by the name axitons \cite{Schiappacasse18,Vaquero19,Eggemeier19}. For the gravitationally bound clumps of axions Kolb and Tkachev introduced the name bose stars. These self-gravitating objects belong to the class of oscillatons that we discuss in Chapter \ref{fejrelgrav}.

A relativistic real scalar field in the non-relativistic limit can be conveniently represented by a complex scalar field \cite{Namjoo18,Braaten18b}. In the non-relativistic limit oscillons become rather similar to complex scalar field Q-balls \cite{Rosen68,Coleman85,Copeland14,Shnir18,Nugaev19}, and they have an approximately conserved adiabatic charge denoted by I in this case \cite{Kasuya03,Mukaida14,Kawasaki15}. In this context oscillons have been generally called \mbox{I-balls}. The decay rate and longevity of oscillons/I-balls have also been studied in this context \cite{Kawasaki14,Mukaida17,Ibe2019,Eby19}.

The formation, evolution and properties of oscillons formed by classical, i.e. non-quantized, scalar fields are described in detail in Sections \ref{secoscidofejl}-\ref{secoscsugnum} of this review. The energy levels, radiation and stability of oscillons considered as quantum mechanical systems have been studied in the papers \cite{dashen-75,Hertzberg10,Saffin14,Olle19}. Oscillons are playing a key role in enhancing resonant tunneling in scalar quantum field theories \cite{Copeland08,Saffin08}.

For some other reviews on oscillons one can consult the PhD theses of Salmi \cite{SalmiThesis2008} and Sicilia \cite{SiciliaThesis2011}.

\section{Physical applications of oscillons}

The initial studies of oscillons (pulsons) were motivated by the search of particle-like solutions of relativistically invariant nonlinear equations \cite{Bogolubsky-77b,bogolyubskii-77,Makhankov-78}. Bound states were first observed by the numerical investigation of the collision of two $1+1$ dimensional kink solutions \cite{kudryavtsev-75,Sugiyama79,Moshir81,Campbell83,Wingate83,Romanczukiewicz10}. Gleiser has reached to the study of oscillons by investigating cosmological phase transitions \cite{gleiser-94,copeland-95,Gleiser06}. In case of an asymmetric $\phi^4$ potential the transition between the two vacuum states can lead to bound states \cite{GleiserRogers08}. Oscillons may have evolved during the inflationary period of the early universe, from the inflaton or from coupled scalar fields, and may have influenced the dynamics of inflation \cite{Broadhead05,Gleiser07b,Amin10,Gleiser11,Amin12a,Amin12b,Sainio12,Gleiser14,Adshead15,Antusch16,Lozanov14,Lozanov18,Hasegawa18,Hong18}. Oscillons may also form as a result of the shrinking and collapse of regions bounded by domain walls \cite{Hindmarsh08}. The collision of planar and spherical domain walls also tend to form oscillons, assuming that the symmetry is slightly broken by small fluctuations \cite{Braden15a,Braden15b,Bond15}. The formation of oscillons at cosmic vacuum bubble collisions have also been considered in \cite{Dymnikova00,Johnson12,Mersini14}. Oscillons also emerge when a scalar system is in contact with an external heath bath \cite{Gleiser96,Gleiser03}. The collision of ultrarelativistic $1+1$ dimensional oscillons have been studied in \cite{Amin14}.

Hormuzdiar and Hsu \cite{Hormuzdiar99a} have shown that certain classical pion states can be described by the ${3+1}$ dimensional sine-Gordon equations, of which there exist long-lived oscillon solutions. Oscillons also appear in the bosonic sector of the standard model, under the assumption that the Higgs mass is double of the $W^{\pm}$ boson mass \cite{Farhi05,Graham07a,Graham07b,Sfakianakis12}. The formation of oscillons has been also observed in Abelian Higgs models during numerical studies of the decay of sphalerons \cite{Arnold88,Rebbi96}. Oscillons can also form in Abelian Higgs models by the collision of vortex antivortex pairs, or as a result of symmetry breaking  \cite{Gleiser07,Gleiser09,Gleiser12,Achilleos13,Diakonos15}. The existence of oscillons has also been shown in supersymmetric theories \cite{Correa18}. Structures analogous to oscillons are also shown to form in various Bose-Einstein condensates \cite{Charukhchyan14,SuAtal15}.

In various cosmological models the inflation is directly followed by the preheating epoch, during which the energy of the inflaton is transferred to the coupled fields, and quickly changing domains are forming with large density differences. These dynamical structures may emit gravitational waves that may become observable by the planned space-based gravitational-wave observatories. During the process the scalar field is likely to form oscillons, which may provide a detectable peak on the spectrum of gravitational waves \cite{Zhou13,Antusch17,Antusch18a,LiuGuo18,Antusch18b,Amin18,Kitajima18,LiuGuo19,Sang19,Lozanov19}. The existence of oscillons may also enhance the formation of fermions \cite{Borsanyi09,Saffin17}. From the oscillons existing after the inflationary epoch a large number of small mass black holes may have formed, which may be still present as dark matter in the universe \cite{Cotner18}.

It is important to clarify the effect of the expansion of the universe on the formation and lifetime of oscillons. As we have seen, on flat $1+1$ dimensional background the radiation of oscillons become gradually extremely weak, and hence these oscillons exist for arbitrarily long time. Numerical simulations show that on one-dimensional expanding background oscillons can be easily formed, and by scalar radiation their amplitude slowly decreases, while their size becomes bigger \cite{Graham06}. However, when their size grows so much that it becomes comparable to the size of the cosmological horizon, these oscillons suddenly decay. By the small-amplitude expansion of $1+1$ dimensional oscillons on de Sitter background it can be shown that their radiation is exponentially small in terms of the cosmological constant \cite{Farhi08}. Oscillons also form on three spatial dimensional expanding background, from a real scalar field, and also in a non-Abelian Higgs model \cite{Gleiser10a}. We present the effect of the expansion of the universe on bound states formed by a real scalar field coupled to gravity (oscillatons) in Section \ref{secosctnlambpos}.

\section{Time-evolution of oscillons} \label{secoscidofejl}

\subsection{The interaction potential and the initial data}

The behavior of oscillons have been studied most thoroughly in the literature for the case of a symmetric fourth-order potential with two minimums. Since the left-hand side of the field equation \eqref{fieldeq2} only depends on the derivatives of $\phi$, by a $\phi\to\phi+\phi_0$ constant shifting of the scalar field we can make the potential symmetric around $\phi=0$. Since the field equation only contains the derivative of the potential, we can take its general form as
\begin{equation}
 U(\phi)=\frac{m^2}{8\lambda^2}\left(\phi^2-\lambda^2\right)^2 \ ,
\end{equation}
where $m$ and $\lambda$ two positive parameters. Since the expansion of this potential around the $\phi=\pm\lambda$ minimums starts with $m(\phi-\lambda)^2/2$ terms, the parameter $m$ determines the mass of the scalar field. In the \eqref{fieldeq2} field equation by the $\phi\to\lambda\phi$ rescaling of the scalar field we move the minimums into the places $\phi=\pm 1$, and after that we can work with the $\lambda=1$ value. Then by the rescaling of the space and time coordinates we can set the scalar field mass into an arbitrary positive value. For the time-evolution studies of oscillons the most widely used choice is $m=\sqrt{2}$, in which case
\begin{equation}
 U(\phi)=\frac{1}{4}\left(\phi^2-1\right)^2
 \quad , \qquad U'(\phi)=\phi\left(\phi^2-1\right) \ .
\end{equation}
In order to have a finite energy state, at infinity the value of the scalar field must tend to one of the $\phi=\pm 1$ vacuum values. In most publications the $\phi=-1$ choice is made, but in some papers $\phi=1$ is used. The two choices are equivalent, one only has to multiply the initial data by $-1$. In order ensure better compatibility with the later small-amplitude calculations, in this review we assume that the field tends to the $\phi=-1$ value at infinity, and by the substitution $\phi\to\phi-1$ we shift this vacuum value into the position $\phi=0$. In this case, the potential and its derivative takes the form
\begin{equation}
 U(\phi)=\frac{1}{4}\phi^2\left(\phi-2\right)^2
 \quad , \qquad U'(\phi)=\phi\left(\phi-1\right)\left(\phi-2\right) \ . \label{eqpotnum}
\end{equation}
Since any symmetric fourth-order potential that has two minimums can be transformed to this form, it is sufficient to study the system for this specific potential. The choice that $\phi$ tends to zero at infinity also enhances the stability and precision of the applied numerical code.

Gleiser and his co-authors have studied the time-evolution of a Gaussian shaped initial data centered on the origin, in case of $3+1$ spacetime dimensions \cite{gleiser-94,copeland-95,GleiserSicilia08,GleiserSicilia09}. Since the field equation is second order, at the moment of time $t=0$ it is necessary to give the value and the derivative of the scalar field
\begin{equation}
 \left.\phi\right|_{t=0} = 
 C\exp(-r^2/r_0^2)\quad , \qquad  
 \left.\frac{\partial\phi}{\partial t}\right|_{t=0} = 0 \ , \label{e:ff}
\end{equation}
where $C$ is the initial value of the scalar field at the center, and $r_0$ is a parameter characterizing the size of the bubble. When $C$ is positive, the initial amplitude is directed towards the other minimum of the potential. The relation between $C$ and the parameter $\phi_c$ used in the papers of Gleiser is $C=\phi_c+1$, as a result of the shift in the interaction potential. The parameters $C$ and $r_0$ agree with the parameters used in \cite{Andersen2012} for the comprehensive lifetime plots presented there. The initial data tends to the $\phi=0$ vacuum value at infinity. Changing the value of $r_0$ and $C$, (\ref{e:ff}) determines a two-parameter family of smooth localized initial data. For a large open subset of the possible $r_0$ and $C$ values, after a short transitional period the system settles into a long-lived nearly periodic oscillon state. At this stage it radiates slowly, but when its total energy drops below a certain value,  the oscillon suddenly decays, radiating all its remaining energy to infinity.

\begin{figure}[!hbt]
\centering
\includegraphics[width=12cm]{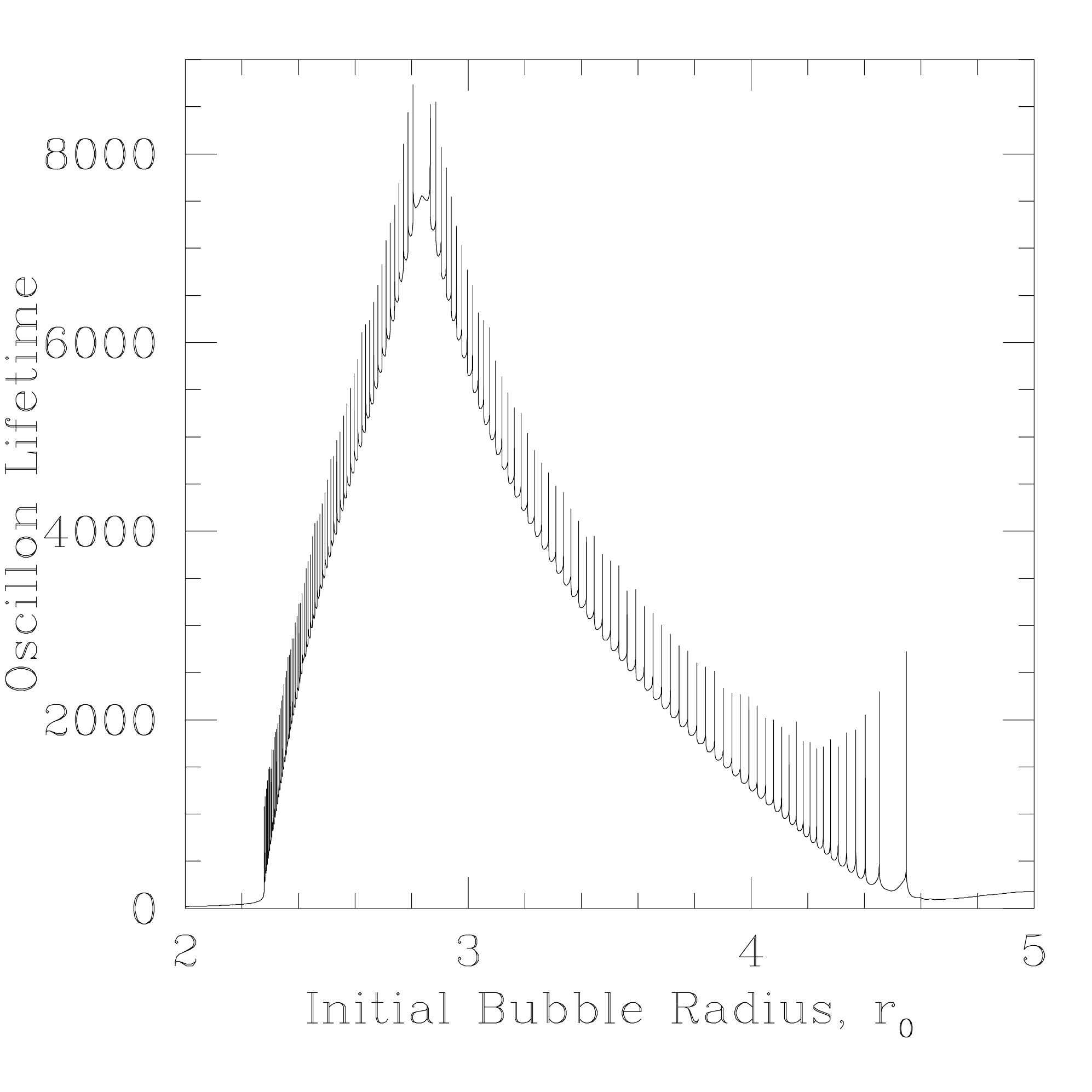}
\caption{The lifetime of oscillons as a function of the initial radius $r_0$ of the bubble, for the case $C=2$. (Source of figure: Ethan Honda \cite{HondaThesis2000}).} \label{fighonda}
\end{figure}

If we fix the initial central value of the field with the choice $C=2$, then initially the scalar field changes between the two vacuum values of the potential. The one-parameter family obtained by fixing $C=2$ and changing $r_0$ is the most thoroughly studied family of solutions in the literature. On Fig.~\ref{fighonda} we reprint the exquisitely detailed lifetime plot of $3+1$ dimensional oscillons from the thesis of Ethan Honda from the year 2000 \cite{HondaThesis2000,HondaChoptuik2002}.
Since the final decay happens very quickly and always in a similar way, the figure does not depend significantly on the precise definition of the decay. We can, for example, consider the lifetime the interval at which the energy decreases to ten percent of the original. On the figure we can see $125$ narrow resonance-like peaks, where the lifetime suddenly increases. In order to obtain the peaks with the heights shown on the figure the parameter $r_0$ had to be tuned with $10^{-14}$ precision. With further tuning the tallness of the peaks can be further increased, but that requires a numerical code that uses more than the usual $16$ digit numbers during the calculations. We will study in details the oscillon states and the reasons for the appearance of the peaks in later sections, when presenting our own results on the subject. Before the publications of our results it was conjectured that the peaks on Fig.~\ref{fighonda} correspond to time-periodic localized breather states. However, as we will show, the oscillons at the peaks also lose energy by emitting radiation, even if extremely slowly, so they cannot be exactly periodic.

\subsection{Numerical methods for studying the time-evolution}

In the initial publications about oscillons (pulsons) there is no information given on the numerical code used to find them, or on the applied boundary conditions \cite{Bogolubsky-77b,bogolyubskii-77,Makhankov-78}. In all papers that followed, the finite difference method was used for the discretization of the differential equations when calculating the numerical solution of the initial value problem. In a few publications the method of characteristics has been applied, but in most cases uniform grid points are used along the $r$ radial and $t$ time coordinates, with initially second order, and later fourth-order interpolation.

In case of spherical symmetry the necessary and sufficient condition for the regularity at the center is that the odd order derivatives of the scalar $\phi$ with respect to the radial coordinate $r$ should vanish at the center $r=0$. Then the function $\phi$ can be extended to negative values of $r$, such that it is smooth and mirror symmetric with respect to $r=0$. It is also usual to require the vanishing of the odd order derivatives for the $1+1$ spacetime dimensional case, instead of the boundary condition at $r=-\infty$ and the requirement of comoving coordinate system. In arbitrary dimensions, because of the field equations, the condition of regularity is equivalent to the requirement that $\phi$ is finite at the origin and its derivative with respect to $r$ is zero. In numerical codes the simplest method for requiring the central regularity is the utilization of the mirror symmetry at the calculation of derivatives, by the addition of virtual grid points with negative radial coordinates.

In the ideal case, the exterior boundary condition should be specified at $r=\infty$, and it should express that the system is isolated, i.e.~there is no incoming radiation from infinity. In actual numerical calculations only finite number of grid points can be used, so the most widely used method is to choose the outer boundary at some large $r=r_{\text{max}}$ value. If the outer boundary surface is chosen sufficiently far away, then it will take very long time until any disturbance reaches the central domain, so it is not important what kind of boundary condition is used there. Since the propagation of the perturbation is by the speed of light, in this way one can only perform simulations that remain valid only for a short time. If instead of a time period $T$ we wish to calculate for $2T$, then we have to take twice as large spatial domain, which means twice as many grid points. Because of this, the necessary computational capacity increases proportionally with $T^2$. It makes the calculation somewhat more efficient if the outer boundary moves outwards by the speed of light, by the constant addition of new grid points \cite{copeland-95}, preventing the outgoing radiation reflecting from there. However, the necessary computational capacity still scales with $T^2$, so this method is only applicable for moderately long time periods.

Since it is the long-time evolution of oscillons which is of primary interest, still it turns out  necessary to choose the outer boundary at a constant finite $r=r_{\text{max}}$ value. The problem becomes even more severe, especially when using higher order discretization, because the  propagation speed of numerical disturbances may be several times that of the speed of light. The boundary condition at the outer surface has to prevent the reflection of the outgoing radiation as well as possible, and therefore avoid the inappropriate disturbance of the evolution of the oscillon in the central region. For a zero mass scalar field the Sommerfeld condition can be applied, the form of which in case of one-dimensional space can be written as
\begin{equation}
 \left(\frac{\partial}{\partial t}
 +\frac{\partial}{\partial r}\right)\phi\ \biggr|_{r=r_{\text{max}}}=0 \ ,
\end{equation}
and for a three-dimensional spherically symmetric field as
\begin{equation}
 \left(\frac{\partial}{\partial t}
 +\frac{\partial}{\partial r}\right)\left(r\phi\right)\, \biggr|_{r=r_{\text{max}}}=0 \ .
\end{equation}
Although oscillons only evolve from massive fields, while the Sommerfeld condition is for the massless case, its application still significantly decreases the disturbance reflected from the outer boundary \cite{geicke-83,geicke-84,geicke-94}.

For the absorption of outgoing radiation and the preventing of reflection Gleiser and Sornborger \cite{GleiserSornborger,Gleiser18} have introduced a damping term $\gamma(r)\frac{\partial\phi}{\partial t}$ into the right-hand side of \eqref{fieldeq2}. Taking the coefficient $\gamma(r)$ to be zero inside a radius $r_0$, and outside of the region proportional to $(r-r_0)^2$, the dissipation arises gradually, hence the approach was named adiabatic damping method. The value of $r_0$ had to be chosen at least $20$ times as large as the radius of the oscillon. The main advantage of this method is that it is not necessary to use more spatial grid points when increasing the simulation time, so the necessary computational capacity only increases proportionally to $T$. The method was mainly used for the study of $2+1$ dimensional oscillons since those are not decaying even after extremely long time periods. Further similar method for the study of massive fields is the application of a sponge filter, in which case a dissipation term proportional to the Sommerfeld condition is added to the field equation in a finite domain before the outer boundary \cite{Israeli81,ChoptuikThesis1986,Marsa96,Balakrishna98}. The disadvantage of adding a dissipation term by the above methods is that it changes the differential equation that we solve. Even when the distance between the neighboring lattice grid points tends to zero, the result is not approaching the solution of the original physical field equation. In order to check the validity of a numerical result, the dependence on the radius of the damping region and the dependence on the strength of the damping has to be investigated separately.

A detailed study of the parameter space of three-dimensional oscillons was carried out by Honda and Choptuik \cite{HondaThesis2000,HondaChoptuik2002}. In their work they used the so called ``monotonically increasingly boosted coordinates'' for decreasing the reflection at the outer boundary. This was achieved by the introduction of a new radial coordinate,
\begin{equation}
 \tilde r= r+f(r)t \ ,
\end{equation}
keeping the time coordinate $t$ and the angular coordinates unchanged. The monotonically increasing $f(r)$ function was chosen in such a way that going outwards it should increase from $0$ to $1$ in a spherical shell shaped domain which is centered around a certain $r_0$ radius and has thickness $\delta$. In this way, in the inner region the coordinate system remains unchanged, while in the transitional region it gradually transforms into the null coordinates used outside. In the transitional region the speed of both the outgoing and the ingoing radiation becomes very slow. Practically, the radiation freezes into this zone. As the propagation speed decreases, the waves become piled up, get blueshifted, so their wavelength decreases. For the small wavelength radiation the dissipation inherent in the numerical method is stronger, so the waves become practically absorbed in this region. Since propagation in both directions is inhibited, the reflection of the outgoing waves from the outer region becomes negligible. The disadvantage of this method is that the chosen coordinate system is not time-independent anymore. The lattice points in the inner region slowly migrate into the more and more wide transitional region, and less and less points remain for the description of the core of the oscillon. Luckily, this decrease is very slow. It can be shown that the necessary computational capacity scales as $T\ln T$. This method was used by Honda and Choptuik for the investigation of $3+1$ dimensional oscillons. In that case the emphasis is not on simulations that are valid for very long time periods, but on large precision runs with several different initial data choices for intermediately long time intervals. The fractal structure emerging at studying the possible end results of the evolution of various initial data for the case of the asymmetric double-well $\phi^4$ potential has been presented in \cite{Honda10}.

\subsection{Our numerical method} \label{secnummodssajat}

In our numerical calculations we have treated the problem of the outer boundary by the compactification of space \cite{Fodor2006,Fodor2008,Fodor2009a,Fodor2009b}. We map the domain of the original radial coordinate $0\leq r<\infty$ by the expression
\begin{equation}
r=\frac{2R}{\kappa(1-R^2)} \label{eqlowupr}
\end{equation}
into the domain $0\leq R<1$ of the new radial coordinate, where $\kappa$ is a constant. The spatial grid points are chosen uniformly with respect to the coordinate $R$. The constant $\kappa$ can be chosen different values depending on the size of the investigated oscillon. It is practical to give $\kappa$ such a value that approximately the same number of grid points are used to the description of the central oscillon and to the description of the radiation in the distant region. The typical value in our computations was $\kappa=0.05$, but for small-amplitude large sized oscillons it is useful to choose the value of $\kappa$ inversely proportional to the size. Since the size of small-amplitude oscillons is inversely proportional to their amplitudes, in these cases we choose $\kappa$ proportionally to the oscillon amplitude.

Using the radial coordinate $R$ the field equation \eqref{fieldeq2} can be written into the following form:
\begin{equation}
\frac{\partial^2\phi}{\partial t^2}=
\frac{\kappa^2(1-R^2)^3}{2(1+R^2)}\left[
\frac{(1-R^2)}{2(1+R^2)}
\frac{\partial^2\phi}{\partial R^2}
-\frac{R(3+R^2)}{(1+R^2)^2}
\frac{\partial\phi}{\partial R}
+\frac{(d-1)}{2R}
\frac{\partial\phi}{\partial R}
\right]
-U'(\phi) \,. \label{eqfieldbigr}
\end{equation}
Introducing the new variables
\begin{eqnarray}
\phi_{t}&=&\frac{\partial\phi}{\partial t} \ , \\
\phi_{R}&=&\frac{\partial\phi}{\partial R} \ , 
\end{eqnarray}
the problem can be interpreted as a system of first order differential equations for the variables $\phi$, $\phi_{t}$ and $\phi_{R}$. The equations to be solved are
\begin{align}
 \frac{\partial\phi}{\partial t}&=\phi_{t} \ , \label{eqphidt} \\
 \frac{\partial\phi_t}{\partial t}&=
  \frac{\kappa^2(1-R^2)^3}{2(1+R^2)}\left[
  \frac{(1-R^2)}{2(1+R^2)}\frac{\partial\phi_R}{\partial R}
  -\frac{R(3+R^2)}{(1+R^2)^2}\phi_R
  +\frac{(d-1)}{2R}\phi_R
  \right]-U'(\phi) \ , \label{eqphitdt} \\
 \frac{\partial\phi_R}{\partial t}&=\frac{\partial\phi_t}{\partial R} \ , \label{eqphirdt} \\
 \frac{\partial\phi}{\partial R}&=\phi_R\ . \label{e:constr}
\end{align}
If \eqref{e:constr} is satisfied at the initial moment $t=0$, then because of the evolution equations \eqref{eqphidt} and \eqref{eqphirdt} it also remains valid later, so it can be considered as a constraint. The third term in the square brackets on the right-hand side of \eqref{eqphitdt} cannot be evaluated directly at the grid point corresponding to the center $R=0$. At this point we calculate this term by the identity
\begin{equation}
 \lim_{R\to0}\left(\frac{1}{R}\frac{\partial\phi}{\partial R}\right)=
 \lim_{R\to0}\frac{\partial^2\phi}{\partial R^2} \ .
\end{equation}

By the compactification of space we avoid the necessity that we have to specify exterior boundary conditions on an arbitrarily chosen distant but finite radius sphere. The applied numerical method is a modified version of a numerical code developed and used earlier together with István Rácz for the time-evolution of magnetic monopoles \cite{FodorRacz08,FodorRacz04}.

For the solution of the initial value problem we discretize the independent variables belonging to the coordinates $t$ and $R$. We introduce uniform grids with steps $\Delta t$ and $\Delta R$. Choosing the Courant factor as $\frac{\Delta t}{\Delta R}=1$ our code turns out to be stable. Spatial derivatives are calculated using symmetric fourth-order stencils. The time integration is performed by the ``method of lines'', using a fourth-order Runge-Kutta method, according to the description in the book of Gustafsson, Kreiss and Oliger  \cite{Gustafsson13}. In order to ensure the stability of the code, to each of the first order equations \eqref{eqphidt}-\eqref{eqphirdt} we add a dissipation term that is proportional to the sixth derivative of $\phi$, $\phi_{t}$ and $\phi_{R}$ with respect to $R$. Since we also take these terms proportional to $(\Delta R)^5$, the dissipation is not decreasing the order of the numerical code. It still remains fourth order, since the effect of the dissipation terms decreases with increasing resolution.

We also add a few non-physical grid points to the lattice, for negative radial coordinates at $R<0$, and for ``beyond infinity'' at $R>1$. Instead of calculating the time-evolution for the $R<0$ grid points we use the mirror symmetry of the scalar field to set the values of the functions. Furthermore, since $\phi$ is a massive field, it tends to zero exponentially at infinity. Because of this, for $R\geq 1$ we fix the values of $\phi$, $\phi_t$ and $\phi_R$ to zero during the whole time-evolution. This solves the problem of spatial infinity in equation \eqref{eqfieldbigr}, since at the grid point corresponding to $R=1$ it is not necessary to calculate values. Because of the addition of non-physical grid points it is possible to use symmetric stencils everywhere to calculate spatial derivatives. Otherwise, at the edge points of the grid the use of asymmetric schemes would be necessary. In our method at the lower boundary we use the mirror symmetry, while at the upper boundary we employ the $\phi=0$ condition.

Although the compactification in the spatial directions restricts the value of the $R$ coordinate into a finite interval, the real physical distance between the neighboring grid points grows larger and larger as we approach $R=1$. This distant region, where the code cannot describe properly the outgoing radiation, is shifted further and further away if we decrease spatial step size $\Delta R$. Several actual numerical simulations that we have performed with increasing number of grid points have shown that wave packets formed by outgoing massive fields are getting absorbed in this transitional domain, without getting reflected back into the central region. In this way, our numerical simulations provide a precise description of the scalar field in the central region, even for very long time periods. The simple but physically not uniformly spaced grid, together with the application of the artificial dissipation term leads to the absorption of outgoing radiation, in a similar way as for the methods in the articles  \cite{GleiserSornborger,HondaChoptuik2002}. Furthermore, because of the very low coordinate speed of the incoming light in the $R\approx 1$ asymptotic domain, the behavior of the central domain can be properly determined for a long time even after numerical errors have appeared at large distances.

The numerical simulation of $3+1$ dimensional oscillons for a time interval up to $t=7000$, which is their typical lifetime, using $2^{13}$ grid points takes about one week on a personal computer. Since it was necessary to make a large number of simulations for studying the parameter space, we have mostly used $2^{12}$ resolution.

By performing convergence tests, we have confirmed that our code actually provides a fourth-order approximation of the evolution equations. We have also monitored the energy conservation and the constraint equation \eqref{e:constr} during the evolution. The time-evolution of massive Klein-Gordon fields, which can also be calculated by a Green function method \cite{FodorRacz03}, was an important cross-check of our code. Time-evolution of various scalar field initial data calculated by the two independent methods agreed extremely well in the central region, even for as long time periods as $t\approx10^4$.

After the publication of our results on the evolution of oscillons \cite{Fodor2006,Fodor2008}, others have also studied the problem by using independent numerical codes, obtaining similar and consistent results \cite{Saffin2007,Salmi12,Andersen2012}. Saffin and Tranberg \cite{Saffin2007} have used artificial dissipation term in a distant spherically symmetric domain, and took the value of the scalar field zero at the outer boundary. Salmi and Hindmarsh \cite{Salmi12} have applied a boundary condition at the distant outer boundary that involves the second derivatives of the scalar field. The method absorbs the radiation very well for massive fields with frequencies larger than $m$. Interestingly, they have observed a power law dependence for the amplitude, frequency and energy. Using the same boundary conditions, Andersen and Tranberg \cite{Andersen2012} have mapped the dependence of the lifetime of oscillons on the parameters $r_0$ and $C$, which describe the width and amplitude of the initial data \eqref{e:ff}. They have published informative spectacularly colored figures for the $d=2,3,4,5,6,7$ spatial dimensional cases. They have shown that the narrow peaks observed earlier by Honda and Choptuik correspond to extremely thin one-dimensional walls in the $(r_0,C)$ space.

\subsection{Time-evolution of Gaussian initial data}

Using our numerical code we have started out by reproducing the results of Gleiser \cite{gleiser-94,copeland-95} and Honda \cite{HondaThesis2000,HondaChoptuik2002}. In the investigated case the self-interaction of the scalar field is determined by a symmetric fourth-order $U(\phi)$ potential, which is transformed into the form \eqref{eqpotnum} by rescaling the field and the coordinates and by shifting one of the minimums to the place $\phi=0$. At the initial moment $t=0$ the value of the field is chosen a Gaussian shape, according to \eqref{e:ff}. The time derivative of the field is set to zero at $t=0$, which corresponds to time-reflection symmetric evolution. In the expression \eqref{e:ff} the scalar tends to the vacuum value $\phi=0$ at infinity, hence it corresponds to finite energy initial data. Fixing the initial central amplitude as $C=2$ and changing the parameter $r_0$ one can obtain Figure \ref{fighonda} with the $125$ peaks, which was published by Ethan Honda.

\begin{figure}[!hbt]
 \centering
 \begin{minipage}[c]{105mm}
  \includegraphics[width=105mm]{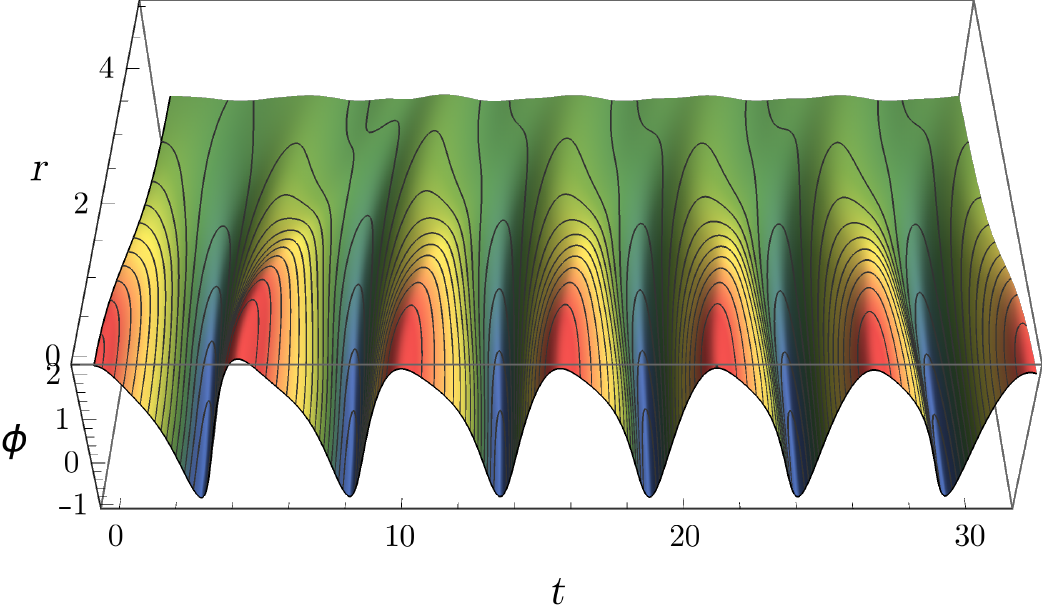}
 \end{minipage}
 \hspace*{3mm}
 \begin{minipage}[c]{15mm}
  \includegraphics[width=15mm]{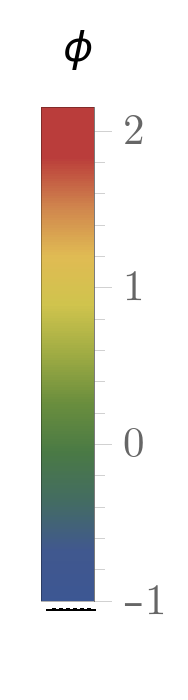}
 \end{minipage}
\caption{\label{figgaussevol} The dependence of the scalar field $\phi$ on the radial coordinate $r$ and time coordinate $t$, for the case $r_0=2.70716$ and $C=2$. The time increases to the right.}
\end{figure}

\begin{figure}[!hbt]
 \centering
 \includegraphics[width=115mm]{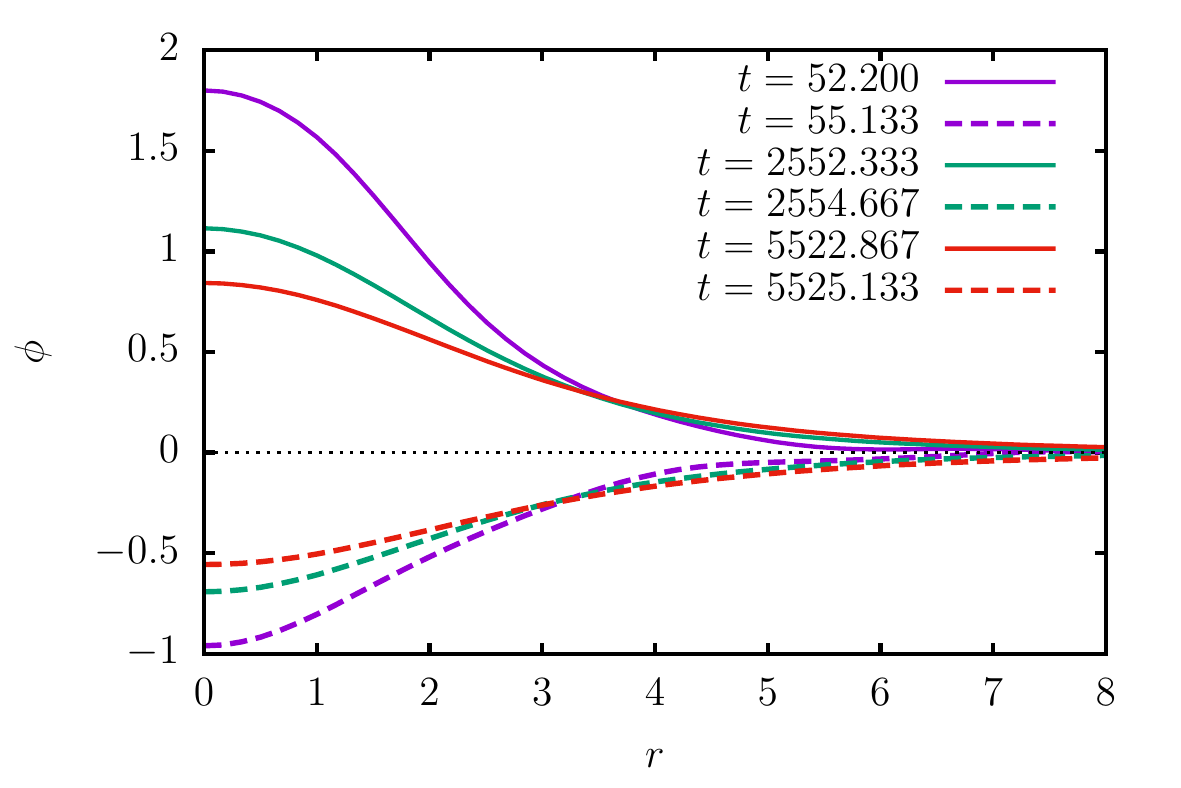}
\caption{\label{figradmax} The shape of the scalar field as a function of the coordinate $r$ in moments of time when the field takes its local maximum or minimum in the center, for the initial data $r_0=2.70716$, $C=2$. The dashed lines represent the directly next minimum after the identically colored maximum.}
\end{figure}

We demonstrate the typical long-time evolution of a Gaussian initial data by the more detailed analysis of the case $C=2$, $r_0=2.70716$, based on our own numerical simulations \cite{Fodor2006}. On Figure \ref{figgaussevol} we show the initial behavior of the scalar field in the central domain.
The field starts to oscillate in a spatial form very similar to a Gaussian shape, with a time period of $T\approx 5.3$, which corresponds to an angular frequency of $\omega=2\pi/T\approx 1.2$. A long-lived oscillon state is formed, which performs approximately $1240$ similar oscillations. Then, at around $t=5800$ the radiation becomes more intense, and the oscillon begins to decay. In this stage, during about $25$ oscillations the amplitude decreases so much that it cannot be considered a localized state anymore. On Fig.~\ref{figradmax} we show the radial dependence of the scalar field at three different oscillations, in an early time, in the middle, and before the decay of the oscillon state.
It can be seen that at around $t=5500$ the amplitude of the oscillation decreases to about half of the original, while the spatial size increases by a relatively small extent. 

\begin{figure}[!hbt]
 \centering
 \includegraphics[width=115mm]{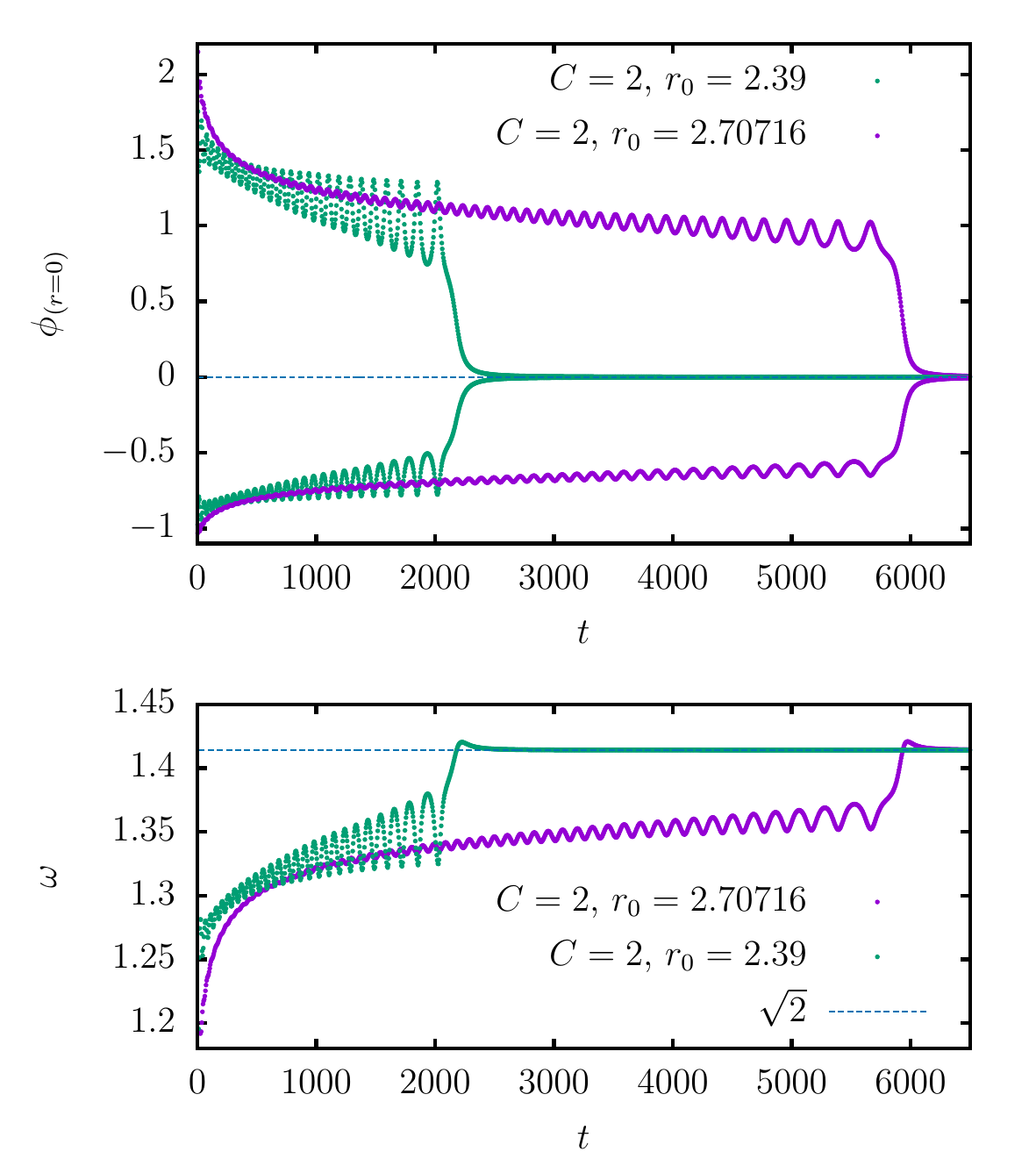}
\caption{\label{fignontune} For two different Gaussian initial data, the upper panel shows the maximum and minimum values of the central scalar field, providing the upper and lower envelope of the oscillations. The lower panel gives time dependence of the frequency.}
\end{figure}

On Figure \ref{fignontune} we show the behavior of the scalar at the center $r=0$ as a function of time.
Since the oscillation period is small compared to the lifetime of the oscillon, we only draw the points corresponding to the local maximums and minimums of the field. The resulting points represent the upper and lower envelope of the central oscillations. On the figure we also give the evolution of another shorter living Gaussian initial data, corresponding to $C=2$ and $r_0=2.39$. The frequency of the oscillations was calculated from the time difference $T=2\pi/\omega$ between two subsequent maximums, determining the precise time of the maximums by a second order interpolation. The frequency of the oscillons is always below the value $\omega=\sqrt{2}$ determined by the scalar field mass, then later during the decay stage it approaches this value. In the decay stage the amplitude of the scalar in the central region decreases proportionally to $(t-t_b)^{-3/2}$, where $t_b\approx 2175$ for the $r_0=2.39$ initial data, and $t_b\approx 5918$ for $r_0=2.70716$. The exponent $-3/2$ is the same as for the radiation of the free linear Klein-Gordon field \cite{FodorRacz03}.

The dependence of the evolution of a Gaussian initial data on the parameters $C$ and $r_0$ was investigated more precisely at first by Copeland, Gleiser and Müller \cite{copeland-95}. The investigated initial configurations generally evolve similarly to those shown on Fig.~\ref{fignontune}, with different lifetimes. Figure 7 of the paper of Copeland, Gleiser and Müller already indicated that in the vicinity of certain initial values resonance-like peaks may appear. Partly this may have motivated the detailed work of Honda and Choptuik \cite{HondaThesis2000,HondaChoptuik2002}, from which we have reprinted Fig.~\ref{fighonda} with the $125$ peaks. The appearance of the narrow peaks is closely related to the low frequency oscillations that can be observed on the upper and lower envelope curves on Fig.~\ref{fignontune}. These were named shape-modes by Honda and Choptuik, and they correspond to the low frequency oscillations in the size and amplitude of the high frequency oscillons. The oscillon states on the two sides of a narrow peak differ in the number of low frequency oscillations appearing on the envelope curve. At each peak the number of oscillations become larger or smaller by one. Proceeding from left to right, let us denote the place of the peaks on Fig.~\ref{fighonda} by $r^{*}_n$\,, where $0\leq n\leq 124$. For the states before the first peak, at $r_0<r^{*}_0$, no low frequency oscillation can be observed on the envelope curve. The number of oscillations is increasing one by one until the $r^{*}_{62}$ peak, with increasing lifetime, and then decreases one by one from peak $r^{*}_{63}$.

\begin{figure}[!hbt]
 \centering
 \includegraphics[width=115mm]{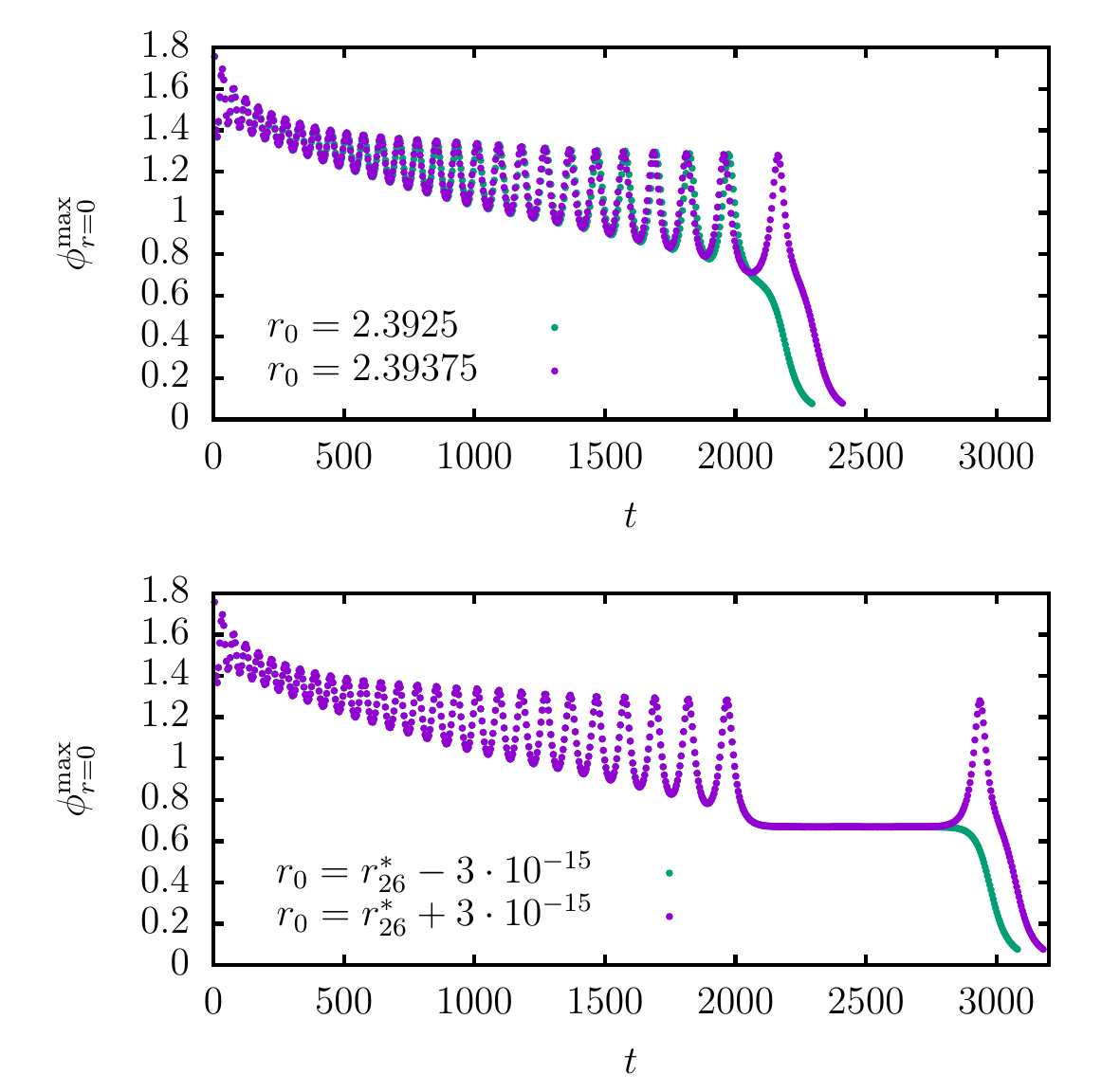}
\caption{\label{figcsx4d} On the upper panel, in case of $C=2$, the points forming the upper envelope of two typical oscillon states are shown on the two sides of the $27$-th peak at $r_{26}^{*}\approx 2.39297$. On the lower panel the evolution of two fine-tuned states can be seen, with parameters extremely close to $r_{26}^{*}$, on the two sides of the peak.}
\end{figure}

On Figure \ref{figcsx4d} we show the evolution of some oscillons close to the $27$-th resonance peak.
In the fine-tuned case, in the time interval $2100<t<2800$, a very closely periodic state forms with almost constant amplitude and frequency \cite{Fodor2006}, which we call \emph{near-periodic state} in the following. The closer is the initial parameter to the critical value the longer the developing near-periodic state becomes. The shape of the lower envelope curve is very similar to the upper one, although its amplitude is different, hence we do not include it in the figure. If the potential determining the self-interaction of the scalar field was symmetric around its minimum, the two envelope curve would be the mirror image of each other. The longer living \emph{supercritical} states arise if we approach $r^{*}_{26}$ from above. In that case, approaching the critical value, on the envelope curve the last low frequency peak moves to a later and later position, making way to a near-periodic state between the last two peaks. At the lower side of the peak, for $r_0<r^{*}_{26}$, close to the critical value, essentially the same near-periodic states are being formed, but at the end they decay without forming a new low frequency peak on the envelope curve. These are called \emph{subcritical} states. The reason for this phenomena is that when the central amplitude of an oscillon decreases below a certain value, the shape-mode that appeared earlier on the envelope curve as a low frequency oscillation becomes unstable. Applying the fine-tuning, we can suppress this unstable shape-mode, and in this way we can observe a smaller amplitude oscillon, which however belongs into the unstable domain.

On Fig.~\ref{fighonda} the near-periodic states belonging to the different peaks oscillate with distinct time periods $T$, but all belong to the interval $4.446<T<4.556$. General oscillon states, including the initial less precisely periodic stages of the tuned oscillons, oscillate with longer periods in the domain $4.6<T<5.5$.

\subsection{Near-periodic states} \label{sec-majdnemper}

It was shown by Honda and Choptuik \cite{HondaChoptuik2002}, that near the resonances the lifetime $\tau$ of the oscillons follow a scaling law,
\begin{equation}
\tau \sim \gamma \ln|r_0-r_0^*|+\delta \label{f:crit}\ ,
\end{equation}
where the scaling exponent $\gamma$ and the constant $\delta$ take different values at each of the resonance peaks, while $\delta$ is distinct even for subcritical and supercritical states. Although it appears that the lifetime of near-periodic states can be increased arbitrarily large by fine-tuning the parameter in the initial data, in practice it is extremely difficult to reach very long lifetimes, since we cannot represent numbers that are very close to the value of the resonance $r^{*}_{n}$ because of the given machine precision of the applied computer and numerical program. It is possible to reach longer lifetimes by using computer programs that calculate with more than the usual $16$ digits, but because of the radical increase in the computation time this method can only be used restrictively. Using the ``long double'' variables of an SGI computer we were able to calculate to $32$ digits instead of the usual $16$, and in this way by improving the fine-tuning we could double the observed lifetime of near-periodic states. Similar increase can be achieved on usual personal computers by using numerical libraries that allow double precision arithmetic \cite{libquadmath,boostmultiprec,Bailey15}.

\begin{figure}[!ht]
\centering
\includegraphics[width=115mm]{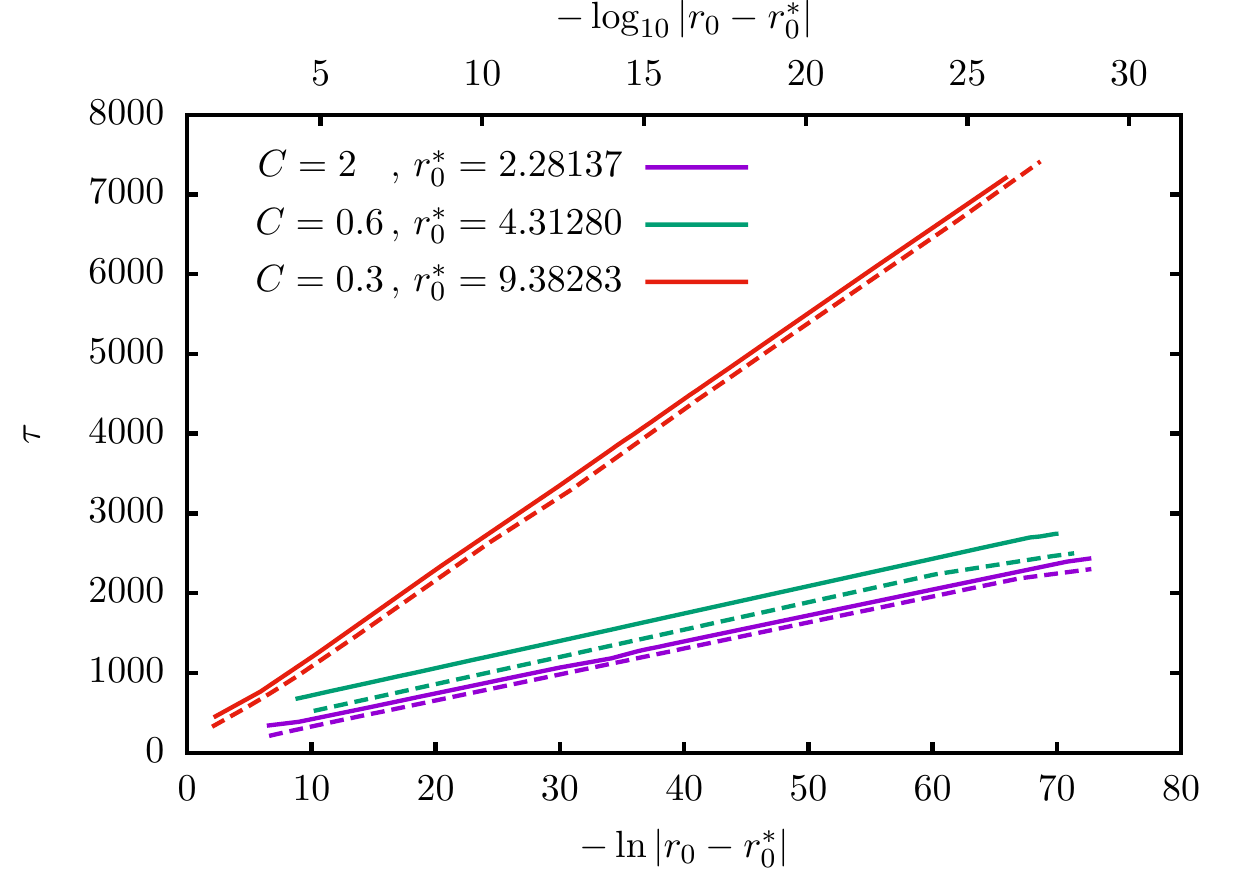}
\caption{\label{f:lt}
The dependence of the lifetime $\tau$ of oscillons on the value of $-\ln|r_0-r_0^*|$ for three different resonances. The lower dashed lines with $r_0<r_0^*$ correspond to subcritical solutions, while the upper lines with $r_0>r_0^*$ correspond to supercritical states.}
\end{figure}

On Figure \ref{f:lt} we show the scaling of the lifetime of oscillons near three different resonance peaks \cite{Fodor2006}.
Instead of studying the resonances belonging to the initial parameter $C=2$ and changing only $r_0$, we have chosen three near-periodic states belonging to the first peaks at different $C$ values. The reason for this choice was that the initial slowly radiating state corresponding to general non-tuned oscillons is the shortest near the first peak. In this case after at most one low frequency modulation the near-periodic state follows, and the computational resources can be concentrated on that.

In order to make it more clear what we understand on the fine-tuning of the parameter $r_0$ to $16$ or $32$ digits, and what is the real error of the various quantities, in Table \ref{tablefirstpeak} we give the position of the first peak in case of $C=2$, for five different numerical resolutions.
\begin{table}[htb]
\centering
\begin{tabular}{c||c|c|c|c|}
$i$ & $n_i$ & $r_0^{*(i)}$ & $\delta_i$ & $c_i$ \\
\hline
\hline
$8$  & $2^{8}$  & $2.281990488596033$ & $6.2\cdot 10^{-4}$ & \\
\hline
$9$  & $2^{9}$  & $2.281392051715203$ & $2.1\cdot 10^{-5}$ &  \\
\hline
$10$ & $2^{10}$ & $2.281371382459355$ & $7.9\cdot 10^{-7}$ & $4.86$ \\
\hline
$11$ & $2^{11}$ & $2.281370625452998$ & $3.1\cdot 10^{-8}$ & $4.77$ \\
\hline
$12$ & $2^{12}$ & $2.281370594875569$ &  & $4.63$ \\
\hline
\end{tabular}
\caption{\label{tablefirstpeak}
The position $r_0^{*(i)}$ of the first peak in case of $C=2$ for five different numerical resolutions. The number of spatial grid points for the resolution labeled by $i$ was $n_i=2^i$ during the fine-tuning. The error is estimated by giving $\delta_i=|r_0^{*(i)}-r_0^{*(12)}|$. The convergence factor is given by the expression
$c_i=\log_2|(r_0^{*(i-2)}-r_0^{*(i-1)})/(r_0^{*(i-1)}-r_0^{*(i)})|$.
}
\end{table}
At each numerical resolution the lifetime of the oscillon increased to approximately $\tau=1100$ when the parameter $r_0$ approached a resolution dependent value to $16$ digits precision. The study of the convergence of the results show that in case of $C=2$ the real position of the first peak is at $r_0^{*}=2.281370594$, with absolute error $10^{-9}$.

\begin{figure}[!ht]
\centering
\includegraphics[width=115mm]{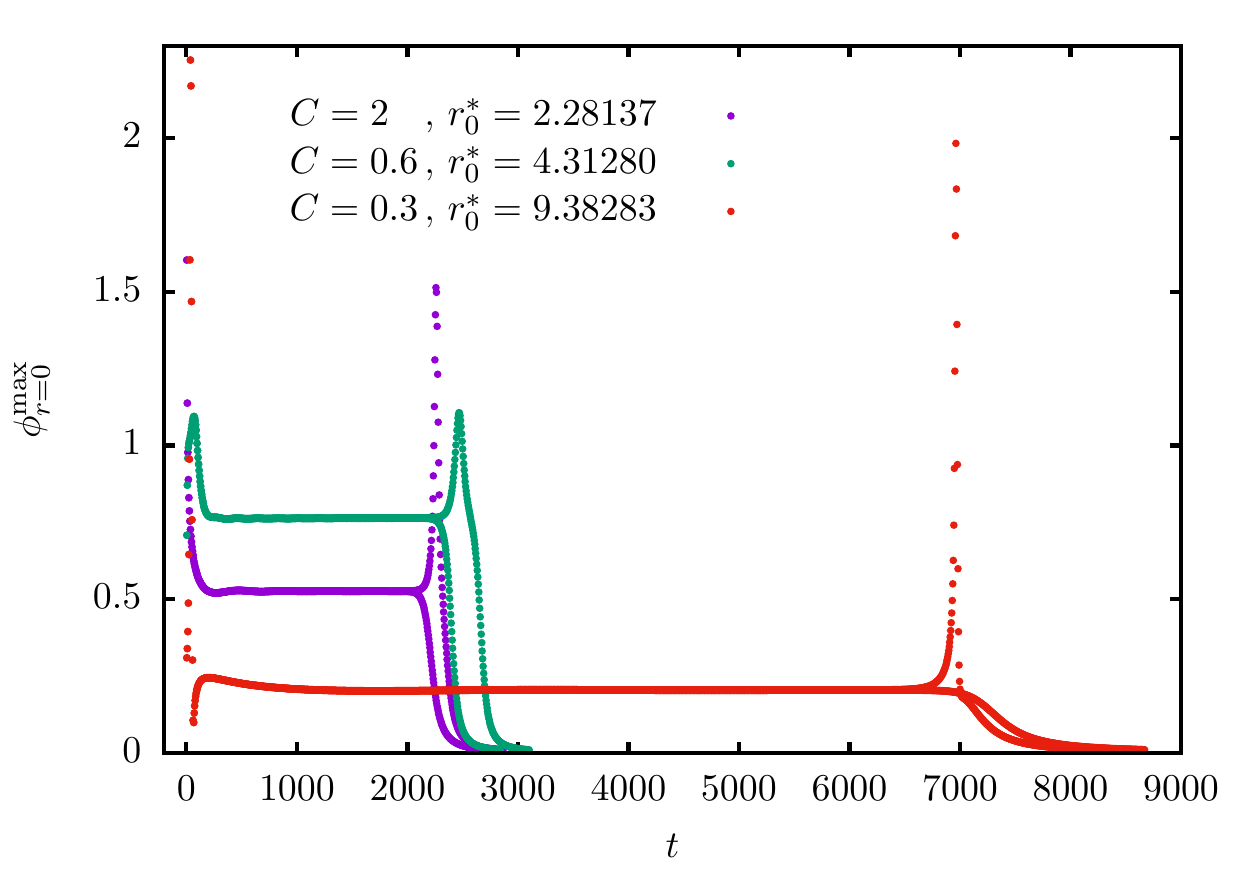}
\caption{\label{figcontour1d}
The points constituting the upper envelope of the central value of the scalar field for oscillons at three fine-tuned first peaks. In all three cases both a subcritical and a supercritical state is shown, with fine-tuning of $32$ digits.
}
\end{figure}

\begin{figure}[!ht]
\centering
\includegraphics[width=115mm]{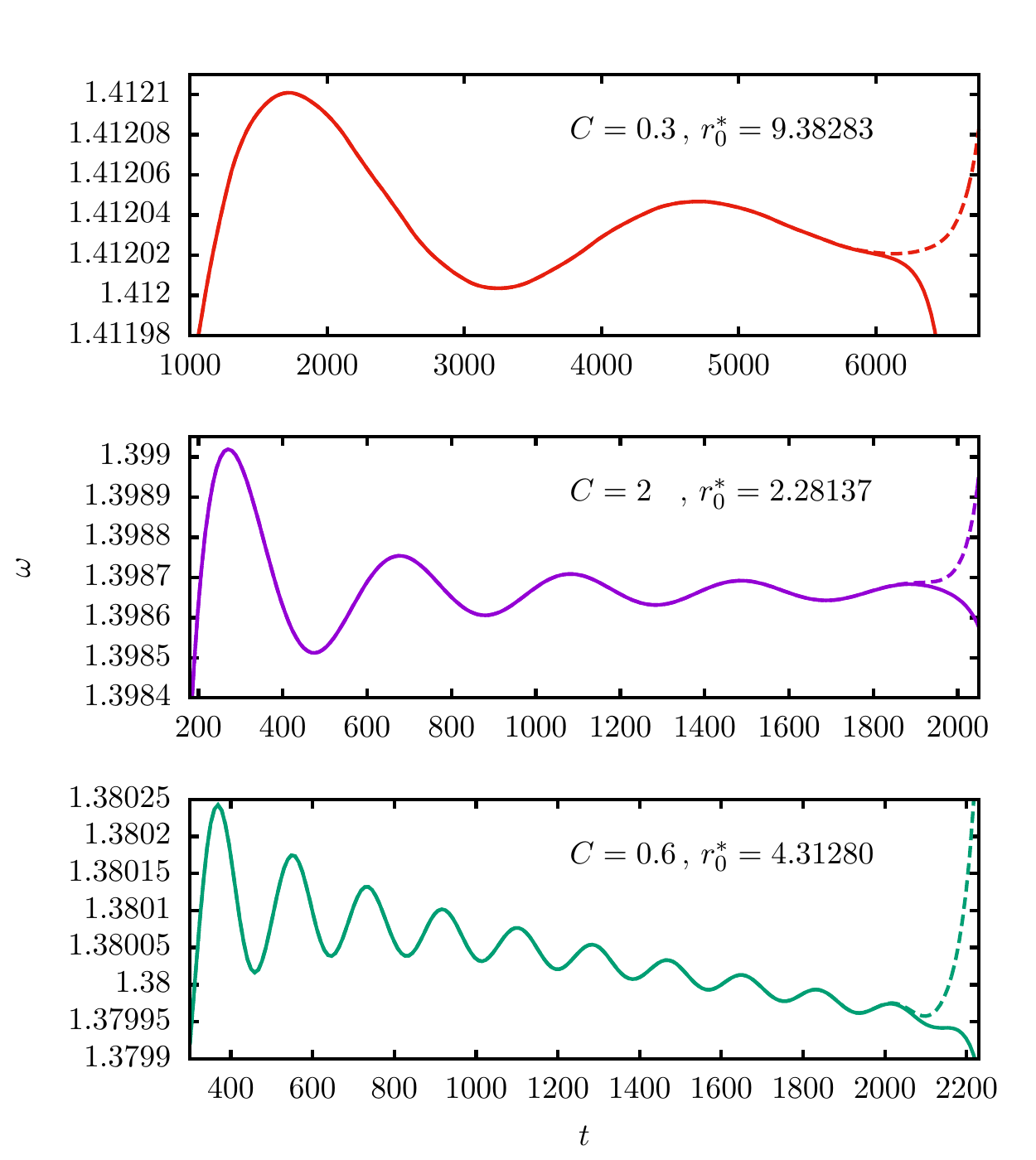}
\caption{\label{figfr1d}
Time dependence of the frequency of the three near-periodic states shown on the previous figure. Dashed lines correspond to subcritical states.
}
\end{figure}

Our numerical simulations clearly show that the different resonance peaks correspond to different near-periodic states. The time dependence of the amplitude of the fine-tuned states corresponding to the three first peaks shown on Fig.~\ref{f:lt} is given on Figure \ref{figcontour1d}.
The time-evolution of the frequency of the corresponding near-periodic states are presented on Fig.~\ref{figfr1d}.
In order to determine the frequency $\omega$ we have minimized the value of the following integral calculated from the numerical results of the time-evolution code
\begin{equation}
\int_{t-t_0}^{t+t_0} \left[\phi(t,\bar
r)-\phi(t+T,\bar r)\right]^2\mathrm{d}t
\end{equation}
for the value of the oscillation period $T=2\pi/\omega$, choosing a certain $r=\bar r$ value of the radial coordinate and an appropriate integration interval $t_0$. For the value of $\bar r$ it is only important that it should be in the core region of the oscillon, while it is practical to choose the value of $t_0$ to be of similar order as the oscillation period $T$. This method using polynomial interpolation gives considerably better precision than the fast Fourier transform (FFT) method.

On all three near-periodic states on Fig.~\ref{figfr1d} a lower frequency modulation can be observed, which corresponds to a stable shape-mode. A similar wavy form could be seen on Fig.~\ref{figcontour1d} as well, if we would show the amplitude of the individual states separately magnified out in the near-periodic domains. Apart from this stable low frequency oscillation there also exists an unstable shape-mode, which was canceled by the fine-tuning.

All near-periodic states that we have found using our numerical code, including the $125$ peaks for the $C=2$ case, are very similar to the just investigated three states. The frequency of all of them belong to interval $[1.379,1.413]$ spanned by the $C=0.6$ and $C=0.3$ first peaks.

\begin{figure}[!ht]
\centering
\includegraphics[width=115mm]{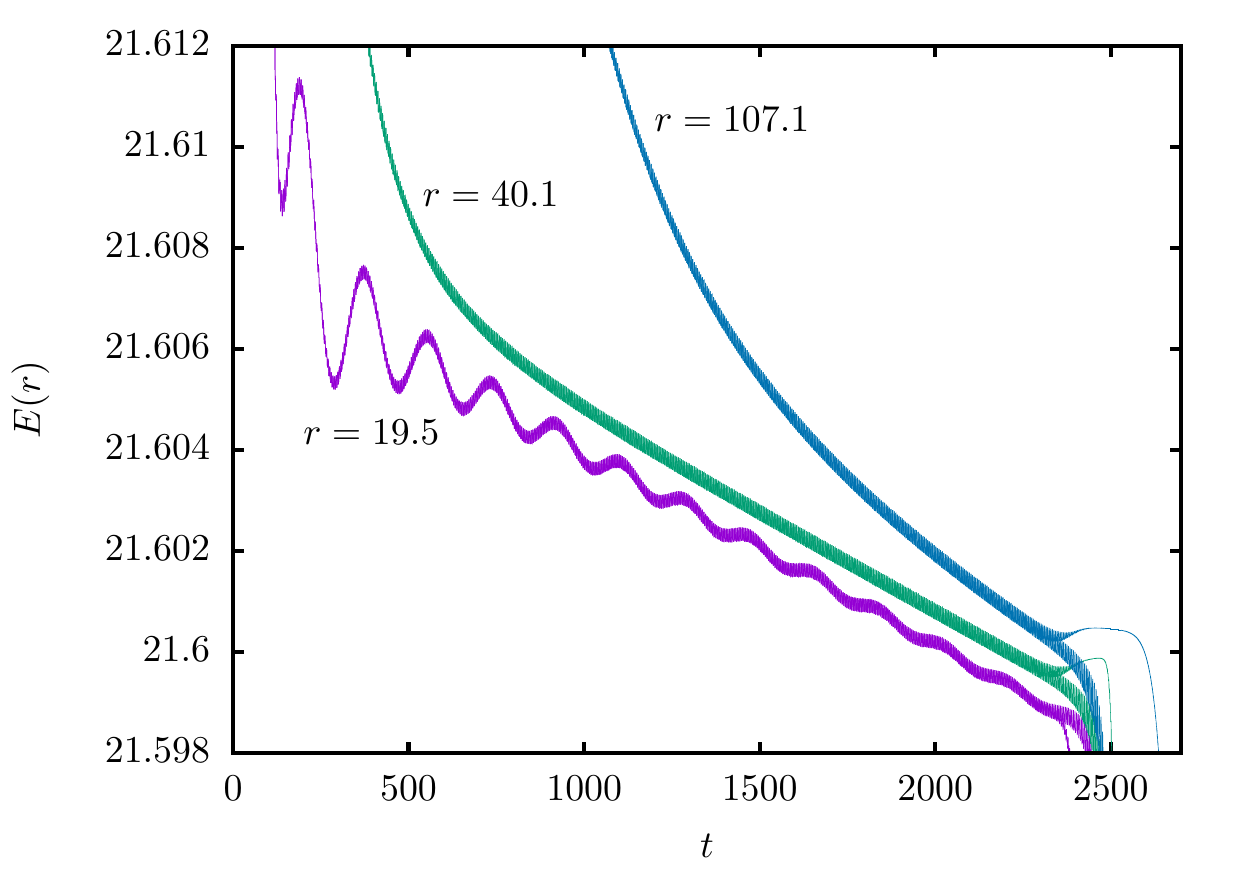}
\caption{\label{figen04d}
The time dependence of the energy inside spheres of radius $r=19.5$, $r=40.1$ and $r=107.1$ for the near-periodic state belonging to the first peak for $C=0.6$.
}
\end{figure}

On the lower panel of Fig.~\ref{figfr1d}, in the case $C=0.6$, it can be clearly seen a slow but uniform decrease in the frequency. At the two other states, which are closer to the frequency $\sqrt{2}$ no such tendency can be seen during the time period that we could follow numerically with the fine-tuning method. Corresponding to this frequency change, there is also a slow decrease in the numerically calculated energy of the $C=0.6$ near-periodic state. The time dependence of the energy $E(r)$ inside balls with given radiuses calculated by the identity \eqref{eqensph} is presented on Figure \ref{figen04d}.
The definitive decrease of the energy shows that the near-periodic state cannot be exactly time-periodic. Based on the observed rate of radiation the time interval at which the energy would decrease to half of the original can be estimated as $\tau_e=2.6\cdot 10^6$. All the other near-periodic states are necessarily losing energy by radiating out scalar field, but the rate of radiation becomes weaker as the frequency gets nearer to the value $m=\sqrt{2}$. We will show later that this decrease is exponential.

The different endings at around $t=2500$ on Fig.~\ref{figen04d} correspond to subcritical and supercritical evolutions. In case of a subcritical decay the field suddenly gets radiated, in such a way that the energy is only flowing outwards. For the supercritical case the energy first becomes concentrated in a smaller region, and then gets radiated out to infinity. Although this is similar to the behavior of an unstable spherical shell, no shell structure can be observed on the plot of the energy density. The largest energy density is always at the center. The two kind of decay mechanism of the near-periodic state provides an explanation of why long-lived states can be achieved by the fine-tuning method.

In paper \cite{Ikeda17} the influence of Einsteinian gravity on $\phi^4$ oscillons have been studied using a time evolution code for spherically symmetric self-gravitating, self-interacting scalar fields. For relatively weak gravitational coupling, similarly to the flat background case, they have observed resonance peaks on the lifetime plots, and near-periodic states appearing close to them. Oscillons coupled to gravity, which are generally called oscillatons, will be discussed in Chapter \ref{fejrelgrav}.

\section{Quasibreathers and nanopterons} \label{seckvaznano}

\subsection{Time-periodic states}

For a real scalar field, apart from the $1+1$ dimensional sine-Gordon breather, there is no exactly time-periodic oscillating finite energy spatially localized configuration. Every oscillon loses energy slowly by radiating out the scalar field in a spherically symmetric way. Because of this, the amplitude and frequency of oscillons must also change slowly. The amplitude of the outgoing radiation is generally extremely small in comparison to the amplitude of the oscillations of the oscillon in the central domain. If we compensate the outgoing radiation by an exactly identical amplitude incoming radiation, then an exactly time-periodic state is formed. Because of the incoming radiation, in the faraway domain a spherically symmetric standing wave forms, that we call \emph{tail}. At the same time, in the inner \emph{core} domain the larger amplitude oscillations remain essentially unchanged. Because of the very small amplitude of the tail with respect to the central amplitude, following the nomenclature of John P. Boyd, we call these solutions \emph{weakly nonlocal} states \cite{Boyd1989a,Boyd-book1998}. 

The study of time-periodic weakly nonlocal solutions is considerably easier, both by analytical and numerical methods, than the direct study of the slowly changing frequency oscillons. For a fixed frequency, still there are many different weakly nonlocal solutions, with different tail-amplitudes and phases. Obviously, the oscillon is best approximated by the weakly nonlocal solution that has the minimal tail-amplitude. This solution is unique for a given frequency, and we name it \emph{quasibreather}.
\begin{figure}[!ht]
\centering
\includegraphics[width=115mm]{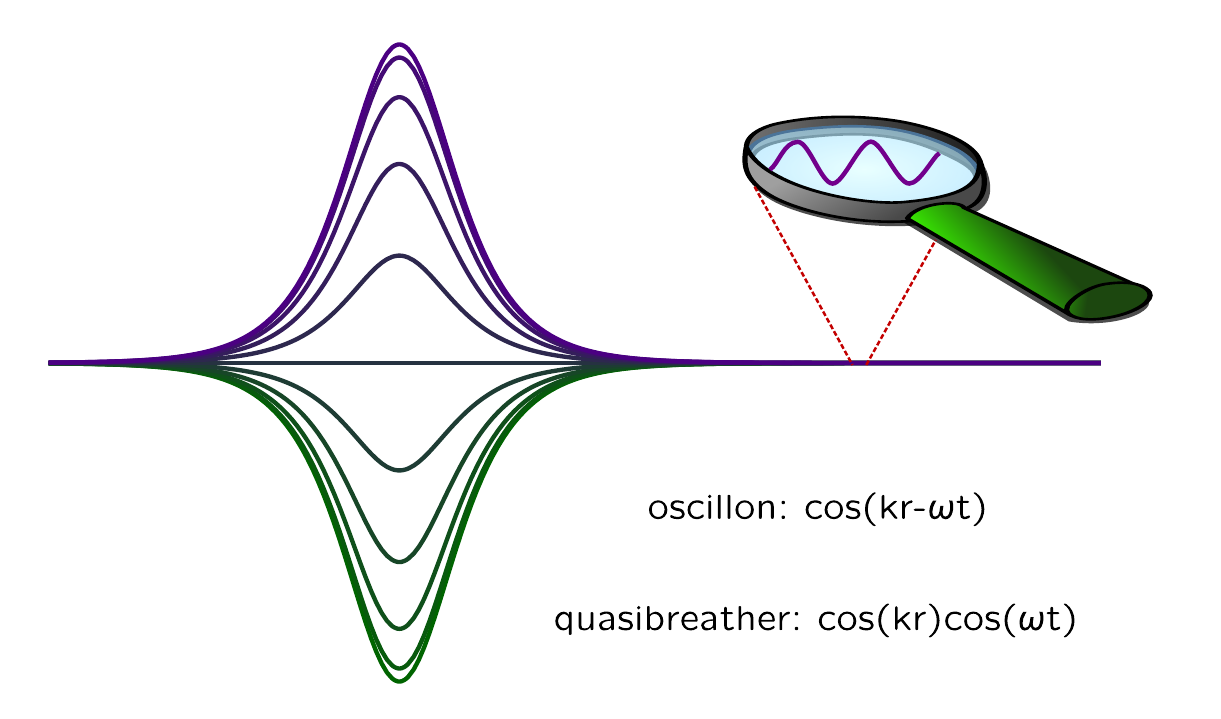}
\end{figure}

At large distances from the core the scalar behaves as a Klein-Gordon field with mass $m$. The standing wave solutions of the Klein-Gordon equation with frequency $\omega_f>m$ at large $r$ distances are
\begin{equation}
 \phi\approx\frac{1}{r^{\frac{d-1}{2}}}\cos[\lambda_f(r-r_0)]\cos[\omega_f(t-t_0)] \label{eqphistandw}
\end{equation}
with arbitrary $r_0$ and $t_0$ phases, and $\omega_f^2=m^2+\lambda_f^2$. The energy $E(r)$ of this wave contained in a sphere of radius $r$ calculated using \eqref{eqensph} grows linearly with $r$ at large distances. This shows that the weakly nonlocal solutions, including the quasibreathers, always have infinite energy. On the other hand, the amplitude of the oscillating tail is generally so small that the energy density there is several order of magnitudes smaller than the energy density in the core region. Considering an intermediately large sized spherical region, most of the energy is still contained in the much smaller central core part, and the object can be considered as a ``quasi'' localized breather. In case of oscillons, in the asymptotic region the scalar field is an outgoing wave of the form $r^{(1-d)/2}\cos(\lambda_f r-\omega_f t)$. The energy $E(r)$ contained in this outgoing wave, assuming that the radiation has begun infinitely long time ago, also diverges for large $r$. That condition obviously cannot be satisfied, since oscillons are evolved from finite energy initial data. Considering the future, in certain cases the oscillons may still exist for infinitely long time, since the magnitude of the radiation decreases quickly as the time passes. The oscillons never decay in case of $d=1$ and $d=2$ spatial dimensions, since in those cases the small amplitude oscillons that emerge as a result of the radiation loss are stable.

The notion of quasibreathers was introduced in our paper \cite{Fodor2006}, and it became adopted in the literature \cite{Saffin2007,SalmiThesis2008,Hertzberg10,Salmi12,Andersen2012}. A closely related concept to quasibreathers is the notion of \emph{nanopterons} introduced by John P. Boyd \cite{Boyd1989a,Boyd1989,Boyd1990,Boyd1995,Boyd-book1998}. The naming comes from the Greek expression ``dvarf-wing''. Nanopteron solutions appear in various hydrodynamic, meteorological, oceanographic, plasma physical and particle physics models. According to Boyd's definition, nanopterons are weakly nonlocal states for which reducing the core amplitude the amplitude of the wing decreases exponentially. As we will see, this is always satisfied for quasibreathers associated to oscillons. For the case when the wing amplitude only decreases according to a power law, Boyd has introduced the naming micropteron. Similarly to quasibreathers, nanopterons also have a large amplitude core region and a very small amplitude standing wave tail, i.e.~wing. Nanopterons are not unique either, different solutions exist depending on the phase of the wing. The quasibreather solution can be considered as the minimal tail-amplitude nanopteron.

We search for quasibreather solutions with frequency $\omega$ in a Fourier series form
\begin{equation}
 \phi=\sum_{n=0}^\infty\Phi_n\cos(n\omega t)  \ , \label{eqphifourier}
\end{equation}
where the functions $\Phi_n$ are time-independent. In general, the frequency $\omega$ of the quasibreather's core is below the mass threshold $m$, and only the higher order modes can have standing wave tails. The scalar field $\phi$ expanded in the form \eqref{eqphifourier} is time-reflection symmetric at the moment $t=0$. The existence of these type of solutions follows from the invariance of the system with respect to time-reflection. The choice is also motivated by the observation that according to numerical simulations weakly radiating oscillons also possess this symmetry to a very high degree. The expansion \eqref{eqphifourier} was first used as a starting point for the search of small amplitude solutions \cite{Kosevich75,dashen-75,bogolyubskii-77c}. We note that the expression
\begin{equation}
 \phi=\Phi_0+\sum_{n=0}^\infty\left\{
 \Phi_{2n+1}\sin[(2n+1)\omega t]+\Phi_{2n+2}\cos[(2n+2)\omega t]
 \right\} \label{eqphifouridashen}
\end{equation}
used in \cite{dashen-75} is also time-reflection symmetric at the moment $t=\pi/(2\omega)$, so it equivalent to \eqref{eqphifourier}. Later, expansion \eqref{eqphifourier} was used for the proof of the nonexistence of exponentially localized breather solutions \cite{Eleonskii1984,Eleonsky1991}. For those type of potentials that are mirror symmetric around their minimum, such as the $U(\phi)=1-\cos\phi$ \, sine-Gordon potential, there are only odd indexed terms in the expansion \eqref{eqphifourier}, which makes the analysis significantly easier.

Substituting the expansion \eqref{eqphifourier} into the field equation \eqref{fieldeq2}, for the functions $\Phi_n$ we obtain the following coupled differential equations,
\begin{equation}
 \Delta\Phi_n+(n^2\omega^2-m^2)\Phi_n=F_n \ , \label{eqphinnonl}
\end{equation}
where the terms $F_n$ are nonlinear functions of $\Phi_0,\Phi_1,\Phi_2,\ldots$, and $m$ is the mass of the scalar field.

We write the expansion of the interaction potential around its minimum in the form
\begin{equation}
 U(\phi)=\frac{1}{2}m^2\phi^2+\sum_{k=2}^\infty\frac{1}{k+1} g_{k}\phi^{k+1} \label{potexpeq}
\end{equation}
in terms of the expansion coefficients $g_{k}$. In this case the derivative of the potential with respect to $\phi$ can be written as
\begin{equation}
 U'(\phi)=m^2\phi+\sum_{k=2}^\infty g_{k}\phi^{k} \ .\label{eqpotdexp}
\end{equation}

The nonlinear terms on the right-hand side of equation \eqref{eqphinnonl} can be obtained in the following way,
\begin{align}
 F_n=\left(1-\frac{1}{2}\delta_{n,0}\right)\biggl[
 &\frac{g_2}{2}\sum_{p,q=0}^\infty \Phi_p\Phi_q\delta_{n,\pm p\pm q}
 +\frac{g_3}{4}\sum_{p,q,r=0}^\infty \Phi_p\Phi_q\Phi_r\delta_{n,\pm p\pm q\pm r} \notag\\
 &+\frac{g_4}{8}\sum_{p,q,r,s=0}^\infty
 \Phi_p\Phi_q\Phi_r\Phi_s\delta_{n,\pm p\pm q\pm r\pm s} \notag\\
 &+\frac{g_5}{16}\sum_{p,q,r,s,t=0}^\infty
 \Phi_p\Phi_q\Phi_r\Phi_s\Phi_t\delta_{n,\pm p\pm q\pm r\pm s\pm t}
 +\cdots
 \biggr] , \label{eqfngnsum}
\end{align}
where
\begin{equation}
 \delta_{n,\pm p\pm q}=\delta_{n,p+q}+\delta_{n,p-q}+\delta_{n,-p+q}+\delta_{n,-p-q} \ ,
 \label{eqdeltapm}
\end{equation}
and we define similarly the other $\delta$ expressions, by adding all possible variations for the signatures. We note that in the definition \eqref{eqdeltapm} the zero values of $p$ and $q$ contribute with double weight with respect to the nonzero values, since in those cases there are more nonvanishing terms in the sum. If the potential is chosen to be the symmetric $\phi^4$ potential in the form \eqref{eqpotnum}, then the mass of the scalar field is $m=\sqrt{2}$, and the nonvanishing expansion coefficients are $g_2=-3$ and $g_3=1$.

For the numerical construction of nanopterons with minimal amplitude tails (i.e.~quasibreathers) equations \eqref{eqphinnonl} were used first by John P. Boyd, in case of $1+1$ dimensional $\phi^4$ theory \cite{Boyd1990}. Spherically symmetric $3+1$ dimensional $\phi^4$ quasibreathers and their radiating tails have been studied first by Richard Watkins. However, the excellent but short and not very detailed report about his important results and ideas only appeared as a preprint, which is not available on the internet \cite{Watkins96}. Alfimov, Evans and Vázquez has calculated spherically symmetric quasibreather configurations in case of $n+1$ dimensional sine-Gordon theory \cite{alfimov2000}. For the description of the observed near-periodic states Honda and Choptuik have also been looking for the solutions of the system \eqref{eqphinnonl} by numerical methods \cite{HondaThesis2000,HondaChoptuik2002}, but they could not observe the standing wave tail part of the functions because of their extremely small amplitude.

In this review we consider only spherically symmetric configurations in detail. Then the Laplacian can be written as in \eqref{eqlapsph}, and the form of equations \eqref{eqphinnonl} are
\begin{equation}
 \frac{\mathrm{d}^2\Phi_n}{\mathrm{d}r^2}
 +\frac{d-1}{r}\frac{\mathrm{d}\Phi_n}{\mathrm{d}r}
 +(n^2\omega^2-m^2)\Phi_n=F_n \ , \label{eqphinsph}
\end{equation}
where the expressions $F_n$ on the right-hand side are still given by \eqref{eqfngnsum}. The behavior of the solutions close to the center, depending on the number of spatial dimensions, constitute of the following two components:
\begin{alignat}{2}
 d=1 \quad : \qquad & \Phi_n\approx \gamma_n r+\delta_n \ , \\
 d=2 \quad : \qquad & \Phi_n\approx \gamma_n \ln r+\delta_n  \ , \\
 d\geq 3 \quad : \qquad & \Phi_n\approx \frac{\gamma_n}{r^{d-2}}+\delta_n \ ,
\end{alignat}
where  $\gamma_n$ and $\delta_n$ are constants. For $d\geq2$ spatial dimensions the condition of the regularity at the center is $\gamma_n=0$ for every $n$. For one spatial dimension we are looking for solutions that are reflection symmetric at the center $r=0$, which also requires $\gamma_n=0$. Because of the central regularity, for any dimensions the functions $\Phi_n$ has to be mirror symmetric at the center $r=0$. The appropriate behavior at the center implies one condition for each Fourier mode $\Phi_n$.

At large distances from the quasibreather's core the functions $\Phi_n$ become small, they decouple, and satisfy the linear left-hand sides of equations \eqref{eqphinsph}. The behavior at large distances depends on the signature of the factor $n^2\omega^2-m^2$. If $n$ is below a certain value then $n^2\omega^2-m^2<0$, and the solutions of the homogeneous equation asymptotically has the form
\begin{equation}
 \Phi_n\approx\frac{1}{r^{(d-1)/2}}\left[\alpha_n\exp(-\hat\lambda_n r)
 +\beta_n\exp(\hat\lambda_n r)\right] \ ,
 \label{eqasymexp}
\end{equation}
where $\hat\lambda_n=\sqrt{m^2-n^2\omega^2}$. The condition for the existence of at least weakly nonlocal solutions is that $\beta_n=0$, with arbitrary $\alpha_n$. Then because of the exponential decay, the contribution of the mode $\Phi_n$ to the total energy, which can be calculated by \eqref{eqensph}, is finite. For each mode satisfying the condition $n^2\omega^2-m^2<0$ we have one condition at the origin and one at infinity, which considering a second order equation can be satisfied generally.

In some exceptional cases for some $\Phi_n$ mode it can happen that $n^2\omega^2-m^2=0$. Then the form of the asymptotic solutions are
\begin{alignat}{2}
 d=2 \quad : \qquad & \Phi_n\approx \alpha_n \ln r+\beta_n \ , \label{eqasylaz1}\\
 d\not=2 \quad : \qquad & \Phi_n\approx \alpha_n r^{2-d}+\beta_n  \ . \label{eqasylaz2}
\end{alignat}
In order to have finite energy the solution has to tend to zero, and hence $\beta_n=0$ must hold, and in case of $d\leq 2$ we also have to require $\alpha_n=0$. From the finiteness of the energy, in case of $d\leq 4$ it also follows that $\alpha_n=0$, but for larger dimensions $\alpha_n$ can be arbitrary. For $d\leq 4$ spatial dimensions, together with the central regularity condition, there are three conditions for the mode $\Phi_n$, which is generally too much. For the quasibreathers that we investigate in the following, we will not consider frequencies for which $n^2\omega^2-m^2=0$ holds for some mode, so the further study of this case is not important.

In general, for any frequency there exist an integer $n_\omega$ such that for $n\geq n_\omega$ the inequality $n^2\omega^2-m^2>0$ holds. Introducing the notation $\lambda_n=\sqrt{n^2\omega^2-m^2}$, the asymptotic behavior of the modes can be written as
\begin{equation}
 \Phi_n\approx\frac{1}{r^{(d-1)/2}}\left[\alpha_n\sin(\lambda_n r)
 +\beta_n\cos(\lambda_n r)\right] \  . \label{eqtailstand}
\end{equation}
Although for $d\geq 2$ the function $\Phi_n$ tends to zero for arbitrary $\alpha_n$ and $\beta_n$, the energy  $E(r)$ contained in a sphere of radius $r$ diverges linearly for any $d$ when  $\alpha_n$ or $\beta_n$ is nonzero. Truly localized, finite energy, time-periodic breather solution can only exist if for every $n\geq n_\omega$ modes $\alpha_n=\beta_n=0$, and the central condition $\gamma_n=0$ also holds. Three conditions for each mode for coupled second order equations is generally too much. This argument shows that genuine breather solutions can only exist in some very exceptional cases, such as for the $1+1$ dimensional sine-Gordon potential, when the system is integrable. In other cases we can find the solution most similar to a localized configuration if we look for the minimal tail-amplitude weakly nonlocal solution, i.e.~the quasibreather.

\subsection{Numerical study of quasibreathers} \label{subsecqbnum}

In the remaining subsections of Section \ref{seckvaznano} we present our own numerical results about quasibreathers, in case of $d=3$ spatial dimensions. For the numerical construction of the minimal standing wave tail quasibreathers we have applied the LORENE numerical library \cite{Lorene-home}. The library was created at Paris Observatory in Meudon, and we carried out our research in collaboration with Philippe Grandclément, one of the developers of the program \cite{Fodor2006}. The LORENE code applies spectral methods and multi-domain decomposition of space. About the spectral method and the application of the LORENE library one can find valuable and practical reference in the online resources of the 2005 Meudon school on this technique \cite{Lorene-school}. We divide the space into concentric spherically symmetric shell shaped domains, and for the radial dependences we employ expansion in terms of Chebyshev polynomials. By this method the solution of the differential equations is reduced into inversion of matrices on the coefficient space in each domain. For the calculation of the weakly nonlocal states we use the method described in detail in the paper \cite{Grandclement-01}, generalizing it to the $\Delta+\lambda_n^2$ Helmholtz operator case. We solve the nonlinear system by iterative method, inverting the linear Helmholtz operator on the left-hand side and applying relaxation. Special care must be taken for the code to avoid the trivial solutions $\Phi_0=0$, $1$ or $2$, and $\Phi_n=0$ corresponding to the vacuum states \cite{Fodor2006}.

The frequency of the observed oscillons and quasibreathers is always in the range $\omega<m$. We note that in case of the $\phi^4$ potential form used in this section the scalar field mass is $m=\sqrt{2}$, but later in order to make the comparison easier with the small-amplitude expansion results in the literature, we will also use the choice $m=1$. The amplitude of the quasibreather's tail can be extremely small for frequencies that are only a little smaller than $m$, but decreasing the frequency the tail-amplitude grows very quickly. When $\omega$ decreases close to the value $m/2$, the tail generally grows so large that we cannot talk about a localized state anymore. Hence in the following we only investigate quasibreathers in the frequency domain
\begin{equation}
 \frac{m}{2}<\omega<m \ ,
\end{equation}
for which the linearized behavior of the modes $\Phi_0$ and $\Phi_1$ is exponentially decaying, according to the first term of equation \eqref{eqasymexp}. The other $\Phi_n$ modes for $n\geq 2$ has standing wave tails according to \eqref{eqtailstand}.

Generally, in applications of the LORENE code the faraway region is mapped into a finite domain by the introduction of a radial coordinate $u=1/r$. Because of the standing wave tail this method cannot be applied in our case for the modes $n\geq 2$. The spectral method, similarly to other generally used numerical approaches, cannot describe functions that cross the zero value infinitely many times in a finite domain. For the $n\geq 2$ modes we match the value of the function $\Phi_n$ at a large $R_{\text{lim}}$ value of the $r$ coordinate to the function
\begin{equation}
 \Phi_n\approx \frac{1}{r}A_n\sin(\lambda_n r+\varphi_n) \ ,
 \label{eqtailstphase}
\end{equation}
which is equivalent to \eqref{eqtailstand} for $d=3$, where the amplitude of the asymptotic tail is $A_n$, and its phase is $\varphi_n$. For the first two modes, for which the linearized behavior is exponentially decaying, we still use a compactified domain for $r>R_{\text{lim}}$, and in this domain on the right-hand side of the differential equations \eqref{eqphinnonl} we only keep the functions $\Phi_0$ and $\Phi_1$, setting the others zero. Since for the chosen potential $m=\sqrt{2}$, $g_2=-3$ and $g_3=1$, based on the expansion \eqref{eqfngnsum} in this exterior domain we solve the following equations:
\begin{align}
&(\Delta -2) \Phi_0= \frac{3}{2} (\Phi_0-1) \Phi_1^2 +  \Phi_0^2 (\Phi_0-3) \ , \\
&(\Delta +\omega^2-2) \Phi_1= 3 \Phi_0(\Phi_0-2)\Phi_1+\frac{3}{4}\Phi_{1}^3 \ .
\end{align}
This approximation is obviously only valid up to intermediately large distances, since for large $r$ the nonlinear source terms that involve the modes $n\geq 2$ will determine the behavior of the modes $\Phi_0$ and $\Phi_1$, and these will also become oscillating instead of exponentially decaying. However, this only happens at significantly larger distances than the $R_{\text{lim}}$ value used by us, and the amplitude of the induced oscillations in the faraway region is so small that it does not reduce the precision of our computations. With careful test we have checked that the functions in the core region and the tail-amplitudes of the oscillating modes are not changing when we further increase the position $R_{\text{lim}}$ of the matching surface.

Allowing the oscillating tails for the time-periodic problem, the solution of the differential equations \eqref{eqphinnonl} is not unique. For arbitrary fixed $\varphi_n$ phases for all $n\geq 2$, we can find a weakly nonlocal solution with the above treatment of the boundary conditions. The asymptotic amplitudes $A_n$ of the modes arise as the result of the numerical calculation. For the approximation of the slowly radiating oscillon solution obviously the minimal tail quasibreather is the most appropriate. For this reason, by modifying the phases $\varphi_n$ we were looking for the unique solution for which the amplitude $A_2$ was minimal for some fixed $\omega$ frequency. Instead of minimizing the amplitude $A_2$, it would be more appropriate to minimize the average energy density in the tail domain, taking into account all the modes for $n\geq 2$ \cite{Saffin2007}. However, since the contribution of each mode to the energy density is proportional to the square of their amplitude, and the amplitude of the higher modes is generally significantly smaller than that of the mode $n=2$, the two approaches lead to the same result with good approximation.

The search for the minimum of the amplitude $A_2$ was performed by the application of the multidimensional minimizer of the GSL Scientific Library \cite{gsl-minim}. Initially all phases were set to the value $\pi/2$, and we have iterated the procedure until we have reached the minimum of $A_2$ to a certain precision, finding the quasibreather among the weakly nonlocal solutions. We have looked for the phases in the interval $0\leq\varphi_n<\pi$, reaching the phases outside of this by changing the signature of the amplitudes $A_n$. The code converges extremely fast because of the smooth behavior of $A_2$.

For our calculations we divide the space into spherically symmetric domains according to the value of the $r$ coordinate. The first three domains were chosen to be the intervals $[0,1]$, $[1,2]$, $[2,4]$, and after these the size of all others was taken to be $4$, continuing up to the value of $r=R_{\text{lim}}$. Increasing the number of the domains the value of $R_{\text{lim}}$ grows as well. During our calculations the typical number of domains was $13$, which corresponds to $R_{\text{lim}}=44$, but we have checked that even if we go up to $R_{\text{lim}}=60$ the amplitudes $A_n$ of the oscillating tails do not change significantly. The number of spectral coefficients in each domain was generally $N_r=33$. For the studied cases it turned out to be sufficient to take into account $6$ of the $\Phi_n$ modes in the Fourier decomposition \eqref{eqphifourier}, but to explore even lower frequencies more modes become necessary.

\begin{figure}[!phtb]
\includegraphics[width=85mm]{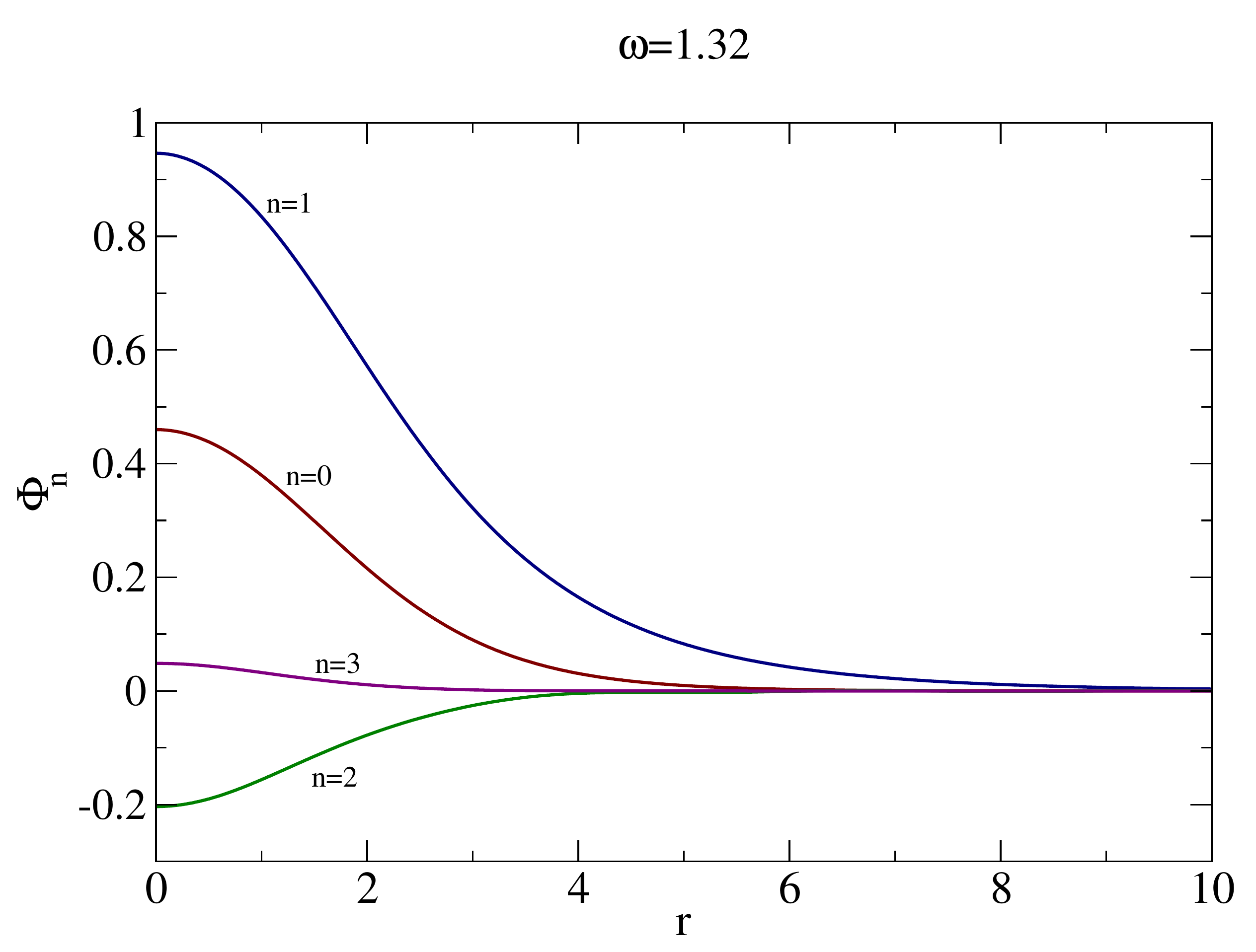}
\includegraphics[width=85mm]{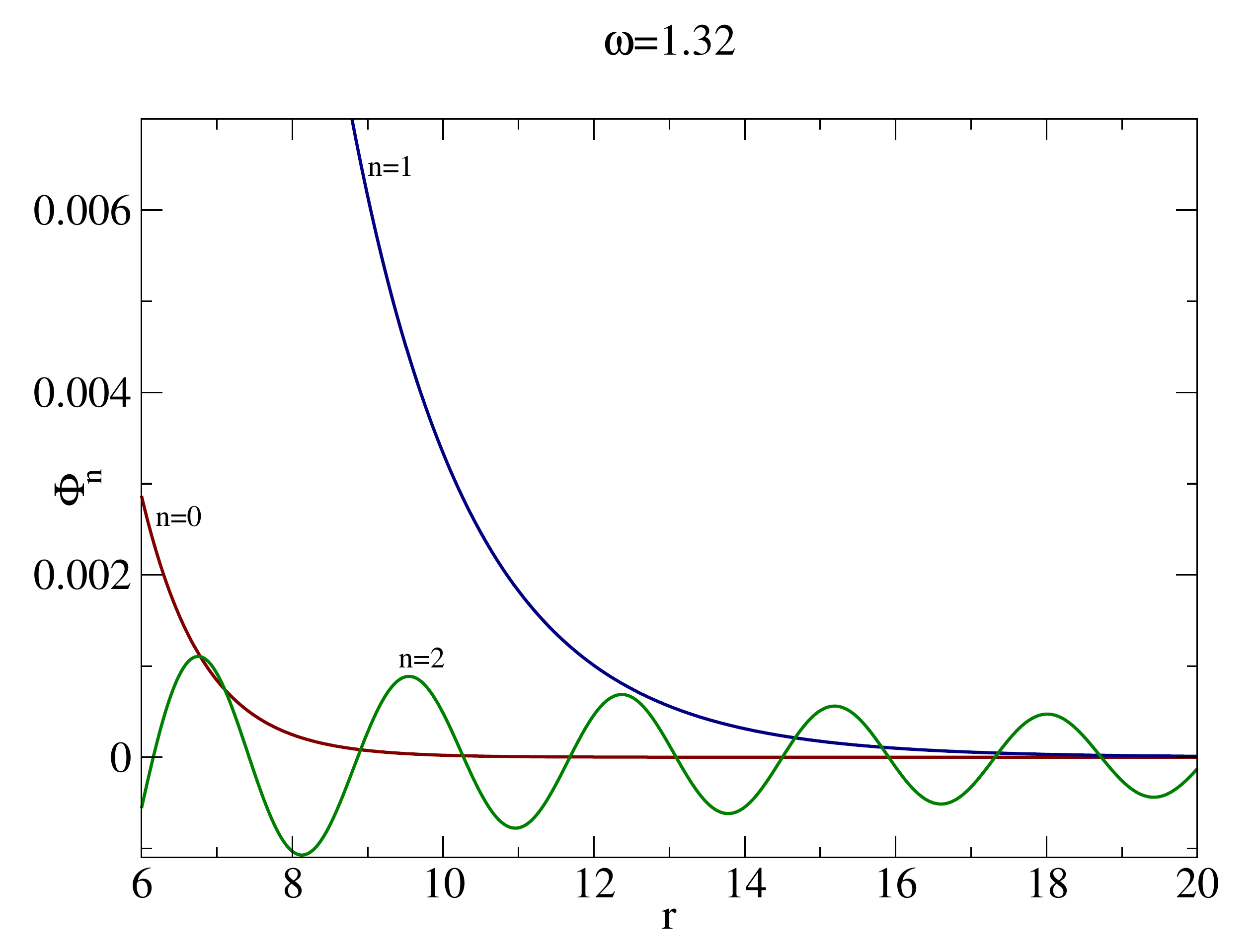}\\
\includegraphics[width=85mm]{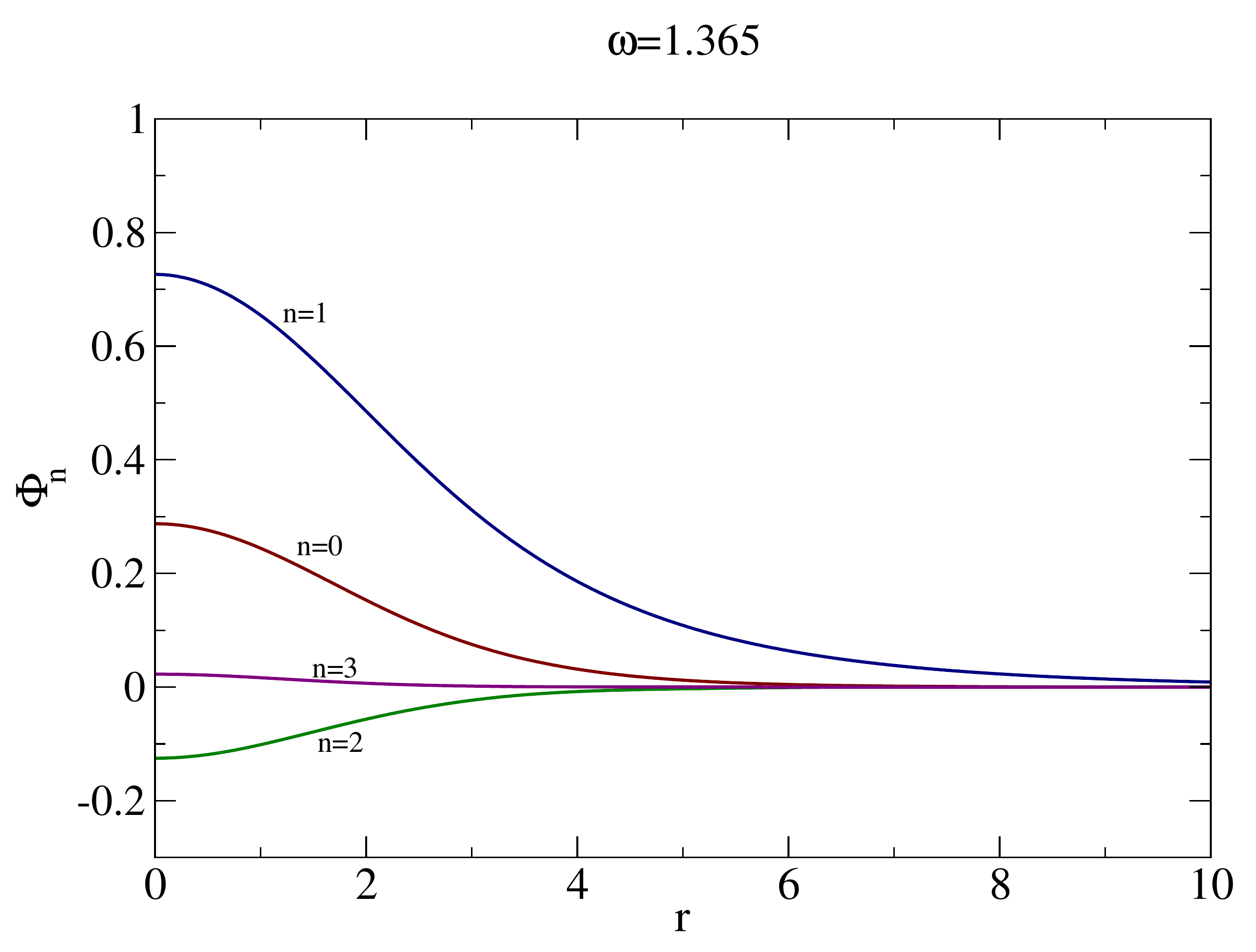}
\includegraphics[width=85mm]{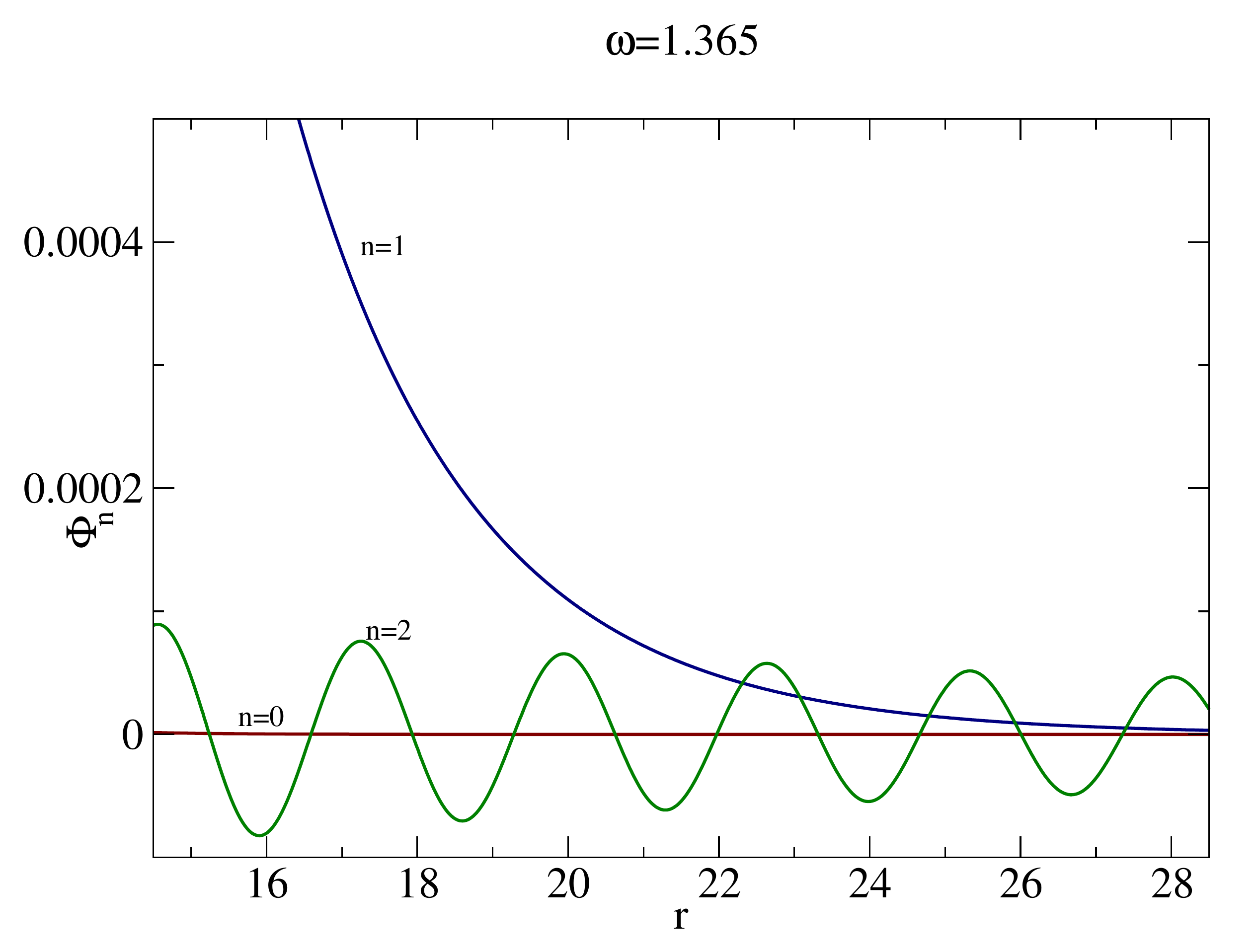}\\
\includegraphics[width=85mm]{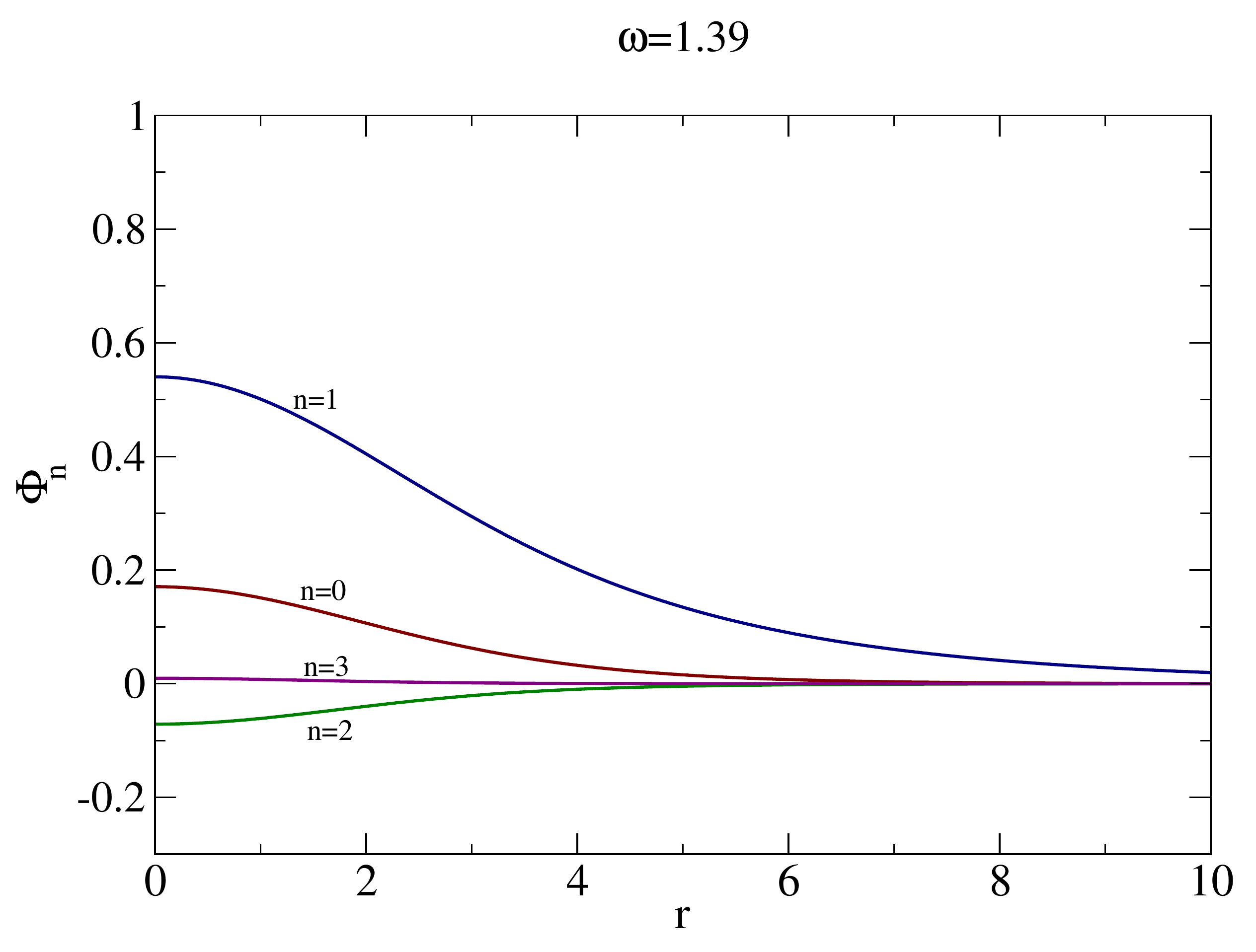}
\includegraphics[width=85mm]{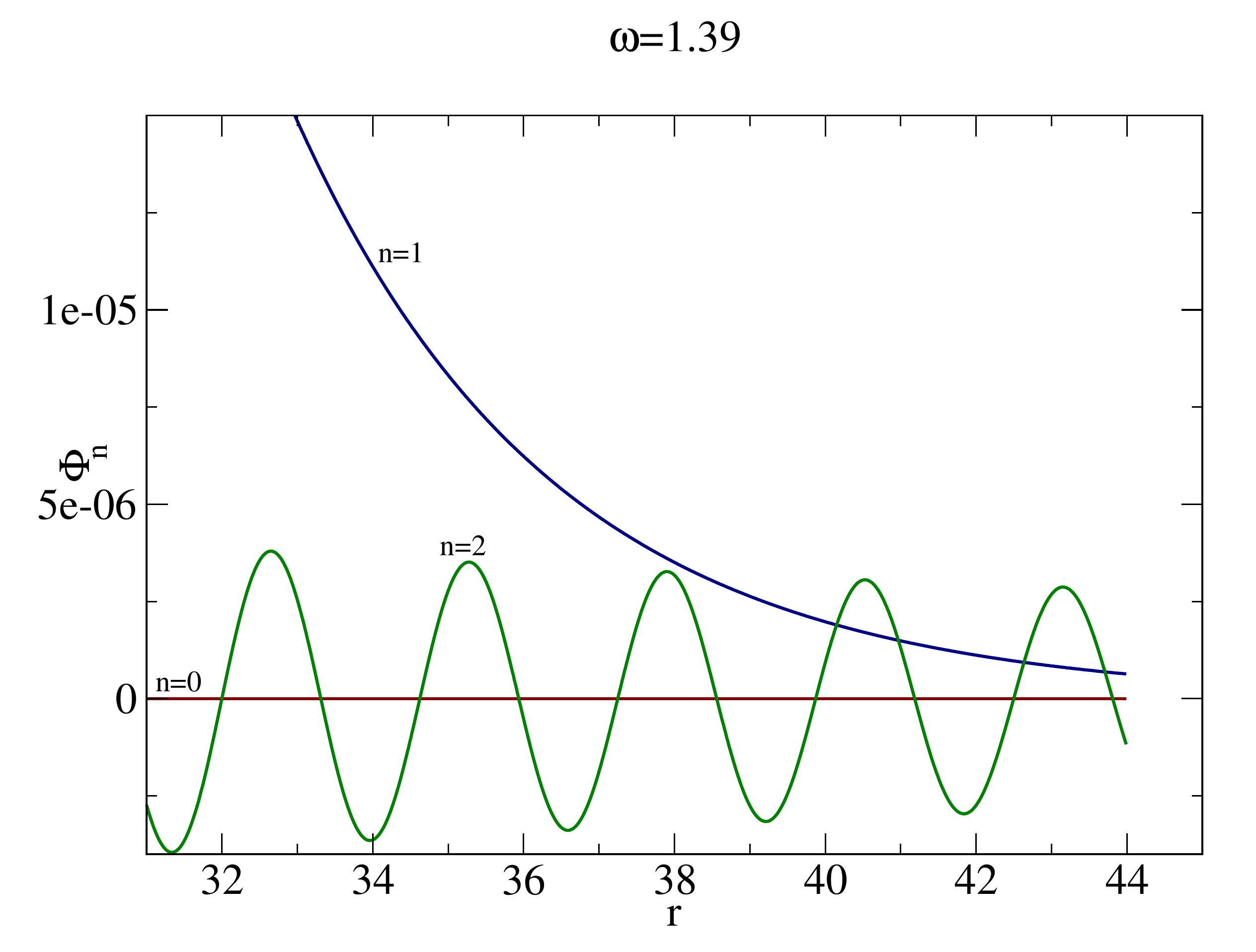}
\caption{\label{f:oscillons}
The $r$ dependence of the Fourier modes $\Phi_n$ of the quasibreather near the core and in the transitional region, for $\omega=1.32$, $1.365$ and $1.39$.}
\end{figure}

On Figure \ref{f:oscillons} we show the radial behavior of the first few $\Phi_n$ modes of the minimized tail quasibreather, in case of three typical frequencies. On the left panels the core domain can be seen, while on the right-hand side panels the region is presented where the tail starts to appear.
It can be seen, that increasing the frequency the central amplitude becomes smaller, and the size of the core-region grows larger. Decreasing the central amplitude the tail-amplitude also becomes smaller, and this decrease of the tail is much faster than the decrease at the center. As we will see later, the relation between the outer and inner amplitudes can be described by an exponential function.

\begin{figure}[!ht]
\centering
\includegraphics[width=115mm]{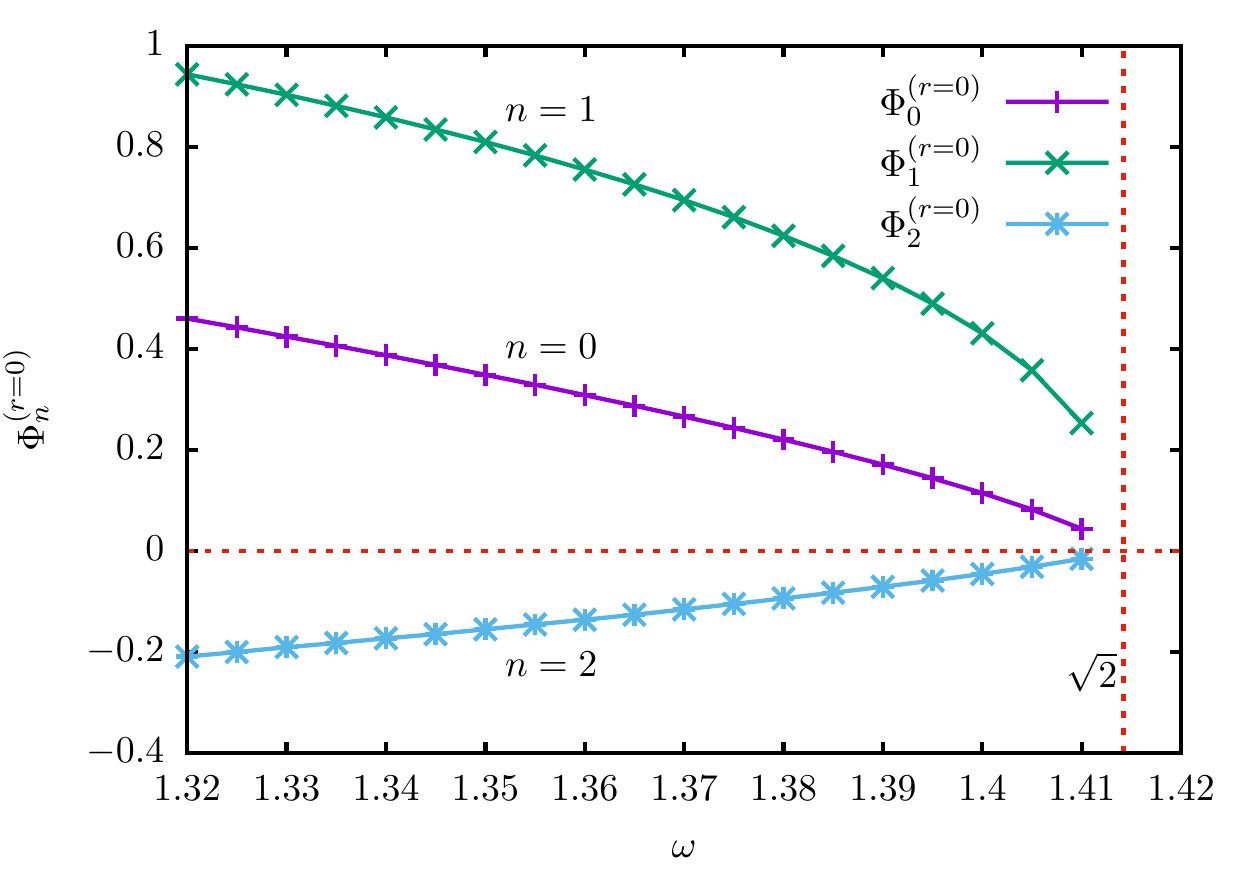}
\caption{\label{conv-ori}
The frequency dependence of the central values of the functions $\Phi_0$, $\Phi_1$ and $\Phi_2$.
}
\end{figure}

\begin{figure}[!ht]
\centering
\includegraphics[width=115mm]{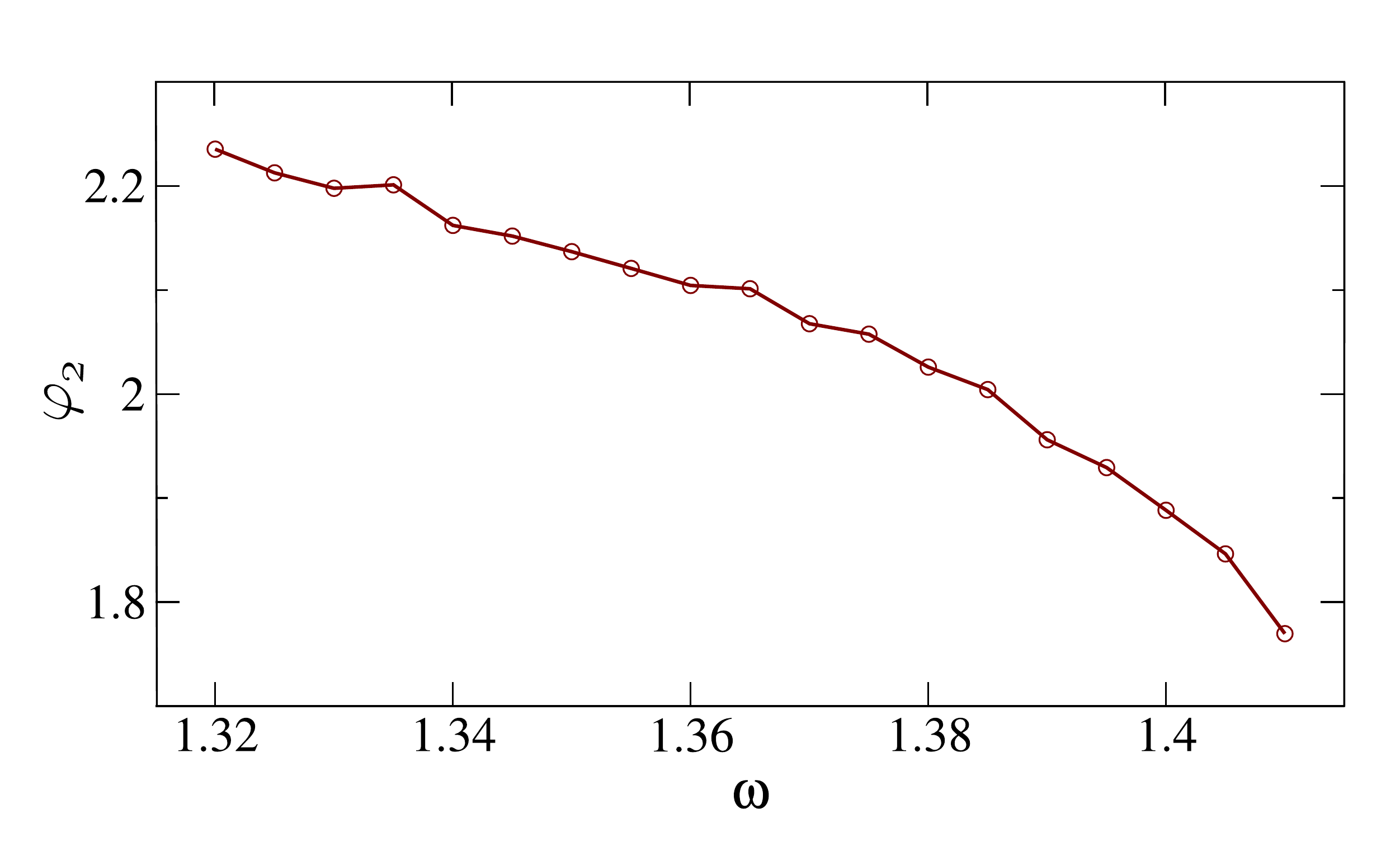}
\caption{\label{conv-phase}
The frequency dependence of the phase $\varphi_2$ of the leading oscillating mode $\Phi_2$.
}
\end{figure}

On Figure \ref{conv-ori} we show the central values of the first three $\Phi_n$ modes as functions of the frequency.
On Fig.~\ref{conv-phase} the phase of the quasibreather's tail in the $\Phi_2$ mode is shown as a function of the frequency.

\begin{figure}[!ht]
\centering
\includegraphics[width=115mm]{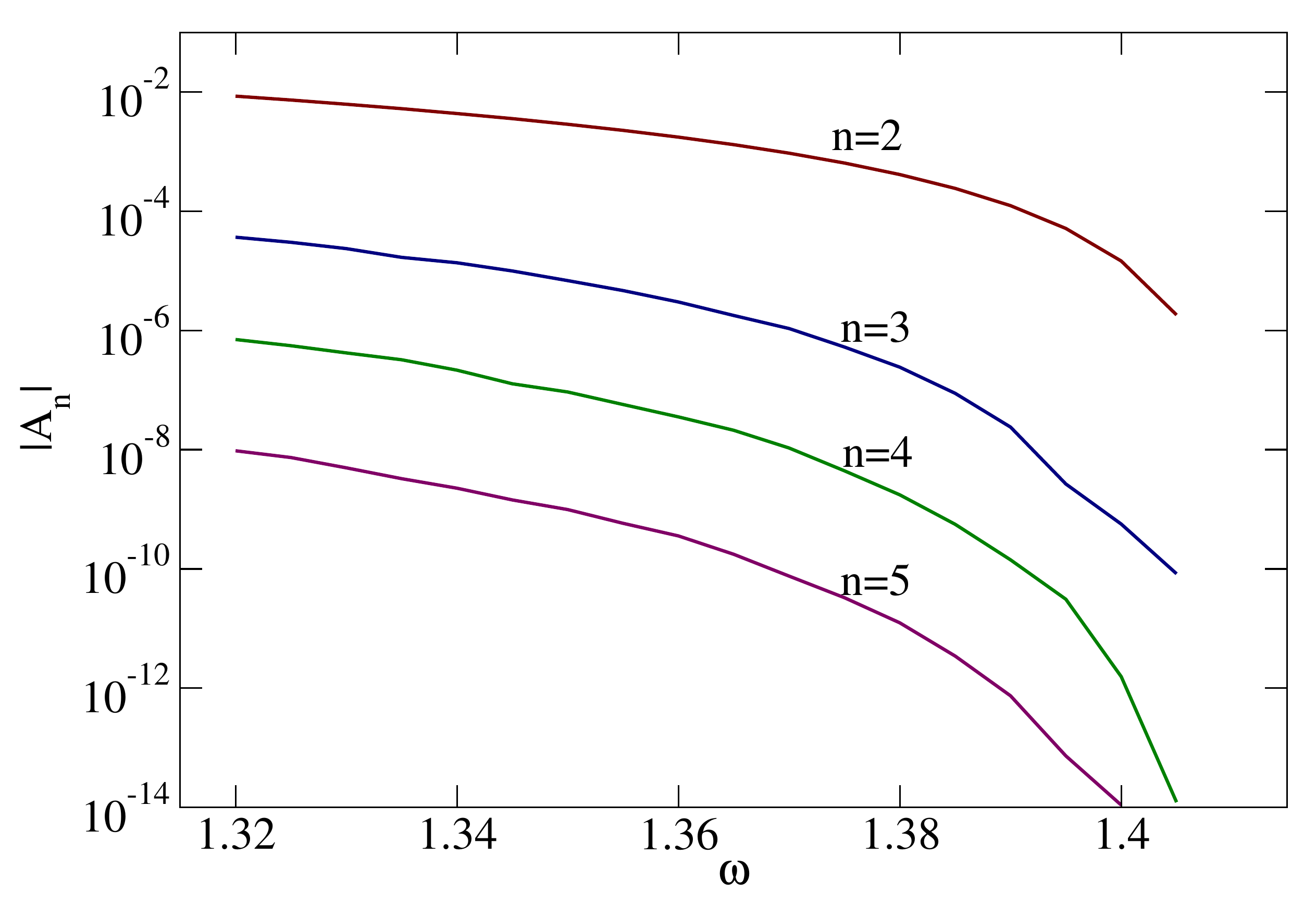}
\caption{\label{coeff-hom}
The frequency dependence of the tail-amplitudes of the first four oscillating modes.
}
\end{figure}

The rapid decrease with respect to the index $n$ of the tails of the oscillating modes is shown on Figure \ref{coeff-hom}.
From the figure it can also be seen that in the studied frequency domain there is no such $\omega$ where the amplitudes tend to zero, hence there is no localized finite energy breather solution of the system. In the $\omega\to m$ limit the amplitude tends to zero even in the central core domain, and the solution approaches the trivial vacuum solution.

\begin{figure}[!htb]
\centering
\includegraphics[width=115mm]{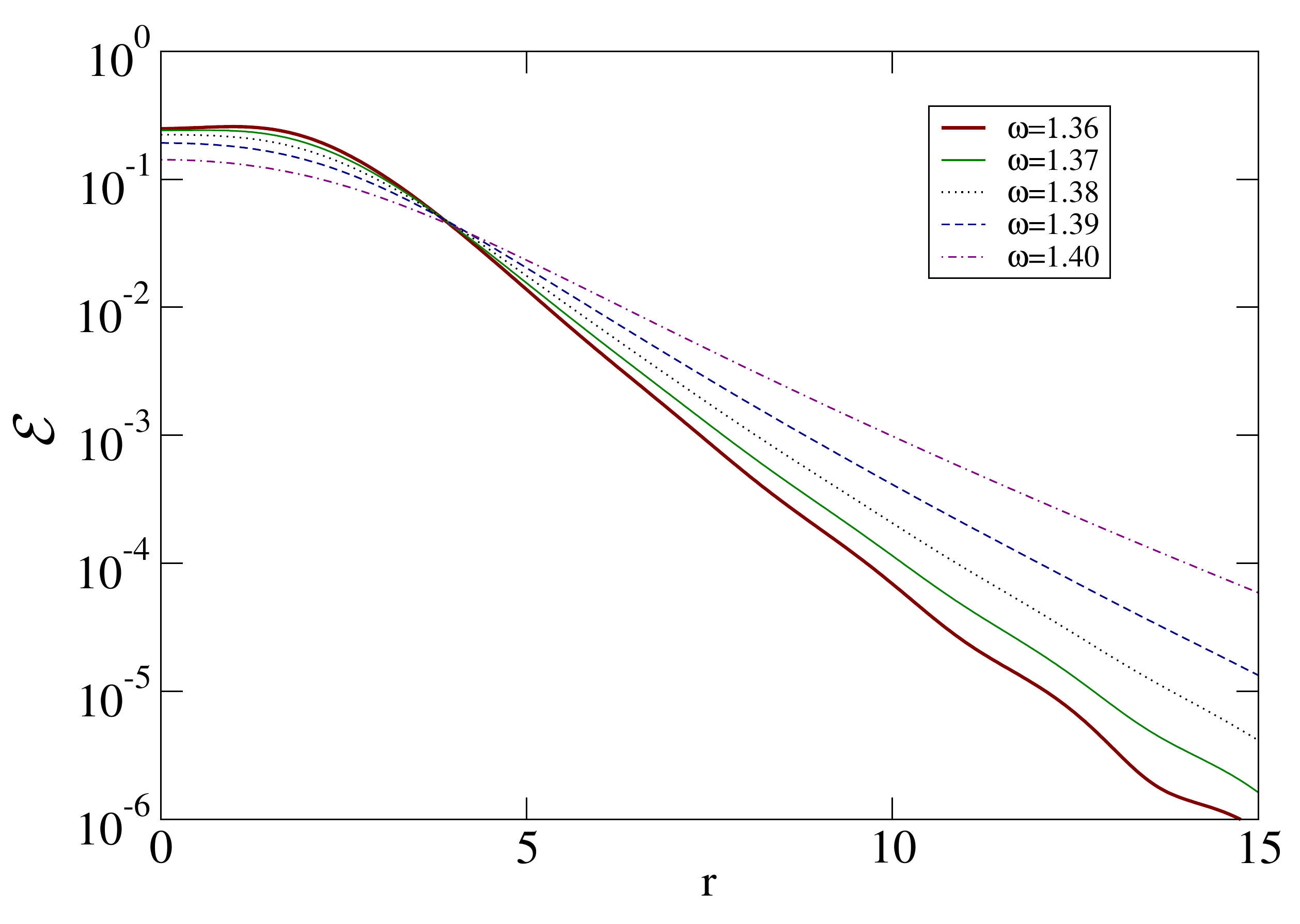}
\caption{\label{energy_ori}
The dependence on the coordinate $r$ of the energy density in the core region of the quasibreather.
}
\end{figure}

\begin{figure}[!htb]
\centering
\includegraphics[width=115mm]{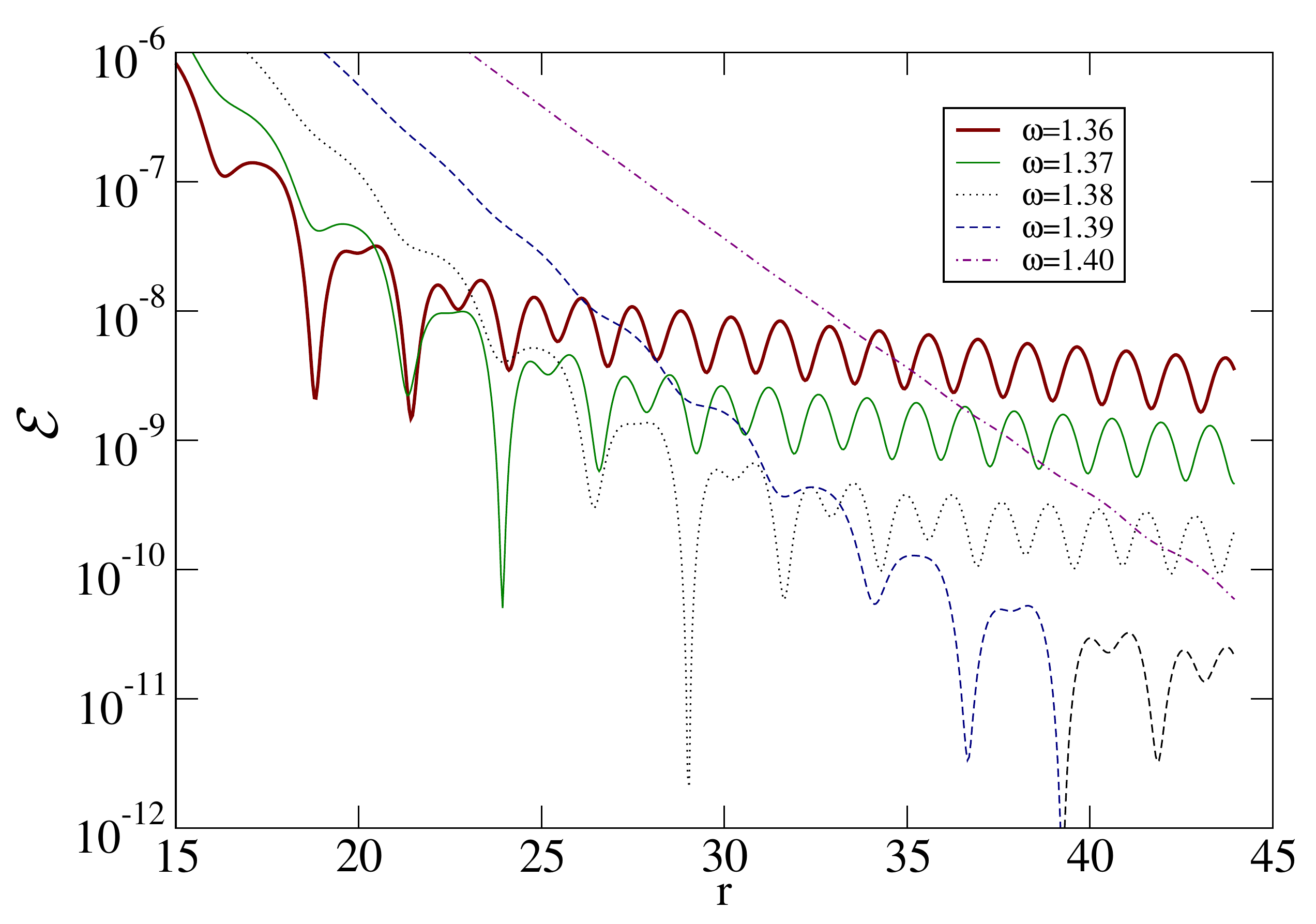}
\caption{\label{energy_end}
Energy density of the quasibreather in the standing wave region.
}
\end{figure}

For the better understanding of the structure of the quasibreathers, on Figures \ref{energy_ori} and \ref{energy_end}
we show the radial dependence of the energy density $\mathcal{E}$, calculated by the expression \eqref{eqendensgen}, in a central and also in an outer domain. For a given frequency, in the tail region the tendency of the decrease of the energy density is according to $1/r^2$, from which it follows that the energy $E(r)$ inside a sphere diverges linearly with increasing $r$. Although the quasibreathers are not finite energy states, still in a large region they approximate with high precision the finite energy but radiating oscillon states. It can also be seen from the figures that in the outer part of the core domain the energy density decreases approximately exponentially, until it reaches the value of the energy density of the tail. The energy density of the tail is essentially proportional to the square of the amplitude $A_2$, because of the relative smallness of the other modes.

We define the radius of the core as the value $r=R_\mathrm{trans}$ where the value of the dominant mode $\Phi_1$ first decreases to the local tail-amplitude of the oscillating mode $\Phi_2$,
\begin{equation}
 \Phi_1\Bigl|_{r=R_\mathrm{trans}} = \frac{|A_2|}{R_\mathrm{trans}} \ .
\end{equation}
This is the distance beyond which the oscillating standing wave tail becomes dominant. 

\begin{figure}[!htb]
\centering
\includegraphics[width=115mm]{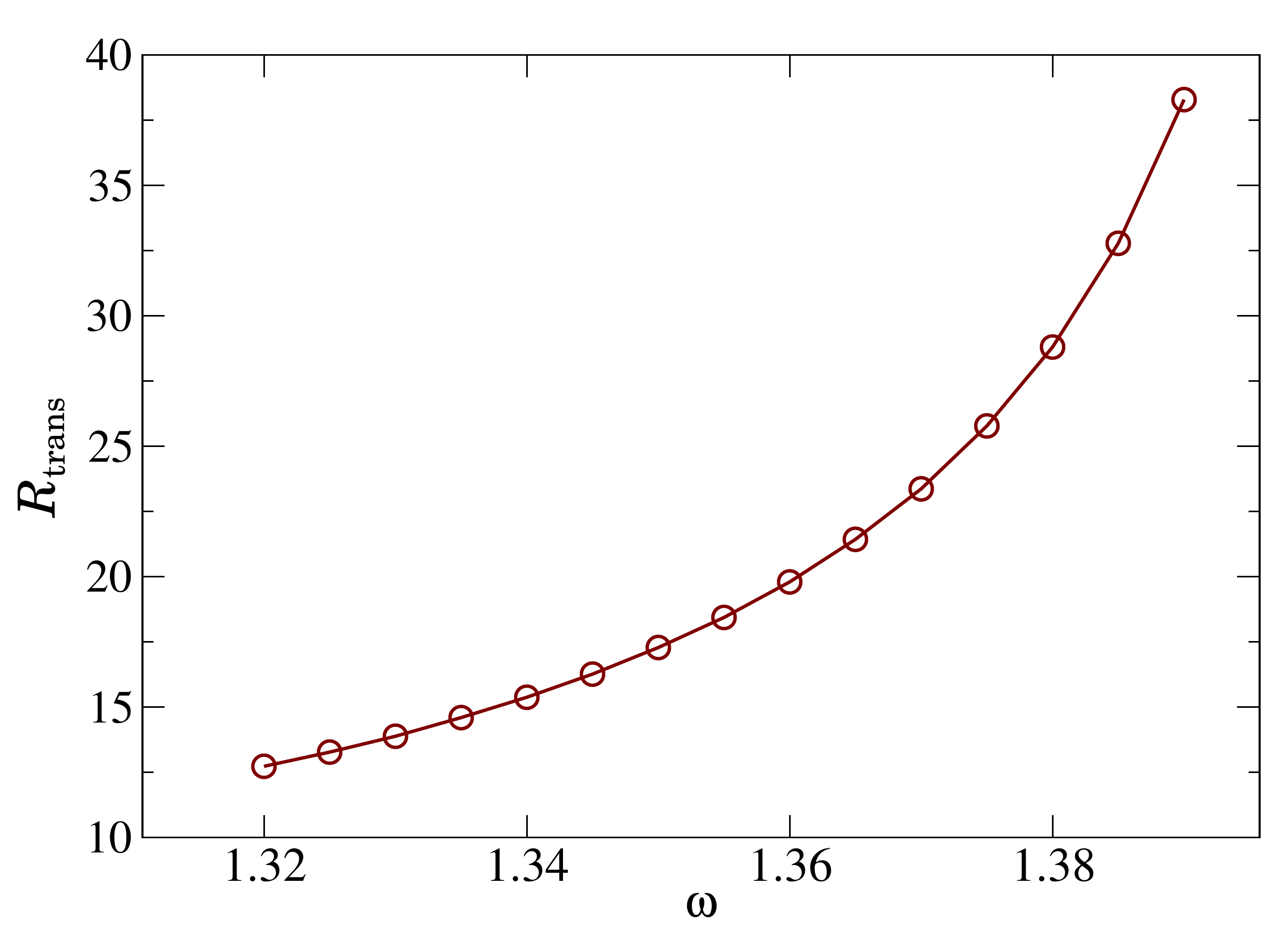}
\caption{\label{figtransr}
The radius $R_\mathrm{trans}$ of the core of the quasibreather as a function of the frequency.
}
\end{figure}

On Figure \ref{figtransr}
we show the change of the radius of the core region as a function of the frequency $\omega$. We have not calculated the core radius for frequencies $\omega>1.39$, because then it becomes larger than the value $R_{\lim} = 44$ used in our code, where we have specified the outer boundary condition. In case of larger $\omega$ it is necessary to increase the value of $R_{\lim}$. In spite of this, we get reliable results for the behavior of the core region even in these cases. We will see later that the size of quasibreathers is inversely proportional to their amplitudes.

\begin{figure}[!htb]
\centering
\includegraphics[width=115mm]{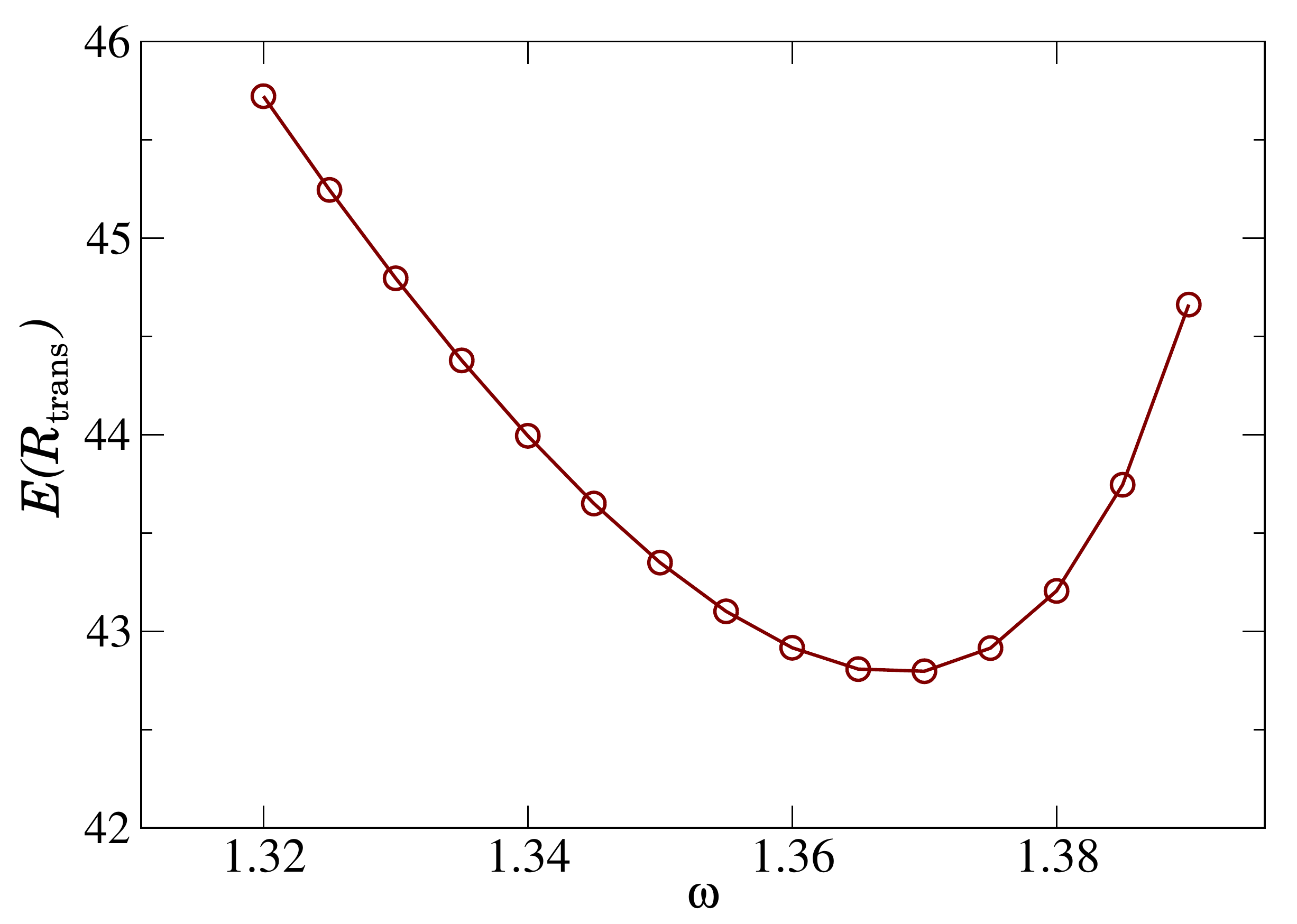}
\caption{\label{figenertot}
The energy $E(R_\mathrm{trans})$ of the quasibreather's core as a function of the frequency.
}
\end{figure}

We consider the total energy of the quasibreather as the $E(R_\mathrm{trans})$ energy inside the core radius $R_\mathrm{trans}$. The $\omega$ dependence of this can be seen on Figure \ref{figenertot}.
In contrast to the core radius the energy is not a monotonic function, it takes a minimal value at the critical frequency $\omega_c\approx 1.368$. The nontrivial behavior is due to two competing effects. Increasing the frequency the core amplitude decreases, but the core size increases. From our analytical calculations it will follow that the energy of the core tends to infinity when the frequency approaches the $m=\sqrt{2}$ value from below.

The critical frequency $\omega_c$ is crucial, since numerical simulations show that oscillons with frequencies $\omega<\omega_c$ are stable, while they are unstable in case of $\omega>\omega_c$. This stability change corresponds to the generally valid astrophysical experience with localized objects similar to stars. If by the increase of the central energy density the full mass-energy is increasing then the configuration is stable, while in the opposite case it is unstable. In case of oscillons and quasibreathers the the central amplitude and central energy density increases monotonically with the decrease of the frequency.

\subsection{The relation between oscillons and quasibreathers} \label{subsecoscqb}

In order to quantitatively compare the oscillons with the quasibreathers, during a certain period of the oscillon's evolution, when the frequency is approximately $\omega$, from the results of the numerical evolution code we calculate the Fourier expansion of the oscillon's scalar field at some radius $r$, and we compare the obtained expansion coefficients with the modes of the frequency $\omega$ quasibreather at radius $r$.

Since the fast Fourier transformation (FFT) is very sensitive to the time step by which the numerical data has been stored, instead of it we use a method based on direct integration, which turns out to be much more precise in the determination of the frequency and the Fourier components of the oscillon. As a first step, we determine the oscillation period by searching for the time $t_1$ and $t_2$ of two consecutive maximums at the central point $r=0$. Since generally both $t_1$ and $t_2$ take places between two saved time slices of the numerical code, we determine the more precise value of these moments by fitting second order polynomials on the data. From our numerical results it can be checked that to a very good approximation the evolution is time-reflection symmetric at the moments $t_1$ and $t_2$, not only at the center but also in a large region around it. We determine the oscillation frequency from the expression $\omega=2\pi/(t_2-t_1)$. After this, we obtain the $n$-th Fourier coefficient at radius $r$ from the function $\phi(t,r)$ written out by the evolution code, by numerically evaluating the following integral,
\begin{equation}
\Phi_n(r)=\int^{t_2}_{t_1}\phi(t,r)\exp(i n\omega t)\mathrm{d}t \ . \label{eqfint}
\end{equation}
We note that the first and last steps of the numerical integration, which are smaller than the others, require special attention. If the function $\phi(t,r)$ was really periodic, and in the moments $t_1$ and $t_2$ it was exactly time-reflection symmetric, then the imaginary part of this integral would be zero for each $n$. We have also computed the imaginary part of the integral, and we have seen that in the core region it is really small compared to the real part.

\begin{figure}[!htb]
\centering
\includegraphics[width=115mm]{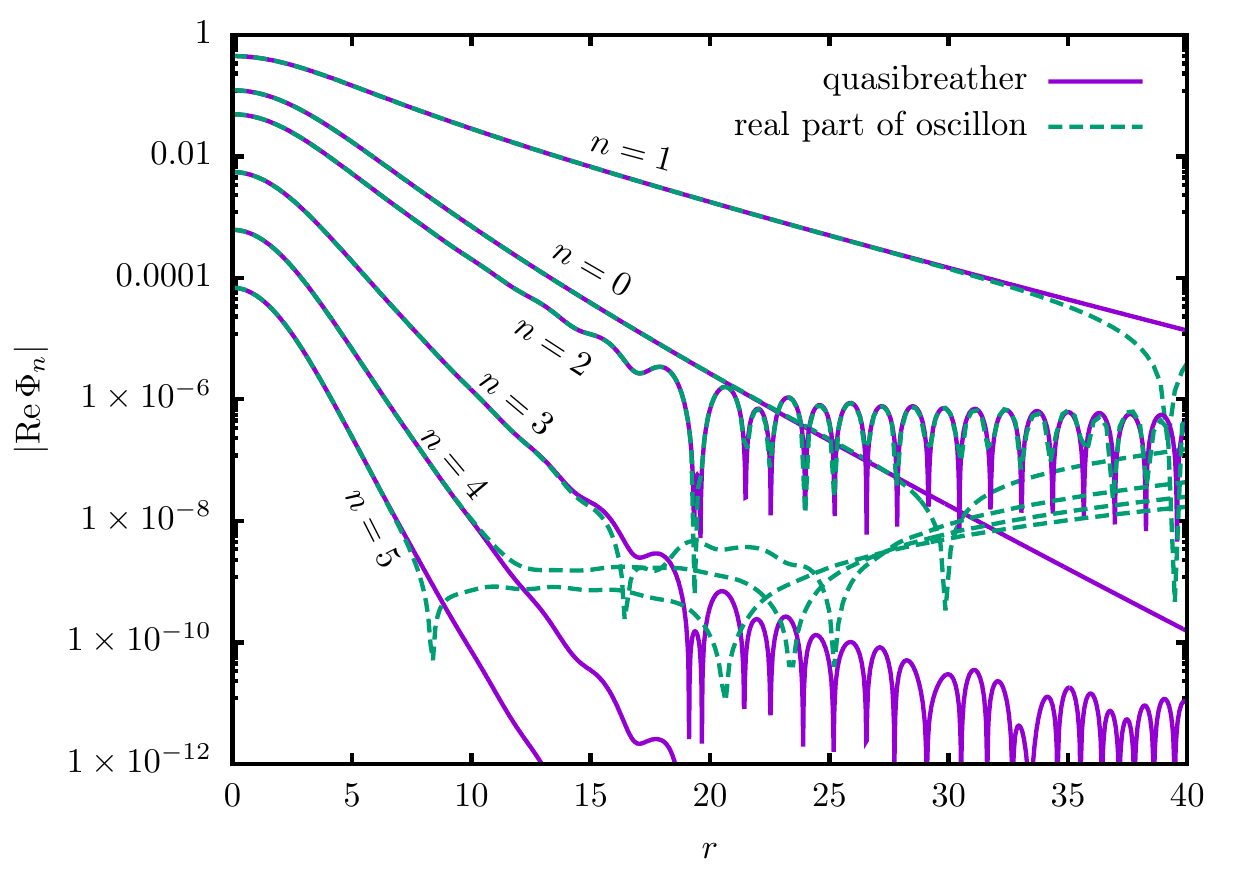}
\caption{\label{figlog1f}
Comparison of the Fourier expansion of the near-periodic state belonging to the firs peak of the $C=2$ initial data to the modes of the $\omega=1.398665$ frequency quasibreather.}
\end{figure}

We first examine the Fourier expansion of the near-periodic states studied in Sec.~\ref{sec-majdnemper}, since these configurations obtained by the fine-tuning of the parameter $r_0$ prove to be periodic and time-reflection symmetric to an extremely high degree. As a first concrete example, we study the near-periodic state belonging to first tuned peak of the initial data \eqref{e:ff} with amplitude $C=2$, of which the time dependence of its frequency was presented on the middle panel of Fig.~\ref{figfr1d}. The Fourier expansion is performed between the consecutive maximums at the moments $t_1=1592.29$ and $t_2=1596.78$. On Figure \ref{figlog1f}
we compare the real part of this expansion to the components of the $\omega=1.398665$ frequency quasibreather, which were obtained by the numerical solution of equations \eqref{eqphinnonl}. On the figure we show the absolute value of the first six components, using a logarithmic scale. The central values of $\Phi_0$ and the odd indexed components are positive, the others are negative. The downward pointing peaks correspond to the zero crossings of the functions. The agreement in the central core region is so good that the relative differences of the values obtained by the two methods are smaller than $10^{-5}$ at the first four modes. Remarkably, the agreement at the mode $\Phi_2$ is surprisingly good even in the oscillating tail region. At the radius where the oscillating tail appears in the mode $\Phi_2$, the amplitude of $\Phi_2$ is approximately $10^{-6}$, and the absolute difference of the curves obtained by the two methods is of the order $10^{-8}$. At this frequency the radius of the quasibreather's core is $r=55.4$, and hence on Fig.~\ref{figlog1f} the mode $\Phi_1$ is still much larger than the oscillating $\Phi_2$ mode. In the $r$ domain shown on the figure the oscillating tail of the quasibreather could not be directly observed on the plot of the full scalar field $\phi$.

\begin{figure}[!htb]
\centering
\includegraphics[width=115mm]{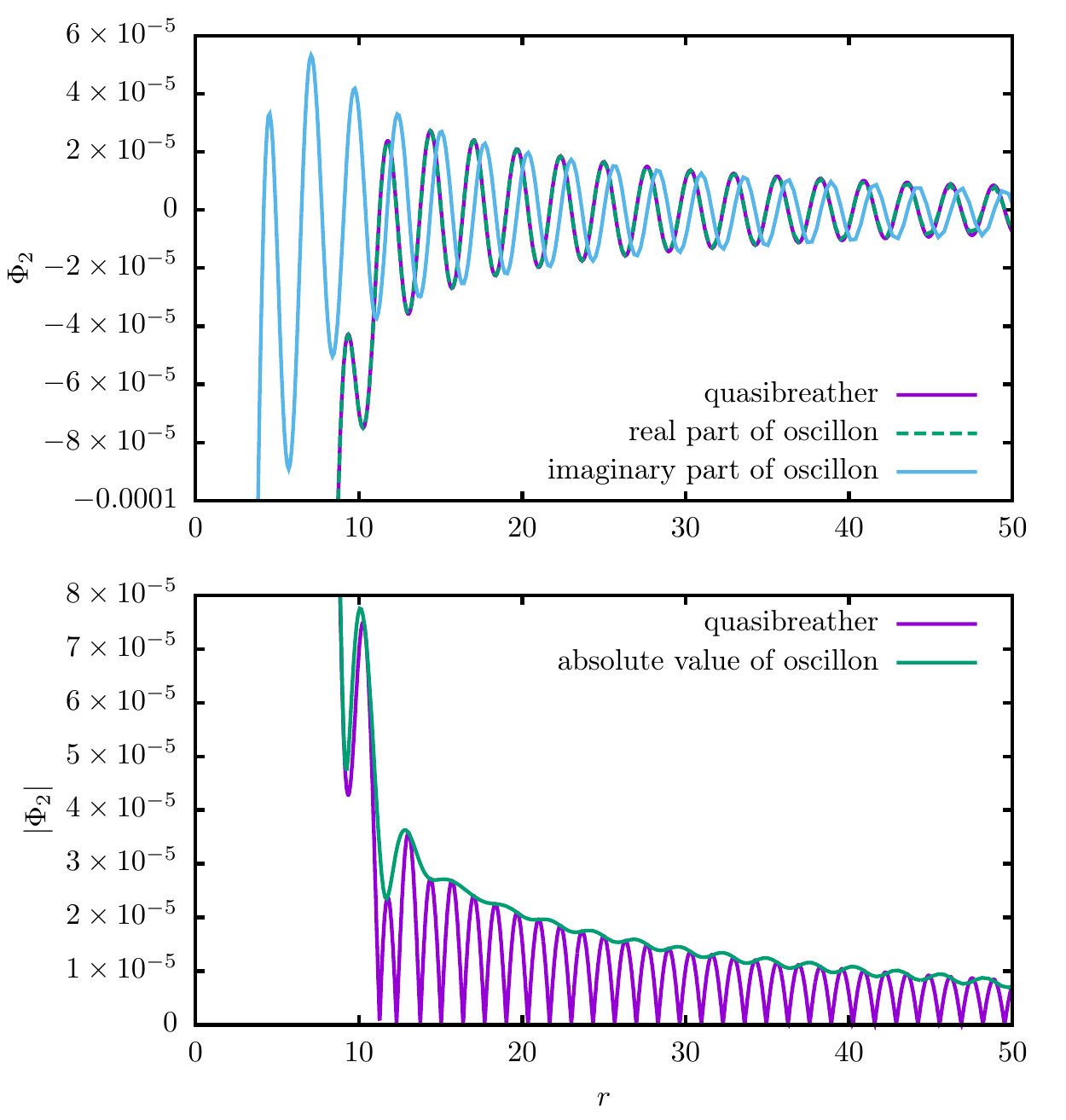}
\caption{\label{figtailc6}
Comparison of the $\Phi_2$ mode of the $\omega=1.38$ frequency quasibreather to the complex $\Phi_2$ function obtained by the Fourier expansion of an identical frequency near-periodic state. On the upper panel we show the real and imaginary parts, while on the lower panel the absolute value.}
\end{figure}

For lower frequencies the core radius becomes smaller, and the agreement between the quasibreather and the oscillon can be checked more easily. As a next example, we investigate the evolution of the $C=0.6$ first peak initial data shown on the lower panel of Fig.~\ref{figfr1d}, during the oscillation period beginning with the central maximum at $t_1=1400.85$. In this case the frequency calculated from the position of the next maximum is $\omega=1.3800002$, and hence we compare it to the $\omega=1.38$ frequency quasibreather. On Figure \ref{figtailc6}
we show the behavior of the $\Phi_2$ Fourier mode in the exterior domain. On the upper part of the figure it can be seen that the real part of the Fourier component of the oscillon agrees to high precision with the corresponding quasibreather's mode, the difference is of the order $10^{-7}$. However, just at the radius $r$ where the oscillating tail appears, the imaginary part of the oscillon's $\Phi_2$ Fourier component becomes the same order as its real part, and hence the time-reflection symmetry fails. The tail of the oscillon can be considered here as an outgoing wave carrying out energy, but its amplitude agrees to good precision with the amplitude of the standing wave tail of the corresponding quasibreather. This can be seen well on the lower part of Fig.~\ref{figtailc6}, where the amplitude of the complex $\Phi_2$ behaves as an envelope curve of the real $\Phi_2$ of the quasibreather. At this frequency the radius of the quasibreather's core is $r=28.8$, and hence the agreement of the amplitudes can also be observed in the tail-domain of the scalar function $\phi$. This agreement is very important, since it allows us to determine the energy loss rate of oscillons by calculating the tail-amplitude of the corresponding frequency quasibreather.

\begin{figure}[!htb]
\centering
\includegraphics[width=115mm]{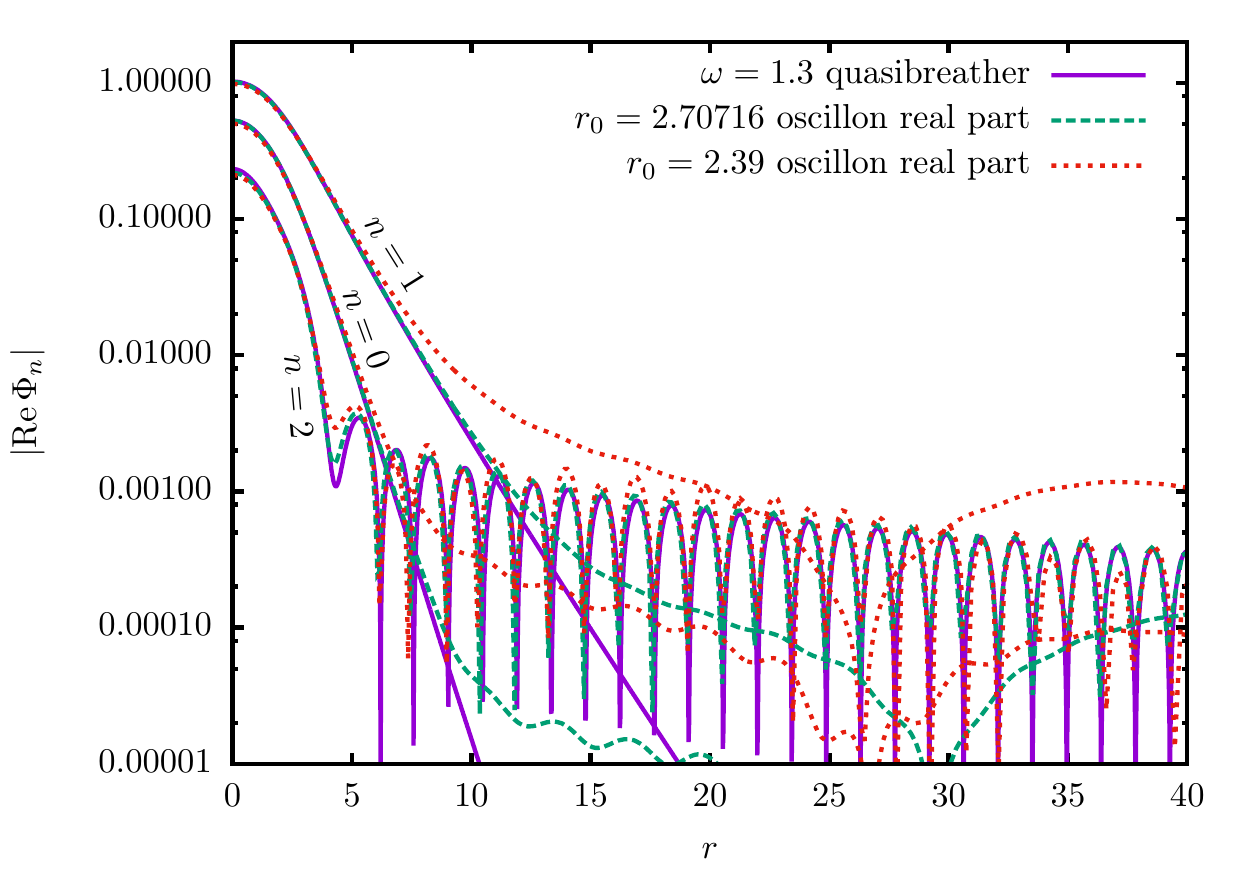}
\caption{\label{figntmode}
Comparison of the Fourier modes of two different oscillon states evolving from a $C=2$ Gaussian initial data with the frequency $\omega=1.3$ quasibreather's modes.}
\end{figure}

We can also carry out the comparison using general, not fine-tuned oscillons. During their evolution both Gaussian initial data shown on Fig.~\ref{fignontune} get into a stage where the frequency is approximately $\omega=1.3$. On Figure \ref{figntmode}
we compare the real part of the first three $\Phi_n$ modes to the identical frequency quasibreather's components. The first oscillon state belongs to the initial data $C=2$, $r_0=2.70716$, and we have carried out the Fourier expansion between the $t_1=471.47$ and consequent maximums, where the frequency turned out to be $\omega=1.2995$. The initial parameters of the other oscillon were $C=2$, $r_0=2.39$, the time of the first maximum was $t_1=298.72$, and the calculated frequency was $\omega=1.3011$. The similarity to the components of the quasibreather in the core region and at the first oscillating mode is clearly apparent, however the agreement is not as good in this case. At the oscillating part of $\Phi_2$ the shape is clearly similar, but the discrepancy can be as large as $30$ percent. At these low frequencies the frequency of the oscillon changes relatively quickly because of the strong radiation. Another factor which makes the agreement less good is the presence of low frequency shape-modes, which can also be seen on Fig.~\ref{fignontune}. According to our experience, the larger is the amplitude of these shape-modes, the worse is the agreement with the modes of the quasibreather. The tail-amplitude of the oscillating mode of the oscillon is generally larger, so the comparison with the identical frequency quasibreather generally underestimates the radiation rate. Studying the oscillon at a later time, when the frequency is already larger and the shape-mode
becomes stronger, the actual radiation amplitude can be twice as large as the one calculated by the quasibreather analogy. According to our understanding, the shape-modes become especially large because of the significant difference between the shape of the Gaussian initial data and the ideal oscillon's shape. We can obtain oscillons with minimally excited shape-modes by using quasibreathers calculated by the spectral method as initial data in the time-evolution code.

\subsection{Time-evolution of the quasibreather initial data}

The connection between oscillons and quasibreathers can be most naturally investigated by using the quasibreather obtained by the high precision spectral code as initial data for the numerical time-evolution code, instead of the earlier used Gaussian initial data. We use the value of the scalar field $\phi$ at the moment $t=0$, when the quasibreather is time-reflection symmetric, and the time derivative of $\phi$ is zero. The grid points in our numerical evolution code are placed uniformly with respect to the radial coordinate $R$ defined in \eqref{eqlowupr}. It may be objected as a principal problem against our procedure, that because of its slowly decaying tail, the quasibreather is an infinite energy state, while the time-evolution code can only describe finite energy systems. This still does not present a serious obstacle, since the physical distances between the grid points, which are distributed uniformly in $R$, are increasing when moving away from the central domain, until the grid becomes unable to represent the given wavelength standing wave tail anymore. In this way, in practice we investigate the time-evolution of a finite energy state obtained by cutting off the tail at a finite distance, where the place of the cutoff in the initial data moves further and further away with increasing numerical resolution. According to our tests, the cutoff surface behaves essentially as an outer boundary with non-reflecting boundary condition implemented.

\begin{figure}[!htb]
\centering
\includegraphics[width=115mm]{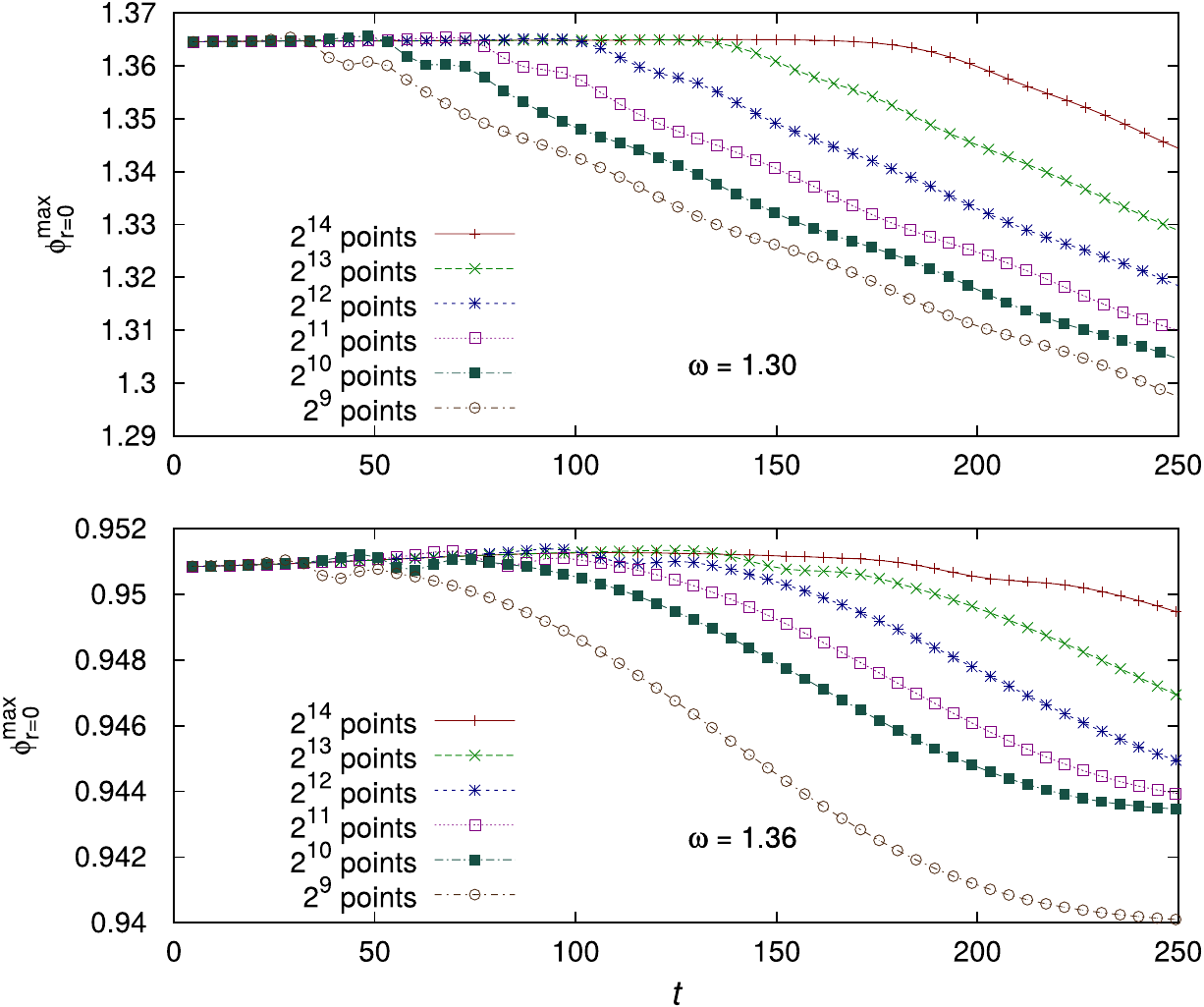}
\caption{\label{figevol3f}
Upper envelopes of the central oscillations in case of quasibreather initial data with frequencies $\omega=1.3$ and $\omega=1.36$, using different number of numerical grid points in the time-evolution code. The symbols on the curves represent the actual maximums of the scalar field $\phi$ in each oscillation period.}
\end{figure}

On Figure \ref{figevol3f}
we can see the time-evolution of two given quasibreathers, with frequencies $\omega=1.3$ and $\omega=1.36$, using different numerical resolutions in the numerical evolution code. It can be seen that the larger the resolution is, the longer the initial almost constant amplitude phase becomes. The reason is that by the increase of the resolution we take into account larger and larger portion of the standing wave tail in the initial data. If we cut the tail at a certain radius, then the disturbance which propagates inwards from there, practically with the speed of light, prevents the solution remaining periodical. Outside of a certain radius the numerical code cannot properly represent the incoming wave still present in the initial data, which would compensate the energy loss of the oscillon. We emphasize that the strong resolution dependence on the figure only follows from the not appropriate representation of the initial data. If we study the evolution of a fixed finite energy localized initial data, the code gives convergent results for many orders longer time periods than the one represented on Fig.~\ref{figevol3f}.

\begin{figure}[!htb]
\centering
\includegraphics[width=115mm]{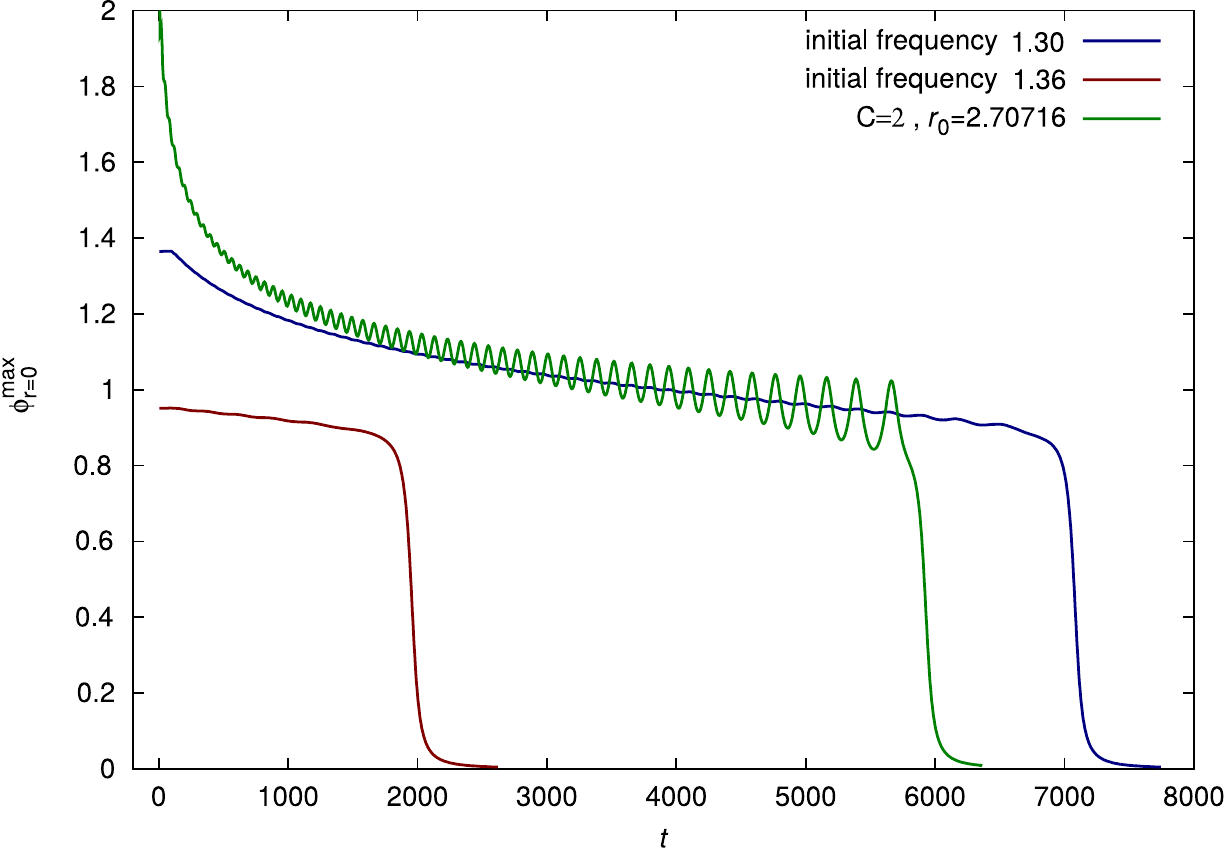}
\caption{\label{figmaxev3f}
The upper envelope curves of the oscillon states forming from the $\omega=1.3$ and $\omega=1.36$ quasibreather initial data, and also from the $C=2$, $r_0=2.70716$ Gaussian initial data.}
\end{figure}

After a certain time the incoming radiation is not present anymore even in that region of the oscillating tail which is otherwise described properly by the evolution code. Since the necessarily present outgoing radiation is not compensated, the oscillation amplitude begins to decrease, very similarly to the case of Gaussian initial data studied earlier. Above a certain resolution the course of this amplitude decrease becomes resolution independent. The curves belonging to different resolutions can be shifted into each other by a shift of time, they only differ in the length of the short initial constant amplitude period. On Figure \ref{figmaxev3f}
we present the long-time evolution of the two quasibreather initial data displayed on the previous figure, for a fixed numerical resolution. In order to make easier the comparison with the evolution of a typical Gaussian initial data, we again present the results for the oscillon already presented on Figures \ref{figgaussevol}-\ref{fignontune}. The evolution of the oscillon obtained from a Gaussian initial data is similar to the evolution of the quasibreather initial data, but the shape-mode, which appears on the amplitude curve as a low frequency oscillation, has significantly larger amplitude in the Gaussian case. The presence of the larger shape-modes appears to be in direct connection with the observation, that the oscillons evolving from a Gaussian initial data radiate more strongly, and their lifetime is shorter.

\begin{figure}[!htb]
\centering
\includegraphics[width=115mm]{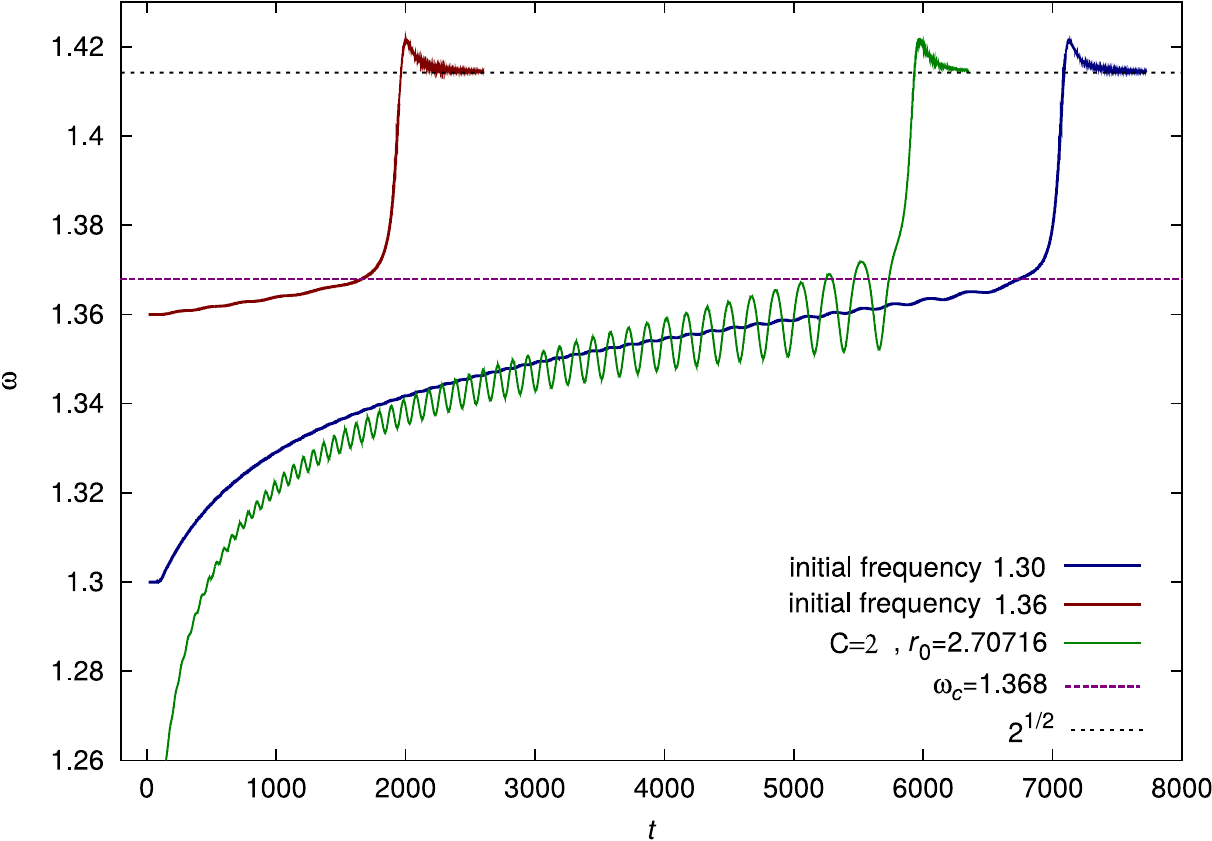}
\caption{\label{figfrev3f}
Time-evolution of the frequency of the three oscillon states presented on the previous figure.}
\end{figure}

On Figure \ref{figfrev3f}
we present the time dependence of the frequency of the three oscillon states shown on the previous figure. During the evolution of oscillons the slow decrease of the amplitude happens simultaneously with the slow increase of the frequency. This slow change continues until the frequency approaches from below the critical frequency $\omega_c=1.368$, which corresponds to the minimum of the energy, according to Fig.~\ref{figenertot}. Oscillons with frequency larger than $\omega_c$ turn out to be unstable. We have constructed near-periodic oscillon states in the domain $\omega_c<\omega<m$ by fine-tuning a parameter in the initial data in Subsection \ref{sec-majdnemper}.

\begin{figure}[!htb]
\centering
\includegraphics[width=115mm]{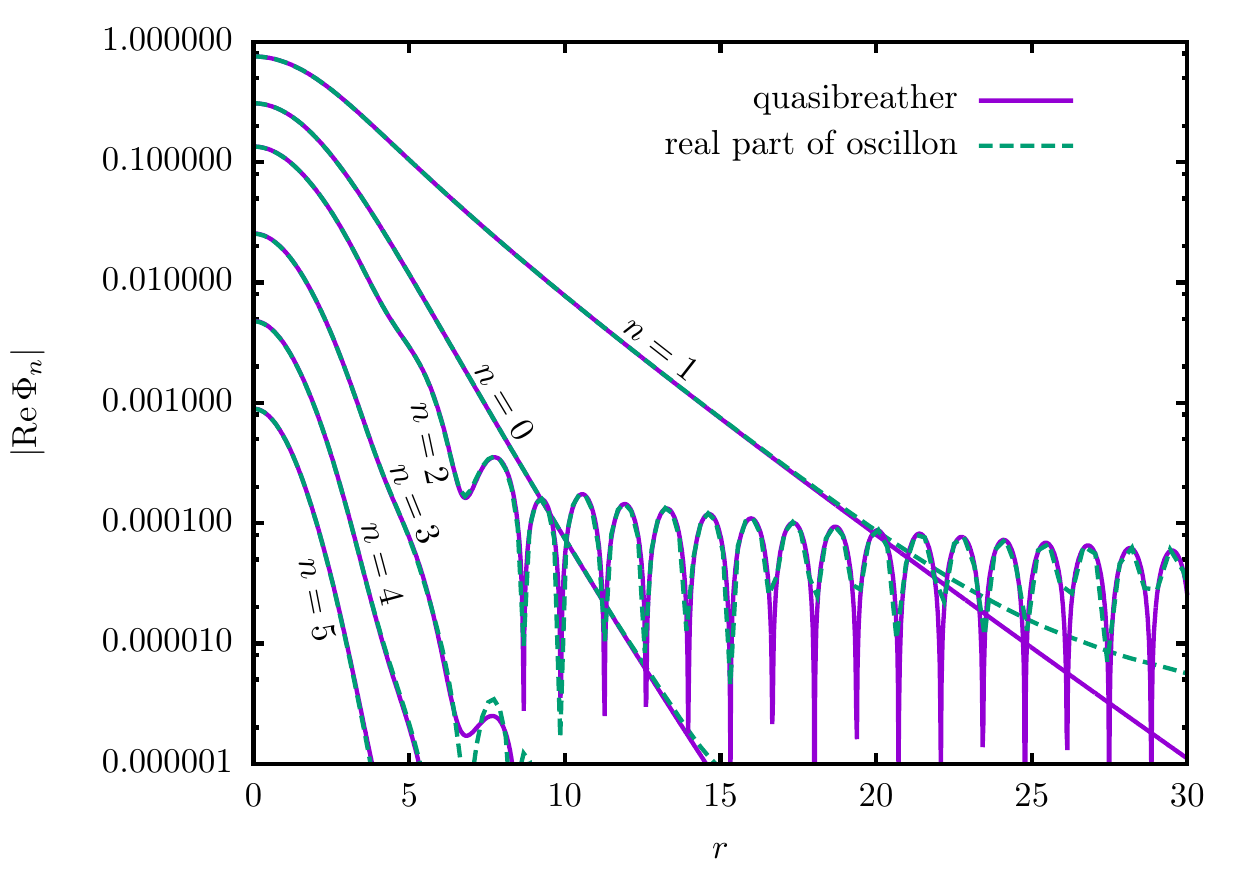}
\caption{\label{figstablef}
Comparison of the Fourier modes of the oscillon evolving from the frequency $\omega=1.3$ quasibreather with the modes of the frequency $\omega=1.36$ quasibreather during a late time period when the frequency of the oscillon has already increased to this value.}
\end{figure}

In order to show how accurately oscillons can be approximated by the appropriate frequency quasibreather, we performed the Fourier expansion of the oscillon evolving from the frequency $\omega=1.3$ quasibreather, at the oscillation period beginning with the maximum at $t_1=5257.45$, when the frequency has already increased to $\omega=1.35997$. On Figure \ref{figstablef}
we demonstrate the particularly good agreement with the modes of the $\omega=1.36$ frequency quasibreather. The fact that the agreement is less good for the oscillons evolving from Gaussian initial data on Fig.~\ref{figntmode}, demonstrates well that the larger discrepancy is likely due to the low frequency shape-modes. The reason for the appearance of the large amplitude shape-modes is the large discrepancy between the Gaussian initial data and the ideal oscillon shape.

\begin{figure}[!htb]
\centering
\includegraphics[width=115mm]{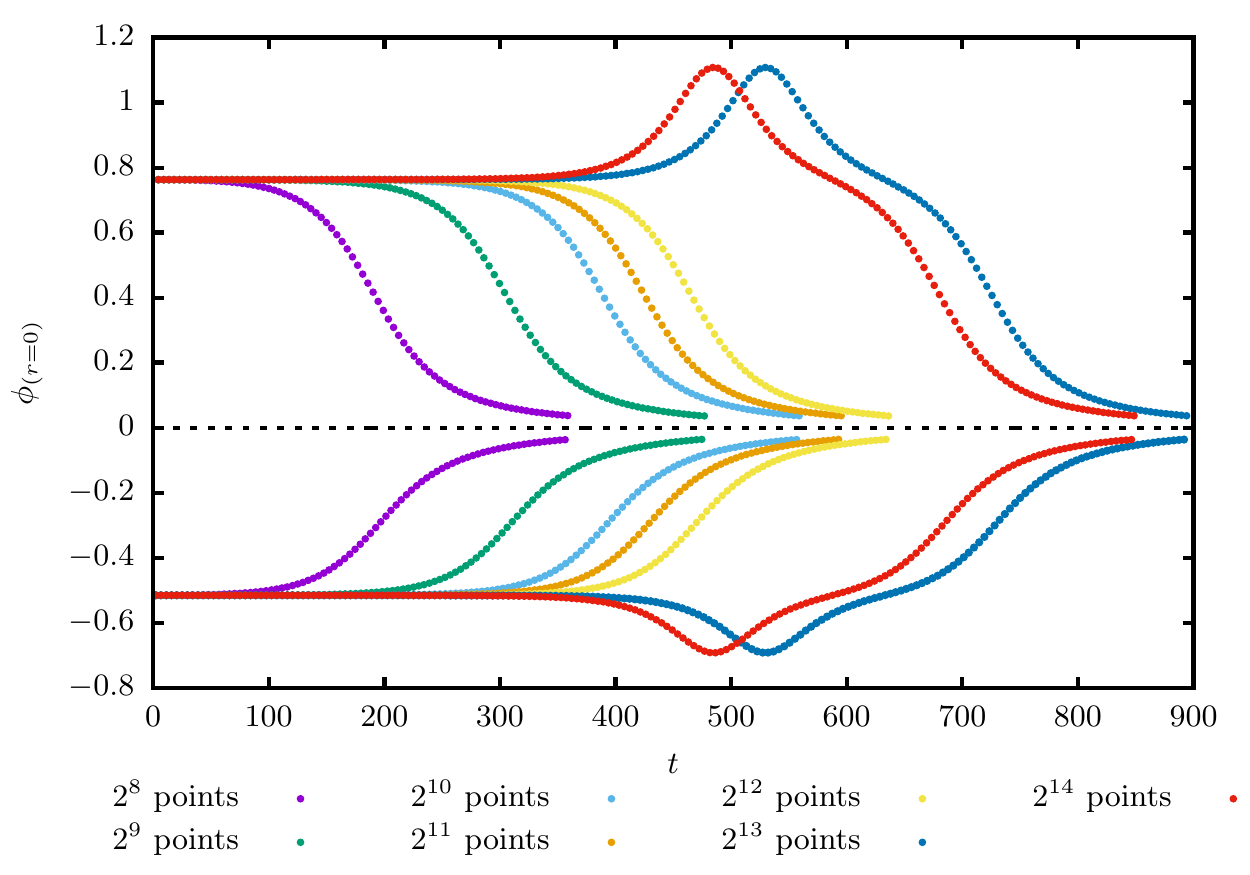}
\caption{\label{figmax38f}
Upper and lower envelopes of the central oscillations of the oscillon state evolving from the quasibreather initial data with $\omega=1.38$ in the unstable frequency domain, for various number of grid points in the evolution code.}
\end{figure}

It is an interesting question that what happens if we use a quasibreather with frequency larger than the critical frequency $\omega_c=1.368$ as initial data in our time-evolution code. On Figure \ref{figmax38f}
we show the evolution of the $\omega=1.38$ quasibreather initial data for various resolutions in the evolution code. It can be seen that the larger the resolution is, generally the longer is the initial near-periodic state. Similarly to the stable case shown on Figures \ref{figevol3f}-\ref{figmaxev3f}, this is because at larger resolutions the grid in the evolution code can represent appropriately up to larger distances the standing wave tail of the initial data. Strangely, at the largest resolution the lifetimes becomes shorter again. This is likely because of a slight numerical error already present in the initial data. After the initial near-periodic state the oscillon decays relatively quickly, similarly to the fine-tuned states shown on Fig.~\ref{figcsx4d}, which were studied in more detail in Subsection \ref{sec-majdnemper}. The decay can be supercritical or subcritical, depending on whether a low frequency oscillation corresponding to a shape-mode occurs or not before the decay. The initial data calculated by the spectral method is so close the ideal, that it depends even on the numerical resolution that the decay happens by a subcritical or supercritical way.

Although the lifetime of a near-periodic oscillon state evolving from a quasibreather with frequency $\omega>\omega_c$ is very long compared to the oscillation period of the oscillon, it is still shorter than the typical lifetime $\tau\approx 2000$ of the fine-tuned states presented in Sec.~\ref{sec-majdnemper}. Actually, it is not surprising that the lifetime of an unstable state calculated from a numerically obtained initial data cannot be as long as the lifetime of an oscillon tuned to $32$ digits. Applying an artificial rescaling of the quasibreather initial data by multiplying it with a number very close to $1$ and fine-tuning this factor, we can again obtain longer living near-periodic states, with frequency slightly different from the original.

Our results presented so far show that the time-evolution of oscillons can be considered as an adiabatic developement through series of quasibreather states with frequencies $\omega(t)$. The core domain of the corresponding quasibreather agrees to high precision with the core of the oscillon. Moreover, the amplitude of the radiative tail responsible for the energy loss of oscillons can also be well approximated by the amplitude of the quasibreather's standing wave tail. The agreement in the tail-amplitude is better in those cases when the shape-mode appearing on the envelope curve is small, i.e.~the oscillon is evolved from a less ``noisy'' initial data. The shape-modes, which correspond to low frequency oscillating deformations of the oscillon, may increase significantly the radiation rate of oscillons and decrease their lifetime.

\section{Small-amplitude expansion} \label{seckisampl}

\subsection{Description of the method}

In this section we describe in detail the expansion procedure in terms of a parameter $\varepsilon$ that represent the amplitude of the oscillons. The expansion gives a very good description of the core region of the oscillons and quasibreathers, even for relatively large amplitudes. Originally, the expansion with respect to the amplitude was applied together with a Fourier expansion with respect to the time coordinate $t$, and the analysis was for one-dimensional space \cite{Kosevich75,dashen-75,Eleonskii1984,SegurKruskal87,Boyd1990,Boyd1995}. The method can be easily generalized for higher dimensions as well. The generalization is valid for arbitrary shaped solutions, but in practice it was only applied for spherically symmetric states \cite{Buslaev1977,bogolyubskii-77c,Makhankov-78,Watkins96}. Kichenassamy \cite{Kichenassamy91} have realized that it is not necessary to apply Fourier expansion together with the small-amplitude expansion. Time-periodicity follows from the weaker assumption that the solution must remain bounded during the whole time evolution. In our paper on the expansion of oscillons we have generalized and worked out in detail this method for $d+1$ dimensional spacetimes \cite{Fodor2008}.

We expand the scalar field $\phi$ with respect to the powers of the parameter $\varepsilon$ in the form,
\begin{equation}
\phi=\sum_{k=1}^\infty\varepsilon^k \phi_k \ , \label{e:sumphi}
\end{equation}
introducing the $\phi_k$ functions that are independent of the parameter. We solve the field equation \eqref{fieldeq2} in case of spherical symmetry, when the Laplacian can be written in the form \eqref{eqlapsph}. In case of one spatial dimension we assume that the scalar $\phi$ is mirror symmetric at $r=0$, which also implies that we choose a system moving together with the center of mass. Although the expansion can be carried out in a similar manner for systems with general symmetry \cite{Fodor2008}, since we will consider spherically symmetric oscillons, for ease of understanding we present the method only for the case of spherical symmetry.

We specify the self-interaction potential $U(\phi)$ by the coefficients $g_k$ in \eqref{potexpeq}. In order to make the expressions simpler we rescale the coordinates $t$ and $r$ so that the mass of the scalar field should be $m=1$. Instead of \eqref{eqpotnum}, the form of the $\phi^4$ potential will have the following form:
\begin{equation}
 U(\phi)=\frac{1}{8}\phi^2\left(\phi-2\right)^2
 \quad , \qquad U'(\phi)=\frac{1}{2}\phi\left(\phi-1\right)\left(\phi-2\right) \ , \label{eqpotone}
\end{equation}
and the nonvanishing expansion coefficients are $g_2=-3/2$ are $g_3=1/2$. The results presented in this section are valid for arbitrary analytic potential. We will only choose a specific potential when comparing with numerical computations.

We can only obtain long-lived localized solutions of the field equation \eqref{fieldeq2} if we choose the characteristic size of the solutions $\varepsilon$ dependent. According to numerical experience, the size of oscillons grows with decreasing $\varepsilon$ amplitude. Because of this, it is advantageous to introduce a new rescaled radial coordinate $\rho$ by the expression
\begin{equation}
\rho=\varepsilon r \ .
\end{equation}
This choice may be also motivated by the exponential decay of the dominant Fourier mode $\Phi_1$ presented in \eqref{eqasymexp}, where the coefficient $\hat\lambda_1$ tends to zero for small $\varepsilon$. Because of the use of the new coordinate the spatial variation of the obtained solutions will be much slower than their change with respect to time. This is always a valid assumption in the core region of the quasibreathers, but in the tail region the derivatives with respect to $t$ and $r$ have the same order, hence the formalism is not able to describe the standing wave tail region. This is not a serious problem in our case, since as it turns out from both numerical and analytical considerations, the amplitude of the tail is $\exp(-1/\varepsilon)$ order small, and hence it can be considered zero for to the power series approximation used in this section.

The frequency of oscillons increases slowly when their amplitude decreases because of the energy loss. If the amplitude $\varepsilon$ of oscillons or quasibreathers approaches zero, their frequency tends to the $\omega_0=m=1$ value. The change in the time-scale can be taken into account most easily if we introduce a rescaled time coordinate
\begin{equation}
\tau=\omega t \ ,
\end{equation}
where $\omega$ is a function of $\varepsilon$. Since during our procedure we keep the frequency of the state at the constant $1$ value with respect to the time coordinate $\tau$, the function $\omega$ gives the physical frequency of the oscillon. We assume that the $\varepsilon$ dependence is analytical, and expand the square of the frequency in the following form:
\begin{equation}
\omega^2=1+\sum_{k=1}^\infty\varepsilon^k\omega_k \ . \label{omegaexp}
\end{equation}
A similar general expansion of the frequency has been published first in the 12-th chapter of the book of Boyd \cite{Boyd-book1998}, in case of one spatial dimension. In earlier articles the analysis was started with imposing the expression $\omega^2=1-\varepsilon^2$, which as we will see soon, can always be achieved by appropriate re-parametrization.

Using the Laplacian
\begin{equation}
 \tilde\Delta\phi=\frac{\partial^2\phi}{\partial \rho^2}
 +\frac{d-1}{\rho}\frac{\partial\phi}{\partial\rho}
 =\frac{1}{\rho^{d-1}}\frac{\partial}{\partial\rho}
 \left(\rho^{d-1}\frac{\partial\phi}{\partial\rho}\right)
 \ , \label{eqlapsphrho}
\end{equation}
with respect to the rescaled coordinate, the field equation \eqref{fieldeq2} can be written into the form:
\begin{equation}
 -\omega^2\frac{\partial^2\phi}{\partial\tau^2} + \varepsilon^2\tilde\Delta \phi=
 \phi +\sum\limits_{k=2}^{\infty}g_k\phi^k \ . \label{e:evoleps}
\end{equation}
Substituting the expansions of $\phi$ and $\omega^2$, the identical $\varepsilon$ power terms determine equations that has to be satisfied separately. The general form of these equations is
\begin{equation}
\frac{\partial^2\phi_k}{\partial\tau^2}+\phi_k=f_k \ ,\label{eqphik}
\end{equation}
where the structure of the left-hand side homogeneous part is always the same, and the inhomogeneous source terms $f_k$ are nonlinear functions of the lower order $\phi_l$ components, for which $l<k$. The expressions $f_k$ depend on the constants $g_l$ for $l\leq k$, and also on the yet unspecified constants $\omega_k$ for $l<k$. In the first equation obviously $f_1=0$, and the next two source terms are:
\begin{align}
f_2&=-g_2\phi_1^2-\omega_1\frac{\partial^2\phi_1}{\partial\tau^2} \ ,\label{e:phi2}\\
f_3&=\ddot\phi_1+\tilde\Delta\phi_1-2g_2\phi_1\phi_2-g_3\phi_1^3
-\omega_1\frac{\partial^2\phi_2}{\partial\tau^2}
-\omega_2\frac{\partial^2\phi_1}{\partial\tau^2} \ .\label{e:phi3}
\end{align}
The homogeneous part is always the expression describing the harmonic oscillator, so at the first glance it may appear that equations \eqref{eqphik} only determine the time dependence of the $\phi_k$ functions. However, as we will see, the requirement that the solutions should not increase without bound with the evolution of time will also fix the spatial dependence of the functions $\phi_k$.

The general solution for $k=1$ of equation \eqref{eqphik} is of the form
\begin{equation}
\phi_1=p_1\cos(\tau+\alpha) \ ,
\end{equation}
where the amplitude $p_1$ and the phase $\alpha$ are arbitrary functions of the radial coordinate $\rho$. All the other equations for $\phi_k$ from \eqref{eqphik} are forced oscillator equations with base frequency $\omega=1$. The inhomogeneous source term $f_k$ on the right-hand side always constitute of sums of terms with time dependence $\cos(n(\tau+\alpha))$ or $\sin(n(\tau+\alpha))$, where $n\geq 0$ integer. When $n\not=1$ the generated solution is always time-periodic, and hence bounded. However, if there are source terms proportional to $\cos(\tau+\alpha)$ or $\sin(\tau+\alpha)$, then they generate solutions proportional to $\tau\sin(\tau+\alpha)$ or $\tau\cos(\tau+\alpha)$, which are unbounded because of the linearly increasing amplitudes. Since we are looking for solutions that remain regular and bounded for a long time, we have to require that for all equations, in the $f_k$ expressions the coefficients of both the $\cos(\tau+\alpha)$ and $\sin(\tau+\alpha)$ terms are zero. In this way, from equation \eqref{e:phi2} determining the form of $f_2$ we can see that necessarily $\omega_1=0$ must hold. This shows that although the potential $U(\phi)$ is generally not symmetric around its minimum, to leading order the frequency is independent of the signature of $\varepsilon$, i.e.~independent of the initial direction of the oscillation.

If $\omega_1=0$ is already satisfied, the solution for $k=2$ of equation \eqref{eqphik} is bounded and periodic in time:
\begin{equation}
\phi_2=p_2\cos(\tau+\alpha)+q_2\sin(\tau+\alpha)
+\frac{g_2}{6}p_1^2[\cos(2(\tau+\alpha))-3] \ ,
\end{equation}
where $p_2$ and $q_2$ are further arbitrary functions of $\rho$. Substituting the obtained solutions for $\phi_1$ and $\phi_2$ into the expression \eqref{e:phi3} of $f_3$, we obtain the following expression,
\begin{align}
&f_3=-\frac{1}{\rho^{d-1}p_1}\frac{\mathrm{d}}{\mathrm{d}\rho}
\left(\rho^{d-1}p_1^2\frac{\mathrm{d}\alpha}{\mathrm{d}\rho}\right)\sin(\tau+\alpha)
+\left[\tilde\Delta p_1+\omega_2p_1+\lambda p_1^3-p_1\left(\frac{\mathrm{d}\alpha}{\mathrm{d}\rho}\right)^2\right]
\cos(\tau+\alpha)\nonumber\\
&-\frac{1}{12}p_1^3(2g_2^2+3g_3)\cos(3(\tau+\alpha))
-g_2p_1\left[q_2\sin(2(\tau+\alpha))+p_2\cos(2(\tau+\alpha))+p_2\right] \ ,\label{e:phi3-1}
\end{align}
where we have introduced the notation
\begin{equation}
\lambda=\frac{5}{6}g_2^2-\frac{3}{4}g_3 \ . \label{lambdadef}
\end{equation}
For the $\phi^4$ potential written in the form \eqref{eqpotone} we have $\lambda=3/2$. For the sine-Gordon potential $U(\phi)=1-\cos\phi$ the constant is $\lambda=1/8$. As we will see soon, small amplitude oscillons can only exist for those potentials for which $\lambda>0$.

For $\phi_3$ we can only obtain a function that is not increasing without bound, if in the form of $f_3$ written in \eqref{e:phi3-1} the coefficient of $\sin(\tau+\alpha)$ is zero, hence
\begin{equation}
 \frac{\mathrm{d}\alpha}{\mathrm{d}\rho}=\frac{C_\alpha}{\rho^{d-1}p_1^2} \ ,
\end{equation}
where $C_\alpha$ is a constant. Clearly, we are looking for solutions for which $p_1$ and $\alpha$ are smooth and bounded at the origin. Since $p_1$ is finite, for $d\geq 2$ spatial dimensions the derivative of $\alpha$, and hence also the function itself, is singular if $C_\alpha\not=0$. In case of $d=1$, if $C_\alpha\not=0$ then because of the mirror symmetry $\alpha$ has a break at $\rho=0$, so it is not differentiable. In this way, for arbitrary dimension $d$, necessarily $C_\alpha=0$, and hence $\alpha$ must be a constant. By this, we have shown that the localized states which are described by the expansion must be time-reflection symmetric. This result was published first in \cite{Fodor2008}. Shifting the time coordinate we can set the value of the constant $\alpha$ to zero, and hence in the following we use the value $\alpha=0$. In earlier literature this was assumed without explanation and proof.

A further condition on the boundedness of the function $\phi_3$ is that in equation \eqref{e:phi3-1} the coefficient of $\cos(\tau+\alpha)$ has to be zero,
\begin{equation}
 \tilde\Delta p_1+\omega_2p_1+\lambda p_1^3=0 \ . \label{p1eqgen}
\end{equation}
Since we are looking for localized solutions that tend to zero, at large distances the cubic term can be neglected. Similarly to the homogeneous part of equation \eqref{eqphinnonl}, in case of positive $\omega_2$ there are only slowly decaying oscillating solutions according to \eqref{eqtailstand}. These have finite energy only if the coefficients of the sine and cosine terms are zero at the same time, which are however too many conditions together with the regularity at the center. The only such solution is the trivial $p_1=0$. Similarly, according to \eqref{eqasylaz1}-\eqref{eqasylaz2}, in case of $\omega_2=0$ there is no nontrivial localized solution. Consequently, $\omega_2<0$, and in this case one of the two asymptotic solutions decays exponentially according to \eqref{eqasymexp} at large distances. According to the expansion \eqref{omegaexp} of the oscillation frequency $\omega$ this means that small amplitude oscillons can only exist below the $m\equiv 1$ mass threshold  \cite{Fodor2008}. This is physically reasonable, since if the frequency $\omega$ is above the base frequency $1$, then there is an oscillating tail associated to the dominating mode according to \eqref{eqtailstand}, which necessarily implies large energy loss. We can expect the existence of long-lived localized states if only $2\omega$, $3\omega$ and higher harmonics are able to radiate.

It is not specified yet which concrete value of the $\varepsilon$ parameter belongs to a physical state with a given amplitude and frequency. So far we have only assumed that for small values of $\varepsilon$ the parameter is proportional to the amplitude of the oscillons. Any $\varepsilon\to\bar\varepsilon(\varepsilon)$ reparametrization which is not violating this assumption is allowed. If we linearly rescale the parameter according to $\varepsilon\to\bar\varepsilon=a\varepsilon$ by a positive constant $a$, then the change of the radial coordinate is $\bar\rho=a\rho$, and the change of the Fourier components is $\bar\phi_k=\phi_k/a^k$. Then $\bar p_1=p_1/a$, and the transformation of the constants describing the $\varepsilon$  dependence of the frequency is $\bar\omega_k=\omega_k/a^k$. The constants $g_k$ and $\lambda$ determining the expansion of the potential remain unchanged. In this case, all terms of equation \eqref{p1eqgen} are only modified by a factor $a^3$, and since $\omega_2<0$, by using $\bar\omega_2=\omega_2/a^2$ one can always achieve $\bar\omega_2=-1$. From now on we assume that from the beginning we have chosen such parametrization that $\omega_2=-1$ holds.

Multiplying equation \eqref{p1eqgen} by $\rho^{d-1}p_1$ we obtain
\begin{equation}
 p_1\frac{\mathrm{d}}{\mathrm{d}\rho}\left(\rho^{d-1}
 \frac{\mathrm{d}p_1}{\mathrm{d}\rho}\right)
 -\rho^{d-1}p_1^2+\lambda\rho^{d-1}p_1^4=0 \ . \label{p1multeq}
\end{equation}
Integrating this between $\rho=0$ and infinity, then integrating by parts in the first term, it can be seen easily that if $\lambda\leq 0$ then only the trivial solution $p_1=0$ exists. By this we have shown that small-amplitude solutions can only exist for potentials for which the scalar $\lambda=\frac{5}{6}g_2^2-\frac{3}{4}g_3$, which was defined from the $g_k$ expansion coefficients according to \eqref{lambdadef}, is positive. In the following we assume that
\begin{equation}
 \lambda>0 \ .
\end{equation}
The coefficient $g_2$ makes the potential $U(\phi)$ asymmetric, and any value of it enhances nonlinearity and helps the appearance of oscillons. For potentials that are symmetric around their minimum $g_2=0$, and then $g_3<0$ is the necessary condition for the existence of oscillons.
\begin{figure}[!htb]
\centering
\includegraphics[width=70mm]{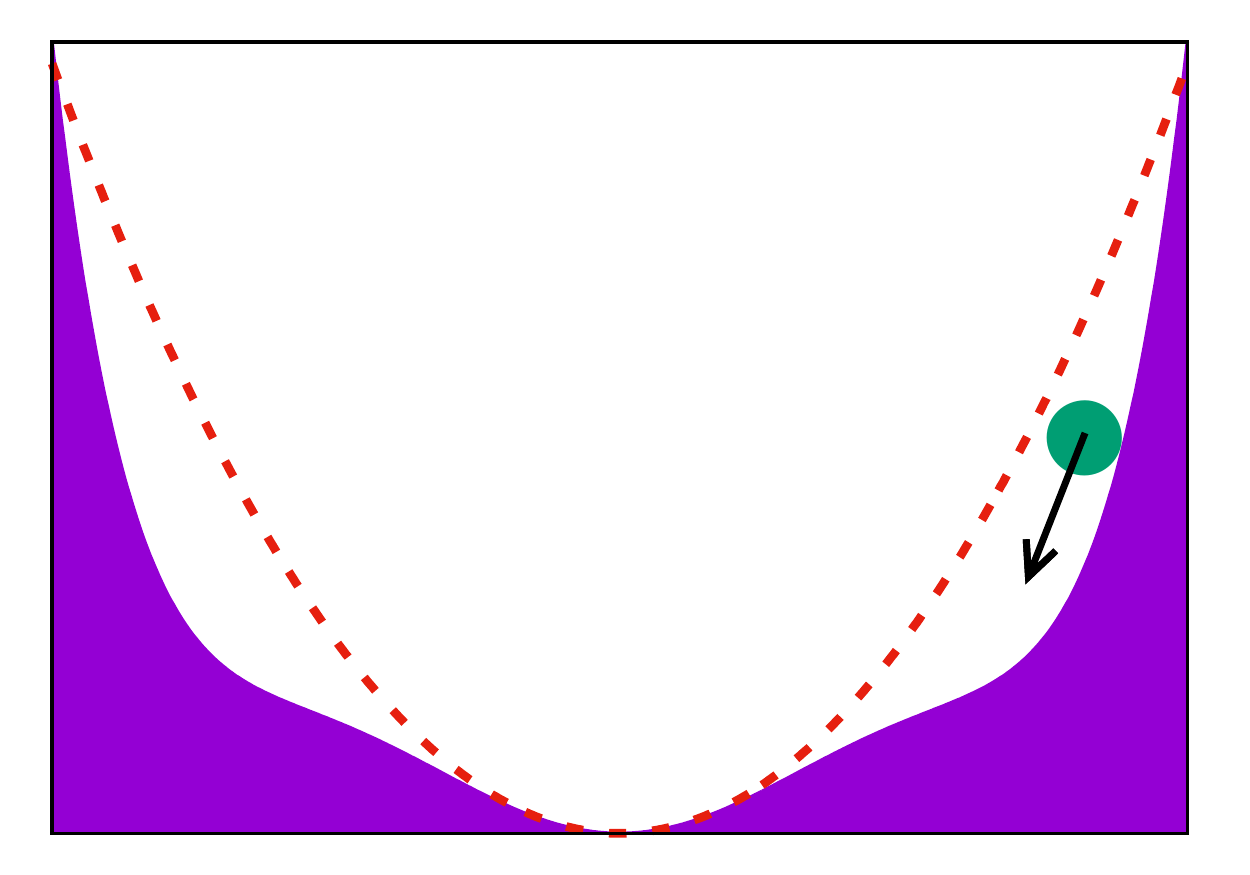}
\end{figure}
In the symmetric case this negative term makes the potential shallower around its minimum than the Klein-Gordon potential, hence increases the oscillation time and decreases the frequency compared to that.

\subsection{The base equation} \label{subsecalapegy}

Introducing the function
\begin{equation}
 S=\sqrt{\lambda}\,p_1 \ ,
\end{equation}
from \eqref{p1eqgen} we obtain an equation which is independent of the parameters,
\begin{equation}
 \tilde\Delta S-S+S^3=0 \ . \label{seqlabel}
\end{equation}
Since we are looking for spherically symmetric solutions, $S$ only depends on the rescaled radial coordinate $\rho$, and the form of the Laplacian is given in \eqref{eqlapsphrho}. It can be seen that the value of the parameter $\lambda$ is in close connection with how nonlinear the system is. The larger $\lambda$ is, the smaller amplitude oscillon is sufficient for a given decrease of the system's frequency. Since \eqref{seqlabel} is appropriate for the description of small amplitude oscillons for arbitrary potentials, following the nomenclature of Buslaev \cite{Buslaev1977}, we refer to it as the \emph{base equation}.

In case of one spatial dimension the solutions of the base equation \eqref{seqlabel} which are symmetric around $\rho=0$ can be written in terms of elliptic functions, and are generally spatially periodic \cite{Boyd-book1998}. The only exception is the localized solution, for which the period becomes infinity. It can be written into the simple form
\begin{equation}
 S=\sqrt{2}\sech\rho \ . \label{seqdegysol}
\end{equation}
Obviously, this multiplied by $-1$ also solves the equation, but it is not necessary to deal with that, since even for potentials not symmetric around their minimum it leads to the same solution, just shifted by a half oscillation period. For one space dimension all functions appearing at the higher orders of the $\varepsilon$ expansion can be expressed as linear combinations of various powers of the right-hand side of \eqref{seqdegysol}, and hence the expansion can be technically much easily performed for $d=1$ than for higher dimensions \cite{SegurKruskal87,Kichenassamy91,Boyd-book1998}. In this review we present a general procedure which is appropriate for arbitrary dimensions.

\begin{figure}[!htb]
\centering
\includegraphics[width=115mm]{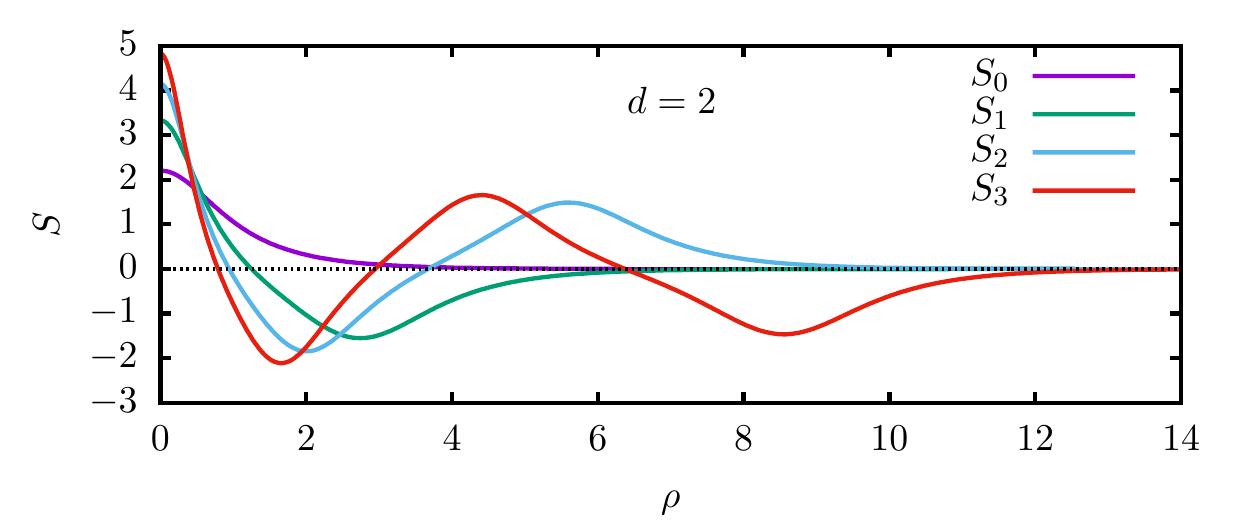}
\includegraphics[width=115mm]{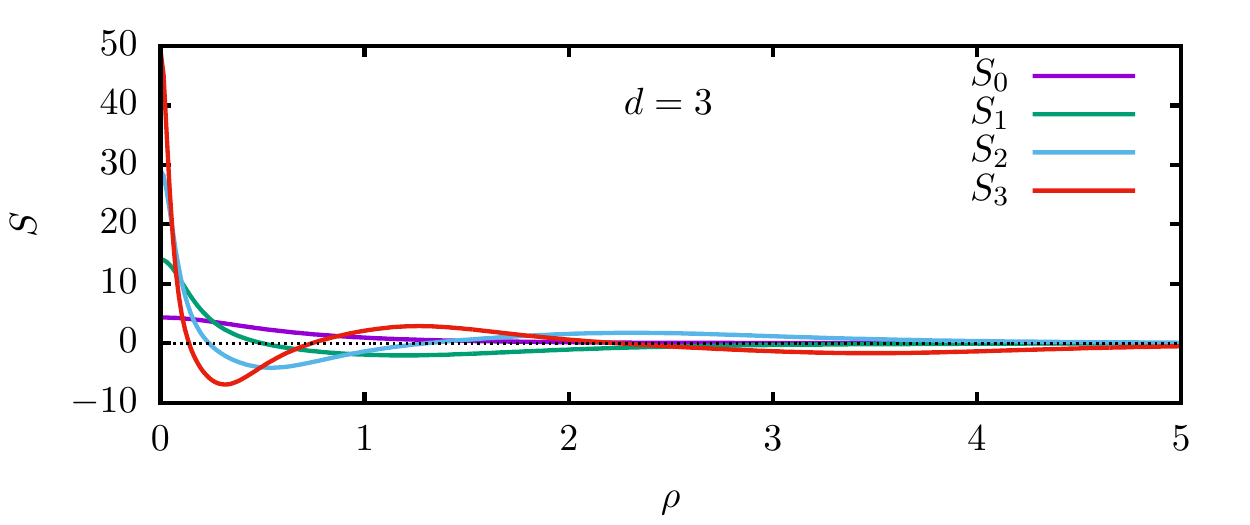}
\caption{\label{figseqsol}
Localized regular solutions of the base equation \eqref{seqlabel} in case of $d=2$ and $d=3$ dimensions.}
\end{figure}

For higher dimensions the solutions are only known in numerical form. The condition for central regularity is that at $\rho=0$ the value of $S$ is finite and its derivative is zero. For $d=2$ and $d=3$ spatial dimensions there are infinitely many localized regular solutions, characterized by an integer $n\geq 0$. The parameter $n$ determines the number of nodes (zero crossings) of the solution denoted by $S\equiv S_n$. On Figure \ref{figseqsol}
we show the first few such functions. By physical intuition it appears reasonable that the solution $S_0$, which is nowhere zero, is the most important. It can be shown that the oscillon solutions corresponding to functions $S$ with nodes have larger energy, are less stable, and have a shorter lifetime \cite{Fodor2008}. In the following we will focus on oscillons belonging to $S_0$, although the expansion procedure is valid for the general case as well. The central value of $S_0$ with nine digits precision is
\begin{equation}
 S_0^{(\rho=0)}=\left
 \{\begin{array}{rl}
  2.20620086 & \text{if } d=2 \ , \\
  4.33738768 & \text{if } d=3 \ . \label{eqscentr}
 \end{array}
\right.
\end{equation}

In this paragraph we show that the base equation \eqref{seqlabel} has no regular localized solutions for $d\geq 4$ spatial dimensions \cite{Fodor2008}. The localized solutions necessarily tend to zero exponentially at infinity. Multiplying the equation by $\rho^{d-1}S$ and integrating, similarly to the procedure at equation \eqref{p1multeq}, we obtain the following virial identity,
\begin{equation}
 \left\langle\left(\frac{\mathrm{d}S}{\mathrm{d}\rho}\right)^2\right\rangle
 +\langle S^2\rangle-\langle S^4\rangle=0 \ , \label{e:viri1}
\end{equation}
where for an arbitrary function $f$ of the coordinate $\rho$
\begin{equation}
 \langle f\rangle:=\int_0^\infty f\rho^{d-1}\mathrm{d}\rho \ .
\end{equation}
One can also obtain a second virial identity, using that the base equation can be derived from the action integral
\begin{equation}
 \mathcal{S}=\int_0^\infty\left[\left(\frac{\mathrm{d}S}{\mathrm{d}\rho}\right)^2
 +S^2-\frac{S^4}{2}\right]\rho^{d-1}\mathrm{d}\rho \ .
\end{equation}
If $S(\rho)$ is a given configuration, then for a constant $\mu$ we can define an elongated configuration $\tilde S(\rho)=S(\rho/\mu)$, and calculate the action integral for it. Introducing the rescaled radial coordinate $\tilde\rho=\rho/\mu$,
\begin{equation}
 \tilde{\mathcal{S}}=\int_0^\infty\left[
 \mu^{d-2}\left(\frac{\mathrm{d}S(\tilde\rho)}{\mathrm{d}\tilde\rho}\right)^2
 +\mu^{d}S(\tilde\rho)^2-\mu^{d}\frac{S(\tilde\rho)^4}{2}\right]
 \tilde\rho^{d-1}\mathrm{d}\tilde\rho \ .
\end{equation}
If $S(\rho)$ is a solution of the base equation, then the derivative of the action with respect to $\mu$ must be zero, from which the second virial identity follows,
\begin{equation}
 (d-2)\left\langle\left(\frac{\mathrm{d}S}{\mathrm{d}\rho}\right)^2\right\rangle
 +d\langle S^2\rangle-\frac{d}{2}\langle S^4\rangle=0 \ . \label{e:viri2}
\end{equation}
From equations \eqref{e:viri1} and \eqref{e:viri2} follows that
\begin{equation}\label{e:crit_dim}
2\langle S^2\rangle+\frac{1}{2}(d-4)\langle S^4\rangle=0 \ ,
\end{equation}
which for a nonzero function $S$ can only hold in case of $d<4$ dimensions. By this we have shown that for $d\geq 4$ spatial dimensions a family of small amplitude oscillons cannot exist for which the amplitude, according to our assumptions, is proportional to a parameter $\varepsilon$, while their size scales according to $1/\varepsilon$. Interestingly, in contrast to this, for $d\geq 4$ it is still possible to find relatively long-lived oscillon states \cite{piette98,Gleiser04,Saffin2007,Andersen2012}. The question of whether there are small-amplitude oscillons among these higher dimensional states, and what are their scaling properties requires further investigation.

\subsection{Higher orders in the expansion}

In order to be able to determine the shape of the oscillon to leading order in $\varepsilon$, we had to proceed by the expansion up to order $\varepsilon^3$, and solve the conditions for the absence of resonances at that order. Similarly, at higher orders, one always have to calculate the expansion two orders higher to ensure that the given order is fully determined. After the elimination of the resonance terms in \eqref{e:phi3-1}, the solution of equation \eqref{eqphik} for $\phi_3$ is
\begin{align}
 \phi_3=&q_3\sin \tau+ p_3\cos \tau \\
 &+\frac{p_1}{3}\left\{\frac{1}{8}
 \left(\frac{4}{3}g_2^2-\lambda\right)p_1^2 \cos(3\tau)
 +g_2 \left[q_2\sin(2\tau)+p_2(\cos(2\tau)-3)\right]\right\} , \notag \label{e:phi3-2}
\end{align}
where $p_3$ and $q_3$ are two new, yet unknown functions of $\rho$. Similarly to this, at higher orders, when solving the equation for $\phi_k$ two new functions appear, $p_k$ and $q_k$, which will be determined by the resonance conditions at two orders higher. For the uniqueness it is also necessary to make a concrete choice for how $\varepsilon$ parametrize the physical states, and set the origin of the time coordinate.

The resonance conditions at fourth order in $\varepsilon$ give two differential equations. Using that $S=\sqrt{\lambda}p_1$, because of the terms proportional to $\sin\tau$ in the source $f_4$\,,
\begin{equation}
 \tilde\Delta q_2-q_2+S^2 q_2=0 \ , \label{e:ph4sin}
\end{equation}
and from the $\cos\tau$ terms,
\begin{equation}
 \tilde\Delta p_2-p_2+\frac{\omega_3}{\sqrt{\lambda}}S+3S^2 p_2=0 \ . \label{e:ph4cos}
\end{equation}

Comparison with the base equation \eqref{seqlabel} shows that equation \eqref{e:ph4sin} for $q_2$ is solved by the function $q_2=a S$, for arbitrary constant $a$. This degree of freedom corresponds to an order $\varepsilon^2$ shift of the time coordinate. By the practical choice $q_2=0$ we ensure that the system remains time-reflection symmetric at $\tau=0$ to higher orders. Similarly all $q_k$ can be chosen zero for $k\geq 2$ as well. That \eqref{e:ph4sin} has no other regular and localized solution, can be shown by the substitution $q_2=f S$, where $f$ is an arbitrary function of $\rho$. Then the equation can be written into the form
\begin{equation}
 \frac{1}{r^{d-1}S}\frac{\mathrm{d}}{\mathrm{d}\rho}
 \left(r^{d-1}S^2\frac{\mathrm{d}f}{\mathrm{d}\rho}\right)=0 \ ,
\end{equation}
from which it follows that the value of $r^{d-1}S^2\frac{\mathrm{d}f}{\mathrm{d}\rho}$ is a constant. If this constant is not zero, then because of the exponential falloff of $S$ the function $q_2$ cannot be bounded, hence it is necessary that $f$ should be a constant, i.e.~$f=a$.

Introducing the function $\bar p_2$ by the substitution
\begin{equation}
 p_2=\bar p_2-\frac{\omega_3}{2\sqrt{\lambda}}\frac{\mathrm{d}}{\mathrm{d}\rho}
 \left(\rho S\right) \label{eqp2traf}
\end{equation}
equation \eqref{e:ph4cos} can be transformed into a homogeneous form
\begin{equation}
 \tilde\Delta \bar p_2-\bar p_2+3S^2\bar p_2=0 \ . \label{eqp2bar}
\end{equation}
This equation corresponds to the linearization of the base equation \eqref{seqlabel} around $S$. Since in the spherically symmetric case the base equation only has discrete solutions, we cannot expect any regular localized solution for $\bar p_2$. That indeed $\bar p_2=0$ is the only such solution is proved in the paper of Buslaev \cite{Buslaev1977}. In the non spherically symmetric case, the spatial translations and rotations of $S$ generate the solutions of \eqref{eqp2bar}.

By changing the $\varepsilon$ parametrization as $\varepsilon\to\bar\varepsilon=\varepsilon(1+a\varepsilon)$, where $a$ is a constant, based on the definition \eqref{omegaexp} it is always possible to set $\omega_3=0$. Similarly to this, at the higher orders of the expansion we can also set $\omega_k=0$ for all $k\geq 3$. By this, we totally fix the freedom in the $\varepsilon$ parametrization, and from now on the relation between $\omega$ and $\varepsilon$ will take the following simple form,
\begin{equation}
 \omega^2=1-\varepsilon^2 \ .
\end{equation}

After setting $\omega_3=0$, according to \eqref{eqp2traf}, the only regular and localized solution that remains is $p_2=0$. Proceeding to higher orders in the expansion, at even orders the same simple equation determines the free functions, and hence it follows that $p_{2k}=0$ for integer $k$. The odd indexed $p_{2k+1}$ functions are determined by further second order differential equations. We note that at the reparametrization $\varepsilon\to\bar\varepsilon=\varepsilon(1+a\varepsilon)$ the change of the function $p_2$ corresponding to \eqref{eqp2traf} comes from the fact that the argument $\rho=\varepsilon r$ of the function $S$ is also $\varepsilon$ dependent.

\begin{figure}[!htb]
\centering
\includegraphics[width=100mm]{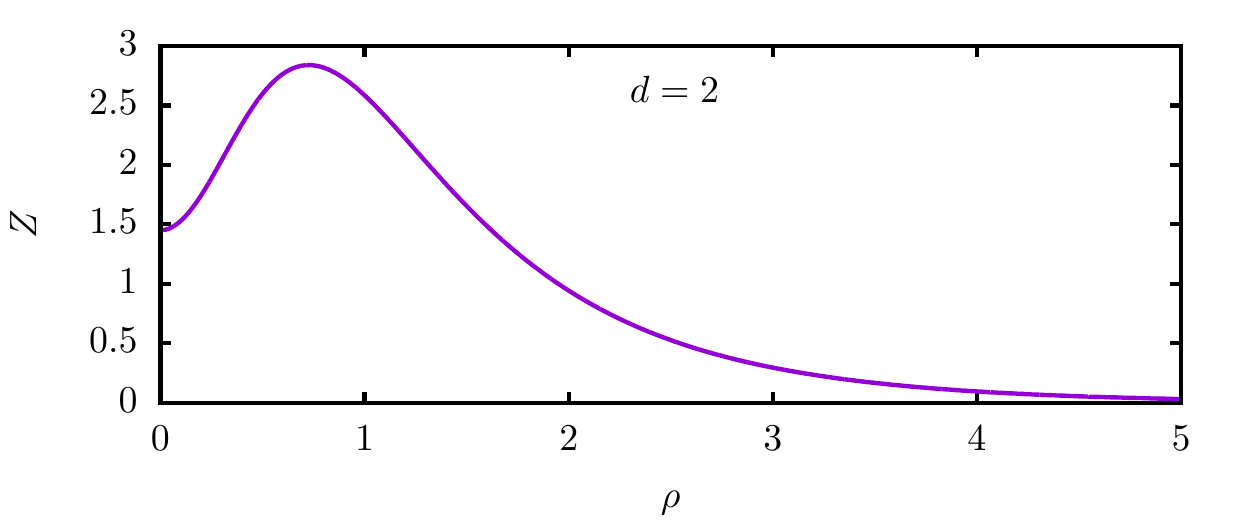}
\includegraphics[width=100mm]{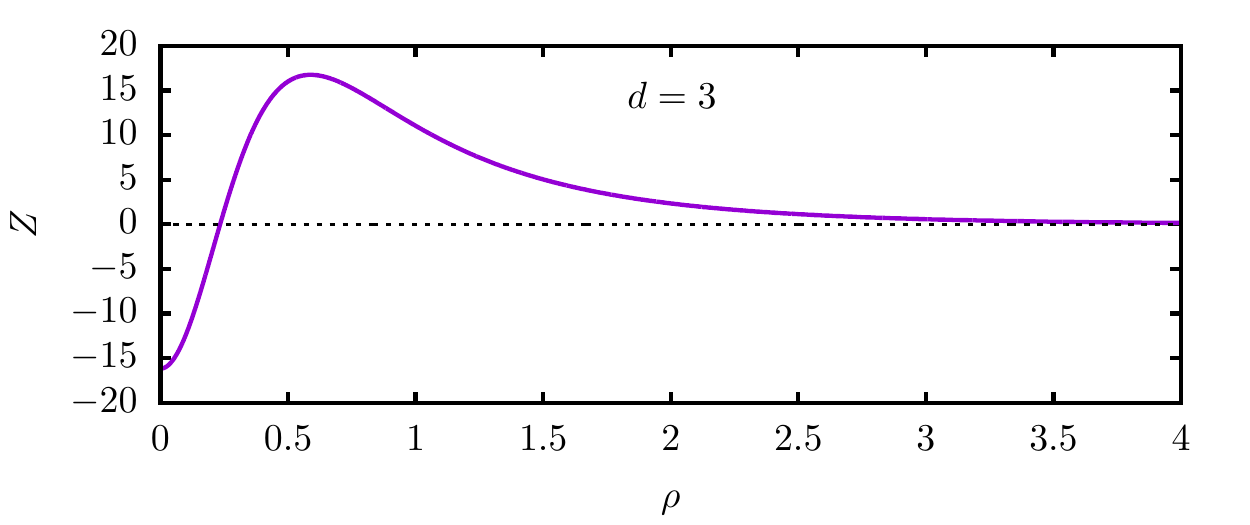}
\caption{\label{figzeqsol}
Solutions of equation \eqref{eqZgen} belonging to the base solution $S_0$ in case of $d=2$ and $d=3$ space dimensions.}
\end{figure}

The $\rho$ dependence of the function $p_3$ is determined at order $\varepsilon^5$ from the cancellation of the terms with time dependence $\cos\tau$. Let us introduce the function $Z$ by the expression
\begin{equation}
p_3=\frac{1}{\lambda^2\sqrt{\lambda}}\left[
\sigma Z-\frac{1}{54}\lambda g_2^2S(32+19S^2)
\right] \ ,  \label{e:p3}
\end{equation}
where
\begin{equation}
 \sigma=\frac{1}{24}\lambda^2-\frac{1}{6}\lambda g_2^2+\frac{5}{8}g_5
-\frac{7}{4}g_2g_4+\frac{35}{27}g_2^4 \ . \label{eqsigmadef}
\end{equation}
Then the condition for $p_3$ can be written into the following much simpler form, 
\begin{equation}
\Delta Z-Z+3S^2Z-S^5=0 \ , \label{eqZgen}
\end{equation}
which is a linear and inhomogeneous equation, independent of the potential $U(\phi)$. The equation is linear in $Z$, and for given $S$ it has a unique regular and localized solution. For $d=1$ spatial dimension this solution is
\begin{equation}
Z=S(4-S^2)/3 \ , \label{eqzdimone}
\end{equation}
where $S$ has the form given in \eqref{seqdegysol}. In case of $d=2$ or $d=3$ spatial dimensions the function $Z$ can be determined numerically. On Figure \ref{figzeqsol}
we show the function $Z$ belonging to the nodeless solution $S_0$ of the base equation. 
The value of the function $Z$ in the center is
\begin{equation}
 Z_0^{(\rho=0)}=\left
 \{\begin{array}{rl}
  1.4507606 & \text{if } d=2 \ , \\
  -16.174027 & \text{if } d=3 \ .
 \end{array}
\right. \label{eqzcent}
\end{equation}
At two orders later, when solving the equation for $\phi_5$, a new $p_5$ function appears, which will become determined at further two orders later.

We summarize the results up to now. We have constructed the $\varepsilon$ expansion of the solution of the field equation \eqref{fieldeq2} up to fourth order. The solution is the sum of terms with harmonic time dependence, and the radial dependence is determined by two universal elliptic partial differential equation, \eqref{seqlabel} and \eqref{eqZgen}. The continuation of the expansion to higher orders is not prevented by any principal obstacle, but the resulting expressions become more complicated, and more coupled differential equations has to be solved numerically. The result of the expansion up to fourth order, for general potential $U(\phi)$, is
\begin{eqnarray}
\phi_1&=&p_1\cos \tau \ , \label{e:phi3b1}\\
\phi_2&=&\frac16 g_2p_1^2\left[\cos(2\tau)-3\right] \ ,\label{e:phi3b2}\\
\phi_3&=&p_3\cos \tau+\frac{1}{72}(4g_2^2-3\lambda)p_1^3\cos(3\tau) \ ,
\label{e:phi3b3}\\
\phi_4&=&
\frac{1}{360}p_1^4\left(3g_4-5g_2\lambda+5g_2^3\right)\cos(4\tau)\nonumber\\
&&-\frac{1}{72}\left[8g_2\left(\frac{\mathrm{d}p_1}{\mathrm{d}\rho}\right)^2
-12g_4p_1^4+16g_2^3p_1^4
-24g_2p_1p_3-23g_2\lambda p_1^4-8g_2p_1^2\right]\cos(2\tau)\nonumber\\
&&-g_2p_1^2-g_2p_1p_3+\frac{1}{6}g_2\lambda p_1^4
-g_2\left(\frac{\mathrm{d}p_1}{\mathrm{d}\rho}\right)^2
+\frac{31}{72}g_2^3p_1^4-\frac{3}{8}g_4p_1^4 \ .\label{e:phi3b4}
\end{eqnarray}
The oscillation frequency of the state is given through the relation $\tau=\omega t$ by the expression $\omega=\sqrt{1-\varepsilon^2}$\,.

For potentials $U(\phi)$ which are symmetric around their minimum, the even coefficients are zero, $g_{2k}=0$. In this case the resulting expressions become considerably simpler, and it is easier to proceed to higher orders with the expansion. Then the even indexed terms are zero in the $\varepsilon$ expansion, namely $\phi_{2k}=0$. Since in the general case $\phi_{2n}$ only contain $\cos(2k\tau)$ terms, and $\phi_{2n+1}$ only consist of $\cos((2k+1)\tau)$ terms, where $k=0\ldots n$, from this also follows that for symmetric potentials even indexed $\Phi_n$ terms do not appear in the Fourier expansion \eqref{eqphifourier}. In this case, according to \eqref{e:p3}, the function $p_3$ is proportional to $Z$, and the equation determining the function $p_5$ becomes relatively simple,
\begin{align}
 \Delta p_5&-p_5+3S^2p_5
 +\frac{SZ}{576\sqrt{\lambda}}(3Z-5S^3)
 \left(\frac{15g_5}{\lambda^2}+1\right)^2 \label{eqp5sym}\\
 &+\frac{S^3}{32\sqrt{\lambda}}
 \left[\left(\frac{\mathrm{d}S}{\mathrm{d}\rho}\right)^2-S^2\right]
 -\frac{S^7}{576\sqrt{\lambda}}\left(\frac{315g_7}{\lambda^3}
 -\frac{60g_5}{\lambda^2}+1\right)=0 \ . \nonumber
\end{align}
Hence for symmetric potentials the fifth order of the expansion is:
\begin{align}
 \phi_5&=p_5\cos\tau
 +\frac{S^5}{1152\sqrt{\lambda}}\left(\frac{3g_5}{\lambda^2}+2\right)
 \cos(5\tau) \label{eqphi5sym} \\
 &-\frac{S}{384\sqrt{\lambda}}\left[
 \left(\frac{30g_5}{\lambda^2}+2\right)SZ+12S^2
 -12\left(\frac{\mathrm{d}S}{\mathrm{d}\rho}\right)^2
 -\left(\frac{15g_5}{\lambda^2}-2\right)S^4
 \right]\cos(3\tau) \ . \nonumber
\end{align}
For $d=1$ spatial dimension the solution of \eqref{eqp5sym} is
\begin{align}
 p_5=&\frac{5S}{648\sqrt{\lambda}}\left(7-6\frac{g_5}{\lambda^2}
 +225\frac{g_5^2}{\lambda^4}+189\frac{g_7}{\lambda^3}\right)
 -\frac{S^3}{2592\sqrt{\lambda}}\left(38+60\frac{g_5}{\lambda^2}
 +1800\frac{g_5^2}{\lambda^4}+945\frac{g_7}{\lambda^3}\right) \notag\\
 &-\frac{S^5}{6912\sqrt{\lambda}}\left(8-120\frac{g_5}{\lambda^2}
 -450\frac{g_5^2}{\lambda^4}+315\frac{g_7}{\lambda^3}\right) \ . \label{eqp5dimone}
\end{align}

\subsection{The energy}\label{ss:energy}

Since the $\varepsilon$ expansion is not able to describe the exponentially small amplitude tail region, as a result of the calculation we necessarily get the energy of the core domain. The form of the energy density according to \eqref{eqendensgen}, using the rescaled coordinates is
\begin{equation}
 \mathcal{E} = \frac{\omega^2}{2} \left(\frac{\partial\phi}{\partial\tau}\right)^2
 + \frac{\varepsilon^2}{2} \left(\frac{\partial\phi}{\partial\rho}\right)^2
 +U(\phi) \ .
\end{equation}
The total energy of the core region, based on \eqref{eqensph} is
\begin{equation}
 E=\frac{1}{\varepsilon^d}\,\frac{2\pi^{d/2}}{\Gamma(d/2)}
 \int_0^{\infty}\rho^{d-1}
 \mathcal{E}\mathrm{d}\rho \ .
\end{equation}
Because of the time-periodicity, it is reasonable to determine the energy density and the energy averaged for one oscillation period,
\begin{equation}
 \bar{{\cal E}}=\frac{1}{2\pi}\int_0^{2\pi}{\cal E}\mathrm{d}\tau \quad , \qquad
 \bar{E}=\frac{1}{2\pi}\int_0^{2\pi}E\mathrm{d}\tau  \ .
\end{equation}
Using the expressions \eqref{e:phi3b1} - \eqref{e:phi3b3} obtained by the $\varepsilon$ expansion, the time averaged energy density can be expanded into the form $\bar{{\cal E}}=\varepsilon^2\bar{{\cal E}}_2+\varepsilon^4\bar{{\cal E}}_4+\mathcal{O}(\varepsilon^6)$, where
\begin{align}
 \bar{{\cal E}}_2=&\frac{S^2}{2\lambda} \ , \label{endenexp2} \\
 \bar{{\cal E}}_4=&\frac{1}{4\lambda}\left(\frac{\mathrm{d}S}{\mathrm{d}\rho}\right)^2
 -\frac{S^2}{216\lambda^2}(64g_2^2+27\lambda)(S^2+2) 
 +\frac{\sigma}{\lambda^3}SZ \ , \label{endenexp4}
\end{align}
and the definition of the constant $\sigma$ is given in \eqref{eqsigmadef}.

Based on these results, the $\varepsilon$ expansion of the time averaged total energy $\bar E$ in case of $d$ dimensions is
\begin{equation}
 \bar{E}=\varepsilon^{2-d}\bar E_2+\varepsilon^{4-d}\bar E_4
 +\mathcal{O}(\varepsilon^{6-d}) \ , \label{e:Eeps-dep}
\end{equation}
where
\begin{equation}
 \bar E_k=\frac{2\pi^{d/2}}{\Gamma(d/2)}
 \int_0^{\infty}\rho^{d-1}
 \bar{{\cal E}}_k\mathrm{d}\rho \ .
\end{equation}
For one space dimension the function $S$ is uniquely given by \eqref{seqdegysol}, and hence $\bar E_2=2/\lambda$. The numerically calculated value corresponding to the nodeless $S_0$ solution of the base equation, for $d=2$ is $\bar E_2=5.85045/\lambda$, and for $d=3$ it is $\bar E_2=9.44863/\lambda$. The dependence of the coefficient $\bar E_4$ on the form of potential $U(\phi)$ is more complicated, hence we will only give its numerical value for the potential $\phi^4$ in the next subsection.

As it can be seen from \eqref{e:Eeps-dep}, the leading order behavior of the time averaged total energy is ${\bar E}\sim\varepsilon^{2-d}$. Because of this, the behavior of the energy for small $\varepsilon$ amplitudes drastically changes at $d=2$ spatial dimensions. For dimensions $d>2$ the energy grows without bound when $\varepsilon$ approaches zero. In case of $d=2$ the energy $\bar E$ tends to a positive constant, while for $d=1$ it tends to zero when $\varepsilon\to 0$. This shows that for $d>2$ the energy of the core of the oscillons (quasibreathers) takes a minimal value at a certain amplitude $\varepsilon_{\rm m}$, which belongs to the oscillation frequency $\omega_{\rm m}=\sqrt{1-\varepsilon_{\rm m}^2}$. In case of $d=3$, calculating the first approximation from expansion \eqref{e:Eeps-dep}, and assuming that $\bar E_4$ is positive, we obtain
\begin{equation}
 \varepsilon_{\rm m}=\sqrt{\frac{\bar E_2}{\bar E_4}} \ . \label{eqepsminen}
\end{equation}

\subsection{Expansion for the case of the \texorpdfstring{$\phi^4$}{phi4} potential}

In this subsection we work out in detail the $\varepsilon$ expansion of oscillons for the case of the $\phi^4$ potential written in the form \eqref{eqpotone}. In this case the nonvanishing expansion coefficients of the potential are $g_2={-3/2}$ and $g_3=1/2$. Because of this, $\lambda=3/2$, and $p_1=S\sqrt{2/3}$. The definition \eqref{e:p3} of the function $Z$ for this potential is
\begin{equation}
 p_3=\frac{\sqrt{2}}{3\sqrt{3}}\left(
 \frac{65}{8}Z-\frac{8}{3}S-\frac{19}{12}S^3
 \right) \ .
\end{equation}
The general equations \eqref{e:phi3b1}-\eqref{e:phi3b4} are valid in this case too, but now it is easier to proceed further with the expansion. At fifth order, when solving for the function $\phi_5$, the new function $p_5$ appears. As an alternative representation of it, we define a function $Y$ by the expression
\begin{equation}
 p_5=\frac{\sqrt 2}{9\sqrt 3}\left(
 Y-\frac{1235}{32}S^2Z+\frac{1503}{16}Z-24S-\frac{17}{3}S^3
 +\frac{11525}{384}S^5\right) \ .  \label{e:Ydef}
\end{equation}
The equation determining $Y$ is
\begin{align}
 \frac{\mathrm{d}^2Y}{\mathrm{d}\rho^2}&+\frac{d-1}{\rho}\,\frac{\mathrm{d}Y}{\mathrm{d}\rho}
 -Y+3S^2Y+\frac{4225}{64}SZ(3Z-5S^3) \label{e:Yeq}\\
 &+\frac{53d(d-1)}{\rho^2}S\left(\frac{\mathrm{d}S}{\mathrm{d}\rho}\right)^2
 +\frac{106(d-1)}{\rho}S^2(S^2-1)\frac{\mathrm{d}S}{\mathrm{d}\rho}
 +\frac{8287}{48}S^7
 =0 \ . \nonumber
\end{align}
For the function $Y$ belonging to the nodeless base solution $S_0$, which is regular at the center and tends to zero at infinity, the central value for $d=2$ is $-87.78183$, and for $d=3$ it is $60356.38$.

In the following equations we give the $\varepsilon$ expansion coefficients $\phi_k$ of the scalar field $\phi$ up to sixth order, at the moment $\tau=0$ when the field is time-reflection symmetric:
\begin{align}
 \phi_1^{(\tau=0)}=&\sqrt{\frac{2}{3}}\, S \ , \label{e:ph01}\\
 \phi_2^{(\tau=0)}=&\frac{1}{3}S^2 \ ,\\
 \phi_3^{(\tau=0)}=&\frac{1}{9}\sqrt{\frac{2}{3}}\left(
  \frac{195}{8}Z-8S-\frac{35}{8}S^3\right) \ , \label{e:ph03}\\
 \phi_4^{(\tau=0)}=&\frac{1}{9}\left[
  \frac{65}{4}SZ+10\left(\frac{\mathrm{d}S}{\mathrm{d}\rho}\right)^2
  +\frac{8}{3}S^2-\frac{125}{12}S^4
  \right] \ ,\\
 \phi_5^{(\tau=0)}=&\frac{1}{9}\sqrt{\frac{2}{3}}\Biggl[
  Y-\frac{2275}{64}S^2Z+\frac{1503}{16}Z
  -\frac{15}{32}S\left(\frac{\mathrm{d}S}{\mathrm{d}\rho}\right)^2
  -24S-\frac{595}{96}S^3+\frac{11285}{384}S^5
  \biggr] \ ,\\
 \phi_6^{(\tau=0)}=&\frac{2}{27}\Biggl[
  SY+\frac{325}{4}\frac{\mathrm{d}S}{\mathrm{d}\rho}\frac{\mathrm{d}Z}{\mathrm{d}\rho}
  +\frac{4225}{128}Z^2-\frac{8125}{48}S^3Z+\frac{6589}{48}SZ\nonumber\\
  &\quad-\frac{9223}{32}S^2\left(\frac{\mathrm{d}S}{\mathrm{d}\rho}\right)^2
  +\frac{88}{3}\left(\frac{\mathrm{d}S}{\mathrm{d}\rho}\right)^2
  +\frac{26d(d-1)}{\rho^2}\left(\frac{\mathrm{d}S}{\mathrm{d}\rho}\right)^2\label{e:ph06}\\
  &\quad+\frac{52(d-1)}{\rho}S(S^2-1)\frac{\mathrm{d}S}{\mathrm{d}\rho}
  +\frac{92}{9}S^2-\frac{35417}{288}S^4
  +\frac{21467}{144}S^6
  \biggr] \ . \nonumber
\end{align}
These expressions are particularly useful because using them we can give initial data for the numerical time evolution code.

\begin{figure}[!htb]
\centering
\includegraphics[width=115mm]{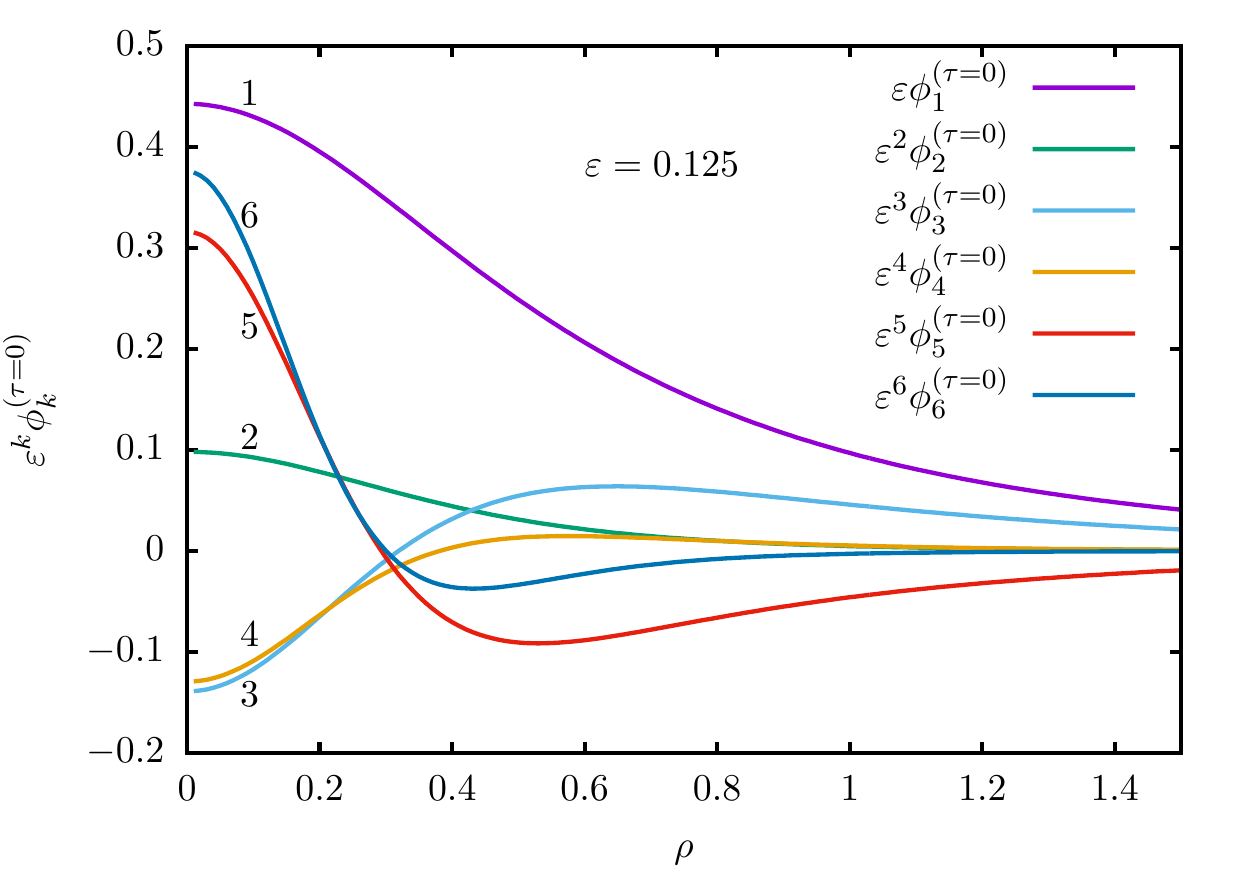}
\caption{\label{figph0d3}
Contributions of various order terms to the scalar field $\phi$, for the choice $\varepsilon=0.125$, in case of $d=3$ spatial dimensions, at the time $\tau=0$.}
\end{figure}

Since the typical values of the functions $\phi_k$ grow quickly with the increase of the index $k$, on Figure \ref{figph0d3}
we show rescaled versions of the functions \eqref{e:ph01}-\eqref{e:ph06}, plotting $\varepsilon^k\phi_k$ for some appropriately chosen value of $\varepsilon$, for the $d=3$ dimensional case. This way we actually show the contributions of the various order terms to $\phi$ at $\tau=0$ for the given choice of $\varepsilon$. The value of the parameter $\varepsilon$ which makes the contribution of an $\varepsilon^{n}$ order term roughly the same as the contribution of the $\varepsilon^{n+1}$ order term can be considered as an upper limit on $\varepsilon$, below which the expansion still gives a useful approximation at the given order $n$. The $\varepsilon$ expansion is actually an asymptotic expansion, and as it is generally true, for a fixed small value of $\varepsilon$, increasing the order of the expansion gives improving approximation, but only up to a certain order. Taking into account even higher orders, we get further away from the true solution, which in our case is the identical frequency quasibreather. Choosing an appropriately small $\varepsilon$, increasing the order $n$ of the expansion, the contributions of the individual $\varepsilon^{n}$ terms become smaller and smaller, but then they start to increase. According to general experience, we can get an ideal approximation if we stop when the contribution of the terms start to increase. In this way, even if the solution that we want to approximate is unknown, we have a good indication on up to what order it is worthwhile to proceed with the expansion.

\begin{figure}[!htb]
\centering
\includegraphics[width=115mm]{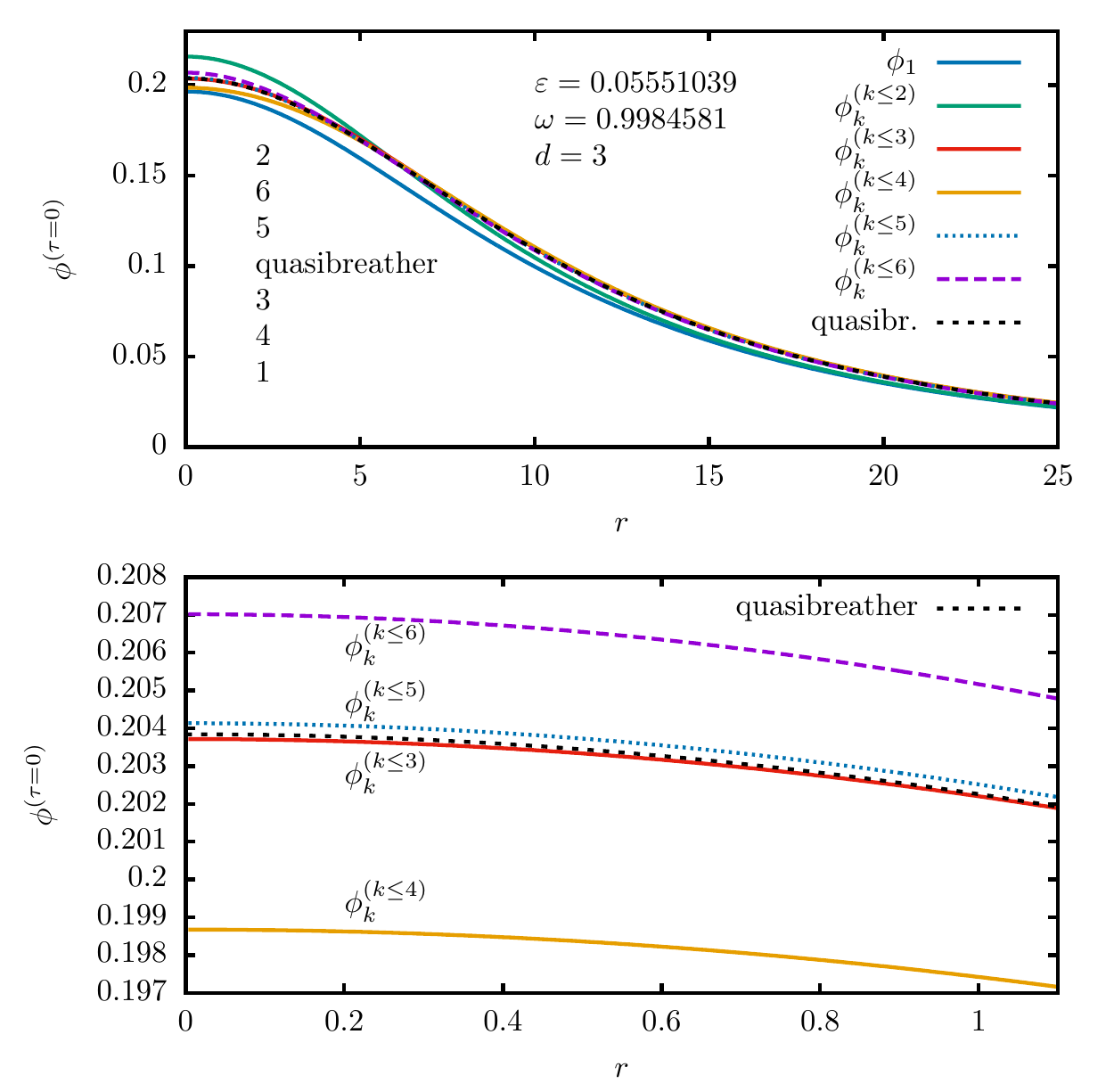}
\caption{\label{figcmp41}
The comparison of the scalar field $\phi$ at the moment $\tau=0$ with various order expansion results, for the $\omega=0.9984581$ frequency quasibreather. The lower panel shows a small region around the center magnified out.}
\end{figure}

Applying the spectral numerical method presented in Subsection \ref{subsecqbnum}, it is possible to determine the quasibreather states to very high precision. On Figure \ref{figcmp41},
for the case of the $\omega=0.9984581$ frequency quasibreather, we compare the numerical value of the scalar field $\phi$ at the moment $\tau=0$ with various order $\varepsilon$ expansion results, for $d=3$ dimensions. The amplitude parameter belonging to this frequency is $\varepsilon=\sqrt{1-\omega^2}=0.05551039$. We note that in Subsection \ref{subsecqbnum} we used a different scaling for the $\phi^4$ potential, and using the conventions there the frequency of this state would be $1.412033$, because of the rescaling of the coordinates $\tau$ and $r$ by $\sqrt{2}$. From the figure it can be seen that in this case we obtain the best approximation if we take into account the first three terms of the expansion. Including higher order terms in $\varepsilon$ only takes us further away from the precise values.

During the rest of the subsection we investigate the $\varepsilon$ dependence of the energy of the configuration for the case of the $\phi^4$ theory. For the form of the $\phi^4$ potential given in \eqref{eqpotone} the time averaged \eqref{endenexp2}-\eqref{endenexp4} expansion coefficients of the energy density are
\begin{equation}
 \bar{{\cal E}}_2=\frac{S^2}{3} \quad , \qquad
 \bar{{\cal E}}_4=\left[\frac{1}{6}\left(\frac{\mathrm{d}S}{\mathrm{d}\rho}\right)^2
 -\frac{41}{108}S^2(S^2+2)+\frac{65}{36}SZ\right]S^4 \ .
\end{equation}
By numerical integration we obtain that the expansion \eqref{e:Eeps-dep} of the time averaged energy of the core domain for the case of $d=2$ space dimension is
\begin{equation}
 {\bar E}\approx 3.9003+26.9618\,\varepsilon^2+\mathcal{O}(\varepsilon^4) \ ,
\end{equation}
while for $d=3$
\begin{equation}
 {\bar E}\approx 6.29908/\varepsilon+264.262\,\varepsilon
 +\mathcal{O}(\varepsilon^3) \ . \label{e:sph-omeps-dep}
\end{equation}
Using this, for the $d=3$ spatial dimensional case, applying \eqref{eqepsminen} we get the estimation $\varepsilon_{\rm m}=0.154$ for the place of the energy minimum, which corresponds to the frequency $\omega_{\rm m}=0.988$. Comparison with the precise numerical results obtained in Subsection \ref{subsecqbnum} shows that this can only be considered as a rough estimation. The frequency dependence of the energy was given on Fig.~\ref{figenertot}. Taking into account of the rescaling of the time coordinate by $\sqrt{2}$, the place of the minimum corresponds to the critical frequency $\omega_c=0.967$, which is considerably smaller than the estimation $\omega_{\rm m}=0.988$. The amplitude parameter belonging to $\omega_c$ is $\varepsilon_c=0.254$ which is much higher than the estimated $\varepsilon_{\rm m}=0.154$.

\subsection{Numerical evolution of the initial data}

The numerical time-evolution code presented in Subsection \ref{secnummodssajat} can be used for the study of the evolution of initial data produced by the small-amplitude expansion procedure. Since for small $\varepsilon$ the size of the oscillon and quasibreather states is inversely proportional to $\varepsilon$, for our computations we chose the value of the coefficient $\kappa$, which was the parameter at the introduction of the coordinate $R$ by \eqref{eqlowupr}, according to the relation $\kappa=5\varepsilon$. We constructed the initial data for $\phi$ at the moment $t=\tau=0$ by calculating the expansion \eqref{e:sumphi} up to some order, applying equations \eqref{e:ph01}-\eqref{e:ph06}. Since the core domain described by the expansion is time-reflection symmetric at $\tau=0$, we set the time derivative of $\phi$ to zero at the initial moment.

\subsubsection{Time evolution for \texorpdfstring{$d=3$}{d=3} spatial dimensions}

In case of three-dimensional space, $\phi^4$ oscillons are unstable for frequencies in the range $1>\omega>\omega_c=0.967$. This frequency domain corresponds to the amplitude parameter range $0<\varepsilon<\varepsilon_c=0.254$, which shows that in this case oscillons become stable only above a relatively large amplitude value. In that case however, at most the first order of the expansion can be used as a valid approximation, and even that differs so much from the precise quasibreather shape that no oscillon state evolves from it with a frequency close to the intended value. Those initial data prepared by the $\varepsilon$ expansion, which approximate better the quasibreather, can be obtained by choosing a relatively small value of the parameter $\varepsilon$, and hence the developing states belong to the unstable domain. In spite of this, these oscillon states can become quite long-lived when we decrease the value of $\varepsilon$, since for small $\varepsilon$ values the higher orders of the expansion approximate very well the shape of the quasibreather.

\begin{figure}[!htb]
\centering
\includegraphics[width=115mm]{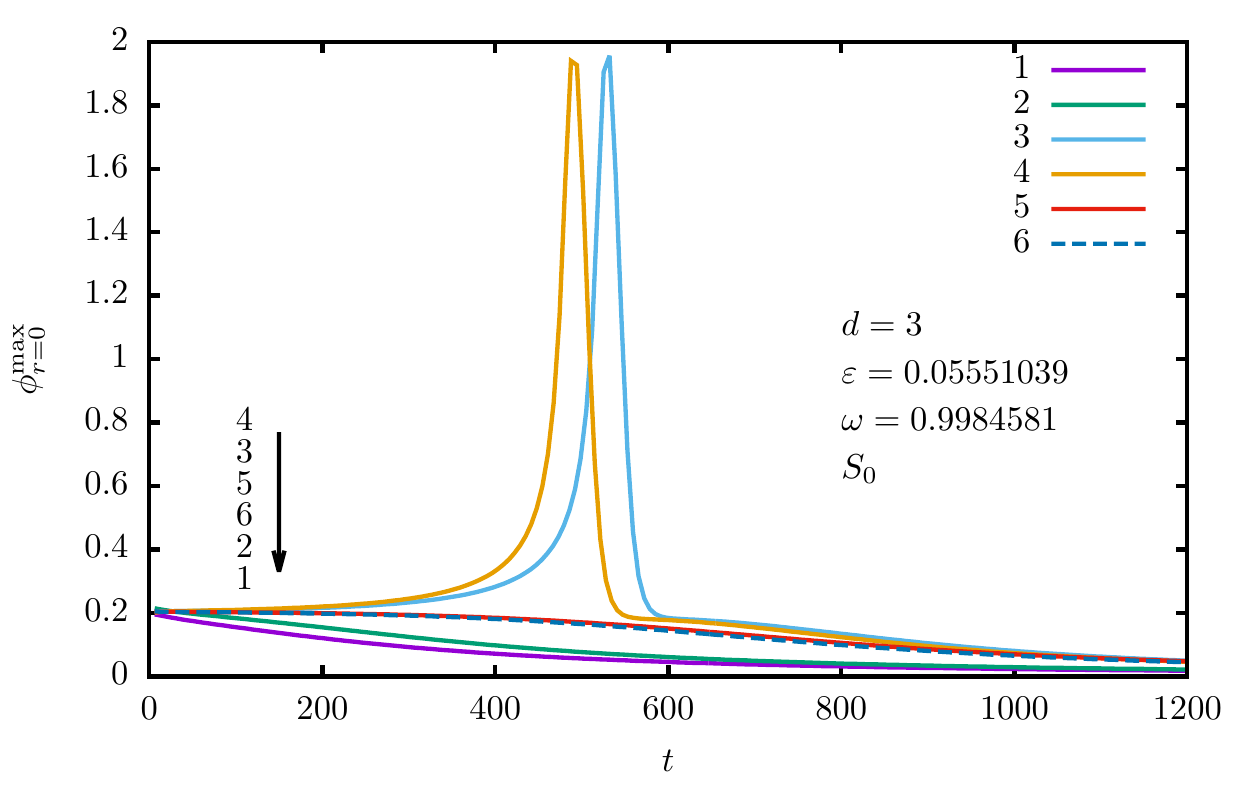}
\caption{\label{figev3d055}
Time evolution of various order $\varepsilon=0.05551039$ initial data obtained by the small-amplitude expansion. We show evolution of initial data calculated up to first to sixth order, for $d=3$ space dimensions.}
\end{figure}

\begin{figure}[!htb]
\centering
\includegraphics[width=115mm]{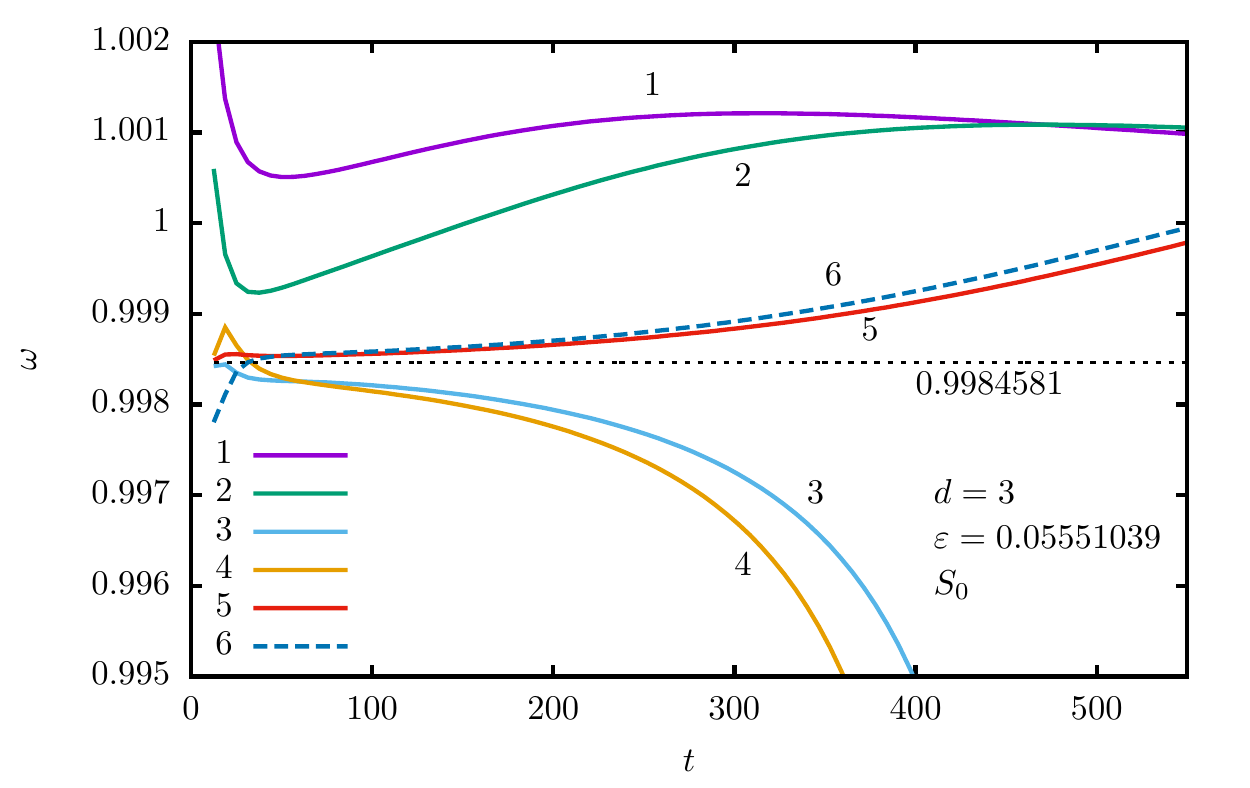}
\caption{\label{figev3d055fr}
Time dependence of the frequency of the oscillons shown on the previous figure.}
\end{figure}

On Figure \ref{figev3d055}
we give the upper envelope curve at the center of oscillons evolving from various order initial data provided by the $\varepsilon$ expansion, for the parameter value $\varepsilon=0.05551039$. Similarly to the near-periodic states shown on Fig.~\ref{figcontour1d}, and to the quasibreather initial data shown on Fig.~\ref{figmax38f}, the unstable states can have two different kind of decay methods. The first is by uniform increase of the oscillon's size, the other by a temporary collapse to a smaller region near the origin. As it can be seen on the figure, that which kind of decay mechanism will actually happen depends even on the order of the initial data. We show the time dependence of the frequency of the evolving states in an initial time period on Figure \ref{figev3d055fr}.
We can see that the first two orders of the expansion give shorter living states that are quite different from the intended oscillons. On the other hand, the third and fifth orders make substantial improvement on the lower order approximations.

\subsubsection{Time evolution for \texorpdfstring{$d=2$}{d=2} spatial dimensions}

\begin{figure}[!htb]
\centering
\includegraphics[width=115mm]{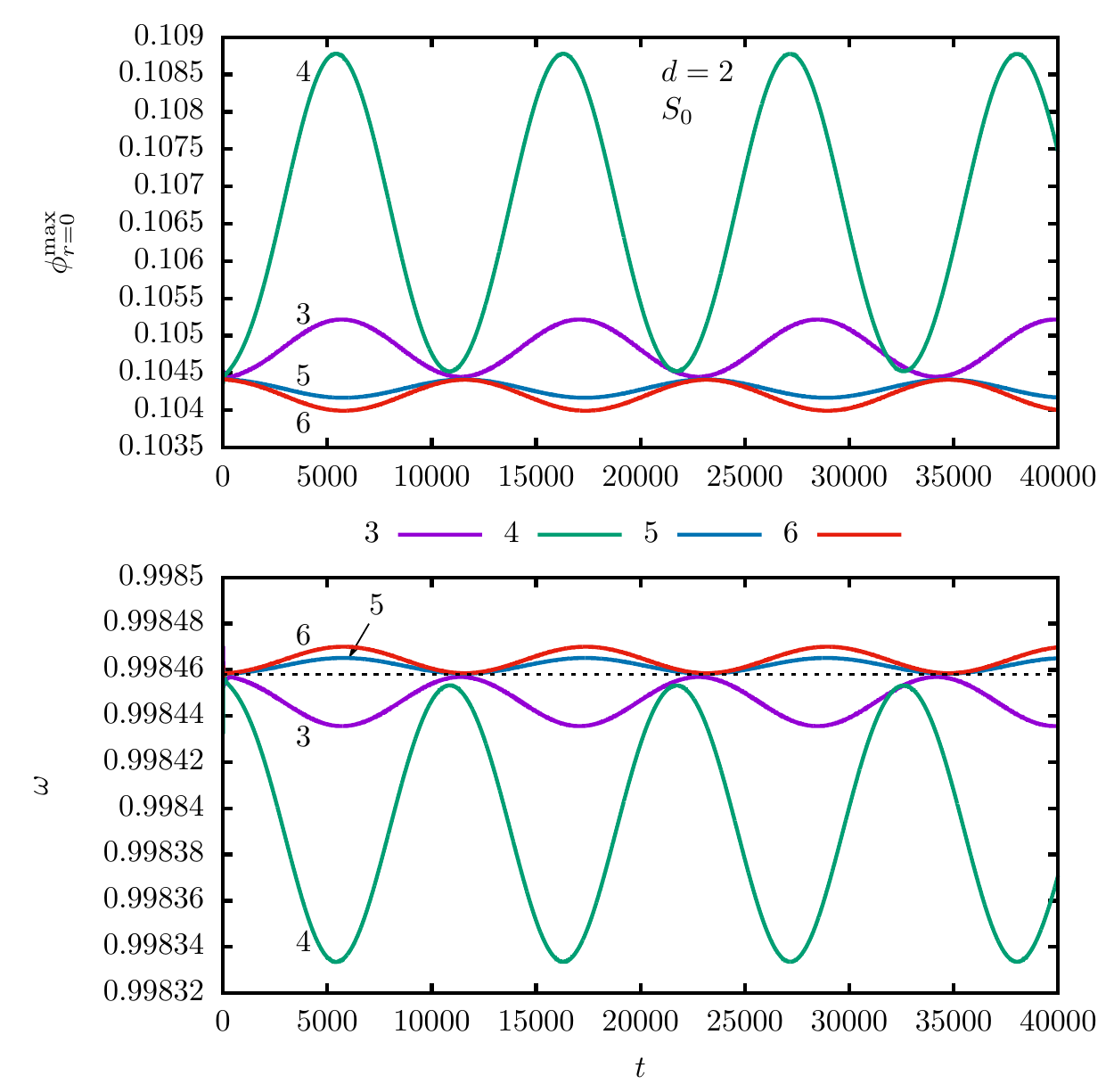}
\caption{\label{figev2d055}
Time evolution of various order initial data with $\varepsilon=0.05551039$, in case of $d=2$ spatial dimensions. On the upper panel we show the upper envelope curve at the center, while on the lower panel we give the frequency. The dashed black line is the theoretically predicted frequency, $\omega=0.9984581$.}
\end{figure}

In case of two spatial dimensions small and moderate amplitude oscillons turn out to be stable. If an oscillon develops from some initial data, its lifetime is infinite, it never decays suddenly. On the other hand, on the amplitude a low frequency modulation appears, corresponding to a shape-mode. It is likely that both for $d=1$ and $d=2$ dimensions there exists some amplitude where the energy becomes maximal, and above which oscillons become unstable, but the value of this limit is currently unknown. On Figure \ref{figev2d055}
we show the time evolution of various order initial data constructed with the choice of the parameter $\varepsilon=0.05551039$, from the third to the sixth order. The initial data provided by the first and second order approximation give relatively quickly decaying states with frequencies far from the intended one, hence we do not show these states on the figure. As it can be observed on the figure, at this relatively low amplitude the third and fifth order correction takes closer to the periodic quasibreather state, while the fourth and sixth order takes us further from it again.

\section{Analytic determination of the radiation amplitude} \label{secsuganal}

\subsection{Oscillons and quasibreathers}

As the comparison of the numerically calculated states in Subsection \ref{subsecoscqb} shows,  the amplitude of the outgoing radiating wave tail of oscillons agree to high precision with the amplitude of the standing wave tail of the identical frequency quasibreather. For $d+1$ dimensional spacetimes the asymptotic form of the $\omega_f$ frequency part of the quasibreather's tail is
\begin{equation}
 \phi=\frac{\alpha}{r^{\frac{d-1}{2}}}\cos(\lambda_f r+\delta)\cos(\omega_f t) \ , \label{eqphitailgen}
\end{equation}
where $\lambda_f=\sqrt{\omega_f^2-m^2}$, and $\delta$ is a constant giving the phase. Adding further contribution with asymptotics $\frac{1}{r^{(d-1)/2}}\cos(\lambda_f r+\delta)\cos(\omega_f t)$, i.e.~the change of $\alpha$, would destroy the regularity of the center (or mirror symmetry for $d=1$). On the other hand, we can freely add a contribution $\frac{1}{r^{(d-1)/2}}\sin(\lambda_f r+\delta)\cos(\omega_f t)$ with arbitrary coefficient, which also solves the linearized equation. That way we get periodic weakly nonlocal solutions which are very similar to the quasibreathers, with essentially the same core region, but with tail-amplitude larger than the minimal one. Violating the time-reflection symmetry by adding a term with asymptotics
$\frac{\alpha}{r^{(d-1)/2}}\sin(\lambda_f r+\delta)\sin(\omega_f t)$, we obtain a wave carrying energy outwards, with identical $\alpha$ amplitude,
\begin{equation}
 \phi=\frac{\alpha}{r^{\frac{d-1}{2}}}\cos(\lambda_f r+\delta-\omega_f t)
 \ . \label{eqphialphatdep}
\end{equation}
Based on \eqref{eqencurr}, the outgoing energy current $\mathrm{S}$ provided by this wave through a sphere of radius $r$, averaged for an oscillation period $2\pi/\omega_f$, is
\begin{equation}
 \bar S=\frac{\pi^{\frac{d}{2}}}{\Gamma\left(\frac{d}{2}\right)} 
 \lambda_f\omega_f\alpha^2 \ . \label{eqsaverage}
\end{equation}
In the following, to obtain the energy loss rate of oscillons we will calculate the amplitude $\alpha$ of the quasibreather's tail.

The proof of the existence of the tail and the determination of its amplitude was achieved first by Segur and Kruskal \cite{SegurKruskal87}. They applied the methods of extension to the complex plane and matched asymptotic expansions for the case of the one-dimensional $\phi^4$ theory. The complex extension method was applied originally by Pokrovskii and Khalatnikov for the calculation of the amplitude of backscattering from a low potential barrier \cite{Pokrovskii61}. The core domain of the quasibreathers can be well described by the expansion with respect to the parameter $\varepsilon$. The amplitude of the core is proportional to $\varepsilon$, while its size is order $1/\varepsilon$. For small values of $\varepsilon$, to this core we need to determine a faraway a standing wave tail correction, with amplitude that is exponentially small in $\varepsilon$. The important advantage of the extension of the equations and their solutions into complex values of the radial coordinate $r$ is that near the singularity which appears on the complex $r$ plane, the contributions of both the core and the tail become order $\mathcal{O}(1)$. Since in the neighborhood of the singularity the functions do not vanish even in the $\varepsilon\to 0$ limit, the correction which is exponentially small on the real axis can be much more easily calculated there.

Segur and Kruskal have solved the differential equations in the neighborhood of the singularity by a numerical method. In our paper \cite{Fodor2009a}, applying Borel summation we have analytically determined the contribution of the oscillating tail near the singularity for potentials $U(\phi)$ that are symmetric around their minimum. This method was applied first by Pomeau, Ramani and Grammaticos for the calculation of the tail-amplitude of the weakly nonlocal solutions of the modified fifth order Korteweg-de Vries (KdV) equation \cite{Pomeau1988}.  In paper \cite{Fodor2009a} we have compared our analytical results for the tail-amplitude to the numerically calculated quasibreathers and to the near-periodic states forming at the numerical time-evolution for different potentials, supporting the correctness of the results. In a subsequent paper we have generalized the method for the calculation of the radiation amplitude of $d$-dimensional spherically symmetric oscillon states \cite{Fodor2009b}.

The method for the calculation of the radiation of oscillons can also be applied for the asymptotic tail of the quasibreather (nanopteron) solutions of the fifth order Korteweg-de Vries (KdV) equation \cite{Pomeau1988}. This problem is considerably simpler than the oscillon case, because we consider time-independent solutions of the KDV equations, so there is no need for the Fourier expansion in time, and instead of the infinitely many Fourier components the study of one function is sufficient. The detailed and pedagogical presentation of the KDV system can be found in the book of Boyd \cite{Boyd-book1998}. As Boyd also points it out, the nonlinearity is not a crucial aspect for the existence of the quasibreather solutions. In his book he reviews an 
analogous simple problem for a function $u(x)$ which is determined by the equation
\begin{equation}
 \frac{\mathrm{d}^2 u}{\mathrm{d}x^2}+u=f(\varepsilon x) \ ,
\end{equation}
where $\varepsilon$ is a small constant, and $f(x)$ is a given localized function, for example $\sech x$. There are solutions of this equation that have a localized core and a tail which is exponentially small in $\varepsilon$. The amplitude of the tail can be calculated by first applying an expansion with respect to $\varepsilon$, and then using complex extension and Borel summation. This simpler problem motivates and illustrates well the sophisticated method which can also be applied for the complicated nonlinear systems \cite{Boyd-book1998}.

\subsection{One-dimensional space with symmetric potential} \label{secegydimszim}

Because of the technical complexity of the method for oscillons, we first present the procedure for a concrete case when the calculation of the radiation is the simplest. In case of a $U(\phi)$ potential that is symmetric around its minimum, in the Fourier expansion \eqref{eqphifourier} there are only odd indexed $\Phi_n$ components, and the equations that have to be solved are also easier to analyze. As we will see later, for $d>1$ spatial dimensions the tail-amplitude of spherically symmetric quasibreathers agree to leading order in $\varepsilon$ with the amplitude of the $d=1$ dimensional states. Because of this, it is necessary to investigate the one-dimensional case first. We will calculate the dimensional correction to this later.

Applying the notation of the expansion \eqref{potexpeq}, the simplest potential which is symmetric around its minimum is $U(\phi)=\frac{1}{2}\phi^2+\frac{g_3}{4}\phi^4$. In this subsection we still keep the mass of the scalar field at $m=1$. Small-amplitude oscillons can only exist if the constant $\lambda=\frac{5}{6}g_2^2-\frac{3}{4}g_3$ defined in \eqref{lambdadef} is positive. Since now $g_2=0$, necessarily $g_3$ must be negative. Because of this, by the rescaling of the scalar field $\phi$, for an arbitrary symmetric potential we can set $g_3=-1$, and then $\lambda=3/4$. To ensure the stability of the system we intend to keep the potential bounded from below, hence we add a sixth degree term to the potential,
\begin{equation}
 U(\phi)=\frac{1}{2}\phi^2-\frac{1}{4}\phi^4+\frac{g_5}{6}\phi^6 \ , \label{eqsympotform}
\end{equation}
where we assume that $g_5>0$. During the following calculations we will use this general symmetric sixth order potential. We will see that the amplitude of the tail depends strongly on the value of $g_5$.

We look for solutions that are time-reflection symmetric at the moment $t=0$, and oscillate with base frequency $\omega$. Fourier expanding the scalar field according to \eqref{eqphifourier}, we introduce the functions $\Phi_n$. Since the potential is symmetric, we choose the even indexed $\Phi_n$ components zero. This choice is consistent with equations \eqref{eqphinsph} for the components, together with their $F_n$ source terms, which are specified in \eqref{eqfngnsum}. Obviously, in this case the summations in the expressions for $F_n$ are only needed for odd integers.

In principle, by the numerical solution of equations \eqref{eqphinsph}, it is possible to calculate the core and tail regions of quasibreathers to high precision. However in the one-dimensional case such results are only known for the $\phi^4$ potential, which is non-symmetric around its minimum \cite{Boyd1990}. The core region is very well described by the small-amplitude expansion, even for moderately large values of the parameter $\varepsilon$, but because of the assumed slow spatial variation it does not give any information about the tail domain. For the solution of the problem we have to calculate an exponentially small correction to the asymptotic series approximation of the core region. For this, we first need to investigate in more detail the results obtained by the $\varepsilon$ expansion.

In Section \ref{seckisampl} we presented the small-amplitude expansion with the weaker assumption that the solution should not increase without bound as the time passes. From this followed the time-periodicity and the time-reflection symmetry to every order in $\varepsilon$. We can obtain the same result by assuming periodicity and time-reflection symmetry, and expanding the Fourier mode equations \eqref{eqphinsph} with respect to powers of the amplitude parameter $\varepsilon$. For this we also have to make the consistent order of magnitude assumption, that the expansion of each $\Phi_n$ starts with power $\varepsilon^n$ term. For odd $n$ the expansion of $\Phi_n$ consists only of odd powers of $\varepsilon$, while for even $n$ there are only even powers. The only small exception for symmetric potentials is $\Phi_0$, which starts with $\varepsilon^2$ instead of $\varepsilon^0$. This behavior can also be seen well from equations \eqref{e:phi3b1} - \eqref{e:phi3b4} and \eqref{eqphi5sym}, which give the first few orders of the expansion. For the expansion one has to use the rescaled radial coordinate $\rho=\varepsilon r$, and the frequency is related to the parameter by $\omega^2=1-\varepsilon^2$. By whichever method we obtain the small-amplitude expansion results, the important point is that the $\varepsilon$ expansion approximates the solution of the Fourier mode equations, in the form of a not convergent asymptotic series.

The $\varepsilon$ expansion becomes significantly simpler for $d=1$ spatial dimension, since in that case the differential equations for the functions $p_{2k+1}$ can be solved analytically. For $d=1$ the solutions can be searched as polynomials in $S=\sqrt{2}\sech\rho$ (see \eqref{eqzdimone}, \eqref{eqp5dimone}). According to equations \eqref{e:phi3b1} - \eqref{e:phi3b4} and \eqref{eqphi5sym}, for the symmetric potential given in \eqref{eqsympotform} the $\varepsilon$ expansion of the first three Fourier components is
\begin{align}
 \Phi_1=&\varepsilon\frac{2\sqrt{2}}{\sqrt{3}}\sech\rho
 +\varepsilon^3\frac{1}{27\sqrt{6}}(80g_5+3)(2\sech\rho-\sech^3\rho) \notag\\
 &+\varepsilon^5\frac{1}{1458\sqrt{6}}\Bigl[5(6400g_5^2-96g_5+63)\sech\rho
 -(25600g_5^2+480g_5+171)\sech^3\rho \notag\\
 &\hspace*{20mm}+3(1600g_5^2+240g_5-9)\sech^5\rho
 \Bigr] +\mathcal{O}(\varepsilon^7) \ , \label{eqpphi1epsexp} \\
 \Phi_3=&-\varepsilon^3\frac{1}{3\sqrt{6}}\sech^3\rho \notag\\
 &+\varepsilon^5\frac{1}{108\sqrt{6}}\left[-2(80g_5+3)\sech^3\rho
 +(200g_5-33)\sech^5\rho\right]+\mathcal{O}(\varepsilon^7) \ , \\
 \Phi_5=&\varepsilon^5\frac{1}{108\sqrt{6}}(8g_5+3)\sech^5\rho
 +\mathcal{O}(\varepsilon^7) \ .  \label{eqpphi5epsexp}
\end{align}
At higher orders it is also true that for odd $k$ the first term of the expansion of $\Phi_k$ is proportional to $\varepsilon^k\sech^k\rho$. For a general symmetric potential the leading order terms of the Fourier components would remain the same. The coefficient $g_7$ of the expansion of the potential would only appear in the third term of \eqref{eqpphi1epsexp}, which is proportional to $\varepsilon^5$. With the setting $g_3=-1$ the other terms given in \eqref{eqpphi1epsexp}-\eqref{eqpphi5epsexp} would remain unchanged.

Although equations \eqref{eqpphi1epsexp}-\eqref{eqpphi5epsexp} cannot give any information about the amplitude of the standing wave tail, they are still very useful, since for small amplitudes they very precisely determine the behavior of the Fourier modes in the whole core domain. For the procedure that we will apply it will be necessary to analytically extend the solution functions $\Phi_k$, which contain both the core and the tail, and the equations \eqref{eqphinsph} which they must satisfy, to the complex $r$ plane. Since the expressions obtained by the $\varepsilon$ expansion contain the variable $\rho=\varepsilon r$, we also have to consider the small-amplitude expansion on the complex $\rho$ plane. The extended functions that we use are complex differentiable at every point of the plane where they can be defined, in other words, they are holomorphic functions. If these functions are known on any open set, they have unique analytic continuation into larger connected open sets on the complex plane. Hence, instead of solving the differential equations on the real axis, we can choose to solve them in some domain on the complex plane, and analytically continue the functions back to the real axis later.

As it can be seen from equations \eqref{eqpphi1epsexp}-\eqref{eqpphi5epsexp}, the behavior of the Fourier mode functions on the complex plane are determined by the function $\sech\rho$. The function $\sech\rho$ has first order poles at the points $\rho=i\frac{\pi}{2}+ik\pi$, where $k$ is arbitrary integer. The radiation rate will be determined by the poles at $\rho=\pm i\frac{\pi}{2}$, which are the closest ones to the real axis. Since our functions take real values on the real axis, because of their symmetry, it is sufficient to study the behavior near $\rho=i\frac{\pi}{2}$. In order to make the generalization for higher dimensions easier later, for the distance of the singularity from the real axis we introduce the notation $P$. In the presently discussed one spatial dimensional case $P=\frac{\pi}{2}$. Introducing the variable $R$ by the equation
\begin{equation}
 \rho=iP+R \ ,
\end{equation}
the Laurent series expansion of the function $\sech\rho$ around the point $\rho=iP$ is
\begin{equation}
 \sech\rho=-\frac{i}{R}+\frac{iR}{6}-\frac{7iR^3}{360}
 +\frac{31iR^5}{15120}+\mathcal{O}(R^7) \ . \label{eqsechrhoexp}
\end{equation}
In each term of the expansion the coefficient of $R^k$ is purely imaginary.

For the Fourier modes $\Phi_k$, as functions of the radial coordinate $r=\rho/\varepsilon$, the singularities closest to the real axis are at the points $r=\pm i\displaystyle\frac{P}{\varepsilon}$, hence they get further and further away when the amplitude $\varepsilon$ is decreased. Let us introduce the variable $y$, which measures the distance from the real axis with respect to the scale corresponding to the original radial coordinate $r$, with the equation
\begin{equation}
 r=i\frac{P}{\varepsilon}+y \ . \label{eqrykapcs}
\end{equation}
Obviously, then $R=\varepsilon y$. The important observation is, that of the expression $\varepsilon\sech\rho$, which appears at several places in \eqref{eqpphi1epsexp}-\eqref{eqpphi5epsexp}, there is a nonvanishing $\varepsilon$ independent part, since
\begin{equation}
 \varepsilon\sech\rho=-\frac{i}{y}+\frac{iy}{6}\varepsilon^2-\frac{7iy^3}{360}\varepsilon^4
 +\frac{31iy^5}{15120}\varepsilon^6+\mathcal{O}(\varepsilon^8) \ .
\end{equation}
Because of this, in case of the \eqref{eqsympotform} symmetric potential, the behavior of the Fourier modes near the singularity is
\begin{align}
 \Phi_1=&-\frac{2\sqrt{2}\,i}{\sqrt{3}\,y}
 -\frac{i}{27\sqrt{6}\,y^3}(80g_5+3)-\frac{i}{486\sqrt{6}\,y^5}(1600g_5^2+240g_5-9)
 +\mathcal{O}(\varepsilon^2) \ , \label{eqpphi1yexp} \\
 \Phi_3=&-\frac{i}{3\sqrt{6}\,y^3}
 -\frac{i}{108\sqrt{6}\,y^5}(200g_5-33)+\mathcal{O}(\varepsilon^2) \ , \\
 \Phi_5=&-\frac{i}{108\sqrt{6}\,y^5}(8g_5+3)
 +\mathcal{O}(\varepsilon^2) \ . \label{eqpphi5yexp}
\end{align}
The first terms of all higher $\Phi_k$ modes are also proportional to $i/y^k$. For a general symmetric potential with $g_3=-1$ the only change in the expansions \eqref{eqpphi1yexp}-\eqref{eqpphi5yexp} is that in the third term in \eqref{eqpphi1yexp} the constant $g_7$ also appears. Since the functions obtained by the $\varepsilon$ expansion are mirror symmetric on the real axis around $r=0$, in every term of the expansion the coefficient of $y^{-k}$ is purely imaginary. Because of this, the contribution of all terms are real on the imaginary $y$ axis.

Since it contains the positive powers of $1/y$, the approximation \eqref{eqpphi1yexp}-\eqref{eqpphi5yexp} can be applied if $|y|$ is large. At first sight this may appear contradictory, since we are now considering places close to the singularity. However, if $\varepsilon$ is small enough, it can be easily satisfied that $y$ is large while the rescaled $R=\varepsilon y$ is still small. The inequality $1\ll |y|\ll 1/\varepsilon$ describes the matching domain of the applied asymptotic matching, which is the overlapping part of the two regions used in the procedure.

The \emph{outer region} is the part of the complex $r$ plane where the small-amplitude expansion is valid, and gives improving approximation with decreasing $\varepsilon$. Obviously, the real  $r$ axis is part of this domain. In this region we solve the Fourier mode equations \eqref{eqphinsph} for a frequency $\omega$ that is just a little smaller than $1$. In our concrete example, with the choice $m=1$ and $d=1$, the equations are
\begin{equation}
 \frac{\mathrm{d}^2\Phi_n}{\mathrm{d}r^2}
 +(\omega^2n^2-1)\Phi_n=F_n \ , \label{eqphouter}
\end{equation}
for odd $n$. We intend to determine the amplitude of the tail when $\varepsilon=\sqrt{1-\omega^2}$ is small.

The \emph{inner region} is the part of the complex $r$ plane in the vicinity of the singular point $r=iP/\varepsilon$, where in the limit $\varepsilon\to 0$ the functions become independent of $\varepsilon$, and hence also can be considered independent of $\omega$. Here we use the shifted, but not rescaled $y$ coordinate. The equation that we have to solve in the inner region can be obtained from \eqref{eqphinsph}, by the substitution $\omega=1$,
\begin{equation}
 \frac{\mathrm{d}^2\Phi_n}{\mathrm{d}y^2}
 +(n^2-1)\Phi_n=F_n \ . \label{eqphinner}
\end{equation}
The functions $F_n$ containing the nonlinear terms are given by \eqref{eqfngnsum} in both regions. For the currently used \eqref{eqsympotform} simple example potential the nonvanishing $g_n$ coefficients are $g_3=-1$ and $g_5$. The crucial advantage of equations \eqref{eqphinner} is that in contrast to \eqref{eqphouter}, it is not necessary to solve them for each $\omega$ frequency separately. The treatment of the limit $\omega\to 1$ is problematic in the outer region, because then on the real axis the amplitudes of both the core and the tail tend to zero.

Substituting the expansion
\begin{equation}
 \Phi_{2n-1}=\sum_{k=n}^\infty a^{(2n-1)}_{2k-1}\frac{1}{y^{2k-1}} \label{eqphinyexpsym}
\end{equation}
into the equations \eqref{eqphinner}, apart from a global $\pm 1$ factor, the constants $a^{(2n-1)}_{2k-1}$ become uniquely determined. In this way we can directly obtain expressions \eqref{eqpphi1yexp}-\eqref{eqpphi5yexp}, and also their generalizations to arbitrary symmetric potentials, without calculating the small-amplitude expansion. The ambiguity in the signature corresponds to the symmetry $\phi\to-\phi$, and related to the choice of the signature of $S$ in \eqref{seqdegysol}. The calculation of the expansion in powers of $1/y$ is technically much easier in this direct way, and can be performed to quite high orders by a software package suitable for symbolic algebraic computations. For our calculations we have used the Maple symbolic computing environment. By direct substitution of the power series, it is possible to determine the coefficients even up to orders $1/y^{60}$. By increasing the $1/y$ order in steps of $2$ and using the already known lower order order results, we were able to proceed up to orders of several hundred in $1/y$, even if we used the same large number of $\Phi_{2n-1}$ Fourier components. According to our experience, the consideration of Fourier components higher than $\Phi_7$ influences the magnitude of the radiating tail only extremely slightly. The essential information that comes from the $\varepsilon$ expansion procedure is that in the inner region the expansion of all $\Phi_n$ Fourier components begins with an $1/y^n$ term, and also that the power of $1/y$ increases in steps of two. An important property of the expansion \eqref{eqphinyexpsym} is that all $a^{(2n-1)}_{2k-1}$ coefficients are purely imaginary.

The understanding of the formalism may be helped by the consideration of the sine-Gordon breather. In this case by the Fourier transformation of the expression \eqref{eqsinegex} of the scalar field $\phi$, the $\Phi_n$ functions can be calculated for any specific $n$. Equation \eqref{eqsgphiepsexp} gives the first few terms of the $\varepsilon$ expansion. Extending the solution $\phi$ into complex values of the radial coordinate, using the substitution $r=i\frac{\pi}{2\varepsilon}+y$ and tending to zero by $\varepsilon$, we obtain the behavior of the scalar field in the inner region:
\begin{equation}
 \phi=4\arctan\left(\frac{\cos t}{iy}\right) \ .
\end{equation}
The functions $\Phi_n$ in the inner region can be easily obtained by Fourier transforming this expression. By expanding the results in powers of $1/y$, we can obtain the coefficients $a^{(2n-1)}_{2k-1}$. For example, at the expansion of $\Phi_3$,
\begin{equation}
 a^{(3)}_{2k-1}=-i\frac{(2k-2)!}{4^{k-2}(k+1)!(k-2)!} \ .  \label{eqsgphi3exp}
\end{equation}

\subsection{Numerical solution in the inner region} \label{secbelnum}

In the inner region the $\Phi_n$ Fourier components are the solutions of \eqref{eqphinner}. The expansion of the solution in the power series \eqref{eqphinyexpsym} is not convergent, but it is an asymptotic series. Cutting the expansion at the appropriate order, for small values of $1/|y|$ it gives a very good approximation. The lack of convergence is connected to the contributions decaying according to $\exp(-|y|)$, which cannot be described by power series expansion. The existence of these exponentially decaying contributions is strongly related to the existence of the exponentially small tail of quasibreathers on the real axis.

Calculating a correction, which is beyond all orders small, to a divergent series may seem to be meaningless at first sight, but following the method of Segur and Kruskal \cite{SegurKruskal87} we can give meaning to the approach by looking for a place where at least the imaginary part of the series is convergent. This place is the imaginary axis ${\rm Re}\,y=0$, where all terms of \eqref{eqphinyexpsym} are real, and hence the imaginary part of the series trivially converges to the value $\mathrm{Im}\,\Phi_n=0$.

We divide the Fourier mode functions $\Phi_n$ into real and imaginary parts by the notation
\begin{equation}
 \Phi_n=\Psi_n+i\Omega_n \ , \label{eqphinpsinomn}
\end{equation}
where $\Psi_n$ and $\Omega_n$ are real valued functions, and substitute into equations \eqref{eqphinner} describing the inner solution. In the linear approximation the equations decouple. As an alternative to $y$ we introduce the coordinate $\tilde y=iy$, which takes increasing real values when coming downwards on the imaginary axis. This is the direction towards the center $r=0$ on the real axis. The linear part of the equations can be written into the form
\begin{equation}
 -\frac{\mathrm{d}^2\Omega_n}{\mathrm{d}\tilde y^2}
 +(n^2-1)\Omega_n=0 \ . \label{omegandiffeq}
\end{equation}
The solution for $n=1$ is $\Omega_1=c_0+c_1 \tilde y$, where $c_0$ and $c_1$ are constants. This inner solution can only be matched to the exterior solution if in the limit $\tilde y\to\infty$ the function $\Omega_1$ tends to zero, which is only possible if $c_0=c_1=0$. All the other $\Omega_n$ must also tend to zero in the $\tilde y\to\infty$ limit, hence their behavior to linear order is
\begin{equation}
 \Omega_n=\nu_n\,\exp\left(-\sqrt{n^2-1}\, \tilde y\right)
 \qquad \text{for} \quad  n\geq 2 \ , \label{eqomnnunlead}
\end{equation}
where $\nu_n$ are constants that should be determined later.

For the study of the behavior of the functions $\Omega_n$ on the imaginary axis the linearized approximation is not enough. Nevertheless, we can get a very precise description if we substitute $\Phi_n=\Psi_n+i\Omega_n$ into \eqref{eqphinner}, and from the terms of $F_n$ given by \eqref{eqfngnsum} we only keep those terms that are linear in the exponentially small variables $\Omega_k$. The right-hand sides of the equations contain $\Omega_j$ terms multiplied by powers of various $\Psi_k$ functions. Since about the imaginary parts we assume that they are exponentially small, in the equations for all $\Psi_n$ we can substitute the approximation of the $\Phi_n$ functions given by the expansion \eqref{eqphinyexpsym}. In this way we get coupled linear differential equations for the functions $\Omega_n$ along the imaginary axis.

In the differential equation for $\Omega_k$ there are source terms in $F_k$ on the right-hand side which are proportional to some other $\Omega_n$. These terms generate terms in $\Omega_k$ which decay as $\exp(-\sqrt{n^2-1}\, \tilde y)/\tilde y^a$ corresponding to the exponential behavior of $\Omega_n$, where $a$ is some positive integer. For example, on the right hand side of the equation determining $\Omega_5$, the imaginary part of $\Phi_1^2\Phi_3$ yields a term $\Psi_1^2\Omega_3$, which generates a behavior $\exp({-\sqrt{8}}\, \tilde y)/\tilde y^2$ in $\Omega_5$ as well. These types of terms also have an influence on the behavior of the original $\Omega_n$, and modify the coefficients of the  $\exp({\sqrt{n^2-1}}\, \tilde y)/\tilde y^a$ terms there.

For large values of $\tilde y$ the less quickly decaying exponential mode will dominate. For symmetric potentials $\Phi_2=0$, and generally the exponential behavior corresponding to $\Omega_3$ dominates for all $\Omega_k$. This results in a leading order behavior for $\Omega_k$ according to $\exp(-\sqrt{8}\, \tilde y)/\tilde y^a$, where $a$ is a positive integer. In this case $\Phi_3$ is the first mode which can radiate, and if the value of $\nu_3$ turns out to be nonzero then $\Omega_3$ provides the dominant exponential correction, which after the continuation to the real axis determines the radiation rate.

For a symmetric potential the leading order behavior of $\Omega_3$ is well described by equation \eqref{eqomnnunlead}, while for other $\Omega_k$ functions the behavior generated by $\Omega_3$ through the nonlinear terms dominate. We can get a better approximation for the behavior of $\Omega_3$ if we look for the solution of the coupled linear equations for the imaginary parts in the form
\begin{equation}
 \Omega_n=\nu_3\,\exp\left(-\sqrt{8}\, \tilde y\right)\left(c_0^{(n)}+c_1^{(n)}\frac{1}{\tilde y}+c_2^{(n)}\frac{1}{\tilde y^2}
 +c_3^{(n)}\frac{1}{\tilde y^3}+\ldots\right)
\end{equation}
where $c_0^{(3)}=1$, and for the real parts $\Psi_n$ we substitute the expression \eqref{eqphinyexpsym}. Substituting into the coupled differential equations for $\Omega_n$, from the vanishing of the coefficients of the powers of $1/\tilde y$ we can obtain the constants $c_k^{(n)}$. In case of the symmetric sixth order potential \eqref{eqsympotform} we obtain the following result:
\begin{align}
 \Omega_1=&\nu_3\,\exp\left(-\sqrt{8}\,
 \tilde y\right)\left(\frac{1}{4\tilde y^2}
 -\frac{1}{4\sqrt{2}\,\tilde y^3}
 -\frac{1520g_5-267}{864\tilde y^4}
 +\ldots\right) , \label{e:om1}\\
 \Omega_3=&\nu_3\,\exp\left(-\sqrt{8}\,
 \tilde y\right)\left(1+\frac{1}{\sqrt{2}\,\tilde y}+\frac{1}{8\tilde y^2}
 -\frac{260g_5+3}{162\sqrt{2}\,\tilde y^3}
 -\frac{520g_5+87}{2592\tilde y^4}
 +\ldots\right) , \label{e:om3} \\
 \Omega_5=&\nu_3\,\exp\left(-\sqrt{8}\,
 \tilde y\right)\left(-\frac{1}{8\tilde y^2}
 -\frac{1}{4\sqrt{2}\,\tilde y^3}
 +\frac{5(128g_5-33)}{864\tilde y^4}
 +\ldots\right) , \\
 \Omega_7=&\nu_3\,\exp\left(-\sqrt{8}\,
 \tilde y\right)\left(\frac{80g_5+21}{1440\tilde y^4}
 +\ldots\right) . \label{e:om7}
\end{align}
We note that without taking into account the back-reaction provided by the other $\Omega_k$ we would get incorrect coefficients in $\Omega_3$ for the $1/\tilde y^3$ and higher terms. The expansion can be relatively easily calculated up to the same order in $1/\tilde y$ as we could proceed in case of \eqref{eqphinyexpsym}. We would also like to point out that the higher expansion coefficients of $\Omega_3$ change if we only take into account the first few $\Phi_n$ modes. For example, if we intend to calculate the value of $\nu_3$ in an approximation using only $\Phi_1$ and $\Phi_3$, setting the other $\Phi_n$ zero, then it is reasonable to use the appropriate modification of \eqref{e:om1} and \eqref{e:om3} for the fast convergence in $\tilde y$.

In this subsection we calculate the coefficient $\nu_3$ by the numerical method which was first applied by Segur and Kruskal \cite{SegurKruskal87} for the $\phi^4$ potential, which is not symmetric around its minimum. In the inner region we solve equations \eqref{eqphinner} for the functions $\Phi_n=\Psi_n+i\Omega_n$ by numerically integrating along a constant $\im y=y_i$ line, where $y_i<0$ real, starting from a large positive value of $\re y=y_r$. Since we are dealing with second order differential equations, as initial value at the point $y=y_r+i y_i$ we use the series \eqref{eqphinyexpsym} calculated up to some order in $1/y$, and also the derivative of the same truncated series expression. Reaching to the imaginary axis $\re y=0$ at the point $y=i y_i$, from the numerically obtained value of $\Omega_3$ there we calculate the coefficient $\nu_3$ using \eqref{e:om3}. Our experience shows that it is worthwhile to calculate \eqref{e:om3} even up to orders $1/\tilde y^{15}$, since by increasing the order of the expansion we can get more precise value for $\nu_3$ even when choosing a relatively small values of $|y_i|$.

This method is based on the assumption that the in the vicinity of the starting point $y=y_r+i y_i$ the series \eqref{eqphinyexpsym} gives a good approximation for $\Phi_n$. Since the series is not convergent, getting further away from the starting point the proper solution starts to become more and more different from the one represented by the series. This discrepancy can be found most easily on the imaginary axis, because there all terms of the series are real. Having a solution that satisfies these conditions in the inner domain, we can match it to a unique solution in the outer domain. However, we do not yet directly get the intended quasibreather solution with the minimal tail. Since according to the chosen initial condition, in the directions ${-\frac{\pi}{2}}<\arg y<0$ the solution is well approximated by the series \eqref{eqphinyexpsym},  the $\varepsilon$ expansion also remains valid for large $\re r$ values of the radial coordinate, hence there is no standing wave tail in any of the $\Phi_n$ components. In exchange for this, the functions $\Phi_n$ cannot be symmetric at the point $r=0$, and if we extend them to negative $r$ values, then there in the distant region a standing wave tail appears, with twice as large amplitude as the minimal one. We call the configurations determined by these type of functions \emph{asymmetric breathers}.

Extending the inner solution along the imaginary $y$ axis downwards and matching to the exterior solution, the asymmetry can be also seen well on the real $r$ axis, since according to the Cauchy-Riemann equations the derivative of the real part cannot be zero. In the paper of Segur and Kruskal \cite{SegurKruskal87} the tail of the asymmetric state vanishes in the directions $r<0$, while in the present review we assume that there is no tail for $r>0$. Because of the symmetry of the system the two choices are equivalent, but the generalization for higher dimensions will be more natural in this way. A choice for the value of the frequency $\omega$ uniquely determines an asymmetric breather, there is no need for the minimization of the tail in the negative $r$ direction. An asymmetric breather is really a breather in the sense that it is not radiating in the positive $r$ direction, however it does not satisfy the boundary condition required by the symmetry at $r=0$.

Naturally, the numerical solution of the differential equations can be calculated only, if we take into account a finite number of $\Phi_n$ functions. According to our experience, by keeping the Fourier modes $\Phi_1$, $\Phi_3$, $\Phi_5$, $\Phi_7$ we can already get extremely precise results. For the initial values of the functions and their derivatives we may easily calculate the expansion \eqref{eqphinyexpsym} even up to order $1/y^{20}$. For an appropriate result at least order $1/y^{7}$ is necessary, but order $1/y^{12}$ already gives high precision. The place of the constant $y_i$ was chosen in the interval $[-5,-22]$, but because of the exponential decay of $\Omega_n$, for $y_i<-9$ we needed to perform the computations using $32$ digit numbers instead of the usual $16$ digits. For the value $y_r$ giving the initial point on the line we used values in the interval $[50,500]$, but generally the choice $y_r=300$ turned out to be appropriate.

For a numerical calculation no parameter can remain in the theory, hence in the symmetric sixth order potential \eqref{eqsympotform} the fixing of $g_5$ is necessary. We present our results with the choice $g_5=1$. As we will see later, the radiation of the oscillon is maximal at a value of $g_5$ not far from $1$. Taking into account only the modes $\Phi_1$ and $\Phi_3$ we obtained the value $\nu_3=-0.910189$. Adding the mode $\Phi_5$ we obtained $\nu_3=-0.909789$, adding $\Phi_7$ we got $\nu_3=-0.90974966$, an also taking into account $\Phi_9$, the result was $\nu_3=-0.90974964$. Based on these, the proper value of $\nu_3$ for seven digits is
\begin{equation}
 \nu_3=-0.9097496 \ . \label{eqnu3g5e2}
\end{equation}
For the determination of the exceptionally small contribution on the imaginary axis the differential equations has to be integrated with extremely high precision. Our calculations were performed using the Maple software package.

The procedure for the calculation of the radiation rate from the value of $\nu_3$ presented in the rest of this subsection is valid for arbitrary symmetric potential. The imaginary part $\Omega_3$ given in equation \eqref{eqomnnunlead} gives a correction to the function $\Phi_3$ on the negative part of the imaginary $y$ axis,
\begin{equation}
 \delta\Phi_3^{(+)}=i\nu_3\exp(-i\sqrt{8}y) \ . \label{eqdeltaphi3}
\end{equation}
In the inner region this $\delta\Phi_3^{(+)}$ solves to leading order the equation that can be obtained by the linearization of \eqref{eqphinner}. Using the relation  \eqref{eqrykapcs} between the $y$ and $r$ coordinates, in the exterior domain this can be matched to the contribution
\begin{equation}
 \delta\Phi_3^{(+)}=\frac{\alpha}{2}i\exp(-i\sqrt{8}r) \ ,
\end{equation}
where we have introduced the notation
\begin{equation}
 \alpha=2\nu_3\exp\left(-\frac{\sqrt{8}\,P}{\varepsilon}\right) \ , \label{eqalphacorr}
\end{equation}
and $P=\frac{\pi}{2}$ is the distance of the singularity from the real axis. The parameter $\varepsilon$ is determined by the frequency of the state through the relation $\varepsilon=\sqrt{1-\omega^2}$. In the exterior domain the function $\delta\Phi_3^{(+)}$ is the solution of the linearization of \eqref{eqphouter} with $\omega=1$. In the $\varepsilon\to 0$ limit the equations remain regular, only the singularities of the studied solutions are getting further and further away from the real axis.

On the negative part of the imaginary $r$ axis, from the singularity at $r=-iP/\varepsilon$, a similar correction appears,
\begin{equation}
 \delta\Phi_3^{(-)}=-\frac{\alpha}{2}i\exp(i\sqrt{8}r) \ .
\end{equation}
The sum of these two contributions already takes real value on the real $r$ axis,
\begin{equation}
 \delta\Phi_3=\delta\Phi_3^{(+)}+\delta\Phi_3^{(-)}=\alpha\sin(\sqrt{8}r) \ . \label{eqdelphi3}
\end{equation}
Adding this $\delta\Phi_3$ contribution to the Fourier modes $\Phi_n$ calculated by the small-amplitude expansion, we obtain the asymmetric breather approximation, which has no standing wave tail for $r>0$, but it is not mirror symmetric at the center $r=0$. We emphasize that the contribution \eqref{eqdelphi3} is only valid in the inner part of the core, outside the core there is no such oscillating tail. Proceeding outwards,  because of the nonlinearity of the system, the existence of the small amplitude but large sized core is able to compensate and cancel this exponentially small contribution. In the core region this exponentially small oscillation cannot be observed on the order $\varepsilon$ amplitude scalar field $\phi$. However, if we look at the spatial derivative at the point $r=0$, this is the only term which gives a contribution and breaks the symmetry of the solution,
\begin{equation}
 \frac{\partial\phi}{\partial r}\biggr\vert_{r=0}=\alpha
 \sqrt{8}\cos\left(3t\right) \ . \label{e:orig}
\end{equation}
In the one spatial dimensional case we are looking for mirror symmetric solutions, ensuring comoving system and uniqueness. For more dimensions the mirror symmetry is the necessary condition for the regularity at the center.

If we consider perturbations around the just constructed asymmetric breather solution, then the  $\omega=3$ frequency component of the perturbation satisfies the left-hand side of equation \eqref{eqphouter}. The choice of the solution $\delta\Phi_3^{(s)}=-\alpha\sin(\sqrt{8}r)$ is necessary in order to compensate the asymmetry at the center. In contrast to $\delta\Phi_3$, this perturbation is valid in both the central and the tail regions. As a result of this, a standing wave tail $\delta\Phi_3^{(s)}$ appears in the exterior domain, with amplitude $-\alpha$. In this way we arrive at the intended quasibreather solution, which has the minimal amplitude tail. To this quasibreather we can add the function $\delta\Phi_3^{(c)}=\beta\cos(\sqrt{8}r)$ with arbitrary amplitude $\beta$, still obtaining symmetric weakly nonlocal states, with tail-amplitude $\sqrt{\alpha^2+\beta^2}$, which is larger than the minimal one.

Summarizing the results, we can state that for symmetric potentials the minimal amplitude tail of the quasibreather is given by
\begin{equation}
 \phi=-\alpha\sin(\sqrt{8}r)\cos(3t) \ ,
\end{equation}
where the amplitude $\alpha$ is determined by the expression \eqref{eqalphacorr} in terms of the constants $\nu_3$ and $P$. The important result is that the tail-amplitude $\alpha$ depends exponentially on the small-amplitude parameter $\varepsilon=\sqrt{1-\omega^2}$.

The constant $\alpha$ corresponds to the amplitude parameter in equation \eqref{eqphitailgen} that describes the general behavior of a standing wave tail. Hence the energy current averaged for an oscillation period, $\bar S$, can be calculated by applying \eqref{eqsaverage}. In the one-dimensional case, for symmetric potentials $\bar S=3\sqrt{8}\alpha^2$, and hence
\begin{equation}
 \bar S=24\sqrt{2}\,\nu_3^2\exp\left(-\frac{2\sqrt{2}\,\pi}{\varepsilon}\right) .
 \label{eqsbarnu3}
\end{equation}
For a symmetric sixth order potential with $g_5=1$ the coefficient in front of the exponential is $24\sqrt{2}\,\nu_3^2=28.0913$.

\subsection{Analytical method for the calculation of the radiation} \label{secborelsum}

For potentials that are symmetric around their minimum, the $\nu_3$ coefficient determining the magnitude of the exponential correction can also be calculated analytically, applying a method based on Borel summation. The constant $\nu_3$ is the coefficient of the correction \eqref{eqomnnunlead} to the power series expansion approximate solution \eqref{eqphinyexpsym} of the differential equations \eqref{eqphinner}. The behavior of the series \eqref{eqphinyexpsym} for large powers of $1/y$ will determine the correction that gives the imaginary part of $\Phi_3$ on the imaginary $y$ axis. The result obtained by the Borel summation will depend only on the asymptotic behavior of the coefficients $a^{(2n-1)}_{2k-1}$ for large $k$. Since the coefficients $a^{(2n-1)}_{2k-1}$ are purely imaginary, no individual term in the series gives any imaginary contribution on the imaginary axis. The result for the correction would be the same for any other series which has the same asymptotic behavior, even if the low order terms are completely different.

\subsubsection{Asymptotic behavior of the \texorpdfstring{$1/y$}{1/y} series}

For the easier understanding of the procedure, we give the form of the first three equations in \eqref{eqphinner}, keeping only those terms which are essential for our calculations:
\begin{align}
 \frac{\mathrm{d}^2\Phi_1}{\mathrm{d}y^2}&=
 \frac{g_3}{4}\left(3\Phi_1^3+3\Phi_1^2\Phi_3+6\Phi_1\Phi_3^2+6\Phi_1\Phi_3\Phi_5
 +6\Phi_1\Phi_5^2+3\Phi_3^2\Phi_5\right) \label{eqddphi1ddy}
 \ , \\
 \frac{\mathrm{d}^2\Phi_3}{\mathrm{d}y^2}+8\Phi_3&=
 \frac{g_3}{4}\left(\Phi_1^3+6\Phi_1^2\Phi_3+3\Phi_3^3+3\Phi_1^2\Phi_5
 +6\Phi_1\Phi_3\Phi_5+6\Phi_3\Phi_5^2\right)
 \ ,  \label{eqddphi3ddy}\\
 \frac{\mathrm{d}^2\Phi_5}{\mathrm{d}y^2}+24\Phi_5&=
 \frac{g_3}{4}\left(3\Phi_1^2\Phi_3+3\Phi_1\Phi_3^2+6\Phi_1^2\Phi_5
 +6\Phi_3^2\Phi_5+3\Phi_5^3\right) \label{eqddphi5ddy}
 \ .
\end{align}
The right-hand sides of the equations were calculated from the general expressions \eqref{eqfngnsum}, which determine the nonlinear terms denoted by $F_n$. Since they contain only higher powers of $1/y$ we have dropped the terms that contain $\Phi_7$ and higher Fourier components. We have also dropped the terms proportional to $g_5$, which are fifth order in the various $\Phi_k$ components. For the same reason, for a general symmetric potential it is not necessary to take into account $g_7$ and higher order $g_k$ coefficients in \eqref{eqddphi1ddy}-\eqref{eqddphi5ddy}. For the symmetric sixth order potential written in the form \eqref{eqsympotform} the coefficients are $g_3=-1$, $g_5$ arbitrary, and $g_k=0$ for $k\geq 7$.

Since for the presently considered symmetric potentials only the odd indexed Fourier components are nonzero, for the purely imaginary constants in the expansion \eqref{eqphinyexpsym} we introduce the notation $a^{(2n-1)}_{2k-1}=iA^{(n)}_{k}$, i.e.~we look for the solution in the form
\begin{equation}
 \Phi_{2n-1}=i\sum_{k=n}^\infty A^{(n)}_{k}\frac{1}{y^{2k-1}} \ , \label{eqphi2nm1exp}
\end{equation}
where $k\geq n$ integers. From the expansion \eqref{eqpphi1yexp} it can be seen that in case of $g_3=-1$ we have $A^{(1)}_{1}=-2\sqrt{2/3}$. The number $n$ in \eqref{eqphinner} corresponds to $2n-1$ now. Taking into account only the linear terms on the left-hand side, since $(2n-1)^2-1=4n(n-1)$, we obtain the equations
\begin{equation}
 (2k-3)(2k-2)A^{(n)}_{k-1}+4n(n-1)A^{(n)}_{k}=0 \ .
\end{equation}
For $n\geq 3$ the solutions are
\begin{equation}
 A^{(n)}_{k}=K_n(-1)^k\frac{(2k-2)!}{\left[4n(n-1)\right]^{k-1/2}} \ ,
\end{equation}
where $K_n$ are some constants. The factor $[4n(n-1)]^{-1/2}$ in the denominator was not absorbed into the constant $K_n$ in order to make the appearance of some future calculations simpler. It can be seen, that with the increase of $k$ the most quickly increasing coefficients are $A^{(2)}_{k}$, which coefficients belong to $\Phi_3$. Hence through the nonlinear terms these will determine the asymptotic behavior of the other coefficients as well.

To leading order the coefficients of $\Phi_1$ are determined by the equation
\begin{equation}
 (2k-3)(2k-2)A^{(1)}_{k-1}=-\frac{3g_3}{4}\left(A^{(1)}_{1}\right)^2 A^{(2)}_{k-1} \ ,
\end{equation}
where the righ-hand side comes from the second nonlinear term of \eqref{eqddphi1ddy}. All other terms give contributions which are smaller at least by a factor $1/k^2$. The solution which is valid for large values of $k$ is
\begin{equation}
 A^{(1)}_{k}=\frac{1}{2k^2}A^{(2)}_{k} \ . \label{eqak1ak2}
\end{equation}
It can be checked that this expression is consistent with the values of the explicitly calculated $A^{(n)}_{k}$ coefficients for large $k$. Increasing the order of the calculation gradually, the coefficients can be determined even up the order $1/y^{300}$ with some software that is capable to carry out algebraic manipulations.

The equation that determines the coefficients of the component $\Phi_5$ is
\begin{equation}
 (2k-3)(2k-2)A^{(3)}_{k-1}+24A^{(3)}_{k}=
 -\frac{3g_3}{4}\left(A^{(1)}_{1}\right)^2 A^{(2)}_{k-1} \ ,
\end{equation}
where the right-hand side expression comes from the first nonlinear term of \eqref{eqddphi5ddy}. The asymptotic solution is
\begin{equation}
 A^{(3)}_{k}=-\frac{1}{4k^2}A^{(2)}_{k} \ . \label{eqak3ak2}
\end{equation}
For the coefficients of the higher $\Phi_{2n-1}$ Fourier components it is also true that for $n\geq 3$ and for large $k$ the coefficients $A^{(n)}_{k}$ grow proportionally to the value of $A^{(2)}_{k}/k^{2n-4}$.

The behavior of the dominant $A^{(2)}_{k}$ can be more precisely determined if we also take into account the first nonlinear term,
\begin{equation}
 (2k-3)(2k-2)A^{(2)}_{k-1}+8A^{(2)}_{k}=
 -\frac{3g_3}{2}\left(A^{(1)}_{1}\right)^2 A^{(2)}_{k-1} \ .
\end{equation}
From this it follows that
\begin{equation}
 A^{(2)}_{k}=K_2(-1)^k\frac{(2k-2)!}{8^{k-1/2}}
 \left(1+\frac{1}{k}+\frac{5}{4k^2}\right) \ , \label{eqa2kprec}
\end{equation}
apart from corrections of order $1/k^3$ and higher. Using this identity we can calculate more precisely the value of the constant $K_2$, even from not too high order $A^{(2)}_{k}$ coefficients. For example, for the sixth order symmetric potential \eqref{eqsympotform}, in case of $g_5=1$,
\begin{equation}
 K_2=0.5791646 \ . \label{eqkk2g5e1}
\end{equation}
The value of the constant $K_2$ is important, because as we will see soon, the simple formula $\nu_3={-K_2\pi/2}$ connects it with the coefficient $\nu_3$ defined in the previous subsection, which determines the radiation loss rate.

\subsubsection{Borel summation}

We intend to Borel sum the non-convergent power series
\begin{equation}
 \Phi_{3}(y)=i\sum_{k=2}^\infty A^{(2)}_{k}\frac{1}{y^{2k-1}} \ , \label{eqphi3ysum}
\end{equation}
which belongs to the Fourier component $\Phi_3$ that gives the dominant radiating tail. Based on \eqref{eqa2kprec}, we assume now that the expansion coefficients are defined by the expression
\begin{equation}
 A^{(2)}_{k}=K_2(-1)^k\frac{(2k-2)!}{8^{k-1/2}} \ , \label{eqa2kcoeff}
\end{equation}
where $K_2$ is a real constant. We consider these as the exact coefficients of a series which has the same asymptotic behavior as the original one. The resulting exponential correction only depends on the behavior of the coefficients for large $k$, since each individual term of the expansion gives only real contribution on the imaginary $y$ axis. The coefficients of $\Phi_{3}(y)$ diverge according to $(2k-2)!$. 

We define the transformed $\widehat\Phi_{3}(z)$ function by the following summation:
\begin{equation}
 \widehat\Phi_{3}(z)=i\sum_{k=2}^\infty \frac{A^{(2)}_{k}}{(2k-1)!}{z^{2k-1}} \ .
\end{equation}
Using the integral representation of the factorial function,
\begin{align}
 \Phi_{3}(y)&=i\sum_{k=2}^\infty \frac{A^{(2)}_{k}}{y^{2k-1}(2k-1)!}
 \int_{0}^\infty\mathrm{d}t\, e^{-t}t^{2k-1} \label{eqphi3yint} \\
 &=\int_{0}^\infty\mathrm{d}t\, e^{-t}\, i \sum_{k=2}^\infty \frac{A^{(2)}_{k}}{(2k-1)!}
 \left(\frac{t}{y}\right)^{2k-1}
 =\int_{0}^\infty\mathrm{d}t\, e^{-t}\, 
 \widehat\Phi_{3}\left(\frac{t}{y}\right) \ . \notag
\end{align}
If the summations and integrals can be appropriately defined, then the Borel summed $\Phi_{3}(y)$ function can be calculated by this method. This way, in certain cases we can get a concrete result for $\Phi_{3}(y)$ even if the original \eqref{eqphi3ysum} series is not convergent. If the coefficients $A^{(2)}_{k}$ are given by \eqref{eqa2kcoeff}, then the series $\widehat\Phi_{3}(z)$ can be summed,
\begin{equation}
 \widehat\Phi_{3}(z)=i\sum_{k=2}^\infty K_2
 \frac{(-1)^k}{2k-1}{\left(\frac{z}{\sqrt{8}}\right)^{2k-1}}
 =\frac{K_2}{2}\left[\ln\left(1-\frac{iz}{\sqrt{8}}\right)
 -\ln\left(1+\frac{iz}{\sqrt{8}}\right)\right] \ . \label{eqphi3tildez}
\end{equation}

Introducing the notation $y=-i\tilde y$, since $z=t/y$ in \eqref{eqphi3yint},
\begin{equation}
 \Phi_{3}(y)=\frac{K_2}{2}\int_{0}^\infty\mathrm{d}t\, e^{-t}
 \left[\ln\left(1+\frac{t}{\sqrt{8}\,\tilde y}\right)
 -\ln\left(1-\frac{t}{\sqrt{8}\,\tilde y}\right)\right] . \label{eqphi3intform}
\end{equation}
On the negative part of the imaginary axis $\tilde y$ is real and positive. Because of this, the argument of the first logarithm is positive, and the integral of the first term is real. The second term has a singularity at the place $t=\sqrt{8}\,\tilde y$, and for $t>\sqrt{8}\,\tilde y$ the value of the logarithm becomes complex. The matching of the exterior and interior domains are performed in the directions ${-\frac{\pi}{2}}<\arg y<0$. In this domain $y$ is of the form $y=a-ib$, where both $a$ and $b$ are positive real numbers. Then $\tilde y=i y={b+ia}$, and hence the singularity $t=\sqrt{8}\,\tilde y$ is located above the real axis. If we approach the negative part of the imaginary $y$ axis by tending to zero by $a$, then the singularity moves to the real $t$ axis from above, and hence in the second term of \eqref{eqphi3intform} we must go around the singularity from below. Since the singularity is only logarithmic, if we tend to zero with the radius of the circle around the singularity the contribution to the integral also approaches zero. The $t$ dependent term in the argument of the logarithm has a negative sign, hence if we go around the singularity from below, then we approach the cut of the logarithm on the negative real axis from above. As a consequence, for $t>\sqrt{8}\,\tilde y$ the imaginary part of the logarithm is $\pi$. Hence on the negative part of the imaginary $y$ axis, i.e.~for $\re y=0$ and $\im y<0$, we can write
\begin{equation}
 \im\Phi_{3}(y)=-\frac{K_2\pi}{2}\int_{\sqrt{8}\,\tilde y}^\infty\mathrm{d}t\, e^{-t}
 =-\frac{K_2\pi}{2}\exp\left(-\sqrt{8}\,\tilde y\right) \ .
\end{equation}
Hence
\begin{equation}
 \im\Phi_{3}(y)=-\frac{K_2\pi}{2}\exp\left(-i\sqrt{8}\, y\right) \ . \label{eqimphi3y}
\end{equation}
The reason for the signature difference with respect to our paper \cite{Fodor2009a} is that there we have chosen the matching region in the directions ${-\pi}<\arg y<{-\frac{\pi}{2}}$.

After integration by parts, the integrals in expression \eqref{eqphi3intform} can be written in terms of the exponential integral function $E_1$ (see \cite{dlmflibrary}),
\begin{equation}
 \int\mathrm{d}t\, e^{-t}\ln(1+\alpha t)=-e^{-t}\ln(1+\alpha t)
 -e^{\frac{1}{\alpha}}E_1\left(t+\frac{1}{\alpha}\right) .
\end{equation}
The function $E_1$ can be written in terms of the exponential integral $\mathrm{Ein}$, which is an entire function, i.e.~holomorphic on all finite points of the complex plane,
\begin{equation}
 E_1(z)=\mathrm{Ein}(z)-\ln(z)-\gamma \ ,
\end{equation}
where $\gamma\approx 0.57721$ is the Euler–Mascheroni constant. Since the entire function $\mathrm{Ein}$ takes real values on the real axis, the cut and the imaginary part of the function $E_1$ on the negative real axis is determined by the complex logarithm function. The result \eqref{eqimphi3y} for the imaginary part of $\Phi_{3}(y)$ on the negative imaginary axis can also be obtained in this way. The whole $\Phi_{3}(y)$ function can also be written in terms of exponential integral functions, but the significance of this is restricted by the fact that the expressions \eqref{eqa2kcoeff}, giving the coefficients $A^{(2)}_{k}$, are only valid asymptotically.

Comparing the equations \eqref{eqimphi3y} and \eqref{eqdeltaphi3} obtained for the exponentially decaying imaginary part of $\Phi_3$, we can get the important relation between the constants $\nu_3$ and $K_2$,
\begin{equation}
 \nu_3=-\frac{K_2\pi}{2} \ . \label{eqnu3kk2}
\end{equation}
This relation is valid for arbitrary symmetric potentials. In case of the sixth order symmetric potential given in \eqref{eqsympotform} and for the choice $g_5=1$, the value of $K_2$ given in \eqref{eqkk2g5e1} corresponds to $\nu_3=-0.9097496$, which agrees to all the given digits with the value written earlier in \eqref{eqnu3g5e2}. The precise value of the radiation rate is obviously not important to so many digits, but the high degree agreement of the results obtained by the two independent calculations gives a strong support for the correctness of the applied methods.

\begin{figure}[!htb]
\centering
\includegraphics[width=110mm]{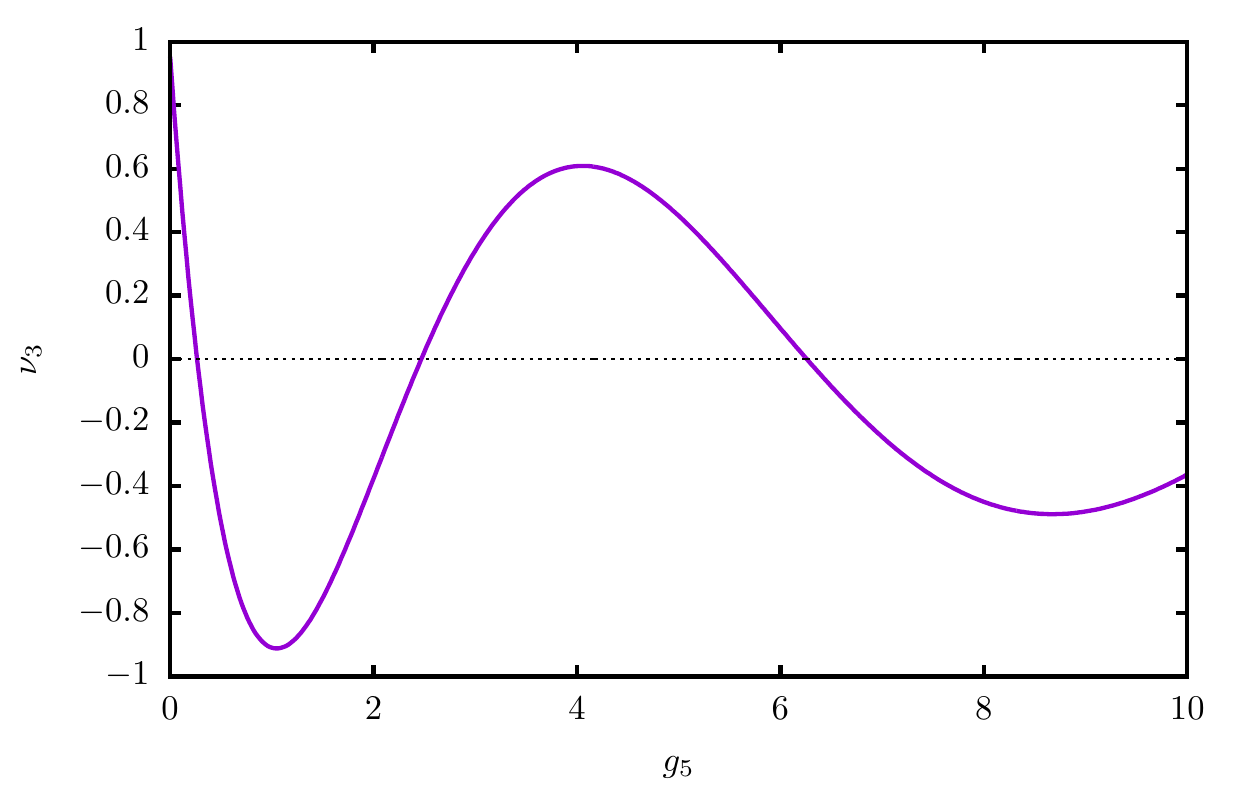}
\caption{\label{fignu3p}
The coefficient $\nu_3$ as a function of the parameter $g_5$ for the symmetric sixth order potential \eqref{eqsympotform}.}
\end{figure}

The calculation of the radiation amplitude from the value of $K_2$ is technically much easier, hence we have used this method to calculate the coefficient $\nu_3$ for other sixth order potentials with different $g_5$ values. On Figure \ref{fignu3p}
we show the dependence of $\nu_3$ on the parameter $g_5$. The first minimum of the function is at $g_5=1.0479$, with value $\nu_3=-0.91202$.

The first place where $\nu_3=0$ is at $g_5=0.26495$, which is quite close to the value $g_5=3/10$ belonging to the sine-Gordon potential, but does not agree with it. From the expansion \eqref{eqsinegoexporig} of the sine-Gordon potential $U(\phi)=1-\cos\phi$, it can be seen that for this potential $g_3=-1/6$. By rescaling the coordinates and the scalar $\phi$ it is possible to set $g_3=-1$, and then the sine-Gordon potential has the form $U(\phi)=\left[1-\cos(\sqrt{6}\,\phi)\right]/6$, and $g_5=3/10$. The value of the curve on Fig.~\ref{fignu3p} at $g_5=3/10$ is $\nu_3=-0.0944$. If we take into account that for the sine-Gordon potential $g_7$ and the higher coefficients are also nonzero, then increasing the order of the expansion of the potential we get smaller and smaller values for $\nu_3$, according to Table \ref{tablenu3sineg}. This is consistent with the fact that in case of the $1+1$ dimensional sine-Gordon breather there is no radiation loss at all.
\begin{table}[!htb]
\centering
\begin{tabular}{r|r @{$\cdot$} l|}
$N_U$ & \multicolumn{2}{c|}{$\nu_3$} \\
\hline
\hline
$6$  & $-9.4$ & $10^{-2}$  \\
\hline
$8$  & $3.4$ & $10^{-3}$  \\
\hline
$10$  & $-4.7$ & $10^{-5}$  \\
\hline
$12$  & $3.0$ & $10^{-7}$  \\
\hline
$14$  & $-1.0$ & $10^{-9}$  \\
\hline
$16$  & $2.0$ & $10^{-12}$  \\
\hline
$18$  & $-2.5$ & $10^{-15}$  \\
\hline
\end{tabular}
\caption{\label{tablenu3sineg}
The value of the coefficient $\nu_3$ for the $U(\phi)=\left[1-\cos(\sqrt{6}\,\phi)\right]/6$ sine-Gordon potential, when the expansion of the potential with respect to $\phi$ is taken into account up to order $\phi^{N_U}$. The results have been obtained by the Borel summation method. We had to use the $\Phi_n$ Fourier components at least up to order $N_U-2$ to obtain the values in the table.
}
\end{table}

\subsection{Non-symmetric potentials} \label{subsecnonsympot}

In case of $d=1$ spatial dimension, if the potential $U(\phi)$ is not symmetric around its minimum, we can still calculate the magnitude of the radiation tail by a method similar to the one presented up to here. In the Fourier expansion \eqref{eqphifourier} of the scalar field there are both even and odd indexed $\Phi_n$ terms in this case. We extend the functions and the equations into the complex $r$ plane, and in the vicinity of the singularity that is closest to the real axis we introduce the coordinate $y$ by the relation \eqref{eqrykapcs}, where $P=\frac{\pi}{2}$ in the one-dimensional case. In the outer region we still solve equations \eqref{eqphouter}, while in the inner region we solve \eqref{eqphinner}, where $F_n$ are given by \eqref{eqfngnsum} in both cases.

Based on the results of the small-amplitude expansion, we look for the solution of the inner equations in the form
\begin{equation}
 \Phi_{n}=\sum_{k=0}^\infty a^{(n)}_{n+2k}\,\frac{1}{y^{n+2k}} \ . \label{eqphinyexpnons}
\end{equation}
From the $\varepsilon$ expansion it also follows that $a^{(0)}_{0}=0$, which means that the expansion of $\Phi_0$ starts with a $1/y^2$ term, while the expansion of all the other $\Phi_n$ starts with a $1/y^n$ term. The power of $y$ increases in steps of two. The coefficients $a^{(n)}_{k}$ follow unambiguously from the $\varepsilon$ expansion, but technically it is much easier to determine the coefficients by the direct substitution of the expansion \eqref{eqphinyexpnons} into the inner equations \eqref{eqphinner}, and by the solution of the equations obtained for the different powers of $y$ for the coefficients $a^{(n)}_{k}$. The calculations can be performed by some algebraic manipulation software. Apart from the signature of $a^{(1)}_{1}$ the solution is unique. The coefficient which corresponds to the choice of the signature of $S$ in \eqref{seqdegysol} is $a^{(1)}_{1}=-i\sqrt{2/\lambda}$, where $\lambda$ is the positive constant defined in \eqref{lambdadef}. The coefficients $a^{(2n)}_{2k}$ belonging to the Fourier components $\Phi_{2n}$ are always purely real, while the coefficients $a^{(2n+1)}_{2k+1}$ belonging to $\Phi_{2n+1}$ are purely imaginary. 

We note that if we allow all even and odd negative exponents of $y$ for all $\Phi_n$ in \eqref{eqphinyexpnons}, starting with $1/y$ in every Fourier component, even then it follows from the solution of the inner equations that the first term of $\Phi_0$ is proportional to $1/y^2$, while all the other $\Phi_n$ start with $1/y^n$. Furthermore if we choose the coefficient of $1/y^2$ in $\Phi_1$ to be zero, then in the expansion of all $\Phi_n$ the powers of $y$ grow in steps of two.

In case of a general potential the first few terms of the expansion \eqref{eqphinyexpnons},  solving the inner equations, are:
\begin{align}
 \Phi_0&=\frac{g_2}{\lambda}\,\frac{1}{y^2}
 +\frac{169g_2^3\lambda+144g_2\lambda^2+72g_2\sigma-81g_4\lambda}{54\lambda^3}\,\frac{1}{y^4}
 +\ldots \ ,\nonumber\\
 \Phi_1&=-\frac{i\sqrt{2}}{\sqrt\lambda}\,\frac{1}{y}
 -\frac{i \sqrt2(19g_2^2\lambda+18\sigma)}{27\lambda^2\sqrt\lambda}\,\frac{1}{y^3}
 +\ldots \ ,  \nonumber\\
 \Phi_2&=-\frac{g_2}{3\lambda}\,\frac{1}{y^2}
 -\frac{220g_2^3\lambda-243g_2\lambda^2+72g_2\sigma-108g_4\lambda}{162\lambda^3}\,\frac{1}{y^4}
 +\ldots \ ,\label{e:1match}\\
 \Phi_3&=\frac{i \sqrt2\,(4g_2^2-3\lambda)}{36\lambda\sqrt\lambda}\,
 \frac{1}{y^3}+\dots \ ,\nonumber\\
 \Phi_4&=\frac{5g_2^3-5g_2\lambda+3g_4}{90\lambda^{2}}\,
 \frac{1}{y^4}+\dots \ ,\nonumber
\end{align}
where the definition of the constant $\lambda$ is in \eqref{lambdadef}, and the constant $\sigma$ is given in \eqref{eqsigmadef}.

In the same way as in \eqref{eqphinpsinomn}, we decompose the Fourier components into real and imaginary parts, $\Phi_n=\Psi_n+i\Omega_n$. Neglecting the nonlinear terms, the imaginary parts satisfy the differential equations \eqref{omegandiffeq}, where $\tilde y=iy$. The solution for the zeroth component is $\Omega_0=c_0\exp(\pm i\tilde y)$, which only tends to zero when going down along the imaginary axis if $c_0=0$. Appropriate solutions exist for $n\geq 2$, which can still be written in the form \eqref{eqomnnunlead}. From the terms of the nonlinear source function $F_n$ given by \eqref{eqfngnsum} we only keep those that are linear in the variables $\Omega_k$. For the functions $\Psi_n$ we substitute the values $\Phi_n$ given by the expansion \eqref{eqphinyexpnons} calculated up to some order, which give real values on the imaginary axis. The asymptotic behavior of the functions $\Omega_n$ is determined by the less slowly decaying exponential contribution, which for non-symmetric potentials is usually generated by $\Omega_2$, and hence this component determines the radiation as well. Solving the coupled linear system, using the coordinate $\tilde y=iy$, we obtain
\begin{align}
 \Omega_0&=\nu_2\,\exp\left(-\sqrt{3}\,\tilde y\right)
 \left(-\frac{2g_2^2-3\lambda}{6\lambda\tilde y^2}
 +\frac{\sqrt{3}(16g_2^4-24g_2^2\lambda-45\lambda^2)}{270\lambda^2\tilde y^3}
 +\ldots\right) \ , \label{eqoom0as} \\
 \Omega_1&=\nu_2\,\exp\left(-\sqrt{3}\,\tilde y\right)
 \left(-\frac{\sqrt{2}g_2}{3\sqrt{\lambda}\tilde y}
 +\frac{8\sqrt{6}g_2^3}{135\sqrt{\lambda^3}\tilde y^2}
 +\ldots\right) \ , \\
 \Omega_2&=\nu_2\,\exp\left(-\sqrt{3}\,\tilde y\right)
 \left(1-\frac{2\sqrt{3}(4g_2^2-15\lambda)}{45\lambda\tilde y}
 +\frac{32g_2^4-282g_2^2\lambda+255\lambda^2}{675\lambda^2\tilde y^2}
 +\ldots\right) \ , \label{e:om2} \\
 \Omega_3&=\nu_2\,\exp\left(-\sqrt{3}\,\tilde y\right)
 \left(\frac{\sqrt{2}g_2}{5\sqrt{\lambda}\tilde y}
 -\frac{8\sqrt{6}g_2(g_2^2-6\lambda)}{225\sqrt{\lambda^3}\tilde y^2}
 +\ldots\right) \ , \\
 \Omega_4&=\nu_2\,\exp\left(-\sqrt{3}\,\tilde y\right)
 \left(\frac{6g_2^2-5\lambda}{30\lambda\tilde y^2}
 +\ldots\right) \ . 
\end{align}
The back-reaction of the $1/\tilde y$ terms generated in $\Omega_1$ and $\Omega_3$ by the leading order term of $\Omega_2$ modifies the second term of $\Omega_2$. The numerical result for $\nu_2$ published in our paper \cite{Fodor2009a} have been calculated without taking into account this back-reaction.

We follow the method which was presented in detail earlier for symmetric potentials. The correction represented by $\Omega_2$ has the form
\begin{equation}
 \delta\Phi_2^{(+)}=i\nu_2\exp(-i\sqrt{3}y) \ . \label{eqdeltaphi2}
\end{equation}
We match this to the correction in the outer region, and then add the similar correction arising from the singularity on the other side of the real axis. On the real axis, inside the core region, similarly to \eqref{eqdelphi3}, we obtain the correction
\begin{equation}
 \delta\Phi_2=\alpha\sin(\sqrt{3}r) \ ,  \label{eqdelphi2}
\end{equation}
where
\begin{equation}
 \alpha=2\nu_2\exp\left(-\frac{\sqrt{3}\,P}{\varepsilon}\right) \ , \label{eqalphcorasym}
\end{equation}
$\varepsilon=\sqrt{1-\omega^2}$, and now in the one-dimensional case $P=\frac{\pi}{2}$. 

In this way we obtain the asymmetric breather, which has no standing wave tail in the positive $r$ direction. In order to compensate the asymmetry in the center $r=0$, and to obtain the quasibreather, we must add to the asymmetric breather the solution $\delta\Phi_2^{(s)}={-\alpha\sin(\sqrt{3}r)}$ of the linearized equations. This determines the minimal amplitude tail of the quasibreather, which oscillates according to $\cos(2t)$. The parameter $\alpha$ defined in \eqref{eqalphcorasym} corresponds to the amplitude parameter in equation \eqref{eqphitailgen}, which describes a general tail. Hence the radiated energy current averaged to an oscillation period, $\bar S$, can be calculated using \eqref{eqsaverage}. In the one-dimensional case, for a not symmetric potential, $\bar S=2\sqrt{3}\alpha^2$, hence
\begin{equation}
 \bar S=8\sqrt{3}\,\nu_2^2\exp\left(-\frac{\sqrt{3}\,\pi}{\varepsilon}\right) .
 \label{eqsbarnu2}
\end{equation}

The coefficient $\nu_2$ that determines the rate of the radiation was calculated first numerically by Segur and Kruskal \cite{SegurKruskal87} for the case of the $\phi^4$ potential, which is not symmetric around its minimum. We have repeated the calculation by taking into account more Fourier modes in our article \cite{Fodor2009a}. We take the potential in the form \eqref{eqpotone}, in which case $m=1$ and the nonvanishing expansion coefficients are $g_2=-3/2$ and $g_3=1/2$. The starting terms of the expansion \eqref{e:1match}, up to order $1/y^8$ are
\begin{align}
 \Phi_0&=-\frac{1}{y^2}-\frac{395}{36y^4}-\frac{19357}{72y^6}
 -\frac{2862666359}{233280y^8}+\cdots \ , \nonumber\\
 \Phi_1&=-\frac{i}{\sqrt{3}}\left(\frac{2}{y}+\frac{103}{18y^3}
 +\frac{19661}{324y^5}+\frac{169114187}{116640y^7}+\cdots\right) \ , \nonumber\\
 \Phi_2&=\frac{1}{3}\left(\frac{1}{y^2}+\frac{47}{9y^4}
 +\frac{26765}{432y^6}+\frac{21535799}{14580y^8}+\cdots\right) \ , \nonumber\\
 \Phi_3&=\frac{i}{6\sqrt{3}}\left(\frac{1}{y^3}+\frac{91}{12y^5}
 +\frac{42767}{432y^7}+\cdots\right) \ , \nonumber\\
 \Phi_4&=-\frac{1}{36}\left(\frac{1}{y^4}+\frac{179}{18y^6}
 +\frac{45883}{324y^8}+\cdots\right) \ , \label{eqphi4yexp}\\
 \Phi_5&=-\frac{i}{72\sqrt{3}}\left(\frac{1}{y^5}+\frac{443}{36y^7}
 +\cdots\right) \ , \nonumber\\
 \Phi_6&=\frac{1}{434}\left(\frac{1}{y^6}+\frac{44}{3y^8}
 +\cdots\right) \ , \nonumber\\
 \Phi_7&=\frac{i}{864\sqrt{3}}\left(\frac{1}{y^7}+\cdots\right) \ , \nonumber\\
 \Phi_8&=-\frac{1}{5184}\left(\frac{1}{y^8}+\cdots\right) \ . \nonumber
\end{align}
For the $\phi^4$ potential the terms of the expansion of the correction \eqref{e:om2}, which appears as the imaginary part of $\Phi_2$, are
\begin{align}
 \Omega_2=\nu_2\,\exp\left(-\sqrt{3}\,
 \tilde y\right)\biggr(&1+\frac{2\sqrt{3}}{5\tilde y}
 -\frac{14}{75\tilde y^2}
 -\frac{43\sqrt{3}}{25\tilde y^3}
 +\frac{4747}{2700\tilde y^4}
 +\frac{3493741}{45000\sqrt{3}\tilde y^5} 
 -\frac{29006059}{675000\tilde y^6} \nonumber\\
 &\left.
 -\frac{1072856045789}{425250000\sqrt{3}\tilde y^7}
 +\frac{15835359963361}{7654500000\tilde y^8}
 +\cdots\right) \ . \label{eqom2phi4}
\end{align}

Apart from a technical difficulty, the method for the numerical determination of the coefficient $\nu_2$ is similar to the technique that was described in detail in the paragraphs after equation \eqref{e:om7}. We solve numerically the differential equations \eqref{eqphinner} for the complex functions $\Phi_n=\Psi_n+i\Omega_n$ along a constant $\im y=y_i$ line, starting from a large value of $\re y=y_r$, until we reach the imaginary axis $\re y=0$. As initial value, we would like to use at the point $y=y_r+iy_i$ the series \eqref{eqphi4yexp} calculated up to a chosen $1/y^n$ order, and its derivative. Reaching the imaginary axis $\re y=0$ at the point $y=iy_i$, from the value of $\Omega_2$ we intend to calculate the coefficient $\nu_2$ using \eqref{eqom2phi4}. Unfortunately, the determination of the numerical solution in this way, as an initial value problem from the point $y=y_r+iy_i$ is problematic, because of the exponential behavior of $\Phi_0$. We were not able to perform the integration up to the imaginary axis in a stable and convergent way. Because of this, following the method of Segur and Kruskal \cite{SegurKruskal87}, we treat the equations for the real and imaginary part of $\Phi_0$ as boundary value problems determined at the two endpoints. Using the series \eqref{eqphi4yexp}, at the point $y=y_r+iy_i$ we only fix the value of $\Phi_0=\Psi_0+i\Omega_0$, but not its derivative. We exchange the missing two requirements by the boundary conditions $\Omega_0=0$ and $\frac{\mathrm{d}}{\mathrm{d}y}\Psi_0=0$ at the point $y=iy_i$ on the imaginary axis. For the other $\Phi_n$ components we still give their values and derivatives at the point $y=y_r+iy_i$. 

After the numerical determination of the solution we can check that the derivatives of the real and imaginary parts of $\Phi_0$ really agree to high precision with the value given by the series \eqref{eqphi4yexp} at the point $y=y_r+iy_i$. For the intended solution the value of $\Omega_0$ at the imaginary axis should be very close to zero. This follows from the observation that the solution $\Omega_0=c_0\exp(\pm i\tilde y)$ of the differential equation \eqref{omegandiffeq} can only tend to zero going down along the imaginary axis if $c_0=0$. We note that for the $\phi^4$ potential the leading $1/\tilde y^2$ order term of $\Omega_0$ in equation \eqref{eqoom0as} is zero.

\begin{figure}[!htb]
\centering
\includegraphics[width=115mm]{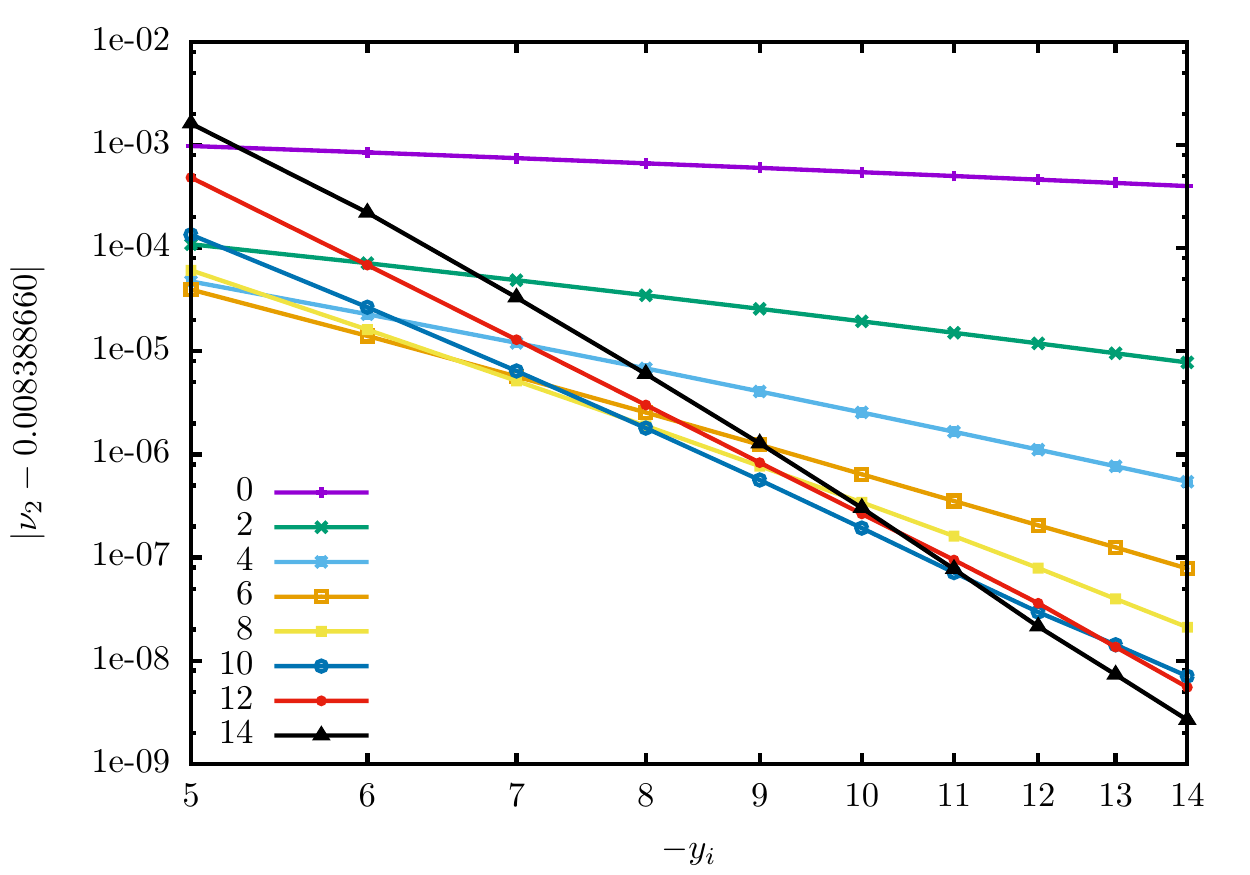}
\caption{\label{fignu2diff}
Logarithmic plot, which shows how $\nu_2$ tends to the value $8.388660\cdot10^{-3}$ with the increase of $|y_i|$, using the modes $\Phi_0$ ... $\Phi_7$. 
For the points indexed with the number $n$, from $0$ to $14$, we took into account the terms in the expression \eqref{eqom2phi4} up to order $1/\tilde y^n$ when calculating $\nu_2$.}
\end{figure}

For our numerical calculations we have used the Maple software package to solve the coupled differential equations as boundary value problems. The position of the constant $\im y=y_i$ line was chosen in the interval $[{-5},{-22}]$, while the starting point $y_r$ was in the interval $[50,500]$. If we chose the constant $y_i$ line below $-13$, then we had to use more than $16$ digits for the calculations, which caused considerable slowdown. The choice $y_r=300$ was appropriate in general. We gave the boundary conditions at the initial point $y=y_r+iy_i$ by the calculation of the expansion \eqref{eqphi4yexp} up to order $1/y^{15}$, although order $1/y^{9}$ already gives three digits precision. In Table \ref{t:n2dep}
\begin{table}[htbp]
\centering
\begin{tabular}{|c|c|}
\hline
$n$ & $\nu_2$\\
\hline
$2$  &  $-8.27\cdot10^{-3}$
\\ 
$3$  &  $7.061\cdot10^{-3}$
\\ 
$4$  &  $8.365\cdot10^{-3}$
\\ 
$5$  &  $8.3885\cdot10^{-3}$
\\ 
$6$  &  $8.3886\cdot10^{-3}$
\\ 
$7$  &  $8.38866\cdot10^{-3}$
\\ \hline
\end{tabular}
\caption{\label{t:n2dep}
The value of the coefficient $\nu_2$ obtained numerically by the use of $n+1$ Fourier modes, $\Phi_0$ ... $\Phi_n$.
}
\end{table}
we show how the radiation coefficient $\nu_2$ changes with the increase of the number of the used $\Phi_n$ Fourier modes. On Figure \ref{fignu2diff}
we show the convergence of $\nu_2$, as a function of the position $y_i$ of the integration line, using various order approximations of equation \eqref{eqom2phi4}, employing $8$ Fourier modes. 

Summing up the results, for the $\phi^4$ potential the value of the radiation coefficient $\nu_2$ is
\begin{equation}
 \nu_2=8.38866\cdot10^{-3} \pm 1\cdot10^{-7} \ . \label{eqnu2phi4}
\end{equation}
The value given here is much more precise than the value which was published earlier in our paper \cite{Fodor2009a}. The reason for this is that at the calculation of the imaginary part $\Omega_2$ we now take into account the back-reaction of the terms generated in the other $\Omega_k$ functions with the same exponential decay. This allows us to apply the high order approximation \eqref{eqom2phi4} to calculate the coefficient $\nu_2$ very precisely even when $|y_i|$ not too large. The signature difference in $\nu_2$ with respect to paper \cite{Fodor2009a} is because in the present review we assume that the asymmetric breather has no tail in the positive $r$ directions. The definition of the constant $\nu_2$ in the paper of Segur and Kruskal \cite{SegurKruskal87} differs in a factor of two from ours, because of the complex description of the Fourier modes there. In our notation their result is $\nu_2=(9.0\pm 2.0)\cdot10^{-3}$.

\subsection{Radiation of spherically symmetric oscillons} \label{secgomboscsug}

The radiation rate of higher dimensional spherically symmetric oscillon states can be calculated similarly to the one-dimensional case, but the procedure becomes technically more complicated. In the following we present the results published in our paper \cite{Fodor2009b}. We look for time-periodic quasibreather solutions with frequency $\omega$, and we intend to determine their tail-amplitude. The components $\Phi_n$ of the Fourier expansion \eqref{eqphifourier} satisfy the equations \eqref{eqphinsph}, which for a scalar field with mass $m=1$ can be written as
\begin{equation}
 \frac{\mathrm{d}^2\Phi_n}{\mathrm{d}r^2}
 +\frac{d-1}{r}\frac{\mathrm{d}\Phi_n}{\mathrm{d}r}
 +(n^2\omega^2-1)\Phi_n=F_n \ . \label{eqnonsphout}
\end{equation}
The $F_n$ expressions on the right-hand side consist of various order products of $\Phi_n$, according to \eqref{eqfngnsum}. The frequency $\omega$ is still related to the amplitude parameter $\varepsilon$ by the relation $\varepsilon^2=1-\omega^2$. We extend the equations \eqref{eqnonsphout} into the complex $r$ plane, and consider the points on the upper half of the imaginary $r$ axis where the solutions $\Phi_n$ are singular. For small values of the $\varepsilon$ parameter we investigate the behavior of the solutions in the vicinity of the singularity which is closest to the real axis. In order to achieve this, we first need to study the small-amplitude expansion results presented in Section \ref{seckisampl} for arbitrary dimensions.

\subsubsection{Small-amplitude expanson results near the singularity}

Based on the expressions \eqref{e:phi3b1}-\eqref{e:phi3b4}, which summarize the results of the first four orders of the $\varepsilon$ expansion, the first few terms of the Fourier components are
\begin{align}
 \Phi_0=&-\varepsilon^2\frac{1}{2} g_2p_1^2 \label{eqphi4fou0}\\
  &+\varepsilon^4\left[
  -g_2p_1^2-g_2p_1p_3+\frac{1}{6}g_2\lambda p_1^4
  -g_2\left(\frac{\mathrm{d}p_1}{\mathrm{d}\rho}\right)^2
  +\frac{31}{72}g_2^3p_1^4-\frac{3}{8}g_4p_1^4
 \right]+\cdots \ , \nonumber\\
 \Phi_1=&\varepsilon p_1+\varepsilon^3 p_3+\cdots \ ,  \\
 \Phi_2=&\varepsilon^2\frac{1}{6} g_2p_1^2 \\
 &-\varepsilon^4\frac{1}{72}\left[8g_2\left(\frac{\mathrm{d}p_1}{\mathrm{d}\rho}\right)^2
 -12g_4p_1^4+16g_2^3p_1^4
 -24g_2p_1p_3-23g_2\lambda p_1^4-8g_2p_1^2\right]
 +\cdots \ ,  \nonumber\\
 \Phi_3=&\varepsilon^3\frac{1}{72}(4g_2^2-3\lambda)p_1^3+\cdots \ , \\
 \Phi_4=&\varepsilon^4\frac{1}{360}p_1^4\left(3g_4-5g_2\lambda+5g_2^3\right)
 +\cdots \ . \label{eqphi4fou4}
\end{align}
The function $p_1$ is connected to $S$ by the relation $S=\sqrt{\lambda}\,p_1$. The function $S$ is determined by the base equation \eqref{seqlabel}, where $\tilde\Delta$ is the Laplacian belonging to the coordinate $\rho=\varepsilon r$, given by \eqref{eqlapsphrho}. The knowledge of the solution $S$ determines the Fourier components $\Phi_n$ up to order $\varepsilon^2$.

The important difference with respect to the one-dimensional case is that for higher dimensions the asymptotically decaying regular, nodeless $S$ solution is only known in a numerical form. As we have seen, for one spatial dimension $S=\sqrt{2}\sech\rho$. From the study of the base equation it also follows that small-amplitude oscillons exist only for $d\leq3$ spatial dimensions. The numerical value of the solution $S$ at the central point was given in \eqref{eqscentr} for $d=2$ and $d=3$ spatial dimensions. Introducing the coordinate $\bar\rho=-i\rho$, by numerical integration of the equation
\begin{equation}
 -\frac{\mathrm{d}^2S}{\mathrm{d}\bar\rho^2}
 -\frac{d-1}{\bar\rho}\frac{\mathrm{d}S}{\mathrm{d}\bar\rho}
 -S+S^3=0
\end{equation}
we can determine the behavior of $S$ on the imaginary axis. Similarly to the one-dimensional case, the singularities which are closest to the real line appear at the places $\rho=\pm iP$, where $P>0$ real. The position of the singularity for one dimension is $P=\frac{\pi}{2}$, otherwise it can be obtained by numerical integration,
\begin{equation}
 P=\left
 \{\begin{array}{rl}
  1.5707963 & \text{for } \ d=1 \ , \\
  1.0925562 & \text{for } \ d=2 \ , \\
  0.6021836 & \text{for } \ d=3 \ . \label{eqpolep}
 \end{array}
\right.
\end{equation}
The value of $P$ can also be calculated by the expansion of the base equation \eqref{seqlabel} around $\rho=0$, and applying Padé approximation.

In the more than one-dimensional case the singularity is not a simple pole as in \eqref{eqsechrhoexp}, but because of the appearing logarithmic terms it is a branch point. Introducing the variable $R$ measuring the distance from the singularity with the relation $\rho=iP+R$,  the expansion of the function $S$ contains the following terms:
\begin{equation}
 S=\sum_{k=0}^\infty\ \sum_{j=4k-1}^\infty S_{j,k} \, R^j\left[\ln(iR)\right]^k \ .
\end{equation}
The constants $S_{j,k}$ are real for even $j$, and purely imaginary for odd $j$, hence $S$ takes real values on the imaginary axis. The first six orders of the expansion are:
\begin{align}
 S&=-\frac{i\sqrt{2}}{R}+\frac{\sqrt{2}(d-1)}{6P}
 -\frac{i\sqrt{2}(d^2-6P^2-8d+7)}{36P^2}R \notag\\
 &-\frac{\sqrt{2}(d-1)(4d^2-18P^2-35d+85)}{216P^3}R^2 \label{eqssrrexp} \\
 &-\frac{i\sqrt{2}(d-1)(2d^3-9dP^2-21d^2+18P^2+72d-80)}{135P^4}R^3\ln(iR)
 +S_{3,0}R^3+\cdots \notag \ .
\end{align}
Only one unspecified coefficient remain in the expansion, the value of $S_{3,0}$ can be any purely imaginary number. After fixing $S_{3,0}$ all the other $S_{j,k}$ are uniquely determined. The value of $S_{3,0}$ is determined by the condition that we study the complex extension of the solution $S$ which is regular at the center, nodeless and asymptotically decaying, of which we gave the central value in \eqref{eqscentr}. This solution has a singularity at the place given in \eqref{eqpolep}. The coefficient $S_{3,0}$ appears at such a high order that it will not influence our subsequent calculations for the determination of the radiation rate. The approximate numerical value for $d=2$ spatial dimensions is $S_{3,0}=-0.0726\, i$, while for $d=3$ it is $S_{3,0}=-0.19\, i$. For $d=1$ dimension, based on \eqref{eqsechrhoexp}, the value is $S_{3,0}=-7\sqrt{2}/360\, i=-0.0275\, i$.

In order to calculate the Fourier components $\Phi_n$ up to order $\varepsilon^4$ using equations \eqref{eqphi4fou0}-\eqref{eqphi4fou4}, it is necessary to know the function $p_3$. This function is related by the expression \eqref{e:p3} to the potential independent function $Z$, which can be obtained by the solution of \eqref{eqZgen}. For one space dimension $Z$ can be obtained by the equations \eqref{eqzdimone} and \eqref{seqdegysol}. For $d\geq2$ the behavior of the function $Z$ on the imaginary axis can be studied by the numerical solution of the differential equation, using the central values given in \eqref{eqzcent}. The singularity of the function $S$ at the point $\rho=iP$ generate a $1/R^3$ type singularity in $Z$ as well. The expansion of $Z$ around the singularity consists of the following terms:
\begin{equation}
 Z=\sum_{j=-3}^\infty Z_{j,0} \, R^j
 +\sum_{k=1}^\infty\ \sum_{j=4k-6}^\infty Z_{j,k} \, R^j\left[\ln(iR)\right]^k \ .
\end{equation}
The first few terms of the expansion are:
\begin{align}
 Z=&-\frac{i2\sqrt{2}}{3R^3}
 -\frac{8\sqrt{2}(d-1)}{15P}\,\frac{\ln(iR)}{R^2}
 +Z_{-2,0}\frac{1}{R^2}
 +\frac{i\sqrt{2}(d^2-45P^2-32d+31)}{45P^2}\,\frac{1}{R} \notag\\
 &+\frac{2\sqrt{2}(d-1)(d^2-6P^2+2d+1)}{135P^3}\ln(iR)
 +\cdots+Z_{3,0}R^3+\cdots \ . \label{eqzzrrexp}
\end{align}
Unlike at $S$, the logarithmic dependence in $Z$ already appears in the order following the leading order term. No $\ln(iR)/R$ appears in the expansion. There are two unspecified parameters, $Z_{-2,0}$ and $Z_{3,0}$. For more than one spatial dimensions the constant $Z_{-2,0}$ can be obtained by numerical integration, requiring that $Z$ should be regular at the center $\rho=0$, and that its central value should be given by \eqref{eqzcent}. The numerical values are
\begin{equation}
 Z_{-2,0}=\left
 \{\begin{array}{rl}
  0 & \text{ for  }\  d=1 \ , \\
  0.49483 & \text{ for  }\  d=2 \ , \\
  3.65482 & \text{ for  }\  d=3 \ . \label{eqzm20}
 \end{array}
\right.
\end{equation}
The constant $Z_{3,0}$ will not appear in our subsequent calculations, and the rate of the radiation will be independent of $Z_{-2,0}$ as well.

In the expressions \eqref{eqphi4fou0}-\eqref{eqphi4fou4} the function $p_1$ always appears with at least one $\varepsilon$ factor, hence we will need the function $\varepsilon S$. Similarly, next to $p_3$ there is at least an $\varepsilon^3$ factor, and hence the product $\varepsilon^3 Z$ will be necessary. According to the expression \eqref{eqrykapcs}, we introduce the $y=R/\varepsilon$ inner coordinate for more than one dimensions as well. Hence up to order $\varepsilon$,
\begin{align}
\varepsilon S=&\frac{-i\sqrt2}{y}
+\varepsilon\frac{\sqrt2(d-1)}{6P} \ , \label{eqep1s}\\
\varepsilon^3 Z=&-\frac{i2\sqrt2}{3}\,\frac{1}{y^3}
-\varepsilon \ln \varepsilon\,\frac{8\sqrt2(d-1)}{15P}
\,\frac{1}{y^2}-\varepsilon\,\frac{8\sqrt2\,(d-1)}{15\,P}\,
\frac{\ln(iy)}{y^2}+ \varepsilon\, Z_{-2,0}\frac{1}{y^2} \ . \label{eqep3z}
\end{align}
The derivative of the function $S$ appears at least with an $\varepsilon^2$ factor,
\begin{equation}
 \varepsilon^2\frac{\mathrm{d}S}{\mathrm{d}\rho}=
 \varepsilon^2\frac{\mathrm{d}S}{\mathrm{d}R}=
 \frac{\mathrm{d}(\varepsilon S)}{\mathrm{d}y}=\frac{i\sqrt2}{y^2} \ . \label{eqep2ds}
\end{equation}
Based on these considerations, substituting into \eqref{eqphi4fou0}-\eqref{eqphi4fou4}, the behavior of the Fourier components $\Phi_n$ near the singularity can be expanded for small $\varepsilon$ in the following way:
\begin{equation}
 \Phi_n=\Phi_n^{(0)}+\varepsilon \ln \varepsilon\,\Phi_n^{(1)}
 +\varepsilon\,\Phi_n^{(2)} + \mathcal{O}(\varepsilon^2 \ln \varepsilon) \ , \label{eqphinepsexp}
\end{equation}
where the functions $\Phi_n^{(k)}$ are already $\varepsilon$ independent.

Using the coordinate $y$ that has the same scale as $r$, the first terms of the expressions \eqref{eqep1s}, \eqref{eqep3z} and \eqref{eqep2ds}, which are $\varepsilon$ independent, are also independent of the dimension $d$. Since the parameter $d$ does not appear in \eqref{eqphi4fou0}-\eqref{eqphi4fou4} either, the functions $\Phi_n^{(0)}$ will be the same as in the earlier discussed one-dimensional case. Hence the first terms of the expansion of $\Phi_n^{(0)}$ are given by \eqref{e:1match}. By this we have shown that the leading order behavior of the Fourier modes near the singularity are dimension independent, and we can apply the one-dimensional results that were presented in detail earlier. Only the position of the singularity is changing for different dimensions, according to \eqref{eqpolep}. We could stop here, and continue with the calculation of the radiating tail to leading order from this information, but since it is relatively easy to calculate the next order correction for small $\varepsilon$, we continue with the study of the next term in \eqref{eqphinepsexp}.

The terms of the functions $\Phi_n^{(1)}$ which can be obtained from equations \eqref{eqphi4fou0}-\eqref{eqphi4fou4} are
\begin{align}
 \Phi_0^{(1)}=-\frac{16ig_2(d-1)\sigma}{15\lambda^3 P}\,\frac{1}{y^3}+\ldots \ , \nonumber\\
 \Phi_1^{(1)}=-\frac{8\sqrt{2}(d-1)\sigma}{15\lambda^2\sqrt{\lambda}\, P}
  \,\frac{1}{y^2}+\ldots \ , \label{eqphi1lnep} \\
 \Phi_2^{(1)}=\frac{16ig_2(d-1)\sigma}{45\lambda^3 P}\,\frac{1}{y^3}+\ldots \ ,\nonumber
\end{align}
where the definition of the constant $\sigma$ was given in \eqref{eqsigmadef}. These and the values of $\Phi_n^{(0)}$ given by \eqref{e:1match} are consistent with the relation
\begin{equation}
 \Phi_n^{(1)}=C\frac{\mathrm{d}}{\mathrm{d}y}\Phi_n^{(0)} \ , \label{eqphin1phin0}
\end{equation}
where the constant $C$ is given by
\begin{equation}
 C=\frac{8i(d-1)\sigma}{15\lambda^2 P} \ . \label{eqcsigma}
\end{equation}
As we will see in the next paragraph, the fact that \eqref{eqphin1phin0} is really a solution follows from the equations that the Fourier modes have to satisfy in the inner region.

Based on the relation $r=i\frac{P}{\varepsilon}+y$ that defines the inner coordinate,
\begin{equation}
 \frac{1}{r}=\varepsilon\frac{1}{iP}+\varepsilon^2\frac{y}{P^2}+\mathcal{O}(\varepsilon^3) \ .
\end{equation}
Using that $\omega^2=1-\varepsilon^2$ and dropping terms of order $\varepsilon^2$ and higher, from \eqref{eqnonsphout} we obtain the following equations in the inner region:
\begin{equation}
 \frac{\mathrm{d}^2\Phi_n}{\mathrm{d}y^2}
 +\varepsilon \frac{d-1}{iP}\frac{\mathrm{d}\Phi_n}{\mathrm{d}y}
 +(n^2-1)\Phi_n=F_n \ . \label{eqnonsphinner}
\end{equation}
Looking for the the $\varepsilon$ expansion of the solution $\Phi_n$ in the form \eqref{eqphinepsexp}, we can see that for the leading order function $\Phi_n^{(0)}$ we get same equation as \eqref{eqphinner}, which was discussed at the one-dimensional case,
\begin{equation}
 \frac{\mathrm{d}^2\Phi_n^{(0)}}{\mathrm{d}y^2}
 +(n^2-1)\Phi_n^{(0)}=F_n\Big|_{\Phi_k=\Phi_k^{(0)}} \ . \label{eqnonsphinnere0}
\end{equation}
The equation determining the $\varepsilon \ln \varepsilon$ term is
\begin{equation}
 \frac{\mathrm{d}^2\Phi_n^{(1)}}{\mathrm{d}y^2}+(n^2-1)\Phi_n^{(1)}
 =\sum_{m=0}^\infty\frac{\partial F_n}{\partial\Phi_m}\Bigg|_{\Phi_k=\Phi_k^{(0)}}\Phi_m^{(1)}
 \ . \label{eqnonsphinnere1}
\end{equation}
This equation is obviously solved by $\Phi_n^{(1)}=C\frac{\mathrm{d}}{\mathrm{d}y}\Phi_n^{(0)}$ for arbitrary constant $C$. However, in order to obtain the value of the constant $C$ given in \eqref{eqcsigma} it is necessary to calculate the first few orders of the small-amplitude expansion.

In the higher dimensional case it is also true that we can obtain much more easily the behavior \eqref{e:1match} of the solution $\Phi_n^{(0)}$ near the singularity if we directly expand the inner equations in powers of $1/y$, instead of applying the small-amplitude expansion procedure, and we can proceed up to much higher orders in this way. From $\Phi_n^{(0)}$ we can obtain the value of $\Phi_n^{(1)}$ easily by applying \eqref{eqphin1phin0} and \eqref{eqcsigma}.

\subsubsection{Correction near the singularity}

As we have already discussed, this expansion is not convergent, but an asymptotic series, which implies that an exponentially small correction appears on the imaginary axis. We would like to obtain an inner solution which can be extended to an outer solution which is non-radiating on the positive part of the real $r$ axis, so it is a higher dimensional generalization of the asymmetric breather considered earlier. In order to achieve this, we assume that in the directions ${-\frac{\pi}{2}}<\arg y<0$ the small-amplitude expansion approximates the solution very well. This implies that in these directions each $\varepsilon$ order components of the Fourier modes \eqref{eqphinepsexp} can be given by the $\varepsilon$ expansion results. The leading order term $\Phi_n^{(0)}$ is  well approximated by a higher order in $1/y$ version of \eqref{e:1match}, the next order $\Phi_n^{(1)}$ contribution by a high order version of \eqref{eqphi1lnep}, and so on.

For a potential $U(\phi)$ which is not symmetric around its minimum, the dominant correction on the imaginary $y$ axis, which determines the radiation, appears in the second Fourier mode, and the imaginary part of $\Phi_2^{(0)}$ is still given by \eqref{eqdeltaphi2}. For a symmetric potential the correction appears in the third mode, and the imaginary part of $\Phi_3^{(0)}$ is given by \eqref{eqdeltaphi3}. We can write the correction for both cases in the following unified way,
\begin{equation}
 \delta\Phi_k^{(0,+)}=i\nu_k\exp(-i\sqrt{k^2-1}\,y) \ , \label{eqdeltaphi20}
\end{equation}
where now and in the rest of this subsection we assume that for symmetric potentials $k=3$, and for those potentials which are not symmetric around the minimum $k=2$. The value of $\nu_k$ is exactly the same as for one-dimension, we can use the numerical or Borel summation results obtained there.

The functions $\Phi_k^{(0)}$ are the actual solutions of \eqref{eqnonsphinnere0}, hence they also include the exponentially small correction on the imaginary axis that is responsible for the radiating tail. These are the solutions which in principle can be calculated very precisely numerically, they are not just an approximation in terms of some expansion. Then the functions $\Phi_n^{(1)}=C\frac{\mathrm{d}}{\mathrm{d}y}\Phi_n^{(0)}$ are precise solutions of \eqref{eqnonsphinnere1}, and they also include the next order contributions to the radiation. Because of the $\varepsilon$ expansion results, in the directions ${-\frac{\pi}{2}}<\arg y<0$ the relation $\Phi_n^{(1)}=C\frac{\mathrm{d}}{\mathrm{d}y}\Phi_n^{(0)}$ holds, where the constant $C$ is given in \eqref{eqcsigma}. Hence \eqref{eqphin1phin0} is also valid on the imaginary $y$ axis, and the contribution to $\Phi_n^{(1)}$ is given by the derivative of \eqref{eqdeltaphi20},
\begin{equation}
 \delta\Phi_k^{(1,+)}=\sqrt{k^2-1}\, C\nu_k\exp(-i\sqrt{k^2-1}\,y) \ , \label{eqdeltaphik1}
\end{equation}
where the value of $C$ can be found in \eqref{eqcsigma}.

We define a rescaled $\widetilde{C}=-i\sqrt{k^2-1}\,C$ constant, which is a real number,
\begin{equation}
 \widetilde{C}=\sqrt{k^2-1}\,\frac{8(d-1)\sigma}{15\lambda^2 P} \ . \label{eqctildedef}
\end{equation}
In this way, taking into account order $\varepsilon \ln \varepsilon$ contributions but dropping order $\varepsilon$ and higher order terms, the correction on the imaginary axis can be written into the form:
\begin{equation}
 \delta\Phi_k^{(+)}=i\nu_k\left(1+
 \widetilde{C}\,\varepsilon \ln \varepsilon\right)
 \exp(-i\sqrt{k^2-1}\,y) \ . \label{eqdelphikp}
\end{equation}
Below the real $r$ axis, near the singularity at the point $-i\frac{P}{\varepsilon}$ an essentially identical correction appears,
\begin{equation}
 \delta\Phi_k^{(-)}=-i\nu_k\left(1+
 \widetilde{C}\,\varepsilon \ln \varepsilon\right)
 \exp(i\sqrt{k^2-1}\,y) \ , \label{eqdelphikm}
\end{equation}
where now $r=-i\frac{P}{\varepsilon}+y$.

\subsubsection{Extension to the outer region}

Up to here we have calculated a correction in the inner region, which domain is described by the coordinate $y$. This generates a correction in the outer region, which is represented by the coordinate $r$. The correction has to be added to the Fourier mode functions $\Phi_n$ which can be obtained by the $\varepsilon$ expansion. Since we assume that $\varepsilon$ is small, in the outer region the correction $\delta\Phi_k$ satisfies the linearization of equation \eqref{eqnonsphout} around $\Phi_n=0$, which can be obtained by dropping the nonlinear terms on the right-hand side, and by substituting $\omega=1$. For $d$ spatial dimensions, the general solution of this can be written in terms of Bessel functions,
\begin{equation}
 \delta\Phi_k=\sqrt[4]{k^2-1}\sqrt{\frac{\pi}{2}}\,r^{1-\frac{d}{2}}
 \left[\alpha Y_{\frac{d}{2}-1}\left(\sqrt{k^2-1}\,r\right)
 +\beta J_{\frac{d}{2}-1}\left(\sqrt{k^2-1}\,r\right)\right] \ , \label{eqphikbessel}
\end{equation}
where $\alpha$ and $\beta$ are constants. The $x\to\infty$ asymptotic behavior of the Bessel functions are (see e.g.~\cite{dlmflibrary}),
\begin{align}
 & J_{\nu}(x)\approx\sqrt{\frac{2}{\pi x}}\,\cos
 \left(x-\frac{\nu\pi}{2}-\frac{\pi}{4}\right) \ , \\
 & Y_{\nu}(x)\approx\sqrt{\frac{2}{\pi x}}\,\sin
 \left(x-\frac{\nu\,\pi}{2}-\frac{\pi}{4}\right) \ .
\end{align}
Based on this, for large $|r|$,
\begin{equation}
 \delta\Phi_k=\frac{1}{r^{\frac{d-1}{2}}}
 \left\{\alpha\sin\left[\sqrt{k^2-1}\,r-\frac{\pi}{4}(d-1)\right]
 +\beta\cos\left[\sqrt{k^2-1}\,r-\frac{\pi}{4}(d-1)\right] \right\} . \label{eqdelphiksin}
\end{equation}
Writing it in terms of exponential functions,
\begin{equation}
 \delta\Phi_k=\frac{1}{2r^{\frac{d-1}{2}}}
 \left[i^{\frac{d-1}{2}}(\beta+i\alpha)\exp(-i\sqrt{k^2-1}\,r)
 +(-i)^{\frac{d-1}{2}}(\beta-i\alpha)\exp(i\sqrt{k^2-1}\,r)
 \right] . \label{eqdelphikexp}
\end{equation}
In the vicinity of the singularity above the real $r$ axis, making the substitution $r=i\frac{P}{\varepsilon}+y$, because of the factor $\exp(-P/\varepsilon)$ the second term of  \eqref{eqdelphikexp} becomes negligible, and since in the matching region $|y|\ll P/\varepsilon$,
\begin{equation}
 \delta\Phi_k=\frac{1}{2}(\beta+i\alpha)
 \left(\frac{\varepsilon}{P}\right)^{\frac{d-1}{2}}
 \exp\left(\sqrt{k^2-1}\,\frac{P}{\varepsilon}\right)
 \exp(-i\sqrt{k^2-1}\,y)  \ .
\end{equation}
Comparing this with the expression \eqref{eqdelphikp} giving the correction in the inner region,
\begin{equation}
 \alpha-i\beta=2\nu_k\left(1+
 \widetilde{C}\,\varepsilon \ln \varepsilon\right)
 \left(\frac{P}{\varepsilon}\right)^{\frac{d-1}{2}}
 \exp\left(-\sqrt{k^2-1}\,\frac{P}{\varepsilon}\right) \ . \label{eqalphahmibeta}
\end{equation}
In the vicinity of the singularity below the real axis, substituting $r=-i\frac{P}{\varepsilon}+y$ into \eqref{eqdelphikexp}, and comparing to \eqref{eqdelphikm}, we similarly obtain that the value of $\alpha+i\beta$ agrees with the value obtained for $\alpha-i\beta$ in \eqref{eqalphahmibeta}. Consequently, $\beta=0$ and the value of $\alpha$ is real,
\begin{equation}
 \alpha=2\nu_k\left(1+
 \widetilde{C}\,\varepsilon \ln \varepsilon\right)
 \left(\frac{P}{\varepsilon}\right)^{\frac{d-1}{2}}
 \exp\left(-\sqrt{k^2-1}\,\frac{P}{\varepsilon}\right)  \ . \label{eqalphanuk}
\end{equation}
From \eqref{eqdelphiksin} only the first term remains, and the correction in the core domain of the oscillon, which is also valid on the real $r$ axis, can be written as
\begin{equation}
 \delta\Phi_k=\frac{\alpha}{r^{\frac{d-1}{2}}}
 \sin\left[\sqrt{k^2-1}\,r-\frac{\pi}{4}(d-1)\right] \ . \label{eqdelphikfin}
\end{equation}

Naturally, for $d>1$ the linear approximation \eqref{eqdelphikfin} becomes invalid close to the origin $r=0$. However, since the size of the oscillon's core is of order $1/\varepsilon$, it still gives a valid approximation in a large region inside the core. Since in the region around the upper singularity we have matched in the directions ${-\frac{\pi}{2}}<\arg y<0$ to the $\varepsilon$ expansion results, the correction \eqref{eqdelphikfin} does not appear outside the core region for large values of the coordinate $r$. In this way, we get to the higher dimensional generalization of the asymmetric breather solution presented in Sec.~\ref{secbelnum}, where for positive $r$ there is no standing wave tail. However, extending the solution by going around the origin on the complex plane, on the negative side of the real $r$ axis a standing wave tail appears, with amplitude twice as large as the minimal one. Since it better represents the essential properties, for $d\geq 2$ spatial dimensions we use the name \emph{singular breather} for this solution.

To leading order, the small perturbations around the above constructed singular breather solution again satisfy the left-hand side of \eqref{eqnonsphout}. The obvious solution, which eliminates the central singularity and cancels the asymmetry, is $\delta\Phi_k^{(s)}=-\delta\Phi_k$, where $\delta\Phi_k$ is given by \eqref{eqdelphikfin}. As a reminder we note, that for a potential which is symmetric around its minimum the mode $k=3$ gives the dominant radiation, otherwise the mode $k=2$. In exchange for the disappearance of the singularity, outside the core region in the positive $r$ directions the standing wave tail $\delta\Phi_k^{(s)}$ appears, with frequency $\omega=k$. The scalar field in the tail has the form
\begin{equation}
 \phi=-\frac{\alpha}{r^{\frac{d-1}{2}}}
 \sin\left[\sqrt{k^2-1}\,r-\frac{\pi}{4}(d-1)\right]
 \cos(kt)\ , \label{eqdphitailmored}
\end{equation}
where the value of $\alpha$ is given by \eqref{eqalphanuk}.

In the expression \eqref{eqalphanuk} giving the radiation amplitude $\alpha$ the coefficient $\nu_k$ is independent of the number of spatial dimensions, hence one can apply the methods that were presented in detail for the one-dimensional case. For symmetric potentials it is advisable to use the Borel summation method presented in Subsection \ref{secborelsum}, applying the identity $\nu_3=-K_2\pi/2$. For potentials that are not symmetric around their minimum only the numerical integration method works, which was discussed in Subsection \ref{secbelnum}. Although $\nu_k$ agrees, but the expression \eqref{eqalphanuk} for the tail-amplitude differs in several ways from the one-dimensional result, which generally results in larger radiation amplitude for higher dimensions. According to \eqref{eqpolep}, the distance $P$ of the singularity from the real axis becomes smaller when increasing the number $d$ of space dimensions, which makes the exponential term less small in \eqref{eqalphanuk}. The factor $\varepsilon^{(d-1)/2}$ in the denominator also increases the radiation amplitude $\alpha$ for the case of $d>1$. Since for $0<\varepsilon<1$ the value of $\varepsilon \ln \varepsilon$ is negative, according to \eqref{eqctildedef}, the correction in the square brackets for $\sigma>0$ decreases, and for $\sigma<0$ increases the radiation with respect to the leading order result. The constant $\sigma$ is defined in \eqref{eqsigmadef} in terms of the constants $g_k$ that determine the expansion of the potential around the minimum.

The parameter $\alpha$ corresponds to the amplitude parameter in equation \eqref{eqphitailgen} that describes a general tail. Hence the energy current averaged for an oscillation period, $\bar S$, can be calculated using \eqref{eqsaverage}, where $\omega_f=k$ and $\lambda_f=\sqrt{k^2-1}$\,,
\begin{equation}
 \bar S=\frac{\pi^{\frac{d}{2}}}{\Gamma\left(\frac{d}{2}\right)}
 4\nu_k^2\, k\sqrt{k^2-1}\,
 \left(1+
 \widetilde{C}\,\varepsilon \ln \varepsilon\right)^2
 \left(\frac{P}{\varepsilon}\right)^{d-1}
 \exp\left(-2\sqrt{k^2-1}\,\frac{P}{\varepsilon}\right) \ . \label{eqbarsgen}
\end{equation}
For a symmetric potential $k=3$, otherwise $k=2$. The constant $\widetilde{C}$ is defined in \eqref{eqctildedef}. The dimension dependent factor is:
\begin{equation}
 \frac{\pi^{\frac{d}{2}}}{\Gamma\left(\frac{d}{2}\right)}=\left
 \{\begin{array}{rl}
  1 & \text{for } d=1 \ , \\
  \pi & \text{for } d=2 \ , \\
  2\pi & \text{for } d=3 \ .
 \end{array}
\right.
\end{equation}

\subsection{Order \texorpdfstring{$\varepsilon$}{epsilon} correction for higher dimensions}

Up to here we have considered the $\varepsilon$ independent and the $\varepsilon \ln \varepsilon$ part of the expansion \eqref{eqphinepsexp} of the inner equations \eqref{eqnonsphinner} for the Fourier modes $\Phi_n$. The equations that $\Phi_n^{(0)}$ and $\Phi_n^{(1)}$ have to solve are \eqref{eqnonsphinnere0} and \eqref{eqnonsphinnere1}, respectively. The part which is linear in $\varepsilon$ is denoted by $\Phi_n^{(2)}$, which is described by the equation
\begin{equation}
 \frac{\mathrm{d}^2\Phi_n^{(2)}}{\mathrm{d}y^2}
 +\frac{d-1}{iP}\,\frac{\mathrm{d}\Phi_n^{(0)}}{\mathrm{d}y}
 +(n^2-1)\Phi_n^{(2)}
 =\sum_{m=0}^\infty\frac{\partial F_n}{\partial\Phi_m}\Bigg|_{\Phi_k=\Phi_k^{(0)}}\Phi_m^{(2)}
 \ . \label{eqnonsphinnere2}
\end{equation}
Using equations \eqref{eqphi4fou0}-\eqref{eqphi4fou4} for the mode functions $\Phi_n$, we have shown earlier that the $\varepsilon$ independent part $\Phi_n^{(0)}$ is the same as for one space dimension, i.e.~it is given by \eqref{e:1match}. We have written the first terms of the part proportional to $\varepsilon \ln \varepsilon$ in \eqref{eqphi1lnep}. For the next order, which is linear in $\varepsilon$, from \eqref{eqphi4fou0}-\eqref{eqphi4fou4} we can obtain:
\begin{align}
 \Phi_0^{(2)}=&\frac{i(d-1)g_2}{3\lambda P}\,\frac{1}{y}
 -\frac{16i(d-1)g_2\sigma}{15\lambda^3 P}\,\frac{\ln(iy)}{y^3}
 +\frac{i\sqrt{2}g_2\sigma Z_{-2,0}}{\lambda^3}\,\frac{1}{y^3} \nonumber\\
 &+\frac{i(d-1)(169g_2^3\lambda+36g_2\lambda^2+18g_2\sigma-81g_4\lambda)}
 {81\lambda^3P}\,\frac{1}{y^3} +\ldots \ ,\nonumber\\
 \Phi_1^{(2)}=&\frac{\sqrt{2}(d-1)}{6\sqrt{\lambda}P}
 -\frac{8\sqrt{2}(d-1)\sigma}{15\lambda^2\sqrt{\lambda}\, P}\,\frac{\ln(iy)}{y^2}
 +\frac{\sigma Z_{-2,0}}{\lambda^2\sqrt{\lambda}}\,\frac{1}{y^2}
 +\frac{19\sqrt{2}(d-1)g_2^2}{54\lambda\sqrt\lambda\,P}\,\frac{1}{y^2}
 +\ldots \ ,  \nonumber\\
 \Phi_2^{(2)}=&-\frac{i(d-1)g_2}{9\lambda P}\,\frac{1}{y}
 +\frac{16i(d-1)g_2\sigma}{45\lambda^3 P}\,\frac{\ln(iy)}{y^3}
 -\frac{i\sqrt{2}g_2\sigma Z_{-2,0}}{3\lambda^3}\,\frac{1}{y^3} \label{eqpphi2yexp} \\
 &-\frac{i(d-1)(220g_2^3\lambda-207g_2\lambda^2+18g_2\sigma-108g_4\lambda)}
 {243\lambda^3 P}\,\frac{1}{y^3}+\ldots \ , \nonumber\\
 \Phi_3^{(2)}=&-\frac{\sqrt2(d-1)(4g_2^2-3\lambda)}{72\lambda\sqrt\lambda P}\,
 \frac{1}{y^2}+\dots \ ,\nonumber\\
 \Phi_4^{(2)}=&\frac{i(d-1)(5g_2^3-5g_2\lambda+3g_4)}{135\lambda^{2}P}\,
 \frac{1}{y^3}+\dots \ . \nonumber
\end{align}

If $\Phi_n^{(2)}$ is a solution of equation \eqref{eqnonsphinnere2}, then $\Phi_n^{(2)}+C_2\frac{\mathrm{d}}{\mathrm{d}y}\Phi_n^{(0)}$ is also a solution for an arbitrary constant $C_2$. From the comparison with the first terms of the expressions \eqref{e:1match} it can be seen that this degree of freedom corresponds to the change of $Z_{-2,0}$. On the other hand, the value of $Z_{-2,0}$ is uniquely determined by the $\varepsilon$ expansion, according to \eqref{eqzm20}.

Equations \eqref{eqpphi2yexp} contain terms proportional to $\ln(iy)$, and the coefficients of these exactly agree with the expressions written in \eqref{eqphi1lnep} for $\Phi_0^{(1)}$. Since all the logarithms are coming from the $\ln(iR)$ terms in \eqref{eqssrrexp} and \eqref{eqzzrrexp}, as a result of the substitution $R=\varepsilon y$ the terms $\ln\varepsilon$ and $\ln(iy)$ necessarily appear in pairs. Introducing a new variable by the substitution
\begin{equation}
 \Phi_n^{(2)}=\Phi_n^{(1)}\ln(iy)+\widetilde\Phi_n^{(2)} \ , \label{eqphinbardef}
\end{equation}
the new function $\widetilde\Phi_n^{(2)}$ will not contain logarithmic terms anymore. Substituting into \eqref{eqnonsphinnere2}, using that $\Phi_n^{(1)}$ satisfies \eqref{eqnonsphinnere1}, and applying the identity \eqref{eqphin1phin0} connecting $\Phi_n^{(1)}$ with the derivative of $\Phi_n^{(0)}$, we can obtain that
\begin{align}
 &\frac{\mathrm{d}^2\widetilde\Phi_n^{(2)}}{\mathrm{d}y^2}
 +(n^2-1)\widetilde\Phi_n^{(2)}
 +\frac{d-1}{iP}\,\frac{\mathrm{d}\Phi_n^{(0)}}{\mathrm{d}y} \notag \\
 &\quad+C\left(\frac{2}{y}\,\frac{\mathrm{d}^2\Phi_n^{(0)}}{\mathrm{d}y^2}
 -\frac{1}{y^2}\,\frac{\mathrm{d}\Phi_n^{(0)}}{\mathrm{d}y}\,\right)
 =\sum_{m=0}^\infty\frac{\partial F_n}{\partial\Phi_m}\Bigg|_{\Phi_k=\Phi_k^{(0)}}
 \widetilde\Phi_m^{(2)} \ . \label{e:epscor}
\end{align}
Looking for the solution of this in the form
\begin{equation}
 \widetilde\Phi_{n}^{(2)}=\sum_{k=0}^\infty b^{(n)}_{n+2k-1}
 \,\frac{1}{y^{n+2k-1}} \ ,\label{eqphinbar2exp}
\end{equation}
where $b^{(0)}_{-1}=0$, and using the earlier obtained results for the expansion \eqref{eqphinyexpnons} of $\Phi_n^{(0)}$, we can obtain \eqref{eqpphi2yexp} much more easily than by the $\varepsilon$ expansion procedure. Furthermore, much higher orders can be achieved in this way. In the solution searched for in the form \eqref{eqphinbar2exp} there remains only one unspecified parameter, $b^{(1)}_{2}$, which is uniquely determined by the parameter $Z_{-2,0}$ according to the second equation in \eqref{eqpphi2yexp}.

The small-amplitude expansion with respect to the powers of $\varepsilon$ is unable to describe the exponentially small tail of the quasibreathers. Corresponding to this, the expansion of the solutions $\Phi_{n}^{(j)}$ of the inner equations with respect to $1/y$ is real to all orders on the imaginary axis. Since we assume that the expansion gives good approximation in the directions ${-\frac{\pi}{2}}<\arg y<0$, on the imaginary axis there appears an $\varepsilon$ independent correction given by \eqref{eqdeltaphi20}, and also an order $\varepsilon \ln \varepsilon$ correction given by \eqref{eqdeltaphik1}. Both are proportional to the parameter $\nu_k$, which can be determined by numerical integration or by Borel summation.

At next order, the correction $\delta\Phi_{k}^{(2,+)}$, which is linear in $\varepsilon$, satisfies the linear left-hand side of equation \eqref{eqnonsphinnere2}. The source term that contains the derivative of $\Phi_{k}^{(0)}$ generates a term in $\Phi_{k}^{(2)}$ which is proportional to $y\exp(-i\sqrt{k^2-1}\,y)$. In the function $\widetilde\Phi_{k}^{(2)}$ defined according to \eqref{eqphinbardef} terms containing $\ln(iy)$ appear. In this way, the determination of the order $\varepsilon$ correction becomes rather complicated in the general case. Luckily, this computation is not so important, since it only gives a second order correction to the leading order result. On the other hand, if at the leading order calculations the coefficient $\nu_k$ turns out to be zero, then at the leading order and at order $\varepsilon \ln \varepsilon$ there is no exponentially small contribution at all, and the order which is linear in $\varepsilon$ will determine the radiation. This not only happens for the sine-Gordon potential, but also for example for the symmetric sixth order potential when the parameter $g_5$ takes one of the values on Fig.~\ref{fignu3p} where $\nu_3=0$.

\subsubsection{Symmetric potential with $\nu_3=0$}

For a potential $U(\phi)$ which is symmetric around its minimum the dominant radiation is provided by the mode $k=3$. If $\nu_3=0$, then considering equation \eqref{e:epscor}, the $\Phi_n^{(0)}$ terms only give real contributions on the imaginary axis. Hence, to leading order, the correction appearing on the imaginary axis is
\begin{equation}
 \delta\Phi_3^{(+)}=\varepsilon\delta\Phi_3^{(2,+)}
 =\varepsilon\delta\widetilde\Phi_3^{(2,+)}
 =i\varepsilon\nu_3^{(2)}\exp(-i\sqrt{8}\,y) \ , \label{eqdeltaphik2p}
\end{equation}
where $\nu_3^{(2)}$ is a constant.

For the determination of the coefficient $\nu_3^{(2)}$ we use the Borel summation method presented in Subsection \ref{secborelsum}. For this we need to study the $1/y$ expansion coefficients of the function $\Phi_3^{(2)}$, for the case of large powers of $1/y$. Using the notation for the expansion coefficients from our paper \cite{Fodor2009b},
\begin{equation}
\widetilde{\Phi}_3^{(2)}=\sum_{k=1}^{\infty} \beta_k\frac{1}{y^{2k}} \ .
\end{equation}
Comparing with \eqref{eqphinbar2exp}, we can see that $\beta_k=b_{2k}^{(3)}$. In agreement with the earlier one-dimensional calculation, the expansion of the leading order functions $\Phi_n^{(0)}$ are of the form \eqref{eqphi2nm1exp}. In the general case the asymptotic behavior for large $k$ is determined by the coefficients $A_k^{(2)}$ belonging to $\Phi_3^{(0)}$, which grow according to \eqref{eqa2kcoeff}. The coefficients $A_k^{(n)}$ belonging to the other $\Phi_n^{(0)}$ are at least $1/k^2$ small with respect to $A_k^{(2)}$ (see \eqref{eqak1ak2} and \eqref{eqak3ak2}). Since the coefficient $A_k^{(2)}$ generally increases according to $(2k-2)!$, the terms proportional to $C$ and the nonlinear terms can at most give $1/k^2$ order contribution in \eqref{e:epscor}. In this way, to leading order we get the following equation:
\begin{equation}
 (2k-2)(2k-1)\beta_{k-1}+8\beta_{k}-(2k-1)\,\frac{d-1}{P}\,A_k^{(2)}=0 \ .
 \label{e:beta}
\end{equation}

We look for the behavior of the coefficients $\beta_k$ for large $k$ in the form:
\begin{equation}
\beta_k = L\,(-1)^k\,\frac{(2k)!}{8^{k}}
+M\,(-1)^k\frac{(2k-1)!}{8^{k}} \ , \label{eqbetakanz}
\end{equation}
where $L$ and $M$ are constants. Substituting into \eqref{e:beta}, the value of $L$ is getting determined by the constant $K_2$ defined in \eqref{eqa2kcoeff},
\begin{equation}
L=\frac{\sqrt2\, (d-1)}{8P}\,K_2 \ .
\end{equation}
According to \eqref{eqnu3kk2}, we have $\nu_3=-K_2\pi/2$, and since we just study the case $\nu_3=0$, it follows that $L=0$. In this way, only the second term remains in \eqref{eqbetakanz}.

For large $k$ we can get a more precise approximation for $\beta_k$, if from the left-hand side of \eqref{e:epscor} we also take into account the term $\frac{3}{2}g_3\left(\Phi_1^{(0)}\right)^2\widetilde{\Phi}_3^{(2)}$  (see \eqref{eqddphi3ddy}). For the case of the $U(\phi)=1-\cos\phi$ sine-Gordon potential the starting term of $\Phi_1^{(0)}$ is $-4i/y$, and hence the part of the term on the right-hand side which is important in the leading orders is $4\widetilde{\Phi}_3^{(2)}/y^2$. Because of this, the equation for $\beta_k$ is
\begin{equation}
 (2k-2)(2k-1)\beta_{k-1}+8\beta_{k}=4\beta_{k-1} \ . \label{eqbetakorr}
\end{equation}
It is not necessary to take into account the $\Phi_3^{(0)}$ terms on the left-hand side of equation \eqref{e:epscor}, since we investigate the $\nu_3=0$ case. For the case of the sine-Gordon potential the exactly calculated coefficients of $\Phi_3^{(0)}$ are given in \eqref{eqsgphi3exp}. Their growth is much slower than that of the constants $\beta_{k}$. From equation \eqref{eqbetakorr} follows that up to $1/k^2$ order,
\begin{equation}
 \beta_k= M(-1)^k\,\frac{(2k-1)!}{8^{k}}
 \left(1+\frac{1}{k}+\frac{3}{4k^2}\right) \ . \label{eqmmbetak}
\end{equation}

\subsubsection{Radiation for the sine-Gordon potential in case of $d\geq2$ dimensions}

We can obtain the value of the constant $M$ to a good approximation if we calculate the coefficients of the expansion \eqref{eqphinbar2exp} of the equation \eqref{e:epscor} up to moderately high orders in $1/y$ by some algebraic manipulation software. Obviously, we can only take into account a finite number of Fourier modes, and we can take into account the expansion of the sine-Gordon potential up to a certain order. It turns out that we can obtain quite precise result if we use the Fourier modes up to $\Phi_9$, the expansion of the potential up to order $\phi^{13}$, and we calculate the expansion coefficients up to order $1/y^{33}$. The values of $\beta_k=b_{2k}^{(3)}$ determined in this way contain terms that are proportional to $b^{(1)}_{2}$, which grow much slower with the increase of $k$ than the other terms. Hence we can neglect these terms and the value of $M$ turns out to be independent of the parameter $Z_{-2,0}$. All the other terms contain a factor $(d-1)/P$, which we can take out by defining the constant $\widetilde{M}$,
\begin{equation}
 M=\frac{d-1}{P}\,\widetilde{M} \ .
\end{equation}
Using the asymptotic formula \eqref{eqmmbetak}, from the calculated $\beta_k$ constants we can obtain the value of $\widetilde{M}$. The result for the sine-Gordon potential, valid up to four digits precision is
\begin{equation}
 \widetilde{M}=0.6011 \ . \label{eqmmtilde}
\end{equation}

As we show in this paragraph, there is a close connection between the value of the constant $M$ and the coefficient $\nu_3^{(2)}$ determining the correction on the imaginary axis by \eqref{eqdeltaphik2p}. The equation connecting them is
\begin{equation}
 \nu_3^{(2)}=-\frac{M\pi}{2} \ . \label{eqnu32mm}
\end{equation}
We show this by Borel summation, similarly to the proof of \eqref{eqnu3kk2}. We have to perform the Borel summation of the series
\begin{equation}
 \widetilde{\Phi}_3^{(2)}(y)=\sum_{k=1}^{\infty} \beta_k\frac{1}{y^{2k}} \quad \ , \qquad
 \beta_k= M(-1)^k\,\frac{(2k-1)!}{8^{k}} \ .
\end{equation}
We define the transformed function by the expression
\begin{equation}
 \widehat{\widetilde{\Phi}}_3^{(2)}(z)=
 \sum_{k=1}^{\infty} \frac{\beta_k}{(2k)!}z^{2k} \ .
\end{equation}
In the same way as at \eqref{eqphi3yint},
\begin{equation}
 \widetilde{\Phi}_3^{(2)}(y)=\int_{0}^\infty\mathrm{d}t\, e^{-t} 
 \ \widehat{\widetilde{\Phi}}_3^{(2)}\left(\frac{t}{y}\right) \ . 
\end{equation}
Similarly to \eqref{eqphi3tildez}, the function $\widehat{\widetilde{\Phi}}_3^{(2)}(z)$ can be summed,
\begin{equation}
 \widehat{\widetilde{\Phi}}_3^{(2)}(z)=\sum_{k=1}^\infty M
 \frac{(-1)^k}{2k}{\left(\frac{z}{\sqrt{8}}\right)^{2k}}
 =-\frac{M}{2}\left[\ln\left(1-\frac{iz}{\sqrt{8}}\right)
 +\ln\left(1+\frac{iz}{\sqrt{8}}\right)\right] \ .
\end{equation}
In this case also, only the second term gives imaginary contribution to the integral for $\re y=0$, and similarly to the proof of \eqref{eqimphi3y}, on the imaginary $y$ axis,
\begin{equation}
 \im\widetilde{\Phi}_3^{(2)}(y)=
 -\frac{M\pi}{2}\exp\left(-i\sqrt{8}\, y\right) \ .
\end{equation}
Comparing with equation \eqref{eqdeltaphik2p} we obtain the intended relation \eqref{eqnu32mm} connecting $\nu_3^{(2)}$ and $M$. 

Continuing the correction down to the real axis into the core domain, and compensating the asymmetry of the singular breather, by the method presented at \eqref{eqdphitailmored} we can obtain, that for dimensions $d=2$ and $d=3$, for the sine-Gordon case, the scalar field tail of the quasibreather is
\begin{equation}
 \phi=-\frac{\alpha^{(2)}}{r^{\frac{d-1}{2}}}
 \sin\left[\sqrt{8}\,r-\frac{\pi}{4}(d-1)\right] \cos(3t) \ ,
\end{equation}
where now
\begin{equation}
 \alpha^{(2)}=-\pi(d-1)\widetilde{M}
 \left(\frac{P}{\varepsilon}\right)^{\frac{d-3}{2}}
 \exp\left(-\sqrt{8}\,\frac{P}{\varepsilon}\right)  \ ,
\end{equation}
and the values of $\widetilde{M}$ and $P$ are given by \eqref{eqmmtilde} and \eqref{eqpolep}.
The radiated energy current averaged over an oscillation period can be obtained in the same way as at \eqref{eqbarsgen},
\begin{equation}
 \bar S=\frac{\pi^{\frac{d}{2}}}{\Gamma\left(\frac{d}{2}\right)}
 6\sqrt{2}\,\pi^2(d-1)^2\widetilde{M}^2\left(\frac{P}{\varepsilon}\right)^{d-3}
 \exp\left(-4\sqrt{2}\,\frac{P}{\varepsilon}\right) \ .
\end{equation}

\section{Numerical investigation of the radiation of oscillons} \label{secoscsugnum}

In this section we apply our numerical time-evolution code, that we have presented in Subsection \ref{secnummodssajat}, for the determination of the radiation rate of oscillons, and we compare the numerical values to the analytical results obtained in the previous section. The localized states formed by the scalar field belong to two large classes, there are stable and unstable oscillons. The unstable states have generally only one decay mode, which we were able to suppress by the fine-tuning of a parameter in the initial data, and we could obtain the near-periodic states by the method presented in Subsection \ref{sec-majdnemper}. These states are generally so closely periodic, that the radiation emitted by them are near the minimal energy loss rate, and hence it can be directly compared to the analytically obtained results.

In case of one and two spatial dimensions, stable oscillon states oscillate for infinitely long time, without ever decaying. In case of three space dimensions, when the energy of an otherwise stable but slowly radiating oscillon decreases below a critical limit determined by the minimum on Fig.~\ref{figenertot}, it gets into the unstable domain and suddenly decays. On Figures \ref{fignontune}, \ref{figmaxev3f} and \ref{figfrev3f} we can see typical examples of this. On these figures we can also observe another important property of oscillons. For any dimensions, on the envelope curve of the oscillations of the scalar field a low frequency modulation can be observed. This shape-mode is caused by a low frequency change of the spatial size and amplitude of the oscillons. These low frequency modes can store considerable amount of energy, and since they may decay relatively quickly, the energy radiated by them can be much larger than that of the minimal radiation by an ideal ``clean'' oscillon. However, by fine-tuning some parameter in the initial data, these shape-modes can be made extremely small. In this way, we were able to study close to ideal oscillons, and we could compare their radiation rate to the theoretically obtained values.

\begin{figure}[!htb]
\centering
\includegraphics[width=115mm]{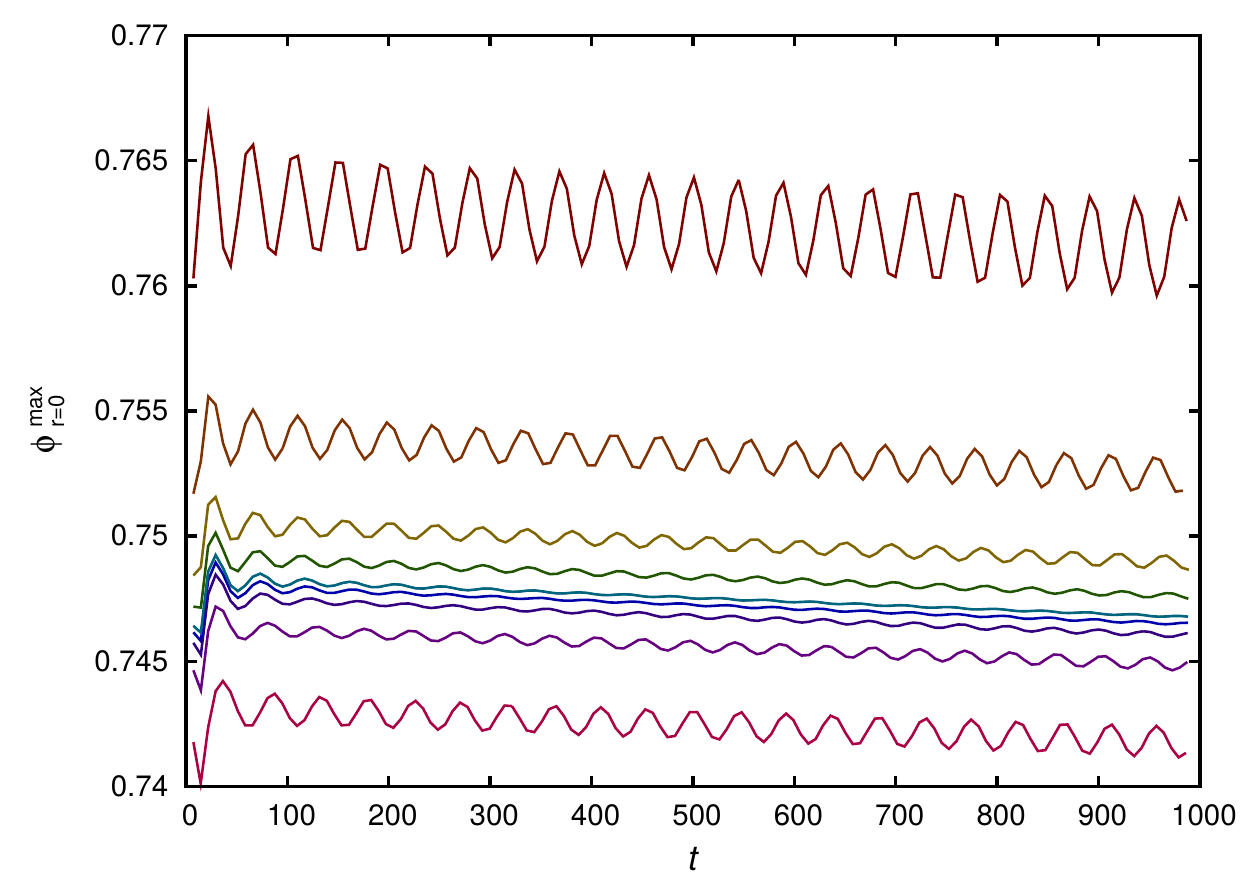}
\caption{\label{figtuneal}
Time evolution of the upper envelope of the central oscillations for the case of $d=1$ dimension and $\phi^4$ potential, for $\varepsilon=0.5$ initial data rescaled by various $\gamma$ factors.
The constant $\gamma$ takes values in the interval $[0,9978,1,0169]$, and the smoothest evolution is obtained for $\gamma=1.00263$.}
\end{figure}

In order to obtain as periodic oscillons as possible, we need to choose initial data which is very precise already before the fine tuning. A large number of various initial states can be most easily prepared by the small-amplitude expansion procedure presented in Section \ref{seckisampl}. According to our experience, in the amplitude range where the radiation of the oscillons is numerically detectable, the order $\varepsilon^3$ initial data gives the most closely periodic time evolution. We construct the initial data $\phi^{(t=0)}$ by using the first three terms of the expansion \eqref{e:sumphi}, based on equations \eqref{e:ph01}-\eqref{e:ph03}. We multiply this initial data by a constant $\gamma$ close to $1$, which we intend to fine-tune. Generally, during the numerical evolution, in a relatively short initial period a smaller part of the energy gets radiated, and after this an oscillon forms which is excited by a shape-mode. Using the least squares method we fit a line on the points corresponding to the maximums of the scalar field, for a relatively longer time period during which a few shape-mode oscillation occurs. We use the sums of the squares of the deviation from this line to characterize the amplitude of the shape-mode. As a function depending on the multiplying factor $\gamma$ we minimize the amplitude of the shape mode using Golden-section search. In order to give an example on how efficient this method can be, on Figure \ref{figtuneal}
we show the effect of the rescaling and tuning for the case of the $\phi^4$ potential written in the form \eqref{eqpotone}, for initial data with $\varepsilon=0.5$. The amplitude of the shape-mode is minimal for $\gamma=1.00263$, in which case the value of the amplitude parameter $\varepsilon$, calculated from the frequency by the relation $\varepsilon=\sqrt{1-\omega^2}$, is slowly changing near the value $0.513$.

\begin{figure}[!htb]
\centering
\includegraphics[width=115mm]{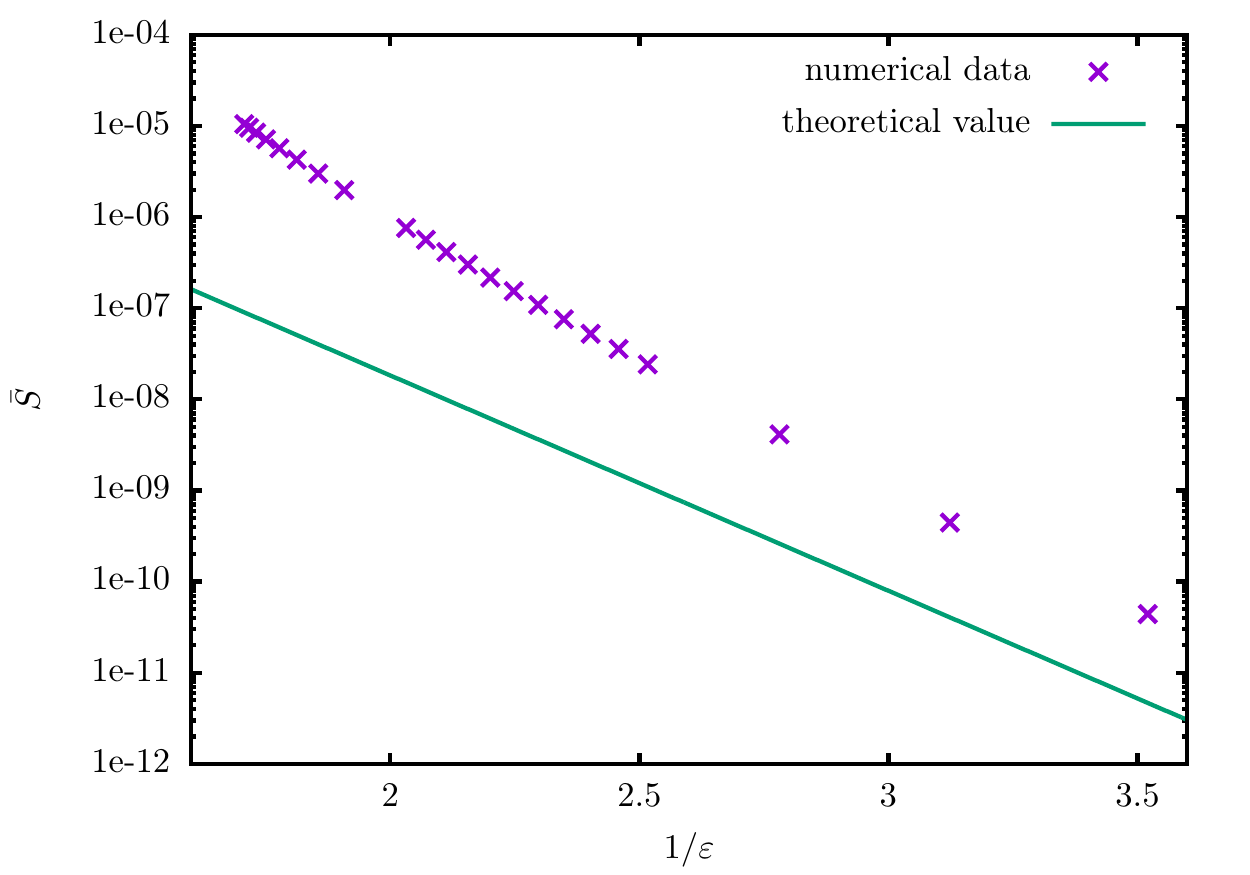}
\caption{\label{figphi41d}
The time averaged energy current $\bar S$ for $\phi^4$ oscillons as a function of the reciprocal of the parameter $\varepsilon$ calculated from the frequency.}
\end{figure}

For the one-dimensional $\phi^4$ oscillon states calculated in this way, on Figure \ref{figphi41d}
we show the numerically calculated energy current averaged over an oscillation period, and compare it to the theoretical result obtained by \eqref{eqsbarnu2} and \eqref{eqnu2phi4}. We show the energy current logarithmically as a function of $1/\varepsilon$, since in this way the theoretically calculated result consists of a straight line. It can be seen that for those relatively large amplitude oscillons for which we were able to determine the energy loss numerically, the radiation is one or two magnitude larger than the theoretically predicted value. The obvious reason for this is that our analytic results are valid only to leading order, for small enough $\varepsilon$ values. If we took into account higher orders in our theoretical calculations, then correction terms containing powers of $1/\varepsilon$ would appear in the expression \eqref{eqsbarnu2} giving the strength of the radiation. Nevertheless, it can be seen that by the decrease of the parameter $\varepsilon$ the numerical data gets closer to the theoretical value. For the $\phi^4$ potential the value of the radiation coefficient is very small, $\nu_2=8.38866\cdot10^{-3}$, which makes the numerical checking of the analytical result especially difficult.

\begin{figure}[!htb]
\centering
\includegraphics[width=115mm]{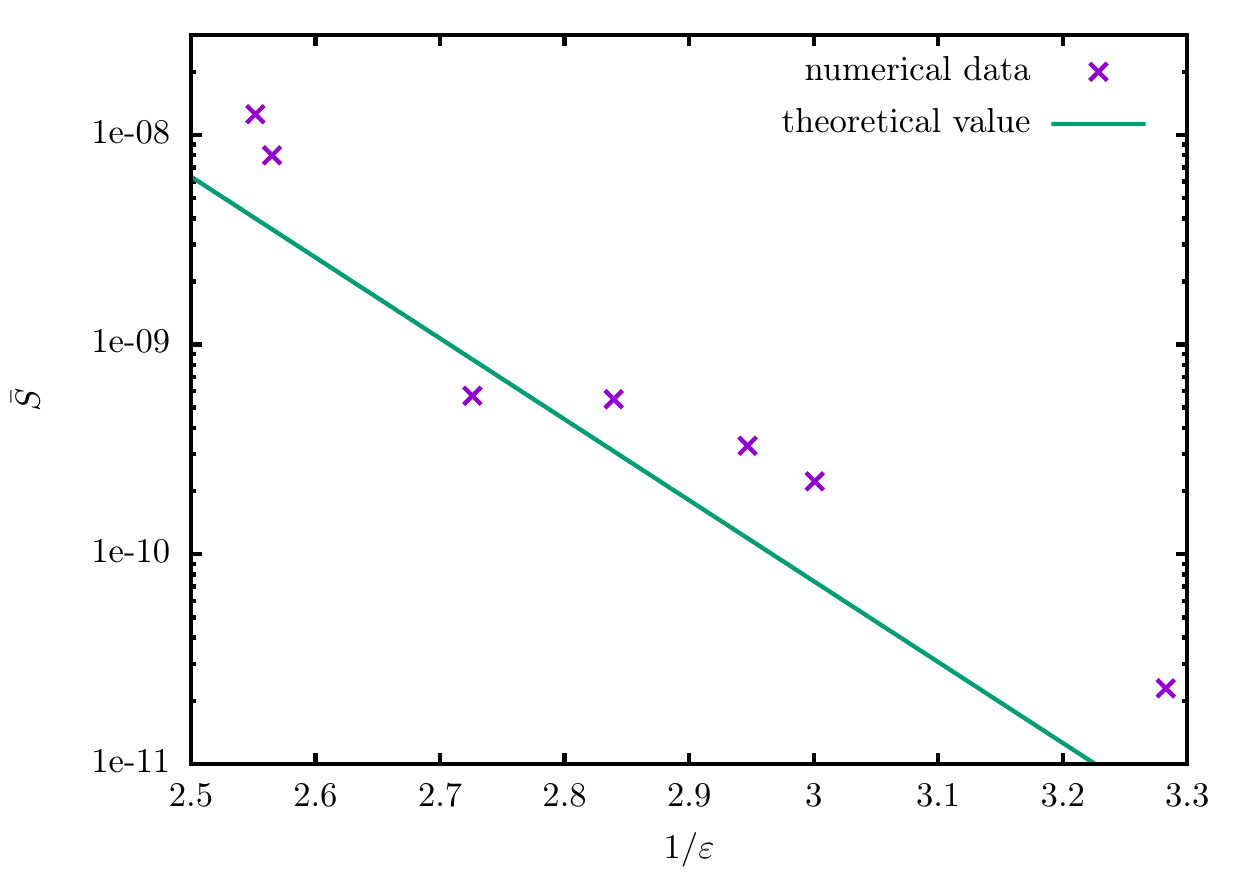}
\caption{\label{figphi6d1}
The value of the time averaged energy current $\bar S$ for the symmetric $\phi^6$ potential in case of $g_5=1$, as a function of $1/\varepsilon$.}
\end{figure}

On Figure \ref{figphi6d1}, still for one spatial dimension, we show the radiation rate for the $\phi^6$ potential given in \eqref{eqsympotform}, for the case $g_5=1$. The analytical result for the time averaged energy current is given in this case by \eqref{eqsbarnu3}, where according to  \eqref{eqnu3g5e2} we have $\nu_3={-0.9097496}$. In this case the difference from the leading order theoretical prediction is much less than for the $\phi^4$ potential.

\begin{figure}[!htb]
\centering
\includegraphics[width=115mm]{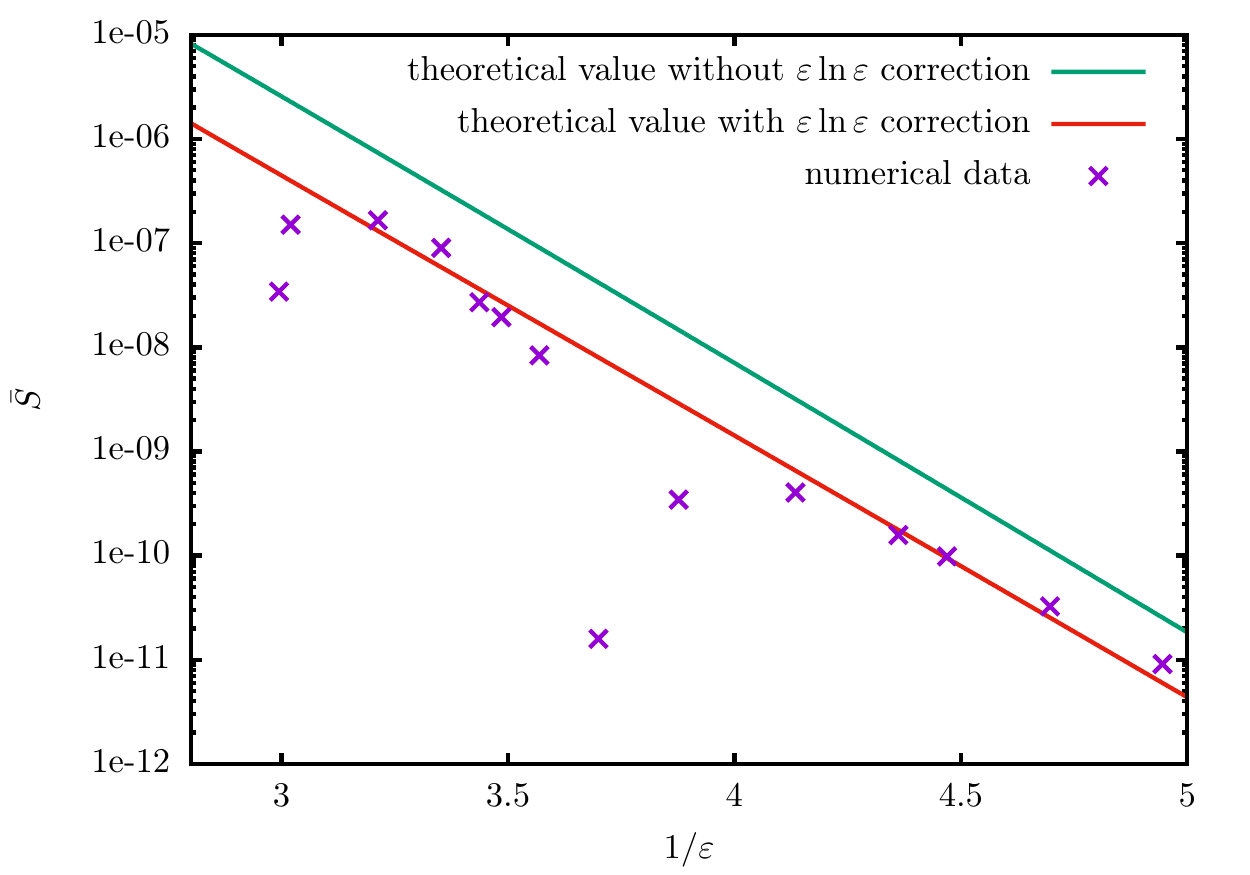}
\caption{\label{figphi6d2}
The time averaged energy current $\bar S$ for $d=2$ spatial dimensions, $\phi^6$ potential, and $g_5=1$.}
\end{figure}

On Figure \ref{figphi6d2}
we show the radiation rate for the same $\phi^6$ potential, but for $d=2$ spatial dimensions. In this case we can get the analytical result from \eqref{eqbarsgen}. We also show the leading order result which can be obtained by dropping the $\varepsilon\ln\varepsilon$ order correction. The $\varepsilon\ln\varepsilon$ contribution takes us closer to the numerical results. However, at such large amplitudes we have to be careful about the validity of the approximation. In equation \eqref{eqbarsgen} the value of the correction term $\widetilde{C}\,\varepsilon \ln \varepsilon$ becomes already very close to $1$. The result that we try to obtain by the expansion of the square $(1+\widetilde{C}\,\varepsilon \ln \varepsilon)^2$ and by dropping the term $(\widetilde{C}\,\varepsilon\ln\varepsilon)^2$ turns out to be invalid.

\begin{figure}[!htb]
\centering
\includegraphics[width=115mm]{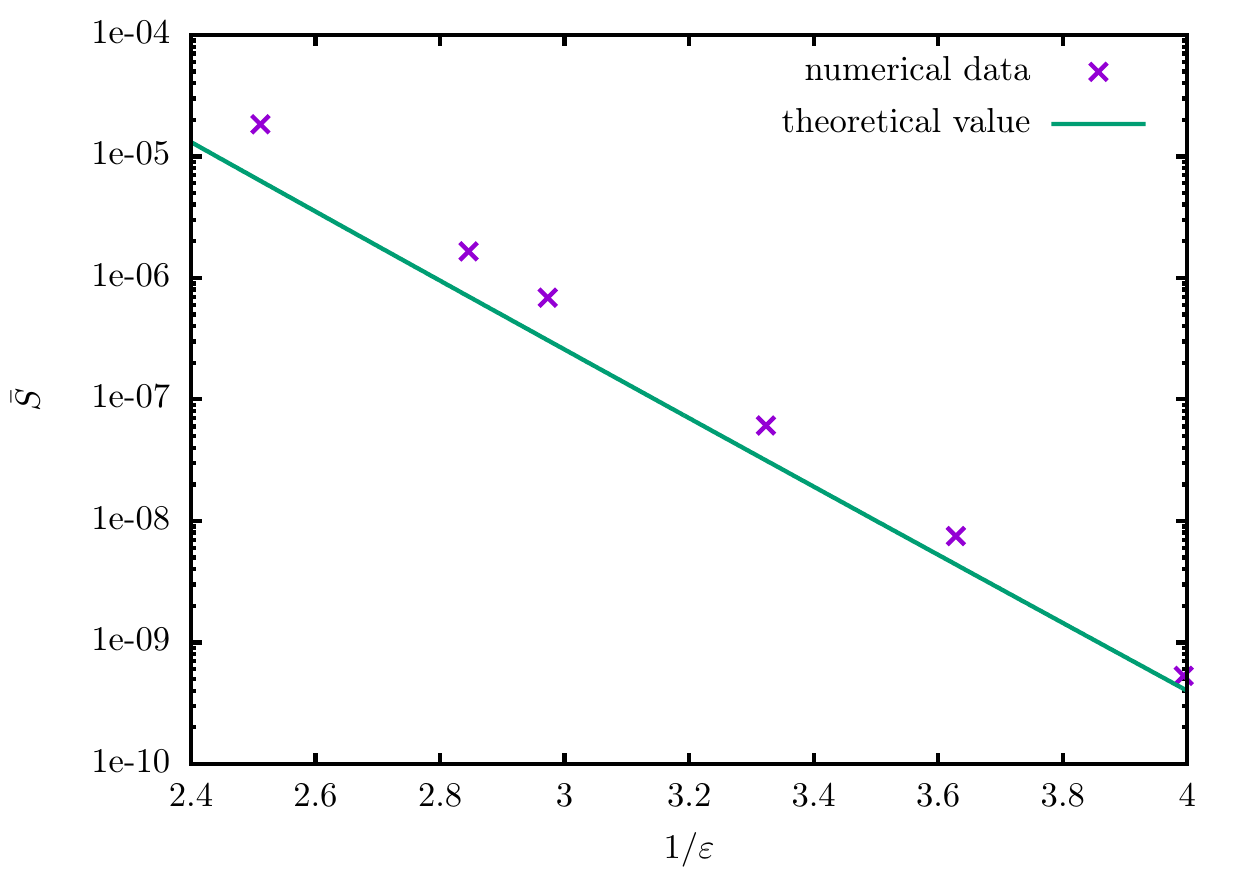}
\caption{\label{figsg2d}
Time averaged energy current for $d=2$ dimensional oscillons with sine-Gordon potential.}
\end{figure}

\begin{figure}[!htb]
\centering
\includegraphics[width=115mm]{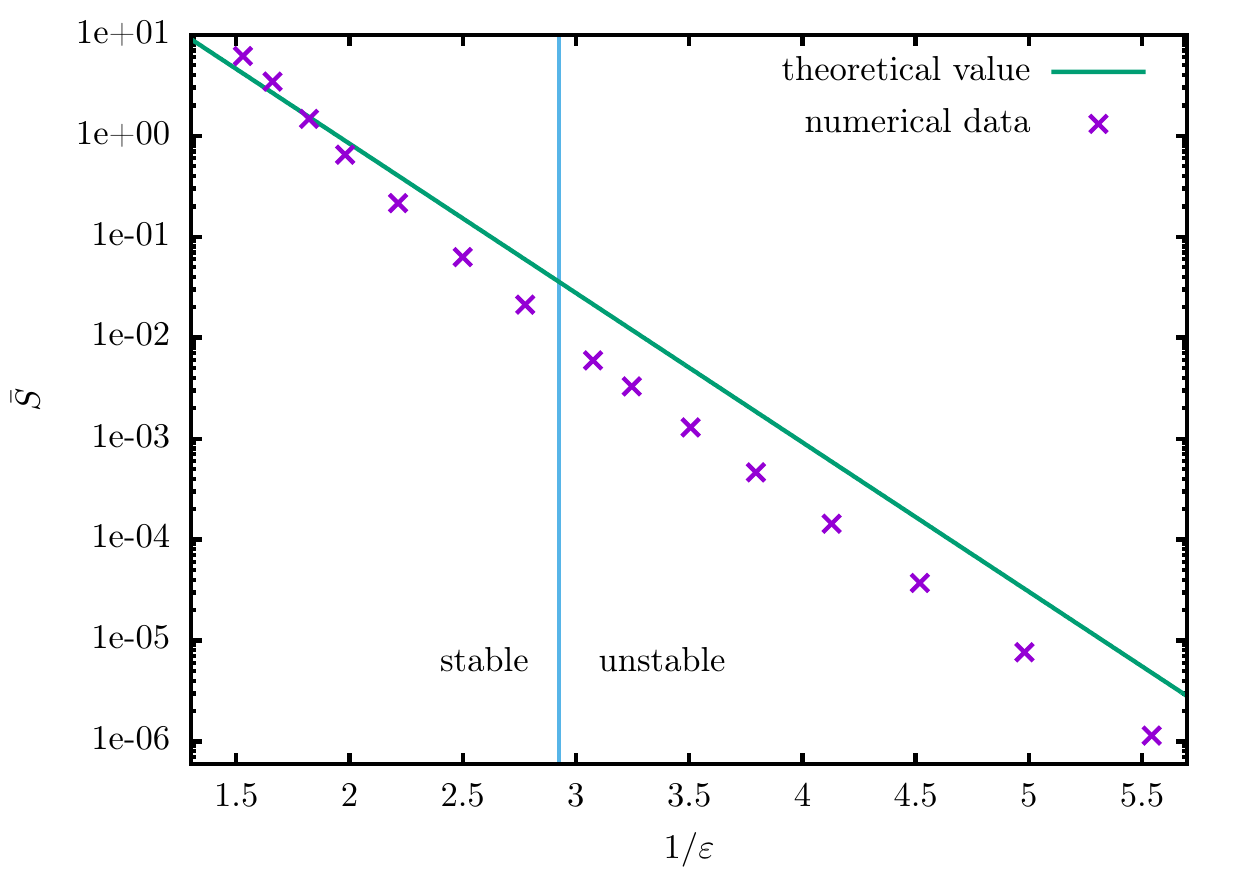}
\caption{\label{figsg3d}
Time averaged energy current for $d=3$ spatial dimensional sine-Gordon oscillons, in the stable and unstable domain.}
\end{figure}

On Figures \ref{figsg2d} and \ref{figsg3d} we compare the numerically calculated radiation of two and three-dimensional sine-Gordon oscillons to the theoretically obtained values.

\chapter{Self-gravitating scalar fields: oscillatons} \label{fejrelgrav}

\section{History of oscillatons}

Since the existence of oscillon (pulson) solutions formed by a real (not complex) nonlinear scalar field on flat Minkowski background is known since 1977 \cite{Bogolubsky-77b,bogolyubskii-77,Makhankov-78}, it would have been natural to study how these configurations change under the influence of Einsteinian gravity. In spite of this, the discovery of the localized states formed by self-gravitating real scalar fields happened completely independently from the studies of oscillons. One of the reasons for this might be that in the gravitational case, because of the nonlinearity of the equations, a non-self-interacting massive Klein-Gordon is already able to form long-lived localized states, while for flat background a nontrivial $U(\phi)$ potential is necessary for this.

It was first shown by Seidel and Suen in 1991, that coupling the Einstein equations to a real Klein-Gordon field localized states can form, which appear time-periodic \cite{seidel-91}. In their first paper they called these objects oscillating soliton stars. Looking for time-periodic solutions, they have solved the system of ordinary differential equations for the first few Fourier modes, and up to the reached precision they have found localized solutions. Using these as initial data in a numerical code developed for the spherically symmetric system, they have indeed obtained very closely periodically evolving states. This have also showed well the stability of the oscillating soliton stars. They have also shown that from general spherically symmetric initial data similar long-lived solutions can evolve, which supported the physical significance of the solutions. Although by the applied numerical precision the solutions appeared to be completely time-periodic, the authors have commented that it is possible that still there is a very slow change in the amplitude and frequency, similarly to how the state of two orbiting black holes changes slowly because of the radiation of gravitational waves.

\begin{figure}[!hbt]
\centering
\includegraphics[width=9cm]{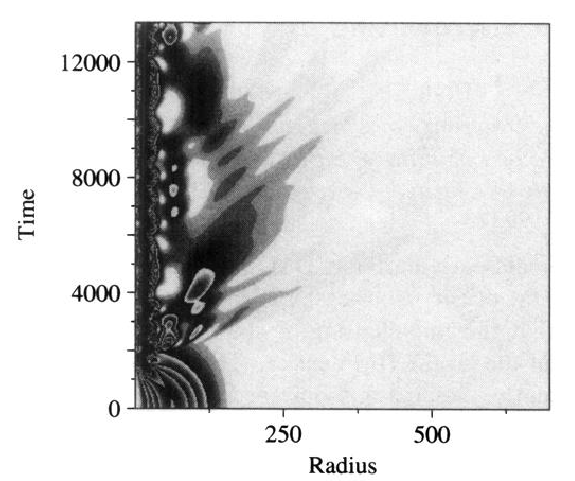}
\caption{\label{f:seidel} Time evolution of the energy density $\rho$ multiplied by the square of the radial coordinate $r$, as a function of $r$, for a self-gravitating spherically symmetric real scalar field. (Source of figure: Seidel and Suen \cite{seidel-94}.)}
\end{figure}

Soliton stars can be formed from both real and complex scalar fields. In case of a complex field the forming objects are known as boson stars \cite{Kaup68,Ruffini69,Jetzer92,Schunck03,Liebling17}. For boson stars only the argument of the complex scalar field $\phi$ depends on time, in the form $e^{i\omega t}$, and because of this the metric of the spacetime remains time-independent. For a real scalar field soliton star both the scalar and the metric oscillate in time, which makes the investigation of this case technically much more complicated. In a subsequent publication \cite{seidel-94} Seidel and Suen have studied the collapse of extended real and complex scalar configurations into smaller domains (see Figure \ref{f:seidel}).
The formation of the soliton stars is made possible by the process that part of the scalar field gets emitted to large distances, which takes away the excess kinetic energy. For the oscillating soliton stars formed from real scalar fields in this second paper Seidel and Suen have introduced the name \emph{oscillaton}. The use of this name became generally accepted in the literature.

\begin{figure}[!hbt]
\centering
\includegraphics[width=9cm]{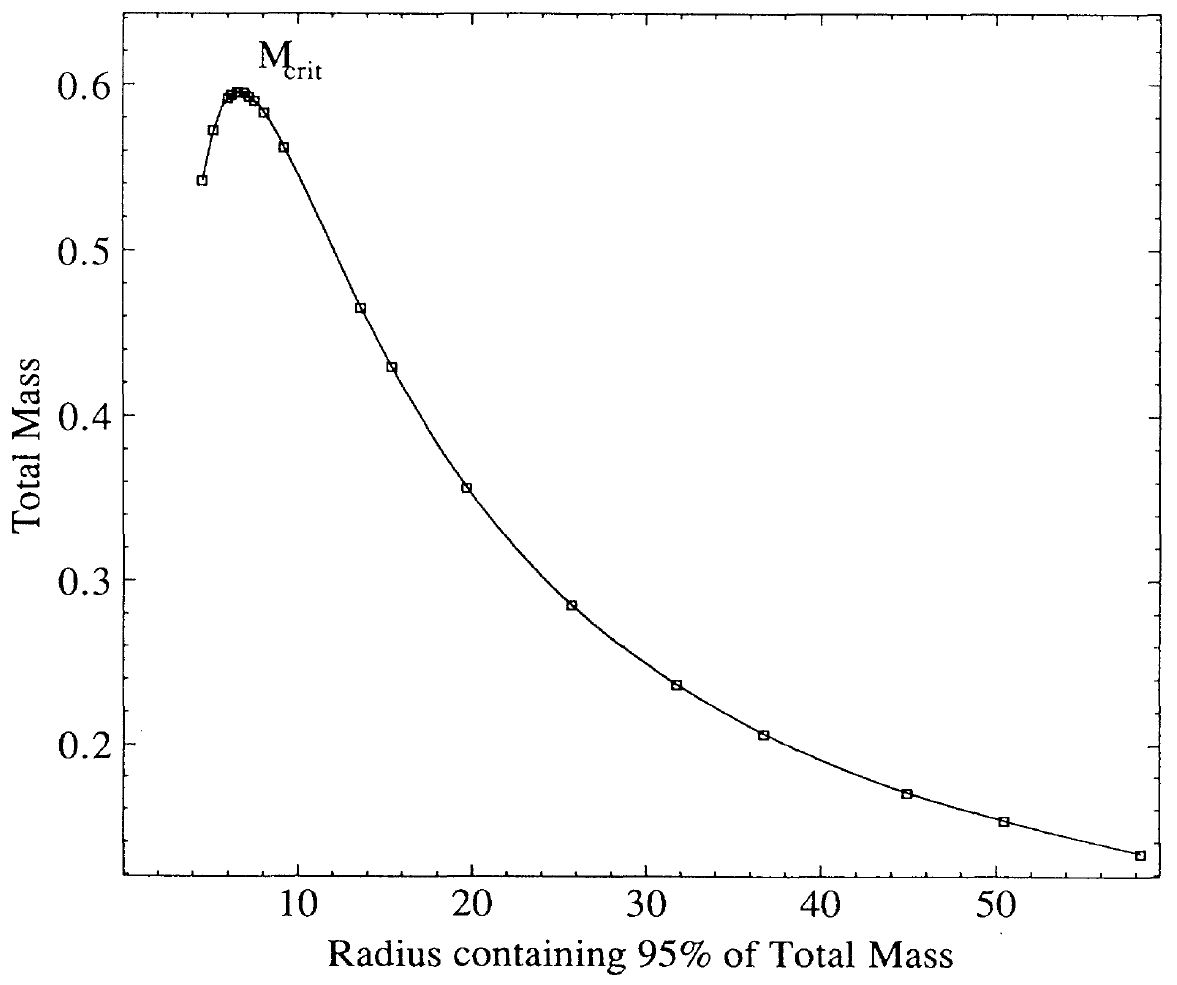}
\caption{\label{f:seidelmass} The total mass of oscillatons as a function of their size. For the size here we use the radius of the sphere which contains $95$ percent of the total mass.  (Source of figure: Seidel and Suen \cite{seidel-91}.)}
\end{figure}

In the first short letter format paper of Seidel and Suen there is already very much important information about oscillatons \cite{seidel-91}. Similarly to the flat background oscillons and to the complex scalar boson stars, the oscillatons are members of a one-parameter family of solutions. Increasing the central amplitude, the size of the oscillatons becomes smaller, and their frequency also decreases. On Figure \ref{f:seidelmass}
we can see how the mass of the oscillaton changes as a function of its size. According to the general behavior, which has been also observed for neutron stars and similar astrophysically important localized states, with the decrease of the size we get to an $\mathrm{M}_{\mathrm{crit}}$ maximal mass state, and further decreasing the size the oscillaton becomes unstable. Since there are two oscillatons belonging to a given mass, applying the terminology used for boson stars, on the figure the stable configurations to the right of the maximum are on the S-branch, while the right-hand side ones constitute the U-branch.

Oscillatons formed from scalar fields in various cosmological models has been considered as dark matter in galaxies \cite{Matos01,Alcubierre02,Susperregi03,Malakolkalami16}. If we take into account only the first term in the Fourier expansion with respect to time, i.e.~we assume that the time dependence of the scalar field is exactly of the form $\cos(\omega t)$, we can already get to the understanding some of the main properties of oscillatons \cite{Urena02a}. The smaller the amplitude of the oscillaton is, the less important are the higher order Fourier components, and the more similar the oscillaton becomes to the same mass boson star configuration. We can find the detailed results obtained by the numerical computations using the Fourier expansion up to the tenth order in the paper of Ureña-López, Matos and Becerril \cite{Urena02b}. However, the numerical precision used there did not make possible the detection of the necessarily existing standing wave tail, which is many orders of magnitude smaller than the typical amplitude in the inner core domain. In the same paper, it was shown that for small amplitudes, i.e.~for large size and weak gravitation, the time-periodic oscillatons and boson stars can be described by a specific system of two coupled second order ordinary differential equations. These are known as Schrödinger-Newton equations in the literature. Without the assumption of time-periodicity, in the small amplitude limit one can obtain the generalization of the Schrödinger-Newton equations for a time dependent complex field, which are also known as Schrödinger-Poisson, or Gross-Pitaevskii-Poisson equations. Applying these equations makes simpler the study of the stability and formation of near-Newtonian oscillatons and boson stars \cite{Guzman03,Guzman04,Guzman06}.

For arbitrary amplitudes, applying a numerical code developed for the time-evolution of a spherically symmetric general relativistic real scalar field, Alcubierre at al.~\cite{Alcubierre03} have studied in detail the behavior of deformed oscillatons. If one applies a relatively small perturbation on oscillatons belonging to the S-branch, they perform low frequency quasinormal oscillations with decreasing amplitude, and they tend to another nearby S-branch oscillaton. Adding a not too large amplitude Gaussian shaped deformation to the initial data calculated by the Fourier expansion method, generally the system still tends to another oscillaton on the S-branch. However, if as a result of the deformation the mass of the system becomes larger than the $\mathrm{M}_{\mathrm{crit}}$ maximal oscillaton mass, then the end result can be a black hole. These investigations not only show that the oscillatons on the S-branch are stable, but also that they correspond to one of the general end states of the system.

On the initial data corresponding to an oscillaton on the U-branch there is necessarily a small numerical perturbation, which increases fast in time, and performing large amplitude quasinormal oscillations the system tends to an S-branch oscillaton. In case of a larger perturbation of an U-branch oscillaton, if the perturbation increases the mass of the system, then the end state will be a black hole. On the other hand, if the perturbation decreases the mass, then the end state is one of the stable oscillatons on the S-branch. Interestingly, during these processes the frequency of the quasinormal oscillation is not changing in time. The frequency remains constant even in those cases when the mass of the system decreases significantly because of the radiated scalar field. The unstable oscillatons on the U-branch correspond to the critical states observed during the study of the collapse of massive scalar fields, which are just on the edge of collapsing to a black hole \cite{Brady97,Okawa14}.

The behavior of oscillatons in case of large $d$ spatial dimensions has been studied in \cite{Rozali18}. Since $f(R)$ gravity is equivalent to a real scalar field coupled to normal Einsteinian gravity, in these alternative theories localized objects corresponding to oscillatons can form even in the vacuum case \cite{Obregon05}.

Most studies of oscillatons have been restricted to the case of the non-self-interacting massive Klein-Gordon potential $U(\phi)=\frac{1}{2}m^2\phi^2$. Not only because this is the simplest and most natural choice, but also because it can be shown that for small amplitudes and large sizes the gravitational interaction dominates above the self-interaction of the scalar field. Ureña-López, Valdez-Alvarado and Becerril \cite{Urena12} have studied, in a similarly detailed way as in the Klein-Gordon case, the system when a self-interaction term $\frac{1}{16}\lambda\phi^4$ is added to the potential, where $\lambda\geq 0$ in order to make the potential bounded from below. In this case too, the solutions can be categorized into S and U branch oscillatons, with similar stability properties as in the Klein-Gordon case. An important difference is that the maximal mass $\mathrm{M}_{\mathrm{crit}}$ is becoming larger with the increase of $\lambda$, similarly as for boson stars. Oscillatons with $\phi^4$ scalar potential have been studied in \cite{Ikeda17} by a spherically symmetric time-evolution code. The possibility of oscillatons with a massless $\phi^4$ potential has been discussed in \cite{Mahmoodzadeh18b}. For cases where gravity can be considered as relatively weak, oscillatons formed by a self-interacting scalar behave much like flat background oscillons. Resonance peaks on the lifetime curve, and near-periodic states have been observed in the gravitational case in \cite{Ikeda17}, similarly to those discussed in Subsection \ref{sec-majdnemper}.

There are infinitely many one-parameter families of time-periodic oscillaton solutions, which are indexed by the number of nodes (crossings of zero) of the scalar field as a function of the radial coordinate \cite{Urena02a,Urena02b}. For the oscillatons belonging to the ground state family, the scalar field $\phi$ has no nodes at the moment $t=0$, which corresponds to the moment of time-reflection symmetry. For the $n$-th excited state oscillatons the scalar has $n$ nodes. It is a natural physical intuition that the excited state oscillatons has larger energy and are probably unstable, hence most of the published papers deal only with ground state oscillatons. Balakrishna at al.~\cite{Balakrishna12} have constructed time-periodic excited states using Fourier expansion, and by a time-evolution code they have demonstrated that the excited states are really unstable. Relatively small amplitude excited oscillatons evolve into ground state oscillatons, while large amplitude ones collapse to black holes. In the same paper \cite{Balakrishna12}, the first results about non spherically symmetric oscillatons can be found, obtained by the use of a $3+1$ dimensional time-evolution code. According to the simulations, non spherically symmetric perturbations quickly decay by the emission of gravitational radiation. A $3+1$ dimensional code is used in \cite{Muia19} for studying the time-evolution of spherically symmetric oscillaton initial data with physically motivated $U(\phi)$ interaction potentials, considering black hole formation, the effect of gravity, and the relation to oscillons.

The real scalar field necessary for the formation of oscillatons may be provided most naturally by axions and similar low mass weakly interacting hypothetical bosonic particles \cite{Tkachev86,Hogan88,Kolb93,Kolb94,Arvanitaki10,Marsh16,Marsh17,Krippendorf18}. If instead of the non-self-interacting $U(\phi)=\frac{1}{2}m^2\phi^2$ Klein-Gordon potential, the self-interaction is determined by a potential $U(\phi)$ describing an axion field, then in several articles the forming oscillatons are named axion stars, see e.g.~\cite{Visinelli18,Braaten18,Chavanis18}. In a few papers oscillatons are used with the spelling oscilloton. Cosmological simulations studying the evolution of axion-like fields show that in the central parts of the diffuse dark matter structures solitonic cores are getting formed, which correspond to oscillatons \cite{Schive14,Veltmaat16}. The dependence on the initial data of the formation of axion stars and black holes have been studied in the articles \cite{Okawa14,Okawa15,Helfer17,Michel18,Widdicombe18}. During the collision of an axion star with a neutron star strong gravitational and electromagnetic waves are radiated out \cite{Iwazaki15,Raby16}.

Using a general $3+1$ dimensional numerical code it was demonstrated that when an axion star and a neutron star is collided, the end result is a black hole or neutron star surrounded by an axion cloud \cite{Clough18}. Fuzzy dark matter formed by very low mass axion-like particles can give an explanation of why we cannot observe dark matter lumps below a certain size in dwarf galaxies \cite{HuAtal00,Marsh15,Schwabe16,HuiAtal17}. The collapse is essentially prevented by the Heisenberg uncertainty principle. The oscillatons formed from ultra light mass scalar dark matter are simply called solitons in several papers. For the study of the formation of the small-amplitude large-sized dark matter structures, instead of the full Einstein equations, it is sufficient to consider the Schrödinger-Poisson (Gross-Pitaevskii-Poisson) equations, which are valid in the weak gravitational case \cite{Edwards18,Madarassy15,Amin19}.

Since oscillatons may be even galactic sized, it is important to investigate how the time-periodically changing metric influences the geodetic motion of stars and other compact objects \cite{Becerril2006,Boskovic18,Mahmoodzadeh18}. The scalar dark matter may be accumulated in the center of stars, and form an oscillaton-like core there. Contrary to initial expectations, the accumulation of dark matter does not enhance the collapse of stars into black holes \cite{Brito15,Brito16}. Massive vector fields can also form localized states. In case of a complex vector field, similarly to boson stars, these configurations have a static metric \cite{Brito16b}, while for a real vector field, similarly to oscillatons, they have a periodically oscillating metric \cite{Garfinkle03,Brito16}. These objects are called Proca stars. Numerical simulations performed by a general numerical code show that colliding oscillatons only form black holes if the mass initially present in the system is considerably larger than the $\mathrm{M}_{\mathrm{crit}}$ maximal mass of oscillatons \cite{Brito15,Brito16}. At the collision of oscillatons the emitted gravitational waves may be quite different from the waves emitted by the collision of black holes \cite{Helfer19}.

In case of boson stars, the real and imaginary parts of the scalar field with time dependence $\phi=p\exp(i\omega t)$ can be considered as two individual real scalars \cite{Hawley00Thesis}. Assuming that the two scalar fields are of the form $\phi_1=p_1\cos(\omega t)$ and $\phi_2=p_2\cos(\omega t+\delta)$, by the choice $p_1=p_2=p$ and $\delta=-\pi/2$ we obtain boson stars. Choosing $p_1=p_2=p/2$ and $\delta=0$ we have $\phi=p\cos(\omega t)$, so we obtain the leading order approximation of oscillatons. In their paper \cite{Hawley03} Hawley and Choptuik have studied the time evolution of initial data, where $p_1=p_2=p$ was chosen corresponding to the boson star value, and the phase difference $\delta$ was arbitrary. As a result, in every case a long-lived localized near-periodic state is formed, with a low frequency modulation in the amplitude of the high frequency oscillations. Interestingly, the initially chosen phase difference  $\delta$ between $\phi_1$ and $\phi_2$ remains constant to a high precision during the whole time evolution. The numerical results make it likely that there exists a two-parameter family of solutions, which contains oscillatons and boson stars as well. Similar results can be obtained for the relation between complex and real Proca stars \cite{Giovanni18}.

Although in their 1991 paper showing the existence of oscillatons Seidel and Suen \cite{seidel-91} have pointed out that it is possible that oscillatons lose energy very slowly by radiating out scalar field, the authors of many subsequent papers could not find any indication supporting this in their numerical simulations. Generally, it was implicitly assumed that oscillatons are really exponentially localized and time-periodic. It was first pointed out by Don Page in 2003 that oscillatons must necessarily radiate, giving the first results for their classical and quantum radiation rate \cite{PageDon04}. That no indication of mass-energy loss was observed in several papers is understandable in view of how much smaller is the amplitude of the tail responsible for the radiation in comparison to the amplitude of the core. The most favorable case to observe the tail is the maximal mass stable oscillaton, when the central amplitude is near $0.5$. For this oscillaton the amplitude of the scalar field decreases exponentially when going outwards, until the oscillating tail appears with magnitude of the order $10^{-8}$. For smaller mass smaller amplitude oscillatons the ratio of the tail and core amplitudes become even smaller, since the magnitude of the tail decreases exponentially as a function of the central amplitude.

In paper \cite{PageDon04} Don Page has presented a calculation for the quantum decay of oscillatons by the annihilation of scalarons into gravitons. For smaller amplitude oscillatons the quantum process turns out to be stronger than the classical radiation. Quantizing a self-gravitating real scalar field may lead to a stationary complex field configuration, which is generally called boson star \cite{Ruffini69}. When the self-interaction of the field is determined by the axion potential, the corresponding localized object is also called axion star \cite{Barranco11,Eby15}. The quantum decay rate of axion stars has been considered in papers \cite{Eby16,Braaten17,Eby18,Eby19}. In this review we only discuss in detail the results for the classical radiation of oscillatons.

\section{Scalar fields in general relativity}

\subsection{The equations describing the system}

We investigate a real scalar field $\phi$ coupled to gravity in the framework of general relativity. The self-interaction of the scalar is determined by the potential $U(\phi)$. We assume that the curved spacetime is $d+1$ dimensional, with coordinates $x^a$, and the signature of the metric $g_{ab}$ is $(-+++...)$. In this review we use $G=c=\hbar=1$ Planck units, and where it is necessary we convert our results into SI units. The Lagrangian density consists of the sum of the term $\mathcal{L}_M$ describing the scalar field according to \eqref{eqlagrmatter}, and the term $\mathcal{L}_G$ characterizing the gravity,
\begin{equation}
\mathcal{L}=\frac{1}{16\pi}\mathcal{L}_G+\mathcal{L}_M \ . \label{eqacttot}
\end{equation}
The Lagrangian density of gravity in case of a cosmological constant $\Lambda$ is
\begin{equation}
\mathcal{L}_G=\sqrt{-g}(R-2\Lambda) \ ,
\end{equation}
where the determinant of the metric $g_{ab}$ is $g$, and the scalar curvature is $R$. By the variation of the action integral obtained from $\mathcal{L}$ in \eqref{eqacttot} we can get the Einstein equations,
\begin{equation}
G_{ab}+\Lambda g_{ab}=8\pi T_{ab} \ , \label{eq:einst}
\end{equation}
where the stress-energy tensor of the scalar field is given by \eqref{streneq} in the general case as well. In case of $d=1$ spatial dimension, the trace of the Einstein tensor $G_{ab}=R_{ab}-\frac{1}{2}g_{ab}R$ is zero, hence by taking the trace of the Einstein equations follows that $U(\phi)$ is constant. Hence from now on we assume that the number of spatial dimensions are at least two, i.e.~$d\geq 2$. Variation with respect to the scalar field $\phi$ leads to the Klein-Gordon equation \eqref{fieldeq1} even in the general case. This wave equation is not independent from the Einstein equations, since it can also be obtained by taking the divergence of \eqref{eq:einst} and using the twice contracted Bianchi identity $\nabla^a G_{ab}=0$.

Considering them as functions of the coordinates $x^c$, if the scalar $\phi(x^c)$ and metric $g_{ab}(x^c)$ solves the Einstein equations with potential $U$ and cosmological constant $\Lambda$, then for arbitrary positive constant $\gamma$,
\begin{equation}
\hat\phi(x^c)=\phi(\gamma x^c) \ , \qquad
\hat g_{ab}(x^c)=g_{ab}(\gamma x^c)   \label{scaleprop}
\end{equation}
is also a solution, with rescaled potential $\gamma^2 U$ and cosmological constant $\gamma^2\Lambda$. We can use this freedom to set the value of the scalar field mass to an arbitrary positive value. In case of spherical symmetry and spherical coordinates the rescaling of the angular coordinates is of course not necessary, one only has to rescale the time $t$ and radial $r$ coordinates. In the absence of a cosmological constant the natural length scale of the system is given by the scalar field mass $m$. Hence it is natural to make the choice $m=1$, which we will assume in the following when studying the $\Lambda=0$ case. At the application of the physically interesting results we will use the rescaling \eqref{scaleprop} to restore the $m$ dependence, and we will use SI units instead of the Planck units.

We assume that the potential $U(\phi)$ has a minimum at the place $\phi=0$, and its value there is $U(\phi)=0$. We assume that the potential is analytic, and write the expansion around the minimum of the potential in the form \eqref{potexpeq}, in term of the mass $m$ of the scalar field and the coefficients $g_k$. In the wave equation \eqref{fieldeq1}, which describes the evolution of the scalar field, the derivative of the potential, $U'(\phi)$, is given by \eqref{eqpotdexp}.

In order to eliminate the $8\pi$ factors appearing at the terms containing the scalar field, we introduce a rescaled scalar field and a rescaled potential,
\begin{equation}
\bar\phi=\sqrt{8\pi}\,\phi \ , \qquad
\bar U(\bar\phi)=8\pi U(\phi) \ .  \label{eq:rscpu}
\end{equation}
Then, according to \eqref{streneq}, the right-hand side of the Einstein equations is
\begin{equation}
 8\pi T_{ab}
 =\nabla_a\bar\phi\nabla_b\bar\phi-g_{ab}\left[\frac{1}{2}\nabla^c\bar\phi\nabla_c\bar\phi
 +\bar U(\bar\phi)\right] \ .
\end{equation}
The expansion of the rescaled potential is
\begin{equation}
 \bar U(\bar\phi)=\frac{1}{2}m^2\bar\phi^2
 +\sum\limits_{k=2}^{\infty}\frac{1}{k+1}\bar g_k\bar\phi^{k+1} \ ,
\end{equation}
where the transformation of the constants determining the expansion are
\begin{equation}
\bar g_k=\frac{g_k}{(8\pi)^{(k-1)/2}} \ .
\end{equation}
Since $\bar U'(\bar\phi)=\sqrt{8\pi}\,U'(\phi)$, the form of the wave equation \eqref{fieldeq1} is unchanged,
\begin{equation}
 \nabla^a\nabla_a\bar\phi=\bar U'(\bar\phi) \ . \label{fieldeq1tr}
\end{equation}

\subsection{Mass in case of spherical symmetry}

We study $d+1$ spherically symmetric spacetime, using coordinates $x^a=(t,r,\theta_1,...,\theta_{d-1})$. We can choose the metric diagonal, and name the components in the following way,
\begin{equation}
 g_{tt}=-A \ , \quad g_{rr}=B \ , \quad
 g_{\theta_1\theta_1}=C \ , \quad
 g_{\theta_n\theta_n}=C\prod_{k=1}^{n-1}\sin^2\theta_k \ , \label{eq:metrgen}
\end{equation}
where $A$, $B$ and $C$ are functions of the time coordinate $t$ and radial coordinate $r$, and if $d>2$ then $2\leq n\leq d-1$.

The function $C$ determines the metric induced on the $d-1$ dimensional symmetry spheres, and hence also gives the size of their surface. The geometry of the surface is the same as for a sphere of radius
\begin{equation}
 \hat r=\sqrt{C}  \label{eq:radfunc}
\end{equation}
on flat background. In case of spherical symmetry, for arbitrary time dependence of the metric, this $\hat r$ function can be defined in a coordinate system independent way. At any point of the spacetime the value of $\hat r$ is determined by the area of the symmetry sphere that the point belongs to, i.e.~that how large would be the radius of the sphere with same geometry in flat space. In terms of the function $\hat r$ in case of arbitrary cosmological constant $\Lambda$ we can define the $\hat m$ Misner-Sharp energy function \cite{Misner64,Nakao95},
\begin{equation}
 \hat m=\frac{(d-1)\pi^{\frac{d}{2}}}
 {8\pi\Gamma\left(\frac{d}{2}\right)}
 \hat r^{d-2}\left(1-g^{ab}\hat r_{,a}\hat r_{,b}-\frac{2\Lambda}{d(d-1)}\hat r^2\right)
 \ . \label{eq:massfunc}
\end{equation}
The function can be calculated locally from the components of the metric and their derivatives. As we will see soon, the value of $\hat m$ can be interpreted as the magnitude of the mass inside the sphere of radius $\hat r$. Its time derivative agrees with the mass-energy current radiated away by the scalar field. For large $r$ its asymptotic value gives the total mass $M$ of the spacetime. At the center $\hat r=0$, for any moment of time $t$, necessarily $\hat m=0$. This can be most easily seen using Schwarzschild coordinate $r=\hat r=\sqrt{C}$, since then at the center $g^{rr}=1$ and the bracketed part of \eqref{eq:massfunc} is zero for $r=0$.

That $\hat m$ represents the mass inside the spheres can be most naturally shown by defining an energy function $E$ in terms of the integral of a conserved quantity, and showing that $E=\hat m$. By the use of the function $\hat r$ we can define the Kodama vector \cite{Kodama80,Hayward96},
\begin{equation}
 K^a=\epsilon^{ab}\nabla_b\hat r \ , \label{eqkodamadef}
\end{equation}
where the antisymmetric tensor $\epsilon_{ab}$ is the volume element in the $(t,r)$ plane. Choosing the orientation that $\epsilon_{rt}=\sqrt{AB}$, the Kodama vector is future pointing, and its nonzero components are
\begin{equation}
 K^t=\frac{\hat r_{,r}}{\sqrt{AB}} \quad  , \qquad
 K^r=-\frac{\hat r_{,t}}{\sqrt{AB}} \ ,
\end{equation}
where comma is used for partial derivatives with respect to $t$ and $r$. Resulting from the definition \eqref{eqkodamadef}, the Kodama vector is divergence free, $\nabla_a K^a=0$. Using the form of the metric \eqref{eq:metrgen}, it can be shown by direct calculation that $G^{ab}\nabla_a K_b=0$. From the Einstein equations follows that the energy current
\begin{equation}
J_a=T_{ab}K^b
\end{equation}
is also divergence free, $\nabla_a J^a=0$, and hence it defines a conserved charge. Integrating on a constant $t$ hypersurface, which has a future pointing normal vector $n^a$, the conserved charge is
\begin{equation}
 E=\frac{2\pi^{\frac{d}{2}}}{\Gamma\left(\frac{d}{2}\right)}
 \int_0^r\hat r^{d-1}\sqrt{B}\,n^aJ_a \mathrm{d}r  \label{eq:er}
 =\frac{2\pi^{\frac{d}{2}}}{\Gamma\left(\frac{d}{2}\right)}
 \int_0^r\frac{\hat r^{d-1}}{A}\left(T_{tt}\hat r_{,r}
 -T_{tr}\hat r_{,t}\right) \mathrm{d}r \ .
\end{equation}
By a longer calculation it can be checked that the derivative of the Misner-Sharp energy function $\hat m$ is
\begin{equation}
 \nabla_a \hat m=-\frac{2\pi^{\frac{d}{2}}\hat r^{d-1}}
 {\Gamma\left(\frac{d}{2}\right)}\epsilon_{ab}J^b \ .
\end{equation}
For the radial derivative it follows that
\begin{equation}
 \hat m_{,r}=\frac{2\pi^{\frac{d}{2}}\hat r^{d-1}}
 {\Gamma\left(\frac{d}{2}\right)A}
 \left(T_{tt}\hat r_{,r}
 -T_{tr}\hat r_{,t}\right) \ .
\end{equation}
Comparing with \eqref{eq:er} follows the intended identity, $E=\hat m$.

The time derivative of the Misner-Sharp energy function is
\begin{equation}
 \hat m_{,t}=\frac{2\pi^{\frac{d}{2}}\hat r^{d-1}}
 {\Gamma\left(\frac{d}{2}\right)B}
 \left(T_{rt}\hat r_{,r}
 -T_{rr}\hat r_{,t}\right) \ . \label{eqhatmtgen}
\end{equation}
If $\Lambda=0$, then the metric becomes asymptotically flat at large distances, $A=B=1$, $C=r^2$ and $\hat r=r$. In this case, using \eqref{streneq} and \eqref{eq:rscpu},
\begin{equation}
 \hat m_{,t}=\frac{2\pi^{\frac{d}{2}}r^{d-1}}
 {\Gamma\left(\frac{d}{2}\right)}
 \phi_{,t}\phi_{,r}=\frac{2\pi^{\frac{d}{2}}r^{d-1}}
 {8\pi\Gamma\left(\frac{d}{2}\right)}
 \bar\phi_{,t}\bar\phi_{,r} \ .  \label{eq:minkmt}
\end{equation}
This expression will be used for the calculation of the mass loss of oscillatons. The result is consistent with the expressions \eqref{eqsrbar} and \eqref{eqencurr} written for the radiated energy on flat background.

Assuming that the scalar field tends to zero fast enough at infinity, the metric of the spacetime must tend to the vacuum Schwarzschild-Tangherlini metric \cite{Tangherlini63}. Using Schwarzschild coordinates, when $C=r^2$, the form of this metric is
\begin{equation}
 \mathrm{d}s^2=-\left(1-\frac{r_0^{d-2}}{r^{d-2}}\right)\mathrm{d}t^2
 +\left(1-\frac{r_0^{d-2}}{r^{d-2}}\right)^{-1}\mathrm{d}r^2
 +r^2\mathrm{d}\Omega_{d-1}^2 \ ,
\end{equation}
where $r_0$ is a constant depending on the mass. In the spatially conformally flat system, when $C=r^2B$, the metric can be written in the form
\begin{equation}
 \mathrm{d}s^2=-\left(\frac{4r^{d-2}-r_0^{d-2}}
 {4r^{d-2}+r_0^{d-2}}\right)^2\mathrm{d}t^2
 +\left(1+\frac{r_0^{d-2}}{4r^{d-2}}\right)^{\frac{4}{d-2}}\left(\mathrm{d}r^2
 +r^2\mathrm{d}\Omega_{d-1}^2\right) \ . \label{eqschwtanghconf}
\end{equation}
Calculated in any of these two coordinate systems, the Misner-Sharp energy function $\hat m$ defined in \eqref{eq:massfunc} turns out to be constant for the Schwarzschild-Tangherlini metric, hence it gives the total mass $M$ of the system,
\begin{equation}
 M=\hat m=\frac{(d-1)\pi^{\frac{d}{2}}}
 {8\pi\Gamma\left(\frac{d}{2}\right)}
 r_0^{d-2} \ .
\end{equation}
For $d=3$ spatial dimensions we have $M=r_0/2$.

For localized systems there exists a further mass concept, the proper mass $M_p$, which we can obtain by the integration of the energy density of the matter on a constant $t$ hypersurface. The energy density is $\mu=T_{ab}u^au^b$, where the components of the future pointing unit vector $u^a$ are $(1/\sqrt{A},0,...,0)$. In case of spherical symmetry, using the rescaled scalar field,
\begin{equation}
 \mu=\frac{1}{8\pi}\left[\frac{1}{2A}\left(\bar\phi_{,t}\right)^2
 +\frac{1}{2B}\left(\bar\phi_{,r}\right)^2+\bar U(\bar\phi)\right]
 \ . \label{eq:massen}
\end{equation}
For a metric in the form \eqref{eq:metrgen} the proper mass of the system can be calculated by the following $d$ dimensional volume integral:
\begin{equation}
 M_p=\frac{2\pi^{\frac{d}{2}}}{\Gamma\left(\frac{d}{2}\right)}
 \int_0^\infty \mathrm{d}r\mu\sqrt{BC^{d-1}} \ . \label{eq:massint}
\end{equation}
This mass is generally larger than the total mass $M$, which characterizes the gravitation felt by faraway observers. As we have seen, the total mass can be obtained as the limit of $\hat m$. The difference $E_b=M_p-M$ yields the gravitational binding energy, which is always positive.

\subsection{Spatially conformally flat coordinates}

The spherically symmetric diagonal metric form \eqref{eq:metrgen} does not yet fixes uniquely the applied coordinate system. The Einstein equations only give sufficient conditions for two of the functions $A$, $B$ and $C$, one function can be chosen freely. The most obvious and hence most often applied method for fixing the diffeomorphism freedom is the choice of the radial coordinate $r=\hat r=\sqrt{C}$, i.e.~the use of Schwarzschild coordinates. The assumption that the metric is diagonal fixes the constant $t$ hypersurfaces as well.

However, as it was pointed out first by Don N.~Page \cite{PageDon04}, it is more reasonable to use a coordinate system in which the $d$ dimensional space is conformally flat. This is satisfied if $g_{\theta_1\theta_1}=r^2 g_{rr}$, which in our notation requires that
\begin{equation}
 C=r^2B \ .  \label{eq:conffl}
\end{equation}
In this case the natural radius function is $\hat r=r\sqrt{B}$. This choice is motivated by the realization that in this case the oscillation of the metric component $g_{tt}$ is smaller. In correlation to this, the oscillation sensed by observers moving on constant $(r,\theta_1,...,\theta_{d-1})$ coordinate lines is much larger in the Schwarzschild case. As we will see, for small amplitude configurations, if the amplitude of the oscillaton is order $\varepsilon^2$, then the periodically oscillating part of the acceleration of these observers is order $\varepsilon^3$ in case of Schwarzschild coordinates, while it is order $\varepsilon^5$ small for the spatially conformally flat case. Spatially conformally flat coordinates may be also more appropriate for the study of boson stars in the Newtonian limit \cite{Friedberg87}. In the remaining part of this review we will use the conformally flat coordinates determined by \eqref{eq:conffl}.

Using the conformally flat coordinate system, the nontrivial linearly independent components of the  Einstein equations can be written in the following form:
\begin{align}
&(d-1)\left[\frac{d}{4B^2}\left(B_{,t}\right)^2
-\frac{A}{r^{d-1}B^{(d+2)/4}}\left(\frac{r^{d-1}B_{,r}}
{B^{(6-d)/4}}\right)_{,r}
\right]-2A\,\Lambda \notag\\
&\qquad=\left(\bar\phi_{,t}\right)^2
+\frac{A}{B}\left(\bar\phi_{,r}\right)^2+2A\,\bar U(\bar\phi) \ , \label{eq:eieq1}\\
&(d-1)\left[
\frac{(d-2)\left(r^2B\right)_{,r}}{4r^4A^{2/(d-2)}B^2}
\left(r^2A^{2/(d-2)}B\right)_{,r}
-\frac{1}{A^{1/2}B^{-1+d/4}}
\left(\frac{B^{-1+d/4}B_{,t}}{A^{1/2}}
\right)_{,t}
-\frac{d-2}{r^2}
\right]+2B\,\Lambda \notag\\
&\qquad=\left(\bar\phi_{,r}\right)^2
+\frac{B}{A}\left(\bar\phi_{,t}\right)^2-2B\,\bar U(\bar\phi) \ , \label{eq:eieq2}\\
&-\frac{d-1}{2}\,A^{1/2}\left(\frac{B_{,t}}
{A^{1/2}B}\right)_{,r}=\bar\phi_{,t}\,\bar\phi_{,r} \ , \label{eq:eieq3}\\
&\frac{rB}{A^{1/2}}\left(\frac{A_{,r}}
{rA^{1/2}B}\right)_{,r}
+(d-2){rB^{1/2}}\left(\frac{B_{,r}}{rB^{3/2}}
\right)_{,r}=-2
\left(\bar\phi_{,r}\right)^2 \ . \label{eq:eieq4}
\end{align}
We have chosen the components such that the right-hand sides of \eqref{eq:eieq1}-\eqref{eq:eieq4} should correspond to $16\pi T_{tt}$, $16\pi T_{rr}$, $8\pi T_{tr}$ and $16\pi (T_{\theta_1\theta_1}/r^2-T_{rr})$, respectively. The wave equation \eqref{fieldeq1tr} in case of spatially conformally flat coordinates is
\begin{equation}
 \frac{\bar\phi_{,rr}}{B}-\frac{\bar\phi_{,tt}}{A}
 +\frac{\bar\phi_{,r}}{2r^{2d-2}AB^{d-1}}\left(r^{2d-2}AB^{d-2}\right)_{,r}
 -\frac{\bar\phi_{,t}}{2B^{d}}\left(\frac{B^{d}}{A}\right)_{,t}
 -\bar U'(\bar\phi)=0 \ . \label{eq:wave3}
\end{equation}
The equations that are valid in arbitrary diagonal coordinate system can be found in the appendix of our paper \cite{fodor2010a}.

\section{Small-amplitude expansion of oscillatons for \texorpdfstring{$\Lambda=0$}{Lambda=0}}
\label{secosctnsmallampl}

The method described in Section \ref{seckisampl} for the description of small amplitude oscillons can be generalized for the case of the self-gravitating scalar field as well. The leading order of the expansion, usually derived by less systematic methods, has been already known in the literature \cite{Urena02b,Guzman03,PageDon04,Kichenassamy08}, and leads to the Schrödinger-Newton equations. At first, we have generalized our method, which was initially applied for oscillons, for a coupled dilaton-scalar system on flat background \cite{Fodor2009c}. This theory also leads to the Schrödinger-Newton equations to leading order, and it behaves very similarly to the gravitational system. In the rest of this section we present in detail our results about the expansion of $d+1$ dimensional oscillatons, which were published in \cite{fodor2010a}. We assume that using the scaling freedom \eqref{scaleprop} we have already set the mass of the scalar field to the value $m=1$.

Similarly to the expansion of oscillons, the expansion with respect to the parameter $\varepsilon$ is not convergent, but an asymptotic series. The expansion gives a very good approximation to the core domain of the oscillatons, even for moderately high amplitudes, but it is unable to describe tail region that is responsible for the radiation, which is exponentially small in terms of $\varepsilon$. According to the approximation represented by this expansion, the scalar field $\phi$ tends to zero at large distances exponentially, to each order in $\varepsilon$.

\subsection{Choice of coordinates}

We look for a family of localized spherically symmetric solutions of the \eqref{eq:einst} Einstein equations and the \eqref{fieldeq1tr} wave equation, which are characterized by the parameter $\varepsilon$ related to the amplitude of the states. We require that in the limit $\varepsilon\to 0$ the scalar field $\phi$ tends to zero, and the metric approaches the Minkowski spacetime. We use the coordinate system determined by \eqref{eq:conffl}, in which case the space is conformally flat. In the applied approximation it also follows from the formalism, that configurations for which the scalar field $\phi$ is not growing without bound in time must necessarily oscillate time-periodically.

Similarly to flat background oscillons, we expect that the smaller the central amplitude of an oscillaton is, the larger its size becomes. Numerical simulations clearly support this expectation. Hence we introduce a new rescaled radial coordinate $\rho$ by
\begin{equation}
 \rho=\varepsilon r \ , \label{eqrhoepsr}
\end{equation}
where $\varepsilon$ denotes the small parameter. At the same time this also means that we are looking for solutions which are slowly varying in space.

We expand the scalar field $\bar\phi$ and the metric components in terms of the powers of $\varepsilon$,
\begin{align}
 \bar\phi&=\sum_{k=1}^\infty\epsilon^{2k}\phi_{2k} \ ,\label{eq:phiexp}\\
 A&=1+\sum_{k=1}^\infty\epsilon^{2k}A_{2k} \ , \\
 B&=1+\sum_{k=1}^\infty\epsilon^{2k}B_{2k} \ . \label{eq:bexp}
\end{align}
Since we intend to use asymptotically Minkowskian coordinates, where far from the oscillaton the coordinate $t$ measures the proper time and $r$ measures the radial distance, we search for such $\phi_{2k}$, $A_{2k}$ and $B_{2k}$ functions which tend to zero if $\rho\to\infty$.

The largest difference with respect to flat background oscillons is that in \eqref{eq:phiexp}-\eqref{eq:bexp} there are only even powers of $\varepsilon$. If we allow any power of $\varepsilon$ in the expansion, then by the method presented in detail below, it can be shown that the coefficients of all odd powers of $\varepsilon$ must be necessarily zero, starting from the terms linear in $\varepsilon$. In order to make this review easily understandable, we have included only the necessary even powers in the expansion \eqref{eq:phiexp}-\eqref{eq:bexp}. For small amplitude configurations, if we assume that with the increase of the parameter $\varepsilon$ the spatial size decreases as $1/\varepsilon$, then for flat background oscillons the amplitude of the core grows proportionally to $\varepsilon$, while for self-gravitating oscillatons the growth is proportional to $\varepsilon^2$. This is a fundamental difference between the scaling properties of the two kind of objects.

The oscillation frequency also depends on the amplitude of the oscillatons. Similarly to the flat background case, the smaller the amplitude is the closer the frequency gets to the value $m=1$. This is also supported by numerical simulations. In view of this, we introduce a rescaled time coordinate $\tau$ by
\begin{equation}
 \tau=\omega t \ ,
\end{equation}
where $\omega$ is a function of the parameter $\varepsilon$. We choose the coordinate $\tau$ in such a way, that in terms of that, the frequency should be exactly $1$, independently of the value of $\varepsilon$. Since we have already set the scalar field mass to $m=1$, the value of $\omega$ gives the oscillation frequency of the oscillaton. We look for the expansion of the square of the function $\omega$ in the following form:
\begin{equation}
 \omega^2=1+\sum_{k=1}^\infty\varepsilon^{2k}\omega_{2k} \ . \label{eqom2expgr}
\end{equation}
We could allow odd powers in this expression as well, but their coefficients would turn out to be zero based on the equations resulting from the expansion formalism. There is a huge freedom in how we parametrize the various states. If for small $\varepsilon$ the amplitude remains proportional to $\varepsilon^2$, then any $\varepsilon\to\bar\varepsilon(\varepsilon)$ reparametrization is allowed. The physical parameter is not $\varepsilon$, but the frequency of the states. We will show, that for $3\leq d\leq 5$ spatial dimensions we can fix the parametrization by the choice $\omega=\sqrt{1-\varepsilon^2}$.

\subsection{Leading order results} \label{subsecvezeredm}

The field equations that we have to solve are the \eqref{eq:eieq1}-\eqref{eq:eieq4} Einstein equations and the \eqref{eq:wave3} wave equation. According to the choice $C=r^2B$, we use spatially conformally flat coordinates. The expansion using $C=r^2$ Schwarzschild coordinates can be found in the appendix of our paper \cite{fodor2010a}. Changing to the new coordinates, the derivatives transform according to the rules
\begin{equation}
 \frac{\partial}{\partial t} \to \omega\frac{\partial}{\partial\tau}
 \ , \qquad
 \frac{\partial}{\partial r} \to
 \varepsilon\frac{\partial}{\partial\rho} \ .
\end{equation}
In the equations only the even powers of the frequency $\omega$ appear, of which we can substitute by the help of \eqref{eqom2expgr}. After this, we can separately solve the parts of the equations proportional to $\varepsilon^n$, in increasing order of $n$.

We get the first conditions at order $\varepsilon^2$. From the Einstein equations follow that $\frac{\partial^2 B_2}{\partial\tau^2}=0$ and $\frac{\partial^2 B_2}{\partial\tau\partial\rho}=0$, and from the wave equation that $\frac{\partial^2\phi_2}{\partial\tau^2}+\phi_2=0$. Since we are looking for solutions that remain bounded as the time passes, we cannot allow terms that grow linearly in $\tau$. Hence the solution of the equations are
\begin{equation}
 \phi_2=p_2\cos(\tau+\delta) \ , \quad B_2=b_2 \ , \label{eqphi2b2}
\end{equation}
where $p_2$, $\delta$ and $b_2$ three new functions that are only dependent on $\rho$. The radial dependence of these functions are determined by the conditions at order $\varepsilon^4$.

From the $\varepsilon^4$ part of the Einstein equation \eqref{eq:eieq1} follows that
\begin{equation}
 \frac{\mathrm{d}^2b_2}{\mathrm{d}\rho^2}+\frac{d-1}{\rho}\,
 \frac{\mathrm{d}b_2}{\mathrm{d}\rho}=
 \frac{1}{d-1}\,p_2^2 \ . \label{eqddb2}
\end{equation}
The $\varepsilon^4$ part of the equation \eqref{eq:eieq4} gives the condition
\begin{equation}
 \frac{\partial}{\partial\rho}\left\{
 \frac{1}{\rho}\frac{\partial}{\partial\rho}\left[
 (d-2)b_2+A_2
 \right]\right\}=0 \ .
\end{equation}
The solution of this is $(d-2)b_2+A_2=\frac{1}{2}\rho f_1+f_2$, where $f_1$ and $f_2$ are two arbitrary functions of the coordinate $\tau$. Since we are looking for solutions for both $b_2$ and $A_2$ that tend to zero at infinity, necessarily $f_1=f_2=0$, and hence $(d-2)b_2+A_2=0$. This also means that $A_2$ is time-independent. In order to remind us to the time independence, we introduce the notation
\begin{equation}
 A_2=a_2 \ ,
\end{equation}
where $a_2$ only depends on $\rho$. Then
\begin{equation}
 (d-2)b_2+a_2=0 \ . \label{eqb2a2ff}
\end{equation}
Using this, from the $\varepsilon^4$ part of the Einstein equation \eqref{eq:eieq2} follows the condition:
\begin{equation}
 \frac{\partial^2 B_4}{\partial\tau^2}=\frac{p_2^2}{d-1}
 \cos\left[2(\tau+\delta)\right] \ .
\end{equation}
The solution, which is not growing without bound as time increases, can be written as
\begin{equation}
 B_4=b_4-\frac{p_2^2}{4(d-1)}\cos\left[2(\tau+\delta)\right] \ , \label{eqbb4timedep}
\end{equation}
where $b_4$ is a further arbitrary function of $\rho$. Here we have dropped a term $\tau f_3$, where $f_3$ is a function of $\rho$. After these considerations, from the $\varepsilon^4$ part of equation \eqref{eq:eieq3} follows that $\frac{\mathrm{d}}{\mathrm{d}\rho}\delta=0$. Shifting the time coordinate we can set
\begin{equation}
 \delta=0 \ .
\end{equation}
This shows that the scalar field oscillates with identical phase at all radii $\rho$.

The part of the wave equation \eqref{eq:wave3} which is proportional to $\varepsilon^4$ gives
\begin{equation}
 \frac{\partial^2\phi_4}{\partial\tau^2}+\phi_4
 +\frac{\bar g_2}{2}p_2^2\left[1+\cos(2\tau)\right]
 -\left[
 \frac{\mathrm{d}^2p_2}{\mathrm{d}\rho^2}
 +\frac{d-1}{\rho}\,\frac{\mathrm{d}p_2}{\mathrm{d}\rho}
 +\omega_2 p_2+(d-2)p_2 b_2
 \right]\cos\tau=0 \ . \label{eqphi4taudep}
\end{equation}
The structure of this equation is the same as that of \eqref{eqphik} written at the flat background case. Solution that remains bounded in time can only exist for $\phi_4$ if the coefficient of the resonant $\cos\tau$ term vanishes, from which it follows that
\begin{equation}
 \frac{\mathrm{d}^2p_2}{\mathrm{d}\rho^2}
 +\frac{d-1}{\rho}\,\frac{\mathrm{d}p_2}{\mathrm{d}\rho}
 +\omega_2 p_2+(d-2)p_2 b_2=0 \ . \label{eqddp2a}
\end{equation}
If $d\not=2$, then this equation and \eqref{eqddb2} forms a coupled system for $b_2$ and $p_2$, of which localized solution may exist. In case of $d=2$ spatial dimensions the equations decouple, \eqref{eqddp2a} becomes linear, and hence obviously no localized solution can exist. Because of this, in the following we assume that $d>2$. Then with the help of \eqref{eqb2a2ff}, the coupled system can be written into the form:
\begin{align}
 \frac{\mathrm{d}^2a_2}{\mathrm{d}\rho^2}
 +\frac{d-1}{\rho}\,\frac{\mathrm{d}a_2}{\mathrm{d}\rho}&=
 \frac{d-2}{d-1}\,p_2^2 \ , \label{eq:x2}\\
 \frac{\mathrm{d}^2p_2}{\mathrm{d}\rho^2}
 +\frac{d-1}{\rho}\,\frac{\mathrm{d}p_2}{\mathrm{d}\rho}&=
 p_2(a_2-\omega_2) \ . \label{eq:p2}
\end{align}
The solutions of this system, together with the relation
\begin{equation}
 b_2=\frac{a_2}{2-d} \ ,  \label{eq:b2}
\end{equation}
resulting from \eqref{eqb2a2ff}, using the equations
\begin{equation}
 \phi_2=p_2\cos\tau \ , \quad A_2=a_2 \ , \quad B_2=b_2
\end{equation}
determine the leading $\varepsilon^2$ order behavior of the functions $\bar\phi$, $A$ and $B$ for small amplitude oscillatons. Up to this order the functions are the same for arbitrary scalar potential, assuming that the mass of the scalar field has been set to the value $m=1$. This means that small amplitude, and hence large sized, oscillatons always behave in the same way as in the Klein-Gordon case. In other words, the gravitational interaction dominates over the self-interaction of the scalar for small-amplitude configurations.

If the resonance condition \eqref{eqddp2a} is satisfied then \eqref{eqphi4taudep} determines the time dependence of $\phi_4$,
\begin{equation}
 \phi_4=p_4\cos\tau+\frac{\bar g_2}{6}p_2^2\left[\cos(2\tau)-3\right] \ , \label{eqphi4p4cos}
\end{equation}
where $p_4$ is a function of $\rho$. The $\phi_4$ given in \eqref{eqphi4p4cos} still remains a solution if we add a further $q_4\sin\tau$ term, where $q_4$ is an arbitrary function of $\rho$. However, from the equations at higher orders in $\varepsilon$ follows that by a small shift in the time coordinate it is always possible to set $q_4=0$. Similarly, at all higher orders of $\varepsilon$ it can be shown that only terms with time dependence $\cos(k\tau)$ appear in the expansions, and hence the oscillaton must be time-reflection symmetric at the moment $\tau=0$. The time dependence of the $B_4$ function is given by \eqref{eqbb4timedep}, while the time dependence of $A_4$ and the spatial dependences will be determined at order $\varepsilon^6$.

\subsection{The Schr\"odinger-Newton equations}\label{sec:SN}

Since we have already seen that for $d=2$ spatial dimensions no small amplitude oscillaton solutions can exist, we assume that $d\geq 3$. Introducing the functions $s$ and $S$ by the relations
\begin{equation}
 s=\omega_2-a_2 \ , \quad S=p_2\sqrt{\frac{d-2}{d-1}} \ ,
\label{eq:sands}
\end{equation}
equations \eqref{eq:x2} and \eqref{eq:p2} can be written into the following form:
\begin{align}
 \frac{\mathrm{d}^2S}{\mathrm{d}\rho^2}
 +\frac{d-1}{\rho}\,\frac{\mathrm{d} S}{\mathrm{d}\rho}
 +s S&=0
 \ , \label{eqsn1}\\
 \frac{\mathrm{d}^2s}{\mathrm{d}\rho^2}
 +\frac{d-1}{\rho}\,\frac{\mathrm{d} s}{\mathrm{d}\rho}
 +S^2&=0
 \ , \label{eqsn2}
\end{align}
which are known as the time-independent Schr\"odinger-Newton (SN) equations  \cite{Moroz98,TodMoroz99,Harrison03,Choquard08}. It is also possible to obtain these equations by the study of the collapse of the quantum mechanical wave function resulting from gravitational interaction \cite{Diosi84,Penrose98}. The near-Newtonian behavior of a small-amplitude boson star is also described by the SN equations \cite{Ruffini69,Friedberg87,Ferrell89}. The generalization of the SN equations is also known as Choquard equation in the literature \cite{Moroz17}.

From any solution of the SN equations \eqref{eqsn1}-\eqref{eqsn2}, using the scale invariance
\begin{equation}
 (S(\rho),s(\rho)) \to
 (\lambda^2S(\lambda\rho),\lambda^2s(\lambda\rho))  \label{snscale}
\end{equation}
we can obtain another solution for any $\lambda>0$.

If the dimension of the space is $3\leq d\leq5$, then for any $n\geq 0$ integer there exist a localized regular solution of the SN equations, for which $S$ has exactly $n$ nodes (zero crossings). The nodeless solution belonging to $n=0$ corresponds to the lowest mass most stable oscillaton, hence in the following we will only study that case. For the solutions of the system, if $\rho\to\infty$ the function $S$ tends to zero exponentially, and $s$ tends to a constant $s_0<0$ in the following way:
\begin{equation}
 s\approx s_0+s_1\rho^{2-d} \ . \label{eq:sasympt}
\end{equation}
We use the scaling freedom \eqref{snscale} to set $s_0=-1$. At the same time we change the $\varepsilon$ parametrization such that $\omega_2=-1$ should hold, which ensures that $a_2$ tends to zero asymptotically. Then the asymptotic behavior of the function $a_2$ is
\begin{equation}
 a_2\approx-s_1\rho^{2-d}  \ , \label{eq:a2asy}
\end{equation}
which is only modified by exponentially decaying contributions. At higher orders of the $\varepsilon$ expansion it is always possible to set that in the expansion \eqref{eqom2expgr} there should be $\omega_i=0$ for every $i>2$, by which we fix the $\varepsilon$ parametrization with respect to the frequency $\omega$,
\begin{equation}\label{eq:omeps}
 \omega=\sqrt{1-\varepsilon^2} \ \ \  {\rm for} \ \  \  3\leq d\leq5 \ .
\end{equation}

If the number of space dimensions is $d=6$, then the asymptotically decaying solutions are known,
\begin{equation}
 s=\pm S=\frac{24\alpha^2}{\left(1+\alpha^2\rho^2\right)^2} \ , \label{6dss}
\end{equation}
where $\alpha$ is an arbitrary constant. In this case both $s$ and $S$ decay at infinity, but the solution is not exponentially localized. Furthermore, to ensure that $A$ should tend to $1$ at infinity we have to require that $\omega_2=0$. In case of $d=6$ we can choose $\omega_4=-1$ as the only nonvanishing coefficient. If the number of spatial dimensions is $d>6$, then there is no solution of the SN equations which describes a finite energy localized state \cite{Choquard08}.

Motivated by the asymptotic behavior of the function $s$, if $d\not=2$ it is useful to introduce the functions $\bar\sigma$ and $\bar\nu$ by the equations
\begin{equation}
 \bar\sigma=\frac{\rho^{d-1}}{2-d}\,\frac{\mathrm{d}s}{\mathrm{d}\rho} \ , \qquad
 \bar\nu=s-\rho^{2-d}\bar\sigma \ . \label{eq:sigmadef}
\end{equation}
For $3\leq d\leq5$ dimensions these tend to the earlier defined constants,
\begin{equation}
 \lim_{\rho\to\infty}\bar\sigma=s_1 \ , \qquad
 \lim_{\rho\to\infty}\bar\nu=s_0 \ . \label{e:s0s1}
\end{equation}
The Schr\"odinger-Newton equations can be written into the following equivalent form:
\begin{align}
&\frac{\mathrm{d}\bar\sigma}{\mathrm{d}\rho}+\frac{\rho^{d-1}}{2-d}S^2=0
\ ,\label{e:dsigma}\\
&\frac{\mathrm{d}\bar\nu}{\mathrm{d}\rho}+\frac{\rho}{d-2}S^2=0 \ ,\\
\frac{\mathrm{d}^2S}{\mathrm{d}\rho^2}
+&\frac{d-1}{\rho}\,\frac{\mathrm{d}S}{\mathrm{d}\rho}
+\left(\bar\nu+\rho^{2-d}\bar\sigma\right) S=0 \ ,
\end{align}
which is more appropriate for the search of high precision numerical solutions.

For the case when $S$ is nodeless, in Table \ref{c1table}
\begin{table}[htbp]
\begin{center}
\begin{tabular}{c|c|c|c|}
  & $d=3$  & $d=4$  & $d=5$ \\
  \hline
  $s_c$ & $0.938323$ & $3.429755$ & $13.90730$ \\
  $S_c$ & $1.021493$ & $3.542143$ & $14.01996$ \\
  $s_1$ & $3.505330$ & $7.694895$ & $10.40384$ \\
  \hline
\end{tabular}
\end{center}
\caption{Numerical results for the central values of the functions $s$ and $S$, and also for the constant $s_1$, in case of $d=3$, $4$ and $5$ dimensions.  \label{c1table}}
\end{table}
we give the central values of $s$ and $S$, and also the constant $s_1$. The value of the constant $s_0$ has been set to $-1$.

\subsection{Higher orders of the expansion}\label{sec:next}

From the order $\varepsilon^6$ components of the field equations follows the time dependence of $A_4$,
\begin{equation}
 A_4=a_4^{(0)}+a_4^{(2)}\cos(2\tau) \ ,
\end{equation}
where $a_4^{(0)}$ and $a_4^{(2)}$ are functions of the radial coordinate $\rho$. The functions $p_4$ and $a_4^{(0)}$ are determined by the following coupled differential equations:
\begin{align}
 \frac{\mathrm{d}^2a_4^{(0)}}{\mathrm{d}\rho^2}
 +\frac{d-1}{\rho}\,\frac{\mathrm{d}a_4^{(0)}}{\mathrm{d}\rho}&=
 \frac{2p_2p_4(d-2)}{d-1}+\left(\frac{\mathrm{d}a_2}{\mathrm{d}\rho}\right)^2
 +\omega_2p_2^2-\frac{2p_2^2a_2}{d-1}
 \ , \label{eq:a40}\\
 \frac{\mathrm{d}^2p_4}{\mathrm{d}\rho^2}
 +\frac{d-1}{\rho}\,\frac{\mathrm{d}p_4}{\mathrm{d}\rho}&=
 p_4(a_2-\omega_2)+\left(a_4^{(0)}-\omega_4\right)p_2
 -\frac{a_2p_2(d-1)(a_2-\omega_2)}{d-2} \notag\\
 &-\frac{dp_2^3}{8(d-1)}-\left(\frac{5}{6}\bar g_2^2
 -\frac{3}{4}\bar g_3\right)p_2^3 \ . \label{eq:p4}
\end{align}
We look for the unique solution for which both $a_4^{(0)}$ and $p_4$ tend to zero if $\rho\to\infty$. If $3\leq d\leq5$, then the function $p_4$ tends to zero exponentially, while for large $\rho$ values
\begin{equation}
a_4^{(0)}\approx\frac{1}{2}s_1^2\rho^{4-2d}+s_2\rho^{2-d}+s_3
\label{eq:a4asympt} \ ,
\end{equation}
where $s_1$ is the constant defined in equation \eqref{eq:sasympt}, moreover $s_2$ and $s_3$ are further constants. If $a_4^{(0)}$ and $p_4$ are solutions of \eqref{eq:a40}-\eqref{eq:p4}, then for any constant $c$
\begin{align}
 \bar a_4^{(0)}&=a_4^{(0)}+c\left[2(a_2-\omega_2)
 +\rho\frac{\mathrm{d}a_2}{\mathrm{d}\rho}\right] , \\
 \bar p_4&=p_4+c\left(2p_2+\rho\frac{\mathrm{d}p_2}{\mathrm{d}\rho}\right) ,
\end{align}
are also solutions. This family of solutions is generated by the scale invariance \eqref{snscale} of the SN equations. If we have an arbitrary solution of the equations \eqref{eq:a40}-\eqref{eq:p4}, then by the appropriate choice of $c$ we can obtain another solution for which $s_3=0$ holds in \eqref{eq:a4asympt}.

The equation determining the function $b_4$ is
\begin{align}
 \frac{\mathrm{d}b_4}{\mathrm{d}\rho}=&
 \frac{1}{2-d}\,\frac{\mathrm{d}a_4^{(0)}}{\mathrm{d}\rho}
 +\frac{1}{4(d-2)^2}\,\frac{\mathrm{d}a_2}{\mathrm{d}\rho}
 \left[\rho\frac{\mathrm{d}a_2}{\mathrm{d}\rho}+4(d-1)a_2\right] \notag\\
 &+\frac{\rho}{2(d-1)(d-2)}
 \left[\left(\frac{\mathrm{d}p_2}{\mathrm{d}\rho}\right)^2
 -p_2^2(a_2-\omega_2)\right] \ . \label{eq:sb4}
\end{align}
For large $\rho$ values $b_4$ tends to zero according to
\begin{equation}
 b_4\approx \frac{6-d}{8(d-2)^2}s_1^2\rho^{4-2d}
 +\frac{s_2}{2-d}\rho^{2-d} \ .
\end{equation}
The equation determining the $\cos(2\tau)$ part of the function $A_4$ is
\begin{equation}
 \frac{\mathrm{d}^2a_4^{(2)}}{\mathrm{d}\rho^2}
 -\frac{1}{\rho}\,\frac{\mathrm{d}a_4^{(2)}}{\mathrm{d}\rho}=
 \frac{(d-2)(a_2-\omega_2)p_2^2}{2(d-1)}
 -\frac{d}{2(d-1)}\,\frac{\mathrm{d}p_2}{\mathrm{d}\rho}
 \left(\frac{\mathrm{d}p_2}{\mathrm{d}\rho}
 +\frac{d-2}{\rho}p_2\right) \ . \label{eq:sa42}
\end{equation}
As a reminder we note that for $3\leq d\leq5$ the natural choice is $\omega_2=-1$.

Summarizing our results so far, up to $\varepsilon^4$ order the components of the scalar field and of the metric are
\begin{align}
 \bar\phi&=\varepsilon^2p_2\cos\tau+\varepsilon^4\left\{
 p_4\cos\tau+\frac{\bar g_2p_2^2}{6}\left[\cos(2\tau)-3\right]\right\}
 +\mathcal{O}(\varepsilon^6) \ , \label{eq:phisum}\\
 A&=1+\varepsilon^2a_2+\varepsilon^4\left[
 a_4^{(0)}+a_4^{(2)}\cos(2\tau)\right]+\mathcal{O}(\varepsilon^6) \ , \label{eq:asum}\\
 B&=1-\varepsilon^2\frac{a_2}{d-2}
 +\varepsilon^4\left[
 b_4-\frac{p_2^2}{4(d-1)}\cos(2\tau)
 \right]+\mathcal{O}(\varepsilon^6)
 \ . \label{eq:bsum}
\end{align}
Proceeding to higher orders, the expressions become considerably longer and more complicated. However, it can be shown that for symmetric potentials, in which case $\bar g_{2k}=0$, the scalar field $\bar\phi$ only contains $\cos(k\tau)$ terms with odd $k$, while $A$ and $B$ has only even Fourier components.

The higher order expressions become considerably simpler for symmetric potentials, when $\bar g_{2k}=0$. Since for symmetric potentials the first radiating mode in $\bar\phi$, which is proportional to $\cos(3\tau)$, appears at order $\varepsilon^6$, we give its higher order expansion in this case,
\begin{align}
 \bar\phi=&\varepsilon^2p_2\cos\tau
 +\varepsilon^4p_4\cos\tau+\varepsilon^6p_6\cos\tau \label{eq:phisum3}\\
 &+\varepsilon^6\left(
 \frac{p_2^3 d}{64(d-1)}
 +\frac{\bar g_3p_2^3}{32}+\frac{p_2a_4^{(2)}}{8}
 \right)\cos(3\tau)
 +\mathcal{O}(\varepsilon^8) \ , \notag
\end{align}
where $p_6$ is a function of $\rho$, which is determined by a long differential equation appearing at higher orders.

\begin{figure}[!ht]
    \begin{center}
    \includegraphics[width=100mm]{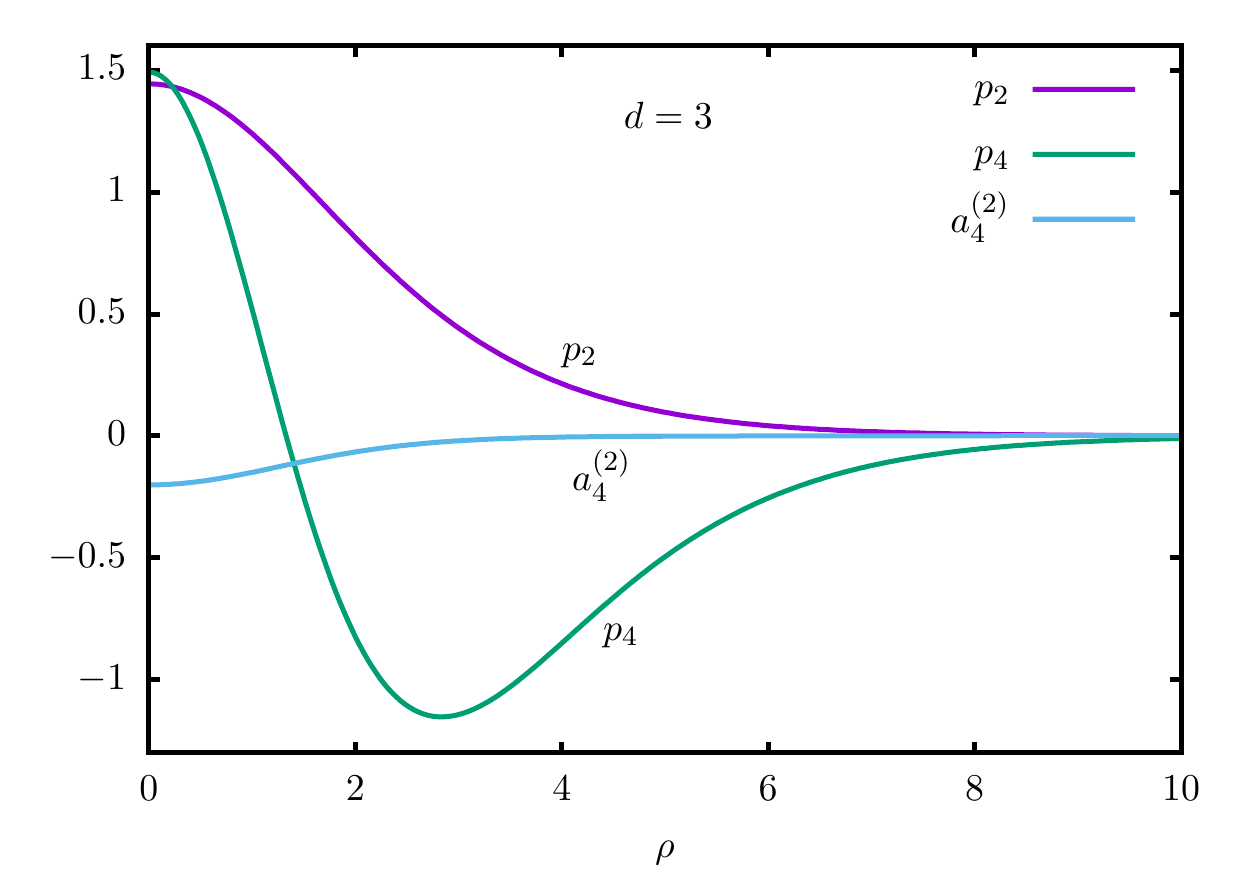}
    \end{center}
    \caption{The exponentially decaying $p_2$, $p_4$ and $a_4^{(2)}$ functions, at the small-amplitude expansion of the Klein-Gordon oscillaton, for $d=3$ spatial dimensions.
      \label{f:exp}}
\end{figure}

\begin{figure}[!ht]
    \begin{center}
    \includegraphics[width=100mm]{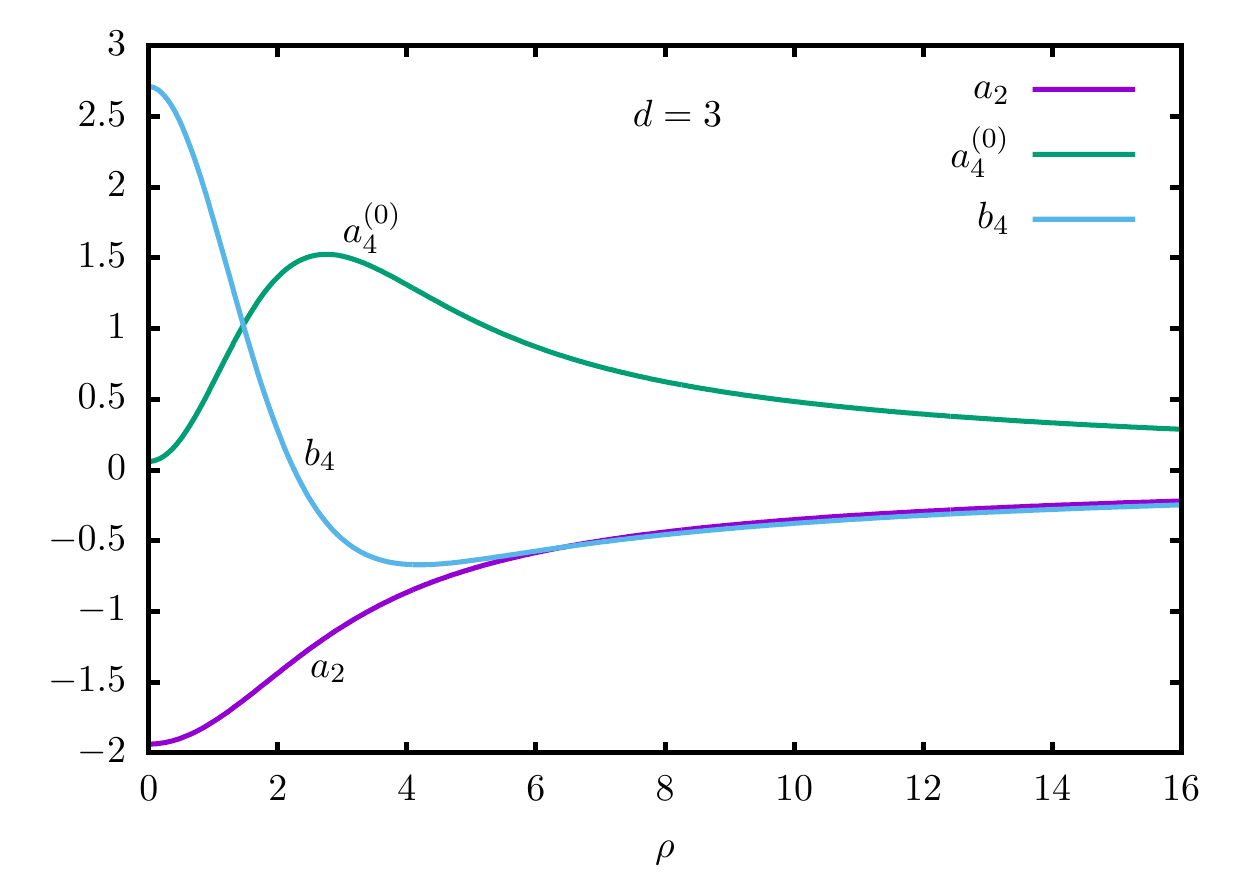}
    \end{center}
    \caption{The functions $a_2$, $a_4^{(0)}$ and $b_4$ for the $d=3$
      Klein-Gordon system. These tend to zero according to a power law when $\rho\to\infty$.
 \label{f:pow}}
\end{figure}

On Figures \ref{f:exp} and \ref{f:pow} we show the numerically calculated $p_2$, $a_2$, $p_4$, $a_4^{(0)}$, $a_4^{(2)}$ and $b_4$ functions for Klein-Gordon potential and $d=3$ spatial dimensions.

Equations \eqref{eq:phisum}-\eqref{eq:bsum} determine a one-parameter family of solutions, which depends on the parameter $\varepsilon$. This family solves the field equations up to order $\varepsilon^4$ for a self-interacting, self-gravitating scalar field with mass $m=1$. By the rescaling \eqref{scaleprop} of the $t$ and $r$ coordinates we can obtain one-parameter families for arbitrary scalar field mass $m$.

Using the spatially conformally flat coordinate system corresponding to the choice $C=r^2B$, according to \eqref{eq:asum}-\eqref{eq:bsum}, up to $\varepsilon^2$ order the metric is not depending on the time coordinate, i.e.~it is static. In contrast to this, using the $C=r^2$ Schwarzschild coordinates, $a_2$ has a $\varepsilon^2$ order part oscillating according to $\cos(2\tau)$ (see Appendix B of our paper \cite{fodor2010a}). This is the most important advantage of spatially conformally flat coordinates over the most widely used Schwarzschild coordinate system. In the Schwarzschild system the constant $r$ observers feel order $\varepsilon^2$ oscillations in the metric. For observers moving on constant $(r,\theta_1,\theta_2...)$ lines, the absolute value of the observer's acceleration vector in the generic diagonal metric \eqref{eq:metrgen} is
\begin{equation}
 \mathrm{a}=\frac{1}{2A\sqrt{B}}\,\frac{\mathrm{d}A}{\mathrm{d}r} \ . \label{eq:acc}
\end{equation}
This has an order $\varepsilon^3$ oscillating component in the Schwarzschild system, while in the spatially conformally flat case the fluctuation in the acceleration is order $\varepsilon^5$ small.

\subsection{Mass, frequency and size of oscillatons} \label{secosctnmass}

At the $\varepsilon$ expansion of oscillatons in the spatially conformally flat system the invariant radius function is $\hat r=r\sqrt{B}$. Into this, we have to substitute the expansion of $B$ in the form \eqref{eq:bexp}. Using the rescaled radial coordinate $\rho=\varepsilon r$, the expansion of the Misner-Sharp energy function defined in \eqref{eq:massfunc} can be written as
\begin{equation}
 \hat m=\varepsilon^{4-d}\hat m^{(1)}
 +\varepsilon^{6-d}\hat m^{(2)}
 +\mathcal{O}\left(\varepsilon^{8-d}\right) \ ,
\end{equation}
where
\begin{align}
 \hat m^{(1)}&=-\frac{(d-1)\pi^{\frac{d}{2}}}
 {8\pi\Gamma\left(\frac{d}{2}\right)}\rho^{d-1}
 \frac{\mathrm{d}B_2}{\mathrm{d}\rho} \ , \label{eqhatm1} \\
 \hat m^{(2)}&=-\frac{(d-1)\pi^{\frac{d}{2}}}
 {8\pi\Gamma\left(\frac{d}{2}\right)}\rho^{d-1}
 \Biggl[\frac{\mathrm{d}B_4}{\mathrm{d}\rho}
 +\frac{\rho}{4}\left(\frac{\mathrm{d}B_2}{\mathrm{d}\rho}\right)^2
 +\frac{d-4}{2}B_2\frac{\mathrm{d}B_2}{\mathrm{d}\rho}\Biggr] \ .
\end{align}
According to \eqref{eqphi2b2} we have $B_2=b_2$, and from \eqref{eq:b2} we get $b_2=a_2/(2-d)$. In view of the definition \eqref{eq:sands} of the function $s$, we have $a_2=-1-s$. The derivative of $s$, based on \eqref{eq:sigmadef}, can be expressed by the function $\bar\sigma$, and hence \eqref{eqhatm1} can be written into the following form:
\begin{equation}
 \hat m^{(1)}=\frac{(d-1)\pi^{\frac{d}{2}}}
 {8\pi\Gamma\left(\frac{d}{2}\right)}\,\bar\sigma \ . \label{eqhatm1b}
\end{equation}

The total mass of the oscillatons is given by the $\rho\to\infty$ limit,
\begin{equation}
M=\varepsilon^{4-d}M^{(1)}
+\varepsilon^{6-d}M^{(2)}
+\mathcal{O}\left(\varepsilon^{8-d}\right) \  , \label{eq:totmass}
\end{equation}
where the constants $M^{(k)}$ are the limits of $\hat m^{(k)}$. According to \eqref{e:s0s1}, the limit of $\bar\sigma$ is the constant $s_1$ defined in \eqref{eq:sasympt}, hence
\begin{equation}
 M^{(1)}=\frac{(d-1)\pi^{\frac{d}{2}}}
 {8\pi\Gamma\left(\frac{d}{2}\right)}\,s_1 \ . \label{eqmm1totm}
\end{equation}
The numerical value of the constant $s_1$ for various dimensions was given in Table \ref{c1table}. Using \eqref{eq:sb4} for the derivative of $b_4$,
\begin{equation}
 M^{(2)}=\lim_{\rho\to\infty}\frac{(d-1)\pi^{\frac{d}{2}}}
 {8\pi\Gamma\left(\frac{d}{2}\right)}\,\frac{\rho^{d-1}}{d-2}
 \Biggl[\frac{\mathrm{d}a_4^{(0)}}{\mathrm{d}\rho}
 -\frac{\rho}{2(d-2)}\left(
 \frac{\mathrm{d}a_2}{\mathrm{d}\rho}\right)^2
 -\frac{3}{2}a_2\frac{\mathrm{d}a_2}{\mathrm{d}\rho}\Biggr] \ ,
\end{equation}
which can be easily calculated numerically, since the expression of which we take the limit tends to a constant exponentially.

The proper mass of oscillatons can be calculated using equation \eqref{eq:massint}. Substituting the $\varepsilon$ expansion results,
\begin{equation}
 M_p=\varepsilon^{4-d}M_p^{(1)}
 +\varepsilon^{6-d}M_p^{(2)}
 +\mathcal{O}\left(\varepsilon^{8-d}\right) \ , \label{eqmpeps}
\end{equation}
where
\begin{align}
 M_p^{(1)}&=\frac{\pi^{\frac{d}{2}}}
 {8\pi\Gamma\left(\frac{d}{2}\right)}
 \int_0^\infty \mathrm{d}\rho\,\rho^{d-1}p_2^2 \ , \label{eq:mp1}\\
 M_p^{(2)}&=\frac{\pi^{\frac{d}{2}}}
 {8\pi\Gamma\left(\frac{d}{2}\right)}
 \int_0^\infty \mathrm{d}\rho\,\rho^{d-1}
 \biggl(2p_2p_4-p_2^2-\frac{3d-4}{2(d-2)}a_2p_2^2\biggr)
 \ . \label{eq:mp2}
\end{align}
Using the expressions \eqref{eq:sands}, \eqref{e:s0s1} and \eqref{e:dsigma}, the leading order coefficient is
\begin{equation}
 M_p^{(1)}=\frac{(d-1)\pi^{\frac{d}{2}}}
 {8\pi\Gamma\left(\frac{d}{2}\right)}\,s_1 \ , \label{eq:propm1}
\end{equation}
which agrees with the leading order coefficient of the total mass $M^{(1)}$ given in \eqref{eqmm1totm}. We show the numerical values of the coefficients of the proper mass $M_p$ and total mass $M$ for the Klein-Gordon case in Table \ref{masstable}.
\begin{table}[htbp]
\begin{center}
\begin{tabular}{c|c|c|c|}
  & $d=3$  & $d=4$  & $d=5$ \\
 \hline
 $M^{(1)}=M_p^{(1)}$ &  $1.75266$ &   $9.06533$ &  $21.7897$ \\
 $M^{(2)}$          & $-2.11742$ & $-43.5347$ & $-555.521$ \\
 $M_p^{(2)}$        & $-1.53319$ & $-39.0020$ & $-533.732$ \\
 \hline
\end{tabular}
\end{center}
\caption{\label{masstable}
 The coefficients of the $\varepsilon$ expansion of the total mass $M$ and the proper mass $M_p$ for a Klein-Gordon scalar with mass $m=1$, for spatial dimensions $d=3,\,4,\,5$.
}
\end{table}
As we have seen, the total mass and the proper mass agree to leading order. However, taking into account the next order of the expansion, according to the physical expectation, the $E_b=M_p{-M}$ bounding energy is always positive.

The results up to here has been obtained with the assumption that the mass of the scalar field has been set to the value $m=1$. For the calculation of the mass in the case $m{\not=}1$ we can use the scaling freedom \eqref{scaleprop}. In this case, a factor $m^2$ appears in the expression \eqref{eq:massen} defining the energy density $\mu$. Since the volume elements in the integrals bring $m^{-d}$ factors, in the expression \eqref{eqmpeps} of the proper mass a factor $m^{2-d}$ appears in each coefficient $M_p^{(k)}$. The same $m^{2-d}$ factor appears in the coefficients $M^{(k)}$ of the total mass in \eqref{eq:totmass}, because of the $\hat r^{d-2}$ factor in the definition \eqref{eq:massfunc}.

Considering the Klein-Gordon potential, if we measure the quantities in SI and electronvolt units, then at $d=3$ spatial dimensions the dependence \eqref{eq:totmass} of the total mass on the parameter $\varepsilon$ and on the scalar field mass  $m$ is
\begin{equation}
 M[\mathrm{kg}]=\varepsilon\left(4.657-5.627\,\varepsilon^2\right)10^{20}
 \frac{1}{m[\mathrm{eV}/c^2]} \ , \label{eqmmepdep3d}
\end{equation}
where the quantities are represented by dimensionless numbers in terms of the units given in the square brackets, instead of the Planck units generally used in this review. Since we have dropped $\mathcal{O}\left(\varepsilon^{4}\right)$ order terms, the result can only be precise for small $\varepsilon$ values. However, as we will see later at the comparison with the numerical results, the expression \eqref{eqmmepdep3d} gives acceptable estimation even for $\varepsilon\approx 0.5$.

The parameter $\varepsilon$ remains dimensionless. However, if the mass of the scalar field is not unity then because of the scaling freedom \eqref{scaleprop} the time scale and hence the frequency is also changing. In case of $m\not=1$, the oscillation frequency of the oscillaton is $\omega=m\sqrt{1-\varepsilon^2}$ in Planck units, while in SI and electronvolt units
\begin{equation}
 \omega[1/\mathrm{s}]=1.519\cdot 10^{15}\,m[\mathrm{eV}/c^2]\,\sqrt{1-\varepsilon^2} \ .
\end{equation}

Because of the rescaling \eqref{scaleprop} the radial coordinate is also changing. This can also be interpreted in the way that in the expression \eqref{eqrhoepsr} the scalar field mass $m$ appears, i.e.~$r=\rho/(\varepsilon m)$. Although the oscillatons are exponentially localized, they do not have a definite outer surface. As a natural definition for their size, we take the radius $r_q$ inside which $q$ percent of the mass can be found. A typical choice is, for example, $q=95$. The mass inside a certain radius is given by the function $\hat m$, which in the leading order of the expansion is given by $\hat m=\varepsilon^{4-d}\hat m^{(1)}$. According to \eqref{eqhatm1b}, the value of $\hat m^{(1)}$ is proportional to the function $\bar\sigma$ introduced in \eqref{eq:sigmadef}. In terms of the radial coordinate $\rho$, we define the radius $\rho_{q}$ of the oscillatons by the following expression:
\begin{equation}
 \frac{q}{100}=\frac{\hat m(\rho_{q})}{\hat m(\infty)}
 \approx\frac{\bar\sigma(\rho_{q})}{\bar\sigma(\infty)} \ .
\end{equation}
In Table \ref{tablerhon} we give the magnitude of the radius $\rho_{q}$ for various choices of $q$.
\begin{table}[htbp]
\begin{center}
\begin{tabular}{c|c|c|c|}
  & $d=3$  & $d=4$  & $d=5$ \\
\hline
$\rho_{50}$   &  $2.240$ &   $1.778$ &  $1.317$ \\
$\rho_{90}$   &  $3.900$ &   $3.013$ &  $2.284$ \\
$\rho_{95}$   &  $4.471$ &   $3.455$ &  $2.652$ \\
$\rho_{99}$   &  $5.675$ &   $4.410$ &  $3.478$ \\
$\rho_{99.9}$ &  $7.239$ &   $5.692$ &  $4.634$ \\
\hline
\end{tabular}
\end{center}
\caption{\label{tablerhon}
 The radius $\rho_{q}$ inside which $q$ percent of the mass can be found, for $d=3,\,4,\,5$ spatial dimensions.
}
\end{table}
The physical size corresponding to the rescaled radius $\rho_{q}$ is $r_{q}=\rho_{q}/(\varepsilon m)$. Measuring the radius in meters and the scalar field mass in electronvolts,
\begin{equation}
 r_{q}[\mathrm{m}]=1.973\cdot 10^{-7}\,
 \frac{\rho_{q}}{\varepsilon\,m[\mathrm{eV}/c^2]} \ . \label{eq:size}
\end{equation}
Although $\bar\sigma$ gives the mass inside a given radius only to leading order, this expression gives an acceptable approximation even for $\varepsilon\approx 0.5$.

The $\varepsilon$ dependence \eqref{eq:totmass} of the total mass of oscillatons provides important information on their stability properties. Similarly to the flat background oscillons studied in Section \ref{chapteroscillon}, with the increase of the parameter $\varepsilon$, the amplitude of the oscillations of the scalar field at the center, and hence also its energy density, increases monotonically. It is a generally observed principle, that if by the increase of the central energy density the total mass $M$ of the system increases, then the configuration is stable, while if $M$ decreases then it is unstable. Numerical simulations support this expectation for oscillatons as well \cite{seidel-91,Alcubierre03}. For the physically relevant $d=3$ spatial dimensional case, according to equation \eqref{eqmpeps}, for small amplitudes the mass increases proportionally to $\varepsilon$, and hence small amplitude oscillatons are stable for any self-interaction potential $U(\phi)$. 

Specializing to the Klein-Gordon potential, since in that case the coefficient of the cubic term is negative, by the increase of $\varepsilon$ we can expect to get to a critical value $\varepsilon_c$, where $M$ becomes maximal, and after that it begins to decrease. The $\varepsilon>\varepsilon_c$ states are unstable. Based on expression \eqref{eqmmepdep3d}, which contains the first two terms of the expansion, the critical amplitude for $d=3$ dimensions can be estimated as
\begin{equation}
 \varepsilon_{\rm m}=\sqrt{\frac{-M^{(1)}}{3M^{(2)}}}= 0.525 \ , \label{eqepscmm}
\end{equation}
which corresponds to the frequency $\omega_{\rm m}=0.851$. Although this is only a leading order result, it is quite close to the precise numerical values $\omega_c=0.8608$ and $\varepsilon_c=0.509$, which will be determined in later in Subsection \ref{subsecnumosctn} by the solution of the Fourier mode equations. According to the expansion, the maximal value of the mass is $M_{\rm m}=\frac{2}{3}\varepsilon_{\rm m}M^{(1)}=0.614$, which is also quite close to the numerically determined precise value $M_{c}=0.60535$. For not unit scalar field mass $M_{c}=0.60535/m$, and in SI and electronvolt units the precise value of the numerically calculated maximal mass is
\begin{equation}
 M_{c}[\mathrm{kg}]=1.6085\cdot 10^{20}\frac{1}{m[\mathrm{eV}/c^2]} \ . \label{eqmcnatunit}
\end{equation}
For example, for a Klein-Gordon scalar field with mass $m[\mathrm{eV}/c^2]=10^{-5}$, a value motivated by axions, the maximal mass is $M_{c}[\mathrm{kg}]=1.6085\cdot 10^{25}$, which is about three times as large as the mass of the Earth. Using the leading order approximation \eqref{eq:size}, the radius of this oscillaton is $r_{95}[\mathrm{cm}]=16.8$. For comparison, the Schwarzschild radius belonging to the maximal mass $M_{c}$ is $2.42\,\mathrm{cm}$.

For a Klein-Gordon field without self-interaction, in case of $d=4$ and $d=5$ spatial dimensions, based on the numbers in Table \ref{masstable}, the mass is a monotonically decreasing function of $\varepsilon$. Based on this approximation, which takes into account the first two leading orders, we can expect that all Klein-Gordon oscillatons are unstable for these dimensions. We are not aware of any numerical investigations which could confirm this.

For scalar fields with a nontrivial $U(\phi)$ self-interaction potential the situation may be completely different. For example, in case of $d=3$, with the choice $\bar g_2=0$ and $\bar g_3=4$, the coefficient $M^{(2)}$ becomes positive, and there is no maximum of the total mass. This indicates that in this case oscillatons with quite large mass may still be stable. Using the same potential for $d=4$ dimensions, at small $\varepsilon$ values the mass is a monotonically increasing function, hence we can expect that the small amplitude oscillatons are stable. If $d=5$, then for the same potential there is a mass minimum, and the oscillatons can be expected to be stable for larger amplitudes only. We can expect that similarly to the three-dimensional flat background oscillons, when the amplitude of this oscillaton decreases below the critical value because of the radiation loss, it suddenly decays.

\section{Radiation law of oscillatons in case of \texorpdfstring{$\Lambda=0$}{Lambda=0}}
\label{secoscsugt}

The method presented in Section \ref{secsuganal} for the determination of the radiation of flat background oscillons can also be generalized for self-gravitating oscillatons. The extension into the complex plane of the Fourier mode equations and the small-amplitude expansion was first applied by Segur and Kruskal for radiation of one-dimensional $\phi^4$ oscillons \cite{SegurKruskal87}. The Borel summation method for the calculation of the small correction near the singularity was introduced by Pomeau, Ramani and Grammaticos \cite{Pomeau1988}. The rest of this section is based on the results published in our paper \cite{fodor2010a}.

\subsection{Fourier expansion}

Since the small-amplitude expansion \eqref{eq:phiexp}-\eqref{eq:bexp} gives exponentially localized states to all orders, it can only be applied for the core region of the oscillaton. Instead of the study of radiating systems with slowly changing frequency, it is considerably simpler to examine exactly time-periodic solutions, which have a large amplitude core and a small amplitude standing wave tail (or wing). These solutions have been named nanopterons by Boyd \cite{Boyd1989a,Boyd-book1998}. The nanopteron with the minimal amplitude tail is named \emph{quasibreather}, and the amplitude of the quasibreather's tail determines the mass loss of the radiating state \cite{Fodor2006}.

For the exactly time-periodic solutions oscillating with frequency $\omega$ we look for the scalar field and metric components in the following Fourier series form:
\begin{align}
 \bar\phi&=\sum_{n=0}^{N_F}\Phi_{n}\cos(n\omega t) \ ,  \label{eq:fourexp1}\\
 A&=1+\sum_{n=0}^{N_F}\bar A_{n}\cos(n\omega t) \ , \label{eq:fourexp2} \\
 B&=1+\sum_{n=0}^{N_F}\bar B_{n}\cos(n\omega t) \ , \label{eq:fourexp3}
\end{align}
where $\phi_{n}$\,, $\bar A_{n}$ and $\bar B_{n}$ only depend on the radial coordinate $r$, and $N_F>0$ integer. Although the exact expansion should consist of infinitely many components, due to the exponential convergence we can already get very good approximation by taking into account the coefficients up to a moderately high $N_F$ order. We assume that the frequency $\omega$ approaches from below the mass threshold $m=1$. In the following calculations we define the parameter $\varepsilon$ by the expression $\varepsilon=\sqrt{1-\omega^2}$. We note that since the definition \eqref{eq:fourexp1} contains $\bar\phi$, while the flat background definition \eqref{eqphifourier} has $\phi$, there is a $\sqrt{8\pi}$ factor difference between the $\Phi_{n}$ functions used here and in Chapter \ref{chapteroscillon}.

The necessary and sufficient condition for the regularity at the center is that the functions $\Phi_n$, $\bar A_n$ and $\bar B_n$ should be finite at the center $r=0$, and that their derivative should be zero. For the $r\to\infty$ asymptotic boundary condition the most natural assumption is that the metric should be asymptotically flat, and the coordinate $t$ should tend to the proper time. As we will see soon, this is a too restrictive condition, but if we do not go out to extreme large $r$ radii, it can still be applied. The asymptotic flatness holds if $\bar A_n\to 0$ and $\bar B_n\to 0$ when $r\to\infty$. The components of the wave equation \eqref{eq:wave3} at large distances can be approximated by \eqref{eqphinsph} which is valid on flat background. If the scalar field tends to zero, the nonlinear source terms $F_n$ become negligible, and the Fourier equations decouple. In the frequency domain $\frac{1}{2}<\omega<1$, which is relevant for the oscillatons, for $n\geq2$ the asymptotic behavior of the functions $\Phi_n$ is still described by \eqref{eqtailstand}, in terms of the amplitude parameters $\alpha_n$ and $\beta_n$. In the same way as at the flat background calculations, the energy density $\mu$, which can be calculated using \eqref{eq:massen}, tends to zero proportionally to $r^{1-d}$. From this follows that if any of the amplitudes $\alpha_n$ or $\beta_n$ is nonzero for any $n\geq2$, then the proper mass becomes infinite, and the spacetime cannot be asymptotically flat. The vanishing of all  $\alpha_n$ and $\beta_n$ for all Fourier modes, together with the central regularity, obviously present too many boundary conditions for the second order differential equations determining $\Phi_n$. Generally, we cannot expect the existence of exactly time-periodic finite mass centrally regular solutions.

If we require that $\alpha_n=\beta_n=0$ for all $n\geq2$, then all $\Phi_n$ tend to zero exponentially, and we obtain a finite mass asymptotically flat solution. However, in the general case this solution will be singular at the center $r=0$. For these type of solutions we have introduced the name \emph{singular breather} (SB) in Subsection \ref{secgomboscsug}. According to an argument based on counting the parameters, after choosing a frequency $\omega$ the singular breather solution is unique. Apart from a small region around the center, inside the core region the SB solution agrees very precisely with the intended \emph{quasibreather} (QB) solution, which has a regular center and a minimal amplitude tail.

It is important to clarify that the quasibreather description is only valid inside a very large but finite radius. However small is the energy density of the oscillating tail, if we consider large enough spheres the contribution of the tail to the total mass will not be negligible anymore. Consequently, the assumption that $\bar A_n$ and $\bar B_n$ tend to zero will not remain valid for arbitrarily large $r$ values, and the form of the wave equation \eqref{eqphinsph} will also change. For large enough $r$ values the metric component $A$ slowly becomes so large that the first radiating Fourier component, which is $\Phi_3$ for symmetric potentials and $\Phi_2$ otherwise, from oscillating mode becomes a decaying mode. Going out for even larger $r$, one by one all the other $\Phi_n$ stop being oscillating. In this way, we obtain exactly time-periodic but infinite mass breather solutions, which has been discussed in more detail in Chapter IX of the paper of Don N.~Page \cite{PageDon04}. The quasibreathers investigated in the present review can be considered as parts of these infinite mass breathers inside a sphere of large radius. The quasibreather contains the core region and also a considerable part of the tail region, where the first radiating mode oscillates, but inside the given radius the contribution of the tail to the mass is still negligible in comparison to the mass of the core. Since the amplitude of the core is order $\varepsilon^2$, and the tail is exponentially suppressed in $\varepsilon$, the quasibreather picture is valid in a sufficiently large volume.

Since the amplitude of the oscillating tail of the quasibreather is very small, apart from a small region around the center, the QB solution differs only a very slightly from the singular breather solution. Although at the origin the scalar field tends to infinity, the amplitude of the singular part of the SB solution is still suppressed exponentially in terms of $\varepsilon$. Hence the identical frequency SB and QB solutions have the same $\varepsilon$ expansions to all orders. At the SB solution the small-amplitude expansion is not valid in a small neighborhood of the origin, while the expansion of the QB solution cannot describe the large distance domain where the standing wave tail dominates. The region $(0,r_{\rm diff})$ around $r=0$ where there is a significant difference between the SB solution and the regular QB solution is very small, its size is $r_{\rm diff}={\cal{O}}(e^{-\delta/\varepsilon})$, where $\delta>0$ is a constant. On the other hand, the total size of the quasibreather's core can be very large, since it is proportional to $1/\varepsilon$.

For potentials $U(\phi)$ which are symmetric around their minimum, i.e.~$\bar g_{2k}=0$ for integer $k$, the Fourier expansion of the scalar contains only odd, while the metric only even indexed components,
\begin{equation}
 \Phi_{2k}=0 \ , \quad \bar A_{2k+1}=0 \ , \quad
 \bar B_{2k+1}=0 \ .
\end{equation}
For symmetric potentials the first radiating mode is $\Phi_{3}$. In the following, we will mainly study symmetric potentials, since the formalism is simpler in that case. Obviously the Klein-Gordon potential, for which $\bar g_k=0$ for $k>1$, is symmetric as well. We will give concrete numerical values for the radiation in the Klein-Gordon case.

For a small amplitude quasibreather or singular breather we can get the relation between the \eqref{eq:fourexp1}-\eqref{eq:fourexp3} Fourier expansion and the \eqref{eq:phiexp}-\eqref{eq:bexp} small-amplitude expansion by using the expressions \eqref{eq:phisum}-\eqref{eq:phisum3}. For symmetric potentials,
\begin{align}
 \Phi_1&=\varepsilon^2p_2+\varepsilon^4p_4
 +\mathcal{O}(\varepsilon^6) \ , \label{eq:bphi1}\\
 \Phi_3&=\varepsilon^6\left(\frac{p_2^3 d}{64(d-1)}
 +\frac{\bar g_3p_2^3}{32}+\frac{p_2a_4^{(2)}}{8}\right)
 +\mathcal{O}(\varepsilon^8) \ ,\\
 \bar A_0&=\varepsilon^2a_2+\varepsilon^4a_4^{(0)}
 +\mathcal{O}(\varepsilon^6) \ ,\\
 \bar A_2&=\varepsilon^4a_4^{(2)}
 +\mathcal{O}(\varepsilon^6) \ ,\\
 \bar B_0&=-\varepsilon^2\frac{a_2}{d-2}+\varepsilon^4b_4
 +\mathcal{O}(\varepsilon^6) \ ,\\
 \bar B_2&=-\varepsilon^4\frac{p_2^2}{4(d-1)}
 +\mathcal{O}(\varepsilon^6) \ . \label{eq:bb2}
\end{align}

\subsection{Singularities on the complex plane} \label{sec:sing}

For the determination of the radiation rate we have to extend to the complex $r$ plane both the small-amplitude expansion and the Fourier expansion, and we have to clarify the connection between them. For the case of the $\varepsilon$ expansion, the functions $\phi_k$, $A_k$ and $B_k$ have symmetrically positioned singularities on the imaginary axis, corresponding to the singularities of the \eqref{eqsn1}-\eqref{eqsn2} Schr\"odinger-Newton (SN) equations. Since their contribution to the radiation decreases exponentially with their distance from the real axis, we only need to consider the nearest pair of singularities.

The behavior of the SN equations on the imaginary axis becomes more understandable if instead of the $s$ and $S$ functions we introduce the new variables
\begin{equation}
 z=\frac{1}{2}(s+S) \quad , \qquad Z=\frac{1}{2}(s-S) \ .
\end{equation}
Then the two SN equations will have the same structure,
\begin{align}
 &\frac{\mathrm{d}^2Z}{\mathrm{d}\rho^2}
 +\frac{d-1}{\rho}\,\frac{\mathrm{d} Z}{\mathrm{d}\rho}
 +Z^2-z Z=0
 \ , \label{eqzn1}\\
 &\frac{\mathrm{d}^2z}{\mathrm{d}\rho^2}
 +\frac{d-1}{\rho}\,\frac{\mathrm{d} z}{\mathrm{d}\rho}
 +z^2-Z z=0
 \ . \label{eqzn2}
\end{align}
Using the central initial values listed in Table \ref{c1table}, and numerically integrating the differential equations along the imaginary axis, it can be observed that $z$ becomes singular at some points $\rho=\pm i P$, where $P>0$ real, while $Z$ tends to zero oscillating
at the same places. For those spatial dimensions for which there exist exponentially localized solutions, the value of the constant $P$ is
\begin{equation}
 P=\left
 \{\begin{array}{rl}
  3.97736 & \text{if } \ d=3 \ , \\
  2.30468 & \text{if } \ d=4 \ , \\
  1.23595 & \text{if } \ d=5 \ . \label{eqpolesn}
 \end{array}
\right.
\end{equation}

Although the solutions of the SN equations are only known numerically, the leading order behavior near the singularities can be determined analytically. Similarly to the approach in Subsection \ref{secegydimszim} for flat background oscillons, in order to describe the distance from the upper singularity, let us introduce the complex $R$ coordinate defined by the relation $\rho=iP+R$. Since on the imaginary axis the functions $z$ and $Z$ take real values, and between the two nearest singularities $z>0$, the leading order behavior near the singularity is necessarily $z=-6/R^2$. In this case, to leading order, the differential equation that $Z$ must satisfy is
\begin{equation}
 \frac{\mathrm{d}^2Z}{\mathrm{d}R^2}+\frac{6}{R^2} Z=0 \ .
\end{equation}
The solution of this on the negative half of the imaginary $R$ axis is
\begin{equation}
 Z=\alpha \sqrt{iR}\,\sin\left(\frac{\sqrt{23}}{2}\ln(iR)+\delta\right) \ , \label{eqzzalphde}
\end{equation}
where $\alpha$ and $\delta$ are real constants that can be determined by the numerical solution of the original equations. The function $Z$ tends to zero, while performing increasing frequency oscillations when approaching the point $\rho=i P$ from below on the imaginary axis. The expansion of the function $z$ to higher orders is
\begin{equation}
 z \approx -\frac{6}{R^2}-\frac{6i(d-1)}{5PR}
 -\frac{(d-1)(d-51)}{50P^2} \ , \label{eq:sexp}
\end{equation}
to which there is also a correction of the form \eqref{eqzzalphde}. The leading terms in the behavior of the functions $s$ and $S$ near the singularity has the same form, given by the right-hand side of \eqref{eq:sexp}. If $d>1$, then continuing the expansion without taking into account the non-analytic terms, logarithmic terms appear, from which the first has the form $R^4\ln R$. Knowing the behavior of $s$ and $S$ according to \eqref{eq:sexp}, equations \eqref{eq:sands} and \eqref{eq:b2} determine the order $\varepsilon^2$ parts of the functions $\phi$, $A$ and $B$ near the singularity.

Substituting into \eqref{eq:a40}, \eqref{eq:p4}, \eqref{eq:sb4} and \eqref{eq:sa42}, the order $\varepsilon^4$ contributions, i.e.~$a_4^{(0)}$, $p_4$, $b_4$ and $a_4^{(2)}$, can also be determined in the neighborhood of the singularity. In the Klein-Gordon case, when $\bar g_k=0$ for $k>1$:
\begin{align}
 &a_4^{(0)}=-\frac{9(25d+208)}{52(d-2)R^4}
 +\frac{324id(d-1)\ln R}{35P(d-2)R^3}
 +\frac{a_{-3}}{R^3}
 +{\cal O}\left(\frac{\ln R}{R^2}\right) \ , \label{eq:a40r}\\
 &p_4\sqrt{\frac{d-2}{d-1}}+a_4^{(0)}=
 \frac{9(43d-104)}{26(d-2)R^4}
 +\frac{9i(d-1)(3d-8)}{5P(d-2)R^3}
 +{\cal O}\left(\frac{1}{R^2}\right) \ ,  \\
 &b_4=\frac{9(333d+832)}{260(d-2)^2R^4}
 -\frac{324id(d-1)\ln R}{35P(d-2)^2R^3}
 -\frac{a_{-3}}{(d-2)R^3}+
 \frac{18i(d-1)}{5P(d-2)R^3}
 +{\cal O}\left(\frac{\ln R}{R^2}\right) \ , \\
 &a_4^{(2)}=-\frac{9(6-d)}{5(d-2)R^4}
 +\frac{6i(d-1)(d-6)}{5P(d-2)R^3}
 +{\cal O}\left(\frac{1}{R^2}\right) \ . \label{eq:a42r}
\end{align}
The value of the constant $a_{-3}$ is fixed by the behavior of the functions on the real axis, namely by the requirement of the exponential decay of $p_4$ for large real $\rho$.

\subsection{Expansion of the Fourier modes near the singularity}

In the $\varepsilon\to 0$ limit the amplitudes of all Fourier components of the oscillaton tend to zero at all points of the real $r$ axis. However, extending to the complex $r$ plane, the Fourier modes have singularities on the imaginary axis. Since the rescaled radial coordinate is $\rho=\varepsilon r$, for small $\varepsilon$ values the singularities closest the the real axis are located at the points $r=\pm i P/\varepsilon$, corresponding to the singularities of the Schr\"odinger-Newton equations. As $\varepsilon$ tends to zero, the singularities get further and further from the real axis, but in their neighborhood the Fourier components $\Phi_{k}$, $\bar A_k$ and $\bar B_{k}$ will not be small, they have $\varepsilon$ independent parts. In agreement with \eqref{eqrykapcs} we introduce the shifted radial coordinate $y$ by
\begin{equation}
 r=\frac{iP}{\varepsilon}+y \ . \label{eq:ry}
\end{equation}
The coordinate $y$ is valid in the \emph{inner region}, and obviously $R=\varepsilon y$. In the neighborhood of the upper singularity we can obtain the leading order behavior of the Fourier modes, if we substitute the results \eqref{eq:sexp}-\eqref{eq:a42r} of the smal-amplitude expansion into the equations \eqref{eq:bphi1}-\eqref{eq:bb2}, and take the $\varepsilon\to 0$ limit. For the case of the Klein-Gordon potential,
\begin{align}
 \Phi_1&=\left(-\frac{6}{y^2}+\frac{999d}{52(d-2)y^4}+...\right)
 \sqrt{\frac{d-1}{d-2}} \ , \label{eq:phi1y}\\
 \Phi_3&=\left(-\frac{27(7d-12)}{40(d-2)y^4}+...\right)
 \sqrt{\frac{d-1}{d-2}} \ , \\
 \bar A_0&=\frac{6}{y^2}-\frac{9(25d+208)}{52(d-2)y^4}+... \ , \\
 \bar A_2&=-\frac{9(6-d)}{5(d-2)y^4}+... \ , \\
 \bar B_0&=-\frac{6}{(d-2)y^2}
 +\frac{9(333d+832)}{260(d-2)^2y^4}+... \ , \\
 \bar B_2&=-\frac{9}{(d-2)y^4}+... \ . \label{eq:b1y}
\end{align}
Since these are expansions with respect to $1/y^2$, they are valid for large $y$ values. On the other hand, we have obtained equations \eqref{eq:sexp}-\eqref{eq:a42r} by assuming that $R$ was small. If $\varepsilon$ is small, both conditions can be satisfied at the same time, since $R=\varepsilon y$.

We can also obtain the expansion \eqref{eq:phi1y}-\eqref{eq:b1y} near the singularity, if in the $\varepsilon\to0$ limit we look for the solution of the Fourier equations in a series form with respect to $1/y^2$,
\begin{align}
 \Phi_{2k+1}&=\sum_{n=k+1}^{\infty}\psi_{2k+1}^{(n)}\frac{1}{y^{2n}}
 \ , \label{eq:yexp1}\\
 \bar A_{2k}&=\sum_{n=k+1}^{\infty}\alpha_{2k}^{(n)}\frac{1}{y^{2n}}
 \ , \\
 \bar B_{2k}&=\sum_{n=k+1}^{\infty}\beta_{2k}^{(n)}\frac{1}{y^{2n}} \ ,
 \label{eq:yexp3}
\end{align}
where $\psi_{2k+1}^{(n)}$, $\alpha_{2k}^{(n)}$ and $\beta_{2k}^{(n)}$ are constants. We can obtain the mode equations that we have to solve from the \eqref{eq:eieq1}-\eqref{eq:eieq4} Einstein equations and from the \eqref{eq:wave3} wave equation, by substituting \eqref{eq:fourexp1}-\eqref{eq:fourexp3}. The equations \eqref {eq:eieq1}-\eqref{eq:wave3} are not independent. The wave equation follows from the Einstein equations by applying the contracted Bianchi identity, and the component $(t,r)$, given in \eqref{eq:eieq3}, is a constraint equation. Because of the cut of the Fourier expansion at a finite $N_F$ order, the mode equations become mutually contradictory. However, if we choose any three equation from \eqref{eq:eieq1}-\eqref{eq:wave3}, the mode equations following from these obviously will have a solution. We have checked that our results for the mass loss of oscillatons are independent on which three equations we choose. We have also checked that the errors at the mode equations coming from two remaining equations tend to zero quickly if $N_F$ grows.

Substituting the definition \eqref{eq:ry} of the coordinate $y$ into \eqref{eq:eieq1}-\eqref{eq:wave3}, and taking the limit $\varepsilon\to0$, near the singularity certain lower order terms in $r$ become negligible. After this, substituting the $1/y^2$ expansion in the form \eqref{eq:yexp1}-\eqref{eq:yexp3} into the mode equations, because of the absence of the odd powers of $1/y$, apart from the signature of $\psi_1^{(1)}$ the coefficients  $\psi_{2k+1}^{(n)}$, $\alpha_{2k}^{(n)}$ and $\beta_{2k}^{(n)}$ will be uniquely determined. The computation of the expansion \eqref{eq:phi1y}-\eqref{eq:b1y} from the Fourier mode equations is technically much simpler than the small-amplitude expansion method, and by the use of an algebraic manipulation software it can be calculated to quite high orders in $1/y$.

Similarly to the flat background case, which was studied in a detailed way in Subsection \ref{secborelsum}, the large $n$ asymptotic behavior of the constants $\psi_{k}^{(n)}$, $\alpha_{k}^{(n)}$ and $\beta_{k}^{(n)}$ will determine the amplitude of the radiative tail of oscillatons. Apart from a common factor, the leading order asymptotic behavior of these constants can be obtained by the study of the structure of the mode equations. It can be shown, that for large $n$ the dominating coefficients are $\psi_{3}^{(n)}$. The result for the third Fourier mode of the Klein-Gordon field is
\begin{equation}
 \psi_{3}^{(n)}=k_d(-1)^n\frac{(2n-1)!}{8^n}
 \Biggl[1+\frac{3(9d-10)}{2(d-2)n} 
 +\frac{3(9d-10)(7d-8)}{2(d-2)^2n^2}
 +{\cal O}\left(\frac{1}{n^3}\right)\Biggr] \ , \label{eq:p3nser}
\end{equation}
where $k_d$ is a factor which depends on $d$ and $N_F$. All other coefficients grow more slowly with $n$ asymptotically. Although the $1/n$ and $1/n^2$ correction terms may depend on the interaction potential of the scalar field, the leading order behavior for any symmetric potential is the same as in \eqref{eq:p3nser}. The value of the constant factor $k_d$ is crucial for the determination of the mass loss rate of the oscillatons. Calculating the coefficients up to order $n=100$, and using the Fourier mode equations up to order $N_F=6$, in the Klein-Gordon case we obtain
\begin{equation}
 k_3=-0.301 \quad , \qquad
 k_4=-0.134 \quad , \qquad
 k_5=-0.0839 \ . \label{eqkdosctn}
\end{equation}

\subsection{The singular breather solution near the singularity}

The expansion \eqref{eq:yexp1}-\eqref{eq:yexp3} is an asymptotic series for the Fourier components $\Phi_k$, $\bar A_k$ and $\bar B_k$. After the determination of the expansion coefficients up to a certain order, we can consider the obtained generalization of \eqref{eq:phi1y}-\eqref{eq:b1y} as boundary conditions to the Fourier mode equations for large $|y|$ in the directions
\begin{equation}
 -\pi/2<\arg\, y<0 \ , \label{eq:matchreg}
\end{equation}
ensuring a unique solution in the inner region. This corresponds to the requirement that the scalar field $\phi$ tends to zero along the positive part of the real $r$ axis for $r\to\infty$, without a standing wave tail. Similarly to the method described in Subsection \ref{secgomboscsug}, we look for the \emph{singular breather} solution described by the functions $\Phi_k^{(SB)}$, $\bar A_k^{(SB)}$ and $\bar B_k^{(SB)}$.

The Fourier components of the wave equation \eqref{eq:wave3} can be written in a form agreeing with \eqref{eqnonsphout} even in the self-gravitating case,
\begin{equation}
 \frac{\mathrm{d}^2\Phi_n}{\mathrm{d}r^2}
 +\frac{d-1}{r}\frac{\mathrm{d}\Phi_n}{\mathrm{d}r}
 +(n^2\omega^2-1)\Phi_n=F_n \ , \label{eq:fourphi}
\end{equation}
where now the source terms $F_n$ are polynomial expressions containing $\Phi_k$, $\bar A_k$, $\bar B_k$ and their derivatives, for $k\leq N_F$. Using the coordinate $y$ in the neighborhood of the singularity, and taking the $\varepsilon\to 0$ limit,
\begin{equation}
 \frac{\mathrm{d}^2\Phi_n}{\mathrm{d}y^2}
 +(n^2-1)\Phi_n=\tilde F_n \ , \label{eq:fourphipole}
\end{equation}
where $\tilde F_n$ is the $\varepsilon\to 0$ limit of the nonlinear source term $F_n$. Apart from the structure of the nonlinear source terms, this equation has the same form as \eqref{eqphinner}, which describes the behavior of one-dimensional oscillons in the inner domain. On the imaginary axis the $1/y^2$ expansion gives real valued functions at all orders. On the other hand, the singular breather solutions of the mode equations must necessarily have a small but nonzero imaginary part on the imaginary axis. Considering the imaginary axis only, the Fourier components $\Phi_k^{(SB)}$ of the scalar field satisfy the linear left-hand side of equation \eqref{eq:fourphipole}. For symmetric $U(\phi)$ potentials the first radiating component is $\Phi_3$, which gives the dominant contribution to the radiation. Consequently, in the same way as at \eqref{eqdeltaphi3} in Subsection \ref{secbelnum}, the SB solution has a part which is exponentially decaying when going down along the imaginary axis,
\begin{equation}
 \im\Phi_3^{(SB)}=\nu_3\exp\left(-i\sqrt{8}\,y\right)
 \qquad \mathrm{for} \qquad \re y=0 \ , \quad \im y<0 \ , \label{eq:phi3}
\end{equation}
where $\nu_3$ is a constant. On the other hand, since the quasibreather solution, which has a small amplitude standing wave tail, is regular at the center $r=0$, its real part is symmetric, while its imaginary part is antisymmetric with respect to the imaginary axis. Hence the component $\Phi_3^{(QB)}$ of the scalar field of the quasibreather has zero imaginary part on the imaginary axis.

For symmetric potentials the value of the constant $\nu_3$ can be calculated by the application of Borel summation \cite{fodor2010a}. The method based on the work of Pomeau, Ramani and Grammaticos \cite{Pomeau1988} has been presented in detail for oscillons in Subsection \ref{secborelsum}. As a first step, associated to the function $\Phi_3$ expanded in the form \eqref{eq:yexp1}, we define the transformed function $\widehat\Phi_3$ by
\begin{equation}
  \widehat\Phi_{3}(z)=\sum_{n=2}^{\infty}\frac{\psi_3^{(n)}}{(2n)!}
  z^{2n} \ .  \label{eq:bortr}
\end{equation}
Similarly to \eqref{eqphi3yint}, we can obtain the Borel summed $\Phi_3$ by the following integral,
\begin{equation}
 \Phi_3^{(SB)}(y)=\int_{0}^{\infty} \mathrm{d}t\,
 e^{-t}\,\widehat\Phi_{3}\left(\frac{t}{y}\right) \ . \label{borelint}
\end{equation}
Let us introduce the coordinate $\tilde y$ by $y=-i\tilde y$. We intend to determine the imaginary part of the function $\Phi_3^{(SB)}(y)$ on the negative part of the imaginary axis, where $\tilde y>0$ real. The argument of the function $\widehat\Phi_3$ is $z=t/y=it/\tilde y$, which is purely imaginary, with a positive imaginary part. Since all terms in \eqref{eq:bortr} contain even powers of $z$, in itself none of the terms in this sum can give a contribution to $\im\Phi_3^{(SB)}$ by \eqref{borelint}. For the Borel summed series, the value of $\im\Phi_3^{(SB)}$ is determined by the large $n$ leading order behavior of the series. We have written the first three terms of this behavior in \eqref{eq:p3nser} for the Klein-Gordon case. Substituting the first leading order term, which is the same for any symmetric potential,
\begin{align}
 \widehat\Phi_3(z)&=\sum_{n=2}^{\infty}k_d\frac{(-1)^n}{2n}
 \left(\frac{z}{\sqrt 8}\right)^{2n}
 =k_d\frac{z^2}{16}+\sum_{n=1}^{\infty}k_d\frac{(-1)^n}{2n}
 \left(\frac{z}{\sqrt 8}\right)^{2n} \notag \\
 &=\frac{k_d}{2}\left[
 \frac{z^2}{8}
 -\ln\left(1-\frac{i z}{\sqrt{8}}\right)
 -\ln\left(1+\frac{i z}{\sqrt{8}}\right)
 \right] \ . \label{e:borelsum}
\end{align}
The result for the Borel summed function is
\begin{equation}
 \Phi_{3}^{(SB)}(y)=\frac{k_d}{2}\int_{0}^\infty\mathrm{d}t\, e^{-t}
 \left[\frac{z^2}{8}
 -\ln\left(1+\frac{t}{\sqrt{8}\,\tilde y}\right)
 -\ln\left(1-\frac{t}{\sqrt{8}\,\tilde y}\right)\right] \ . \label{eqphi3intfosctn}
\end{equation}
Similarly to equation \eqref{eqphi3intform}, only the third term gives imaginary contribution on the imaginary $y$ axis. By the method explained in detail there, we can obtain that
\begin{align}
 \im\Phi_3^{(SB)}(y)=-\frac{k_d\pi}{2}\exp\left(-i \sqrt8 \,y\right)  \ .  \label{e:borel}
\end{align}
Comparison with equation \eqref{eq:phi3} gives the relation to the constant $\nu_3$ which determines the radiation amplitude,
\begin{equation}
 \nu_3=-\frac{1}{2} k_d\pi \ .  \label{eq:nu3kd}
\end{equation}
We have given the numerical value of the dimension dependent constant $k_d$ for the Klein-Gordon scalar in \eqref{eqkdosctn}.

For potentials $U(\phi)$ that are not symmetric around their minimum, the leading order radiating component is $\Phi_2$. Similarly to \eqref{eqdeltaphi2}, the imaginary part on the imaginary axis is
\begin{equation}
 \im\Phi_2^{(SB)}(y)=\nu_2\exp\left(-i\sqrt{3}y\right)
 \quad \mathrm{for} \qquad \re y=0 \ , \quad \im y<0 \ . \label{eq:phi3ns}
\end{equation}
Since in this case the large $n$ dominant behavior of the $1/y$ series is determined by $\Phi_0$, the coefficient $\nu_2$ cannot be determined by the Borel summation method. Its value can be obtained by the numerical integration of the Fourier mode equation near the singularity, by the method presented in papers \cite{SegurKruskal87}  \cite{Fodor2009a}, and also in Subsection \ref{secbelnum} and \ref{subsecnonsympot} of this review.

\subsection{Construction of the quasibreather and the oscillaton} \label{seckbroscelo}

Near the upper singularity \eqref{eq:phi3} and \eqref{eq:phi3ns} represent the small exponentially decaying corrections in the imaginary values. We can write these two expressions in a unified form,
\begin{equation}
 \delta\Phi_k^{(+)}=i\nu_k\exp(-i\sqrt{k^2-1}\,y) \ , \label{eqdeltaphiosctn}
\end{equation}
where $k=3$ for potentials symmetric around their minimum, and $k=2$ for non-symmetric potentials. This expression agrees with equation \eqref{eqdeltaphi20} for flat background oscillons, hence we can follow the procedure presented there for the extension of the correction to the real $r$ axis. We note, that because of the rescaling \eqref{eq:rscpu} of the scalar field $\phi$, the Fourier components $\Phi_k$ contain a factor $\sqrt{8\pi}$ with respect to their flat background correspondents, and this difference also appears in the value of the constant $\nu_k$. We do not consider now the order $\varepsilon \ln \varepsilon$ dimension dependent small corrections to the leading order results.

In the \emph{outer region}, described by the coordinate $r$, the correction $\delta\Phi_k$, which is actually the difference of the singular breather and the quasibreather, satisfies the left-hand side linear part of equation \eqref{eq:fourphi}. The general solution of this can be written in terms of Bessel functions in the form \eqref{eqphikbessel}, which contains two amplitude parameters, $\alpha$ and $\beta$. Due to the contribution of the singularity positioned symmetrically below the real $r$ axis, we have $\beta=0$, and from the matching to the solution in the inner region, similarly to \eqref{eqalphanuk}, it follows that
\begin{equation}
 \alpha=2\nu_k
 \left(\frac{P}{\varepsilon}\right)^{\frac{d-1}{2}}
 \exp\left(-\sqrt{k^2-1}\,\frac{P}{\varepsilon}\right)  \ . \label{eqalphaosctn}
\end{equation}
The scalar field in the tail domain of the quasibreather is
\begin{equation}
 \bar\phi^{(QB)}=-\frac{\alpha}{r^{\frac{d-1}{2}}}
 \sin\left[\sqrt{k^2-1}\,r-\frac{\pi}{4}(d-1)\right]
 \cos(kt)\ , \label{eqdphitailqbosctn}
\end{equation}
corresponding to \eqref{eqdphitailmored}. Adding the time-shifted cosine form solutions of the linearized equation, we obtain the expression for the radiating tail of the oscillaton,
\begin{equation}
 \bar\phi^{(\mathrm{osc})}=-\frac{\alpha}{r^{\frac{d-1}{2}}}
 \sin\left[\sqrt{k^2-1}\,r-\frac{\pi}{4}(d-1)-kt\right]
 \ , \label{eqdphitailosctn}
\end{equation}
where for symmetric potentials $k=3$, otherwise $k=2$. Because of the $\bar\phi=\sqrt{8\pi}\,\phi$ rescaling of the scalar field in \eqref{eq:rscpu}, the real physical amplitude is $\alpha/\sqrt{8\pi}$. Since the transformation \eqref{scaleprop} rescale the coordinates, for not unit scalar field mass $m$, in expressions \eqref{eqdphitailqbosctn} and \eqref{eqdphitailosctn} next to the amplitude $\alpha$ a factor $m^{(1-d)/2}$ appears.

For symmetric potentials $k=3$, and the value of $\nu_3$ can be calculated from the constant $k_d$ according to \eqref{eq:nu3kd}. For the Klein-Gordon potential, in case of $d=3$ spatial dimensions, substituting the numerical values of the constants, the $\varepsilon$ dependence of the amplitude $\alpha$ in \eqref{eqalphaosctn} can be written as
\begin{equation}
 \alpha=\frac{3.761}{\varepsilon}
 \exp\left(-\,\frac{11.2497}{\varepsilon}\right)  \ . \label{eqalphotnkg}
\end{equation}

\subsection{Mass loss rate at small amplitudes} \label{secmasslosseps}

Apart from notational differences, expression \eqref{eqdphitailosctn} agrees with the general flat background radiating tail given in \eqref{eqphialphatdep}. As we have seen, equation \eqref{eq:minkmt}, giving the change of mass in time, is consistent with equations \eqref{eqsrbar} and \eqref{eqencurr} written for the flat background case. In this way, equation \eqref{eqsaverage}, giving the energy current averaged for an oscillation period, can also be applied now, where $\omega_f=k$ and $\lambda_f=\sqrt{k^2-1}$\,. As a result of this, the time averaged mass-energy loss rate of oscillatons is
\begin{equation}
 \bar S=\frac{\pi^{\frac{d}{2}-1}}{2m^{d-3}\,\Gamma\left(\frac{d}{2}\right)}
 \nu_k^2\, k\sqrt{k^2-1}\,
 \left(\frac{P}{\varepsilon}\right)^{d-1}
 \exp\left(-2\sqrt{k^2-1}\,\frac{P}{\varepsilon}\right) \ . \label{eqbarsgenosctn}
\end{equation}
The factor $1/(8\pi)$ difference with respect to the flat background \eqref{eqbarsgen} equation is because of the $\bar\phi=\sqrt{8\pi}\,\phi$ rescaling of the scalar field, which results in a factor in $\nu_k$ as well.

For symmetric potentials, when the mode $k=3$ is responsible for the radiation, for the value of $\nu_3$ we can apply the relation \eqref{eq:nu3kd}. Hence the averaged mass loss during a unit time period is
\begin{equation}
 \bar S=-\,\frac{\overline{\mathrm{d}M}}{\mathrm{d}t}=
 \frac{c_1}{m^{d-3}\varepsilon^{d-1}}
 \exp\left(-\frac{c_2}{\varepsilon}\right) \ , \label{e:symradlaw}
\end{equation}
where the constants depending on the number of space dimensions are:
\begin{equation}
 c_1=3\sqrt{2}\, k_d^2\,P^{\,d-1}\,
 \frac{\pi^{d/2+1}}{4\Gamma\left(\frac{d}{2}\right)}
 \quad , \qquad
 c_2=4\sqrt{2}P \ .
\end{equation}
We give the numerical values of the constants $c_1$ and $c_2$ in Table \ref{ctable}.
\begin{table}[htbp]
\begin{center}
\begin{tabular}{|c|c|c|c|}
\hline
  & $d=3$  & $d=4$  & $d=5$ \\
\hline
$c_1$ & $30.0$ & $7.23$ & $0.720$ \\
$c_2$ & $22.4993$ & $13.0372$ & $6.99159$ \\
\hline
\end{tabular}
\end{center}
\caption{\label{ctable}
  The constants $c_1$ and $c_2$ in the expression \eqref{e:symradlaw} of the averaged mass loss rate, for $d=3,\,4,\,5$ spatial dimensions, in case of a Klein-Gordon field.
}
\end{table}
The constant $c_2$ is the same for all symmetric potentials, but the numbers given for $c_1$ are only valid for the Klein-Gordon field.

Because of the exponential dependence, the larger $\varepsilon$ is, the more chance we have for the observation of the presumably extremely small mass loss. As we have discussed in Subsection \ref{secosctnmass}, for $d=3$ the oscillatons are stable if the amplitude parameter remains below a critical value, i.e.~$\varepsilon<\varepsilon_{c}$. According to equation \eqref{eqepscmm} the approximate numerical value of this is $\varepsilon_{c}\approx\varepsilon_{\rm m}=0.525$. The mass of the oscillaton is maximal when $\varepsilon=\varepsilon_{c}$. Also taking into account the scalar field mass $m$, the estimated value of the maximal mass resulting from the $\varepsilon$ expansion is $M_{c}\approx M_{\rm m}=0.614/m$. Substituting into \eqref{e:symradlaw}, the relative mass loss rate of the maximal mass, most strongly radiating oscillaton can be estimated as
\begin{equation}
 \left(\frac{1}{M}\,\frac{\overline{\mathrm{d} M}}{\mathrm{d} t}
 \right)_{M=M_{c}}=
 -4.33\cdot 10^{-17} m \ .  \label{eq:maxloss1}
\end{equation}

According to equation \eqref{eq:totmass}, for $d=3$ the spatial dimensions, to leading order the total mass is proportional to the amplitude, $M=\varepsilon M^{(1)}/m$. According to Table \ref{masstable}, in this case $M^{(1)}=1.75266$. Substituting into equation \eqref{e:symradlaw}, for the Klein-Gordon potential we obtain
\begin{equation}
 \frac{\overline{\mathrm{d} M}}{\mathrm{d} t}=
 -\frac{c_3}{m^2M^2}\,
 \exp\left(-\frac{c_4}{mM}\right) \ . \label{eq:linmassloss}	
\end{equation}
where
\begin{equation}
 c_3=92.2 \quad , \qquad
 c_4=39.4337 \ . \label{eq:c3c4}
\end{equation}
Equation \eqref{eq:linmassloss} has the same form as the mass loss expression $(122)$ in the paper \cite{PageDon04} of Don Page. However, the constant corresponding to $c_3$ is many magnitudes larger in that paper, with value $3797437.776$. This means that the constant $\alpha$, which determines the amplitude of the radiative tail according to \eqref{eqdphitailosctn}, is estimated to be $202.9$ times larger in paper \cite{PageDon04}. In our paper \cite{fodor2010a} we discuss in detail the difference between the two approaches. The reason for the discrepancy is that in \cite{PageDon04} only the first term of an infinite series has been taken into account, although all terms of the series give contributions which are the same order in $\varepsilon$.

\subsection{Tail-amplitude}

The standing wave tail calculated in the previous subsection is so small that it is not surprising that it has not been observed in papers \cite{seidel-91} and \cite{Urena02b}. In order to compare the central and the tail-amplitudes, in the core domain we approximate the scalar field by the leading order expression $\phi=\varepsilon^2p_2\cos\tau$, while in the tail region we use \eqref{eqdphitailqbosctn}. The tail starts becoming dominant at the radius $r=r_f$ where
\begin{equation}
 \bar\phi(t=0,r)=\varepsilon^2p_2(\varepsilon
 r)=\varepsilon^2S(\varepsilon r)\sqrt{\frac{d-1}{d-2}}
\end{equation}
decreases to a value equal to $\alpha\,r^{(1-d)/2}$. Since according to \eqref{eq:sasympt}, for large $\rho$ we have $s\approx -1+s_1\rho^{2-d}$, for those dimensions which allow the existence of oscillatons, the asymptotic behavior of the function $S$ is
\begin{equation}
 S=\left
 \{\begin{array}{ll} \displaystyle
  S_te^{-\rho}\rho^{s_1/2-1}\left[1-\frac{s_1(s_1-2)}{8\rho}
+\mathcal{O}\left(\frac{1}{\rho^2}\right)\right] & \text{for } d=3 \ , \\[4mm]
  \displaystyle S_t\frac{e^{-\rho}}{\rho^{3/2}}\left[1-\frac{4s_1-3}{8\rho}
+\mathcal{O}\left(\frac{1}{\rho^2}\right)\right] & \text{for } d=4 \ , \\[4mm]
  \displaystyle S_t\frac{e^{-\rho}}{\rho^{2}}\left[1+\frac{1}{\rho}
-\frac{s_1}{4\rho^2}
+\mathcal{O}\left(\frac{1}{\rho^4}\right)\right] & \text{for } d=5 \ ,
 \end{array} \label{eqssasympt}
\right.
\end{equation}
where the numerical value of the constant $s_1$ has been given in Table \ref{c1table}, and we list $S_t$ in Table \ref{sttable}.
\begin{table}[htbp]
\begin{center}
\begin{tabular}{c|c|c|c|}
  & $d=3$  & $d=4$  & $d=5$ \\
  \hline
  $S_t$ & 3.495 & 88.24 & 23.39 \\
  \hline
\end{tabular}
\end{center}
\caption{The numerical value of the constant $S_t$ for various space dimensions. \label{sttable}}
\end{table}
In Table \ref{rttable}
\begin{table}[htbp]
\begin{center}
\begin{tabular}{|c|c|c|c|c|c|c|}
\hline
  & \multicolumn{2}{|c|}{$d=3$}  & \multicolumn{2}{|c|}{$d=4$}
     & \multicolumn{2}{|c|}{$d=5$} \\ \cline{2-7}
  $\varepsilon$ & $r_f$ & $\phi_f$ & $r_f$ & $\phi_f$ & $r_f$ & $\phi_f$\\
  \hline
  $0.1$ & $1160$ & $8.96\cdot10^{-52}$ & $648$ & $2.76\cdot10^{-32}$
    & $346$ & $4.42\cdot10^{-20}$\\
  $0.2$ & $302$ & $4.63\cdot10^{-27}$ & $168$ & $1.06\cdot10^{-17}$
    & $92.6$ & $6.01\cdot10^{-12}$\\
  $0.3$ & $140$ & $9.28\cdot10^{-19}$ & $78.2$ & $9.45\cdot10^{-13}$
    & $45.0$ & $3.83\cdot10^{-9}$\\
  $0.4$ & $81.6$ & $1.40\cdot10^{-14}$ & $46.4$ & $3.07\cdot10^{-10}$
    & $27.9$ & $1.03\cdot10^{-7}$\\
  $0.5$ & $54.3$ & $4.68\cdot10^{-12}$ & $31.5$ & $1.03\cdot10^{-8}$
    & $19.8$ & $7.56\cdot10^{-7}$\\
  $0.6$ & $39.2$ & $2.30\cdot10^{-10}$ & $23.2$ & $1.09\cdot10^{-7}$
    & $15.1$ & $2.87\cdot10^{-6}$\\
  $0.7$ & $29.9$ & $3.76\cdot10^{-9}$ & $18.0$ & $5.91\cdot10^{-7}$
    & $12.2$ & $7.42\cdot10^{-6}$\\
  $0.8$ & $23.8$ & $3.08\cdot10^{-8}$ & $14.6$ & $2.12\cdot10^{-6}$
    & $10.3$ & $1.51\cdot10^{-5}$\\
  \hline
\end{tabular}
\end{center}
\caption{The radius $r_f$ where the standing wave tail becomes dominant, and the amplitude $\phi_f$ of the scalar field there. \label{rttable}}
\end{table}
we give the radius $r_f$ where the tail appears, and also show the amplitude $\phi_f=\alpha r^{(1-d)/2}_f/\sqrt{8\pi}$ of the tail there, for various values of $\varepsilon$. The radius $r_f$ is obviously much larger than the typical size of the core, which can be estimated as $r_{q}=\rho_{q}/(\varepsilon m)$, where $\rho_{q}$ has been given in Table \ref{tablerhon}.

The amplitude of the tail should be compared to the central value of the field, which is
\begin{equation}
 \phi_c=\frac{1}{\sqrt{8\pi}}\varepsilon^2 \phi_{1c} \quad , \qquad
 \phi_{1c}=S_c\sqrt{\frac{d-1}{d-2}} \ ,
\end{equation}
where the value of $S_c$ can be found in Table \ref{c1table}. Obviously, there is only chance to numerically observe the radiating tail for quite large $\varepsilon$ values.

\section{Numerical investigation of quasibreathers for \texorpdfstring{$\Lambda=0$}{Lambda=0}}
\label{seckvbrnum}

In this section we study exactly time-periodic quasibreather states for a self-gravitating real Klein-Gordon field with mass $m=1$, in case of $d=3$ spatial dimensions. We apply a high precision spectral numerical method to determine the structure of the quasibreathers, together with their standing wave tail. The amplitude of the tail is many orders of magnitude smaller than the amplitude in the core domain. We use a spatially conformally flat coordinate system, for which $C=r^2B$, and the equations to be solved are \eqref{eq:eieq1}-\eqref{eq:wave3}. For each chosen oscillation frequency $\omega$, we look for solutions in the Fourier series form \eqref{eq:fourexp1}-\eqref{eq:fourexp3}. Because of the symmetry of the Klein-Gordon potential, the scalar field has only odd, while the metric functions only even Fourier components. This section is based on the results published in our paper \cite{grandclement2011}.

\subsection{Asymptotic behavior} \label{ss:asymptot}

In the moderately distant region, where the mass belonging to the standing wave tail is still much smaller than the mass of the quasibreather's core, the spacetime can be considered asymptotically flat, and hence it tends to the Schwarzschild metric. In a spatially conformally flat coordinate system, the form of the Schwarzschild metric is given in \eqref{eqschwtanghconf}. For three-dimensional space the asymptotic behavior of the metric functions are
\begin{align}
 A &= 1 - \frac{r_A}{r} \ , \label{e:asymptotA} \\
 B &= 1 + \frac{r_B}{r} \ , \label{e:asymptotB}
\end{align}
where $r_A$ and $r_B$ are constants. Obviously, for the solution of the whole system it must be satisfied that $r_A = r_B\equiv r_0$, where now for $d=3$ dimensions we have $r_0=2M$. However, we do not require the relation $r_A = r_B$ in our numerical code, instead, we use it to check the precision of our method.

For large distances, to leading order, the Fourier components $\Phi_n$ of the scalar field still satisfy the left-hand side linear part of \eqref{eqphinsph}, where now $m=1$. Since for the frequency we can assume that $\frac{1}{3}<\omega<1$, there are two kind of behaviors for the Fourier modes. If $n=1$, then the asymptotically decaying solution is
\begin{equation}
 \Phi_1 = C_1 \displaystyle\frac{\exp\left(-\varepsilon r\right)}{r} 
 \ , \label{e:shphi1}
\end{equation}
where $\varepsilon = \sqrt{1-\omega^2}$ and $C_1$ is a constant. If $n>1$, then the solution tends to zero slowly while oscillating,
\begin{equation}
 \Phi_n = C_n \displaystyle\frac{\cos\left(\lambda_n r 
 + \alpha_n\right)}{r} \ , \label{e:shphin}
\end{equation}
where $\lambda_n = \sqrt{n^2\omega^2 - 1}$, furthermore $C_n$ and $\alpha_n$ are constants. The coefficient $C_n$ corresponds to the amplitude parameter $\alpha$ used in \eqref{eqdphitailqbosctn}, in case of general dimensions.

To the leading order approximation the background is the Minkowski spacetime, and the phase $\alpha_n$ is constant. We can obtain a better description if we choose the Schwarzschild spacetime as the background. Then a slow dependence on the radial coordinate appears in the phase. In order to obtain the next order of the approximation we allow that $\alpha_n$ are slowly changing functions of the coordinate $r$, such that for their derivatives  $\alpha_{n,r}\ll\lambda_n$ holds. More precisely, we assume that \eqref{e:shphin} holds, and the order of $\alpha_{n,r}$ is $\lambda_n/r$. Keeping the first two orders in $1/r$,
\begin{align}
 \Phi_{n,r} &= -C_n\left(\lambda_n+\alpha_{n,r}\right)
 \frac{\sin\left(\lambda_n r+\alpha_n\right)}{r} - C_n\frac{\cos\left(\lambda_n r
 + \alpha_n\right)}{r^2} \ , \label{Pder} \\
 \Phi_{n,rr} &= -C_n\left(\lambda_n^2+2\lambda_n\alpha_{n,r}\right)
 \frac{\cos\left(\lambda_n r+\alpha_n\right)}{r} + 2C_n \lambda_n
 \frac{\sin\left(\lambda_n r+\alpha_n\right)}{r^2} \ . \label{Pdder}
\end{align}
Substituting the expressions \eqref{e:asymptotA}-\eqref{e:asymptotB} of the metric functions into the wave equation \eqref{eq:wave3}, setting $r_A = r_B=r_0$, then dropping quadratic and higher order terms in $1/r$, we get
\begin{equation}
 \left(1-\frac{r_0}{r}\right) \Phi_{n,rr} + \frac{2}{r}\Phi_{n,r} + n^2 \omega^2
 \left(1+\frac{r_0}{r}\right) \Phi_n  - \Phi_n = 0 \ . \label{e:wave}
\end{equation}
Substituting the expressions \eqref{e:shphin}-\eqref{Pdder}, from the second order terms follows a differential equation for the phase $\alpha_n$\,,
\begin{equation}
 - 2 \lambda_n \alpha_{n,r} + \frac{r_0}{r} \left(\lambda_n^2 + n^2 \omega^2\right) = 0 \ .
\end{equation}
The solution of this is
\begin{equation}
 \alpha_n = \frac{r_0}{2} \left(\frac{2\lambda_n^2+1}{\lambda_n}\right) \log r
 + \delta_n \ , \label{e:delta}
\end{equation}
where $\delta_n$ is a constant. It can be easily checked, that according to our starting assumption, $\alpha_{n,r}$ is really the same order as $\lambda_n/r$. As outer boundary conditions, for the Fourier components of the scalar field we will use the expression \eqref{e:shphin}, with the substitution of \eqref{e:delta}.

\subsection{Spectral numerical method}\label{s:numerical}

We have obtained the numerical solutions of the self-gravitating scalar system by the use of the KADATH spectral library, which was developed by Philippe Grandclément \cite{kadathwebp, grandclement2010}. The code is applied for the description of functions defined on a two-dimensional space, depending on the coordinates $(t,r)$. For the time coordinate we use only one domain. The physical time coordinate $t$ is connected to the numerical time coordinate $t^\star$ by the relation $t^\star=\omega t$. A spectral expansion is performed with respect to $t^\star$, using even cosines for $A$ and $B$, and odd cosines for $\phi$, according to the Fourier expansion \eqref{eq:fourexp1}-\eqref{eq:fourexp3}.

For the radial coordinate a multidomain decomposition is used, similarly to section $(2.2)$ of \cite{grandclement2010}. We define a core domain, which contains the center of symmetry, and several spherical shell shaped domains, situated between two finite radii. For strictly asymptotically flat spacetimes the infinitely large outer domain outside of the last spherical shell can be naturally studied by the compactification of space. In our case, because of the presence of the standing wave tail that oscillates infinitely many times, we cannot employ the method of compactification. Instead, we solve the equations only up to a large $R_{\rm max}$ value of the coordinate $r$, where we match to the asymptotic solutions discussed in the previous subsection. In each domain the physical coordinate $r$ is connected to the numerical coordinate $r^{\star}$ by an affine law. In the core domain we use $r=R_{\rm nuc}\, r^\star$, where $r^\star\in\left[0,1\right]$ and $R_{\rm nuc}$ is the radius of the domain. The expression used in the shells is
\begin{equation}
 r=\left(\displaystyle\frac{R_{\rm outer}-R_{\rm inner}}{2}\right) r^\star + \left(\displaystyle\frac{R_{\rm outer}+R_{\rm inner}}{2}\right) \ ,
\end{equation}
where $r^\star \in \left[-1, 1\right]$, and the inner and outer radii of the domain are $R_{\rm inner}$ and $R_{\rm outer}$, respectively. The spectral expansion is performed with respect to  $r^{\star}$. In the core domain, since the functions are even with respect to the origin, we use only even Chebyshev polynomials. In the spherical shell domains we expand using standard Chebyshev polynomials. For example, in a given shell the approximation of the metric function $A$ is
\begin{equation}
 A=\sum_{j=0}^{N_t}
 \sum_{i=0}^{N_r} A_{ij} \cos\left(2j t^\star\right) T_i\left(r^{\star}\right) \ ,
\end{equation}
where $N_t$ and $N_r$ are the number of coefficients with respect to $t^{\star}$ and $r^{\star}$. The order $i$ Chebyshev polynomial is denoted by $T_i$, and the constants $A_{ij}$ are the spectral coefficients of the function $A$.

\subsection{Matching conditions on the outer boundary} \label{ss:matching}

We determine the outer boundary conditions based on the asymptotic behavior of the fields discussed in Subsection \ref{ss:asymptot}. Let us consider a function $f(r)$ that we intend to calculate numerically, and assume that we intend to match it to an asymptotic behavior determined by a given function $g(r)$. We require that the function $f(r)$ should be matched smoothly to the function $C g(r)$, where $C$ is a yet unknown constant. The continuity of the function and its derivative give the following conditions:
\begin{align}
 f\left(R_{\rm max}\right) &= C g\left(R_{\rm max}\right) \ , \label{e:matchf} \\
 f_{,r} \left(R_{\rm max}\right) &= C g_{,r}\left(R_{\rm max}\right) \ . \label{e:matchdf}
\end{align}
We can eliminate the constant $C$, and obtain a boundary condition for $f$ which only contains the function $g$,
\begin{equation}
 \left[f g_{,r} - f_{,r} g \right] \left(R_{\rm max}\right) = 0 \ . \label{e:bc}
\end{equation}
We apply this method for the zeroth Fourier components of the metric functions $A$ and $B$, which are matched to the functions \eqref{e:asymptotA}-\eqref{e:asymptotB}. The higher Fourier components of $A$ and $B$, which describe the time dependent parts of these functions, are matched to the identically zero function at the boundary. The first Fourier component of the scalar field is matched to the exponentially decaying \eqref{e:shphi1} function. We match all the other components of the scalar to the oscillating form \eqref{e:shphin}, where in the expression \eqref{e:delta} of the phase $\alpha_n$ we use the value $r_0=r_B$, since both $r_A$ and $r_B$ tend to $r_0$ when the matching radius $R_{\rm max}$ grows.

\subsection{The numerical system} \label{ss:system}

When we use the KADATH spectral numerical library, the partial differential equations describing the system are converted into a system of coupled algebraic equations, where the unknowns are the expansion coefficients of the various fields. The nonlinear system is solved by the Newton-Raphson method. Starting from an initial approximate configuration, the solution is found by iteration. At each step the linearized system, with respect to the unknowns, is getting inverted \cite{grandclement2010}.

The system of equations to be solved, \eqref{eq:eieq1}-\eqref{eq:wave3}, are redundant, containing more equations than unknowns. For our calculations we solve the three equations \eqref{eq:eieq1}, \eqref{eq:eieq2} and \eqref{eq:wave3}, and after obtaining the solutions we confirm that the remaining two conditions are appropriately satisfied.

As we have seen at the small-amplitude expansion in Subsection \ref{secosctnsmallampl}, the size of oscillatons become larger and larger when their frequency $\omega$ tends to the scalar field mass $m$, which is assumed to be scaled to $1$ now. Because of this, we have introduced a rescaled radial coordinate $\rho=\varepsilon r$, where $\varepsilon=\sqrt{1-\omega^2}$. In terms of $\rho$, the geometry of the solutions become rather similar. Taking this into account, we choose the radial domains with respect to the coordinate $r$ in the following way. The first three domains are given by the intervals $\left[0,\,1/\varepsilon\right]$,  $\left[1/\varepsilon,\,2/\varepsilon\right]$ and $\left[2/\varepsilon,\,4/\varepsilon\right]$. The additional domains, indexed by the integer $i$ satisfying \mbox{$4\leq i\leq N_{\rm d}$\,,} are chosen as $\left[4(i-3)/\varepsilon,\,4(i-2)/\varepsilon\right]$. The size of all the outer domains are chosen as $4/\varepsilon$, so that we should be able to describe well the standing wave tail appearing in the scalar field $\phi$. The size of the domains cannot be too large with respect to the wavelength of the oscillations. The place of the outer matching surface is the upper boundary of the last interval, $R_{\rm max}=4(N_{\rm d}-2)/\varepsilon$. The value of $R_{\rm max}$ can be varied by changing the number $N_{\rm d}$ of the used domains. For our computations, generally the number of domains $N_{\rm d}=20$ turned out to be appropriate, which corresponds to $R_{\rm max}=72/\varepsilon$.

Considering the number of spectral coefficients, we have used three kinds of resolutions, a low one with $N_r=13$ radial and $N_t = 5$ time coefficients, a medium one with $N_r = 17$ and $N_t=7$ coefficients, and a high resolution with $N_r = 33$ and $N_t=9$. More technical details about the application of the spectral numerical method can be found in Section III of our paper \cite{grandclement2011}.

For the start of the Newton-Raphson iteration we have obtained the necessary initial approximate solution from the first order of the small-amplitude expansion, with the parameter choice  $\varepsilon\approx0.1$. If a solution is already known for which the frequency $\omega$ is not far from $1$, we can use it as an initial configuration for the search of new solutions, by changing the frequency $\omega$ in small steps. Higher resolution calculations require large number of spectral coefficients, and because of the inversion of the resulting huge matrices, the computations may take considerable amount of time even on computer clusters with several dozens of cores.

\subsection{Minimization of the oscillating tail} \label{ss:minimization}

At the asymptotic behavior \eqref{e:shphin} of all $n>1$ indexed Fourier components of the scalar field there is a phase $\alpha_n$. The expression \eqref{e:delta} of each $\alpha_n$ contains an unspecified constant $\delta_n$. We are looking for the solution, which is as well localized as possible. We intend to construct the quasibreather solution, for which the total energy contained in the tails of all Fourier modes is minimal. In principle, this would mean that we need to minimize the sum of the tail energy densities resulting from all modes, as a function of all $\delta_n$ phases. 

Fortunately, the task is substantially simplified by the fact that the tail-amplitude of the Fourier mode $\Phi_3$ is many orders of magnitude larger than the tail-amplitude of the other modes, and hence it is sufficient to minimize the constant $C_3$ in \eqref{e:shphin}. Further major simplification arises from the observation that the behavior of the higher Fourier modes only very slightly influence the much larger dominant $\Phi_3$ mode. In other words, $C_3$ is almost independent of the $n>3$ indexed $\delta_n$ phases. This means that the minimization of $C_3$ as a function of only $\delta_3$ already gives a very good approximation for the true minimum. This expectation has been also confirmed by our detailed numerical investigations. In all the considered examples, the change of $\delta_5$ and some higher indexed phases induced only negligibly small modifications in the value of the minimum of $C_3$ and in the place of the minimum in terms of $\delta_3$. Based on these considerations, in our calculations generally we have chosen the phases $\delta_n$ to be zero for $n\geq 5$, and we have obtained $\delta_3$ by the minimization of the amplitude $C_3$. We have performed the minimization by golden-section search. 

\begin{figure}[!hbt]
\centering
\includegraphics[width=100mm]{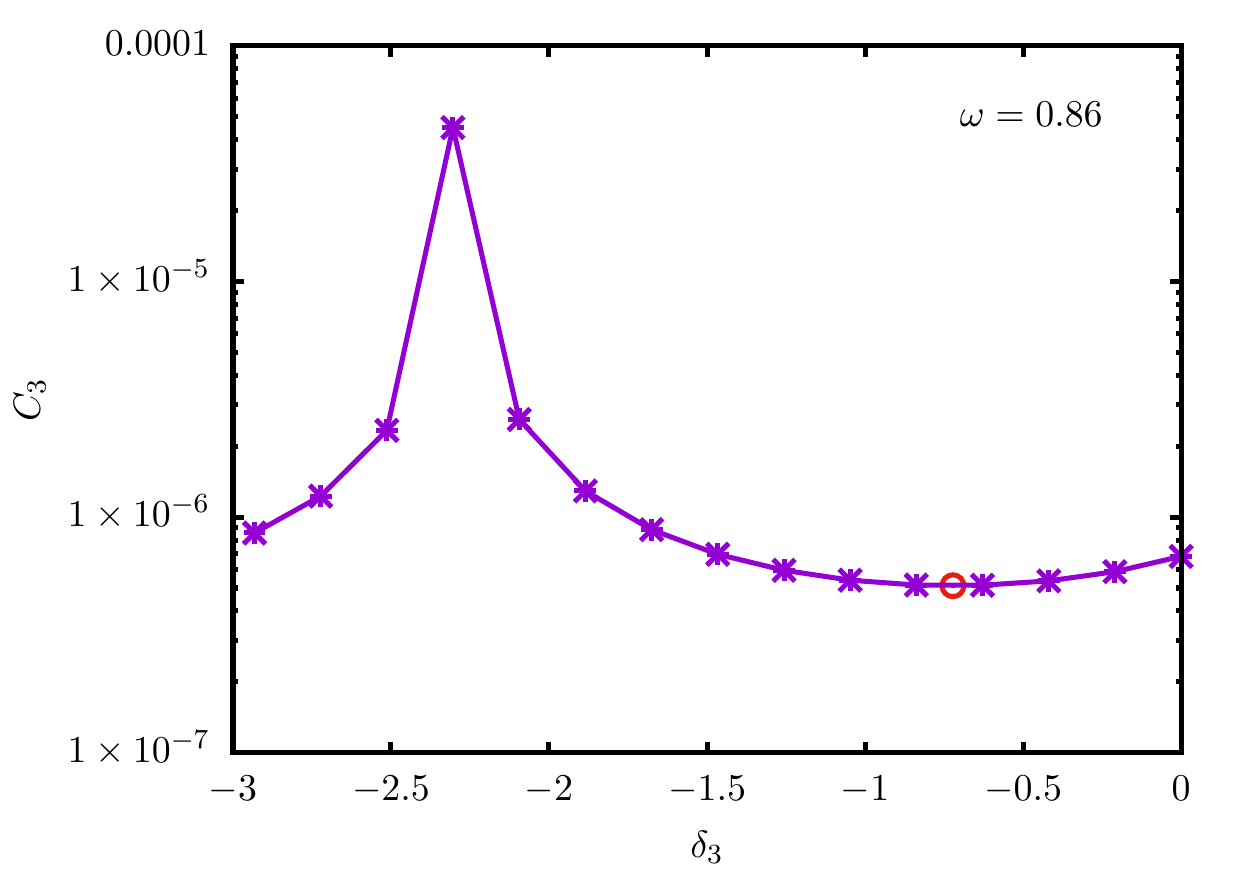}
\caption{\label{f:d3eff} The dominant amplitude $C_3$ as a function of the phase $\delta_3$, in case of $\omega=0.86$. The red circle denotes the place of the minimum, at $\delta_3=-0.72$ and $C_3=5.118\cdot10^{-7}$.}
\end{figure}

On Figure \ref{f:d3eff}
we can see the phase dependence of the amplitude $C_3$ for the frequency choice $\omega=0.86$. It can be seen that in the vicinity of the minimum the function changes very slowly, hence it is not necessary to determine the value of $\delta_3$ very precisely in order to get an accurate value for the amplitude. We have tested, that if we chose various nonzero values for $\delta_5$, the obtained result for $C_3$ only changed extremely slightly, its relative change remained under $10^{-6}$, even though its absolute value was around $10^{-6}$.

\subsection{Numerical results} \label{subsecnumosctn}

\begin{figure}[!hbt]
\centering
\includegraphics[width=70mm]{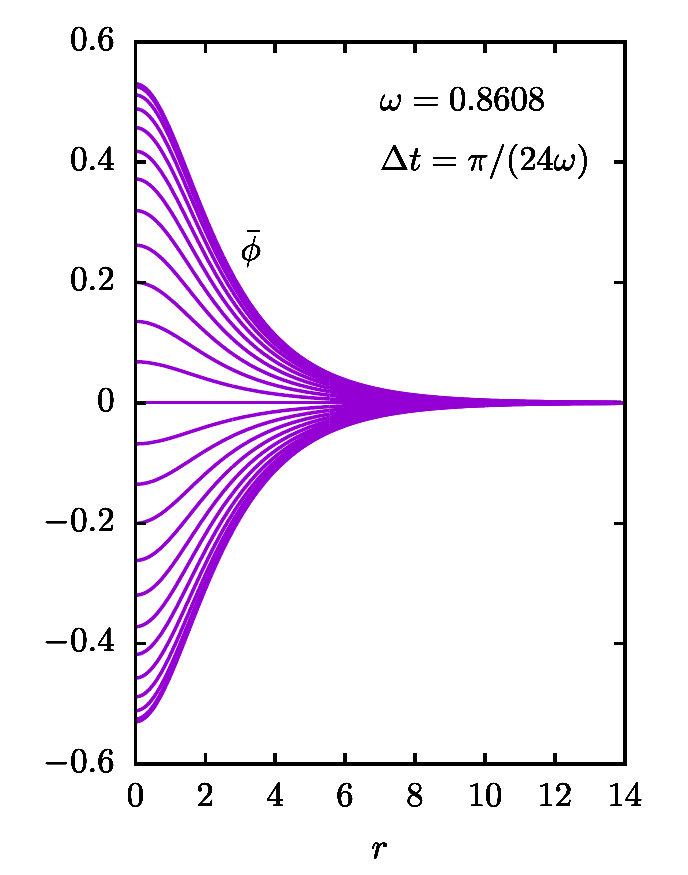}
\includegraphics[width=70mm]{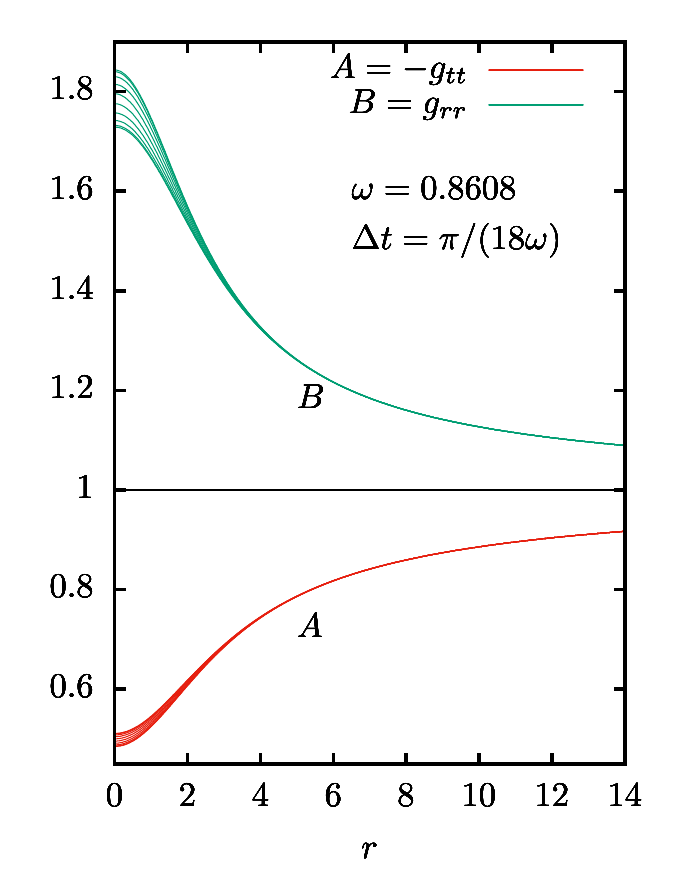}
\caption{\label{f:oscphiaabb} Time evolution of the scalar field $\bar\phi$ and the metric components $A$ and $B$ during one oscillation period, for an oscillaton with frequency $\omega=\omega_c=0.8608$.}
\end{figure}

On Figure \ref{f:oscphiaabb}
we show the change of the scalar field and the metric functions during one oscillation period in the core domain, for the Klein-Gordon oscillaton with the critical frequency $\omega=\omega_c=0.8608$. Oscillatons with lower than the critical frequency are unstable, while for $\omega_c<\omega<1$ they are stable. It can be seen that for $A$ and $B$ the oscillations around the static component $1+\bar A_0$ and $1+\bar B_0$ in \eqref{eq:fourexp2}-\eqref{eq:fourexp3} are relatively small. Studying higher frequency states, it turns out that approaching the  value $1$ with the frequency $\omega$ the relative magnitude of the oscillation in $A$ and $B$ become even smaller, and the state becomes more and more similar to the complex field boson star configuration.

\begin{figure}[!hbt]
\centering
\includegraphics[width=110mm]{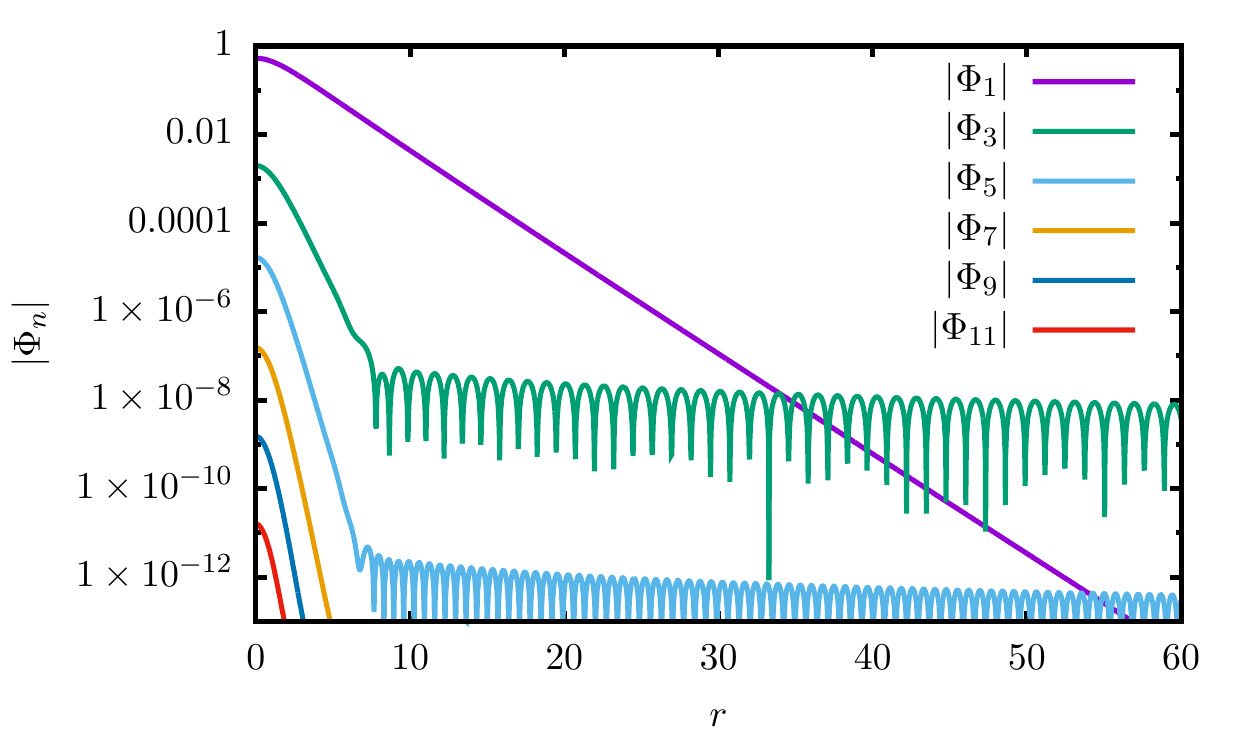}
\caption{\label{f:phincd} The Fourier components $\Phi_n$ of the scalar field $\bar\phi$ at the frequency $\omega=0.8608$.}
\end{figure}

\begin{figure}[!hbt]
\centering
\includegraphics[width=110mm]{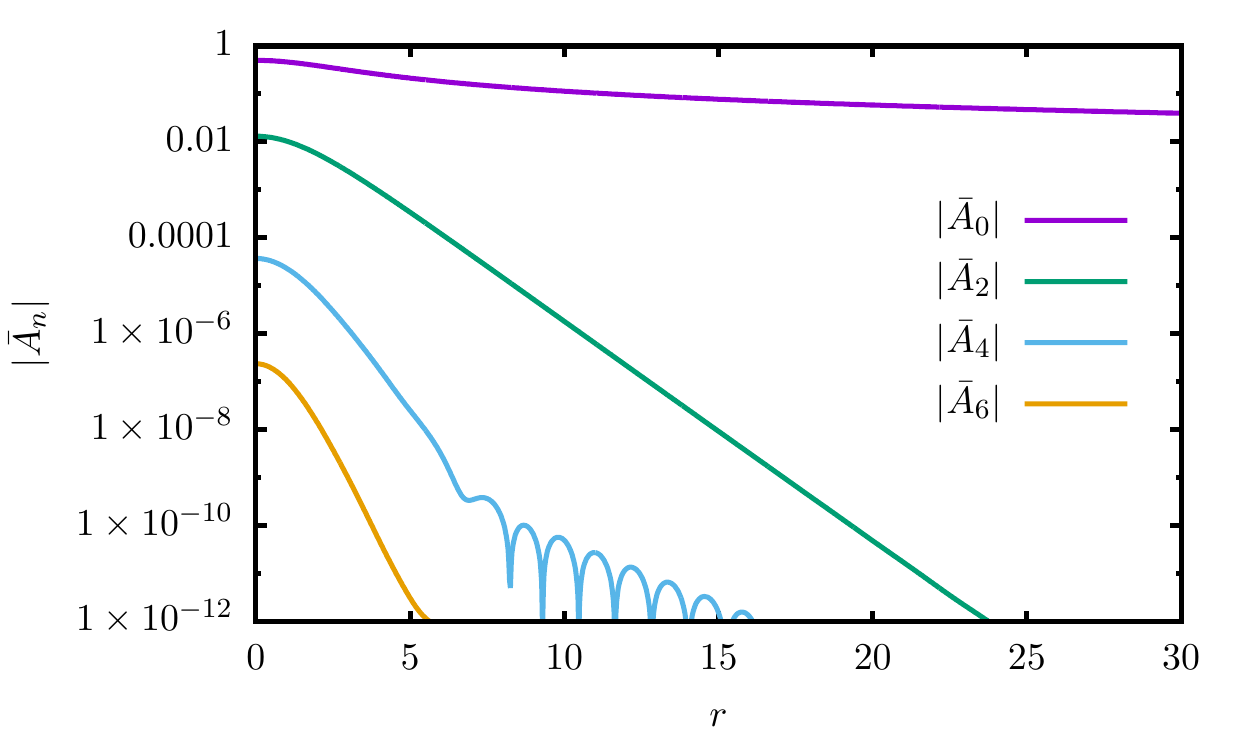}
\caption{\label{f:aanc}  The Fourier components $\bar A_n$ of the metric function $A$, in case of $\omega=0.8608$. }
\end{figure}

\begin{figure}[!hbt]
\centering
\includegraphics[width=110mm]{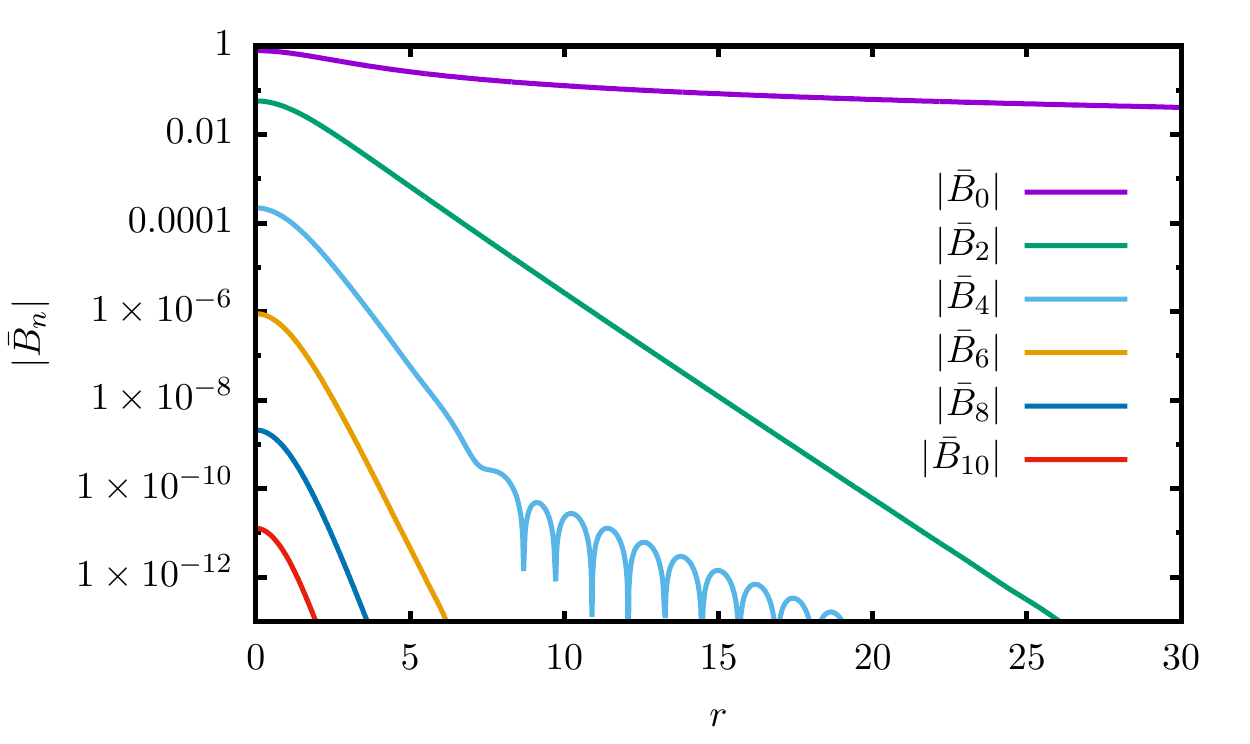}
\caption{\label{f:bbnc} The Fourier components $\bar B_n$ of the function $B$, for $\omega=0.8608$.}
\end{figure}

On Figure \ref{f:phincd},
for the same critical frequency oscillaton, we give the radial dependence of the first few Fourier modes. Because of the several magnitude variation of the functions we show their absolute value in logarithmic scale. The downwards pointing peaks correspond to zero crossings. On Figure \ref{f:aanc}
we can see the components of the metric function $A$, while on Figure \ref{f:bbnc}
we show the components of $B$. We can obtain similar figures for other frequencies as well. With the increase of the frequency the tail-amplitude decreases very quickly. As the frequency approaches the threshold value $1$, the relative magnitude of the higher Fourier modes with respect to the base mode also decreases, and hence in those cases we can already get precise approximation by taking into account only fewer Fourier modes.

\begin{figure}[!hbt]
\centering
\includegraphics[width=110mm]{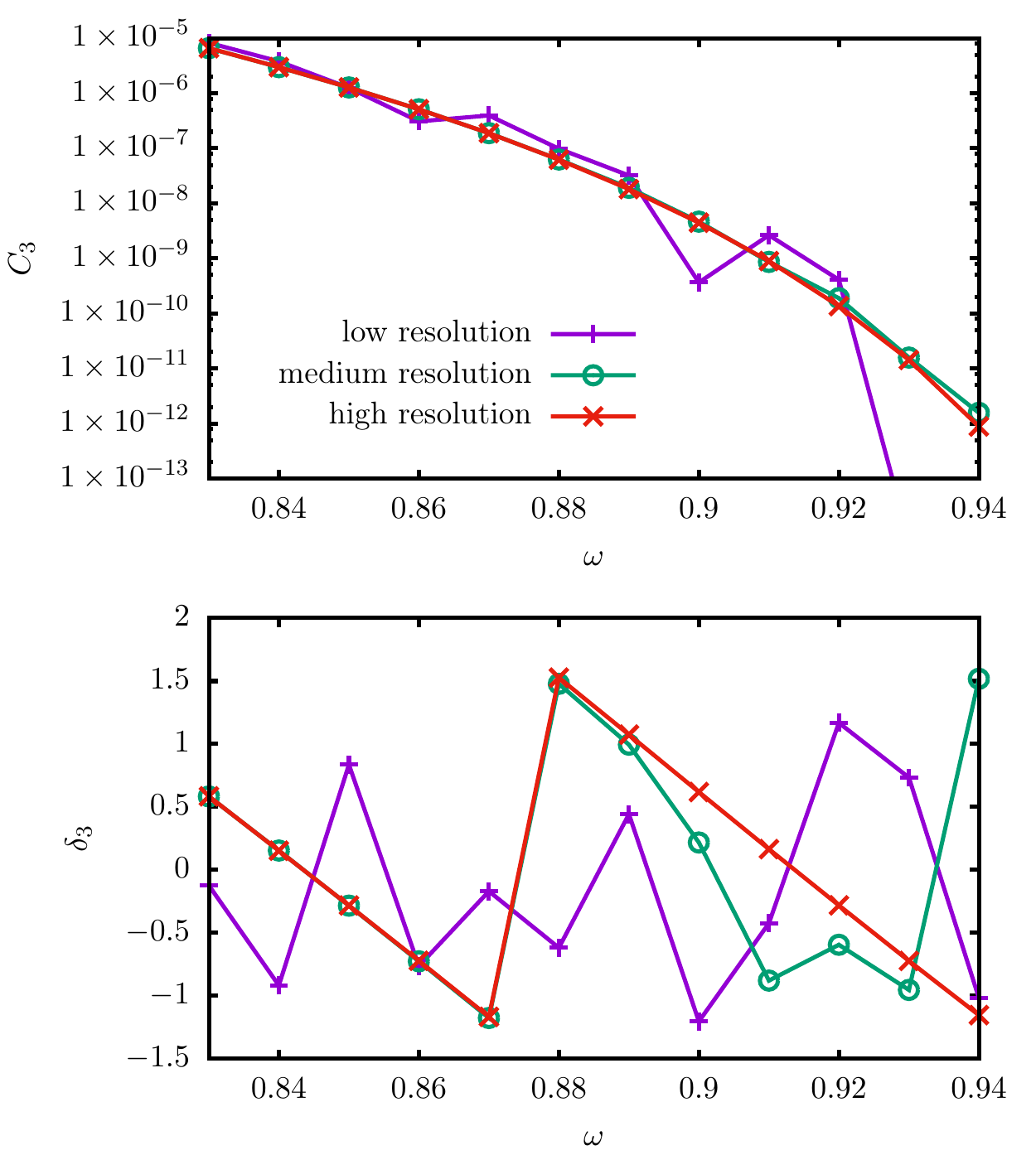}
\caption{\label{f:c3phasec} The minimized amplitude $C_3$ determining the radiation amplitude of oscillatons, and the phase $\delta_3$ belonging to it, for three different numerical resolutions.}
\end{figure}

A crucial result of our numerical simulations is the demonstration that there is no such frequency $\omega$ for which the oscillating tail responsible for the radiation vanishes.
On the upper panel of Figure \ref{f:c3phasec}
we give the frequency dependence of the minimized amplitude $C_3$ in the domain where the tail is large enough so that we are able to compute it numerically. The number of spectral coefficients are, at the low resolution $N_r=13$, $N_t = 5$, at the medium one $N_r = 17$, $N_t=7$, and at the high one $N_r = 33$, $N_t=9$. On the lower panel of the figure we also give the phase $\delta_3$ for the three kind of resolutions. The observation that there is no such  $\omega$ for which $C_3=0$ means that there is no exactly periodic localized breather solution of the system. The mass of all oscillaton solutions decrease very slowly in time because of the energy radiated away by the scalar field. From the lower panel of Fig.~\ref{f:c3phasec} it can be seen that the low resolution is not enough for the determination of the phase, and the medium resolution is appropriate only at lower frequencies. In contrast to this, because of the flatness of the curve, for the value of the minimum amplitude $C_3$ even these lower resolutions give acceptable approximations.

\begin{figure}[!hbt]
\centering
\includegraphics[width=110mm]{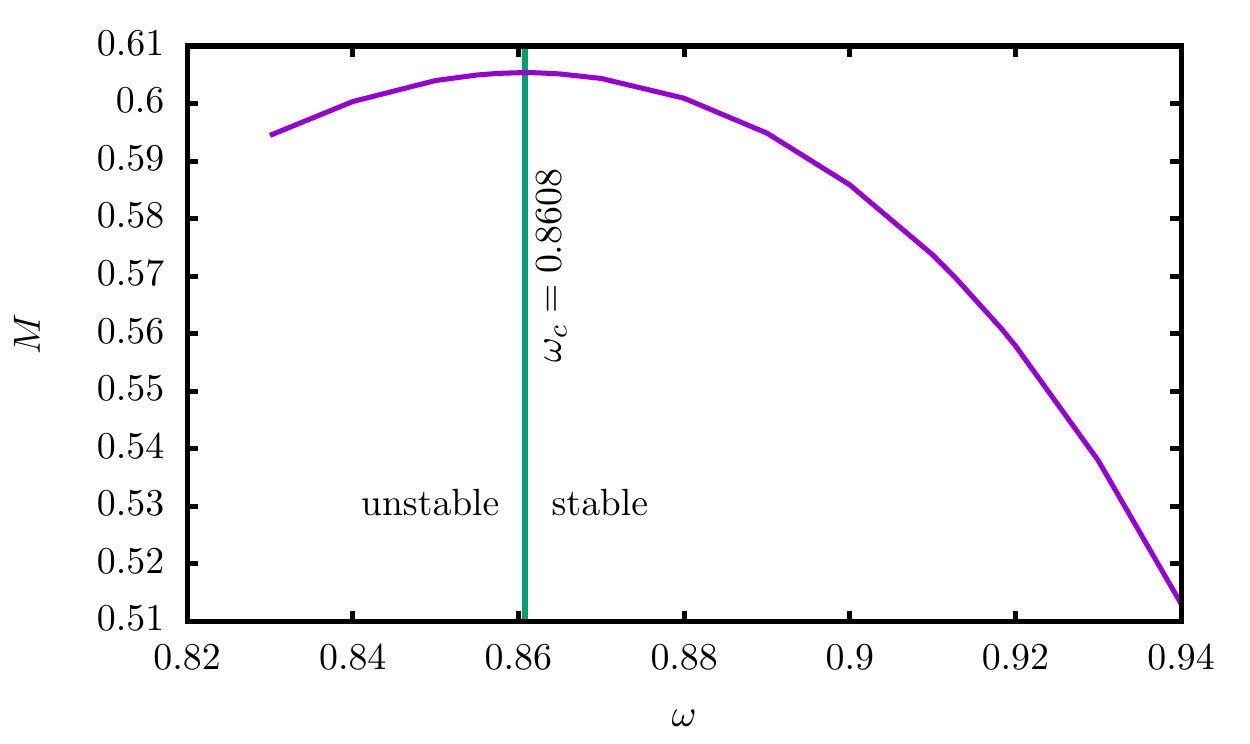}
\caption{\label{f:otnmeps} The mass of Klein-Gordon oscillatons as a function of their frequency.}
\end{figure}

On Figure \ref{f:otnmeps}
we plot the total mass represented by the core domain, in the previously used frequency domain where the tail is still numerically detectable. The mass here is calculated as the mean value $M=r_0/2=(r_A+r_B)/4$ from the asymptotic behavior \eqref{e:asymptotA}-\eqref{e:asymptotB} of the functions $A$ and $B$. The mass takes its maximum at the frequency $\omega_c=0.8608$, where its value is $M_c=0.60535$. This critical state separates from each other the domain of the stable and the unstable oscillatons. Although the mass has a maximum, the central values of the Fourier modes $\Phi_1$ and $\Phi_3$, and hence also the central energy density, are all monotonically decreasing functions of the frequency $\omega$.

\begin{figure}[!hbt]
\centering
\includegraphics[width=110mm]{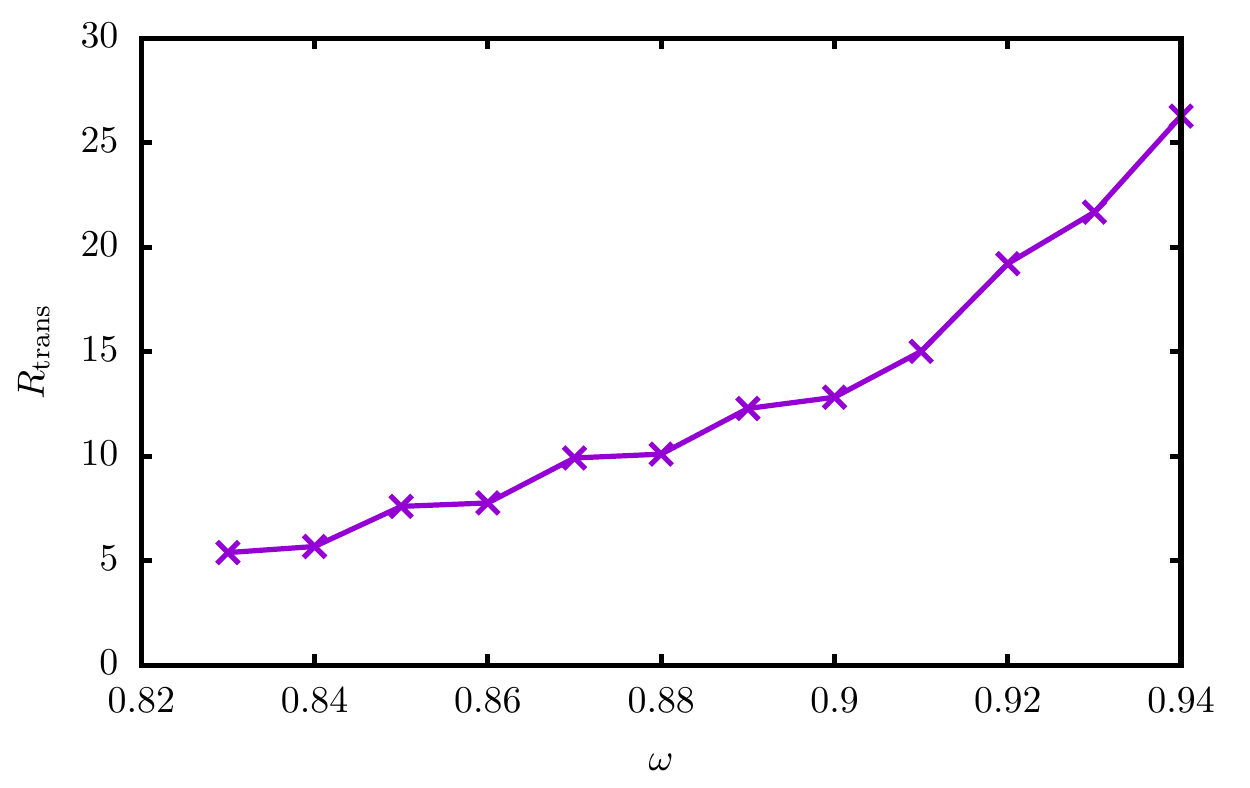}
\caption{\label{f:rtransc} The value of the radius $R_{\mathrm{trans}}$ where the standing wave tail appears in the mode $\Phi_3$, as a function of the frequency.}
\end{figure}

On Figure \ref{f:rtransc}
we show the frequency dependence of the radius $R_{\mathrm{trans}}$, where the radiation in the mode $\Phi_3$ becomes dominant. More precisely, we define $R_{\mathrm{trans}}$ as the lowest radius $r$ where the value of $\Phi_3$ becomes zero, and hence the oscillating tail starts. The small oscillation on the curve appears because of the change in the phase of the tail. Apart from this oscillation, the function $R_{\mathrm{trans}}$ increases monotonically, and according to the small-amplitude expansion, it tends to infinity at $\omega=1$. The radius where the standing wave tail visibly appears in the scalar field $\phi$ is still much larger than this value, since at the radius $R_{\mathrm{trans}}$ the exponentially decaying $\Phi_1$ mode is still much larger than $\Phi_3$ (see e.g.~Fig.~\ref{f:phincd}).

\subsection{Comparison with the small-amplitude expansion}

\begin{figure}[!hbt]
\centering
\includegraphics[width=110mm]{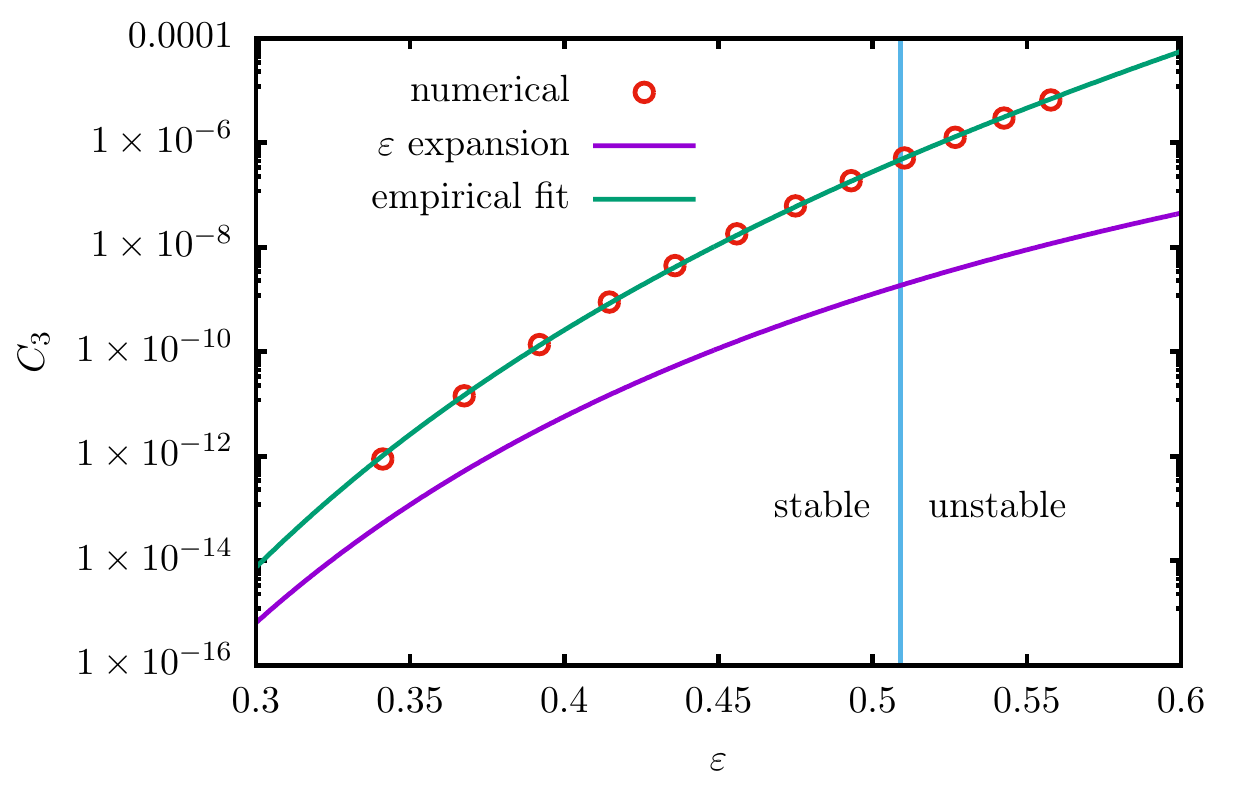}
\caption{\label{f:c3epscmp3d} Comparison of the numerically obtained values of the tail-amplitude $C_3$ of the Fourier mode $\Phi_3$ with the leading order theoretical result shown by the purple curve. The green curve was obtained by a fit to the data points.}
\end{figure}

For the Klein-Gordon potential and three spatial dimensions, the leading order result for the $\varepsilon$ dependence of the tail-amplitude parameter $C_3\equiv\alpha$ has been given in \eqref{eqalphotnkg}. On Figure \ref{f:c3epscmp3d}
we again give the numerically computed $C_3$ amplitudes already presented on Fig.~\ref{f:c3phasec}, but in order to facilitate the comparison, now as a function of the parameter $\varepsilon=\sqrt{1-\omega^2}$. The theoretical curve described by \eqref{eqalphotnkg} lies much below the data points. The following empirical function,
\begin{equation}
 C_3^{\text{emp}}=\frac{3.761}
 {\varepsilon}\left(1+\varepsilon^2\right)^{16.63}\,\exp\left[
 -\frac{11.2497}{\varepsilon}\left(1-0.2990\,\varepsilon^2
 \right)\right] \ ,  \label{eq:c3emp}
\end{equation}
fits very well the numerical results, furthermore at small $\varepsilon$ values it approaches the theoretical result \eqref{eqalphotnkg}. It is a natural expectation, that if we were able to determine the radiation to higher orders in $\varepsilon$, then the place of the pole and the coefficient in front of the exponential would depend polynomially on $\varepsilon$, and hence we would obtain an expression similar to \eqref{eq:c3emp}. Since it approximates very well all our numerical and theoretical results, this empirical formula has a great importance. We can expect that in the whole stable domain, for any frequency in the range $\omega_c<\omega<1$, without further numerical calculation \eqref{eqalphotnkg} gives a good estimate for the amplitude of the tail, and hence for the strength of the radiation. For large $\varepsilon$ values the numerical values shown on Fig.~\ref{f:c3epscmp3d} are several hundred times larger than the analytical results, but with the decrease of $\varepsilon$ the relative difference decreases exponentially. The smallest $\varepsilon$ for which we could numerically calculate the tail is $\varepsilon=0.341$, where the magnitude of the tail-amplitude $C_3$ is $10^{-12}$. In this case the numerical result is still twenty times larger than the leading order theoretical one. The parameter $\varepsilon=0.341$ in a power series expansion generally cannot be considered small. However, because of the exponential decrease of the relative difference it appears convincing that the two approaches would give more and more agreeing results for smaller $\varepsilon$ values.

\begin{figure}[!hbt]
\centering
\includegraphics[width=110mm]{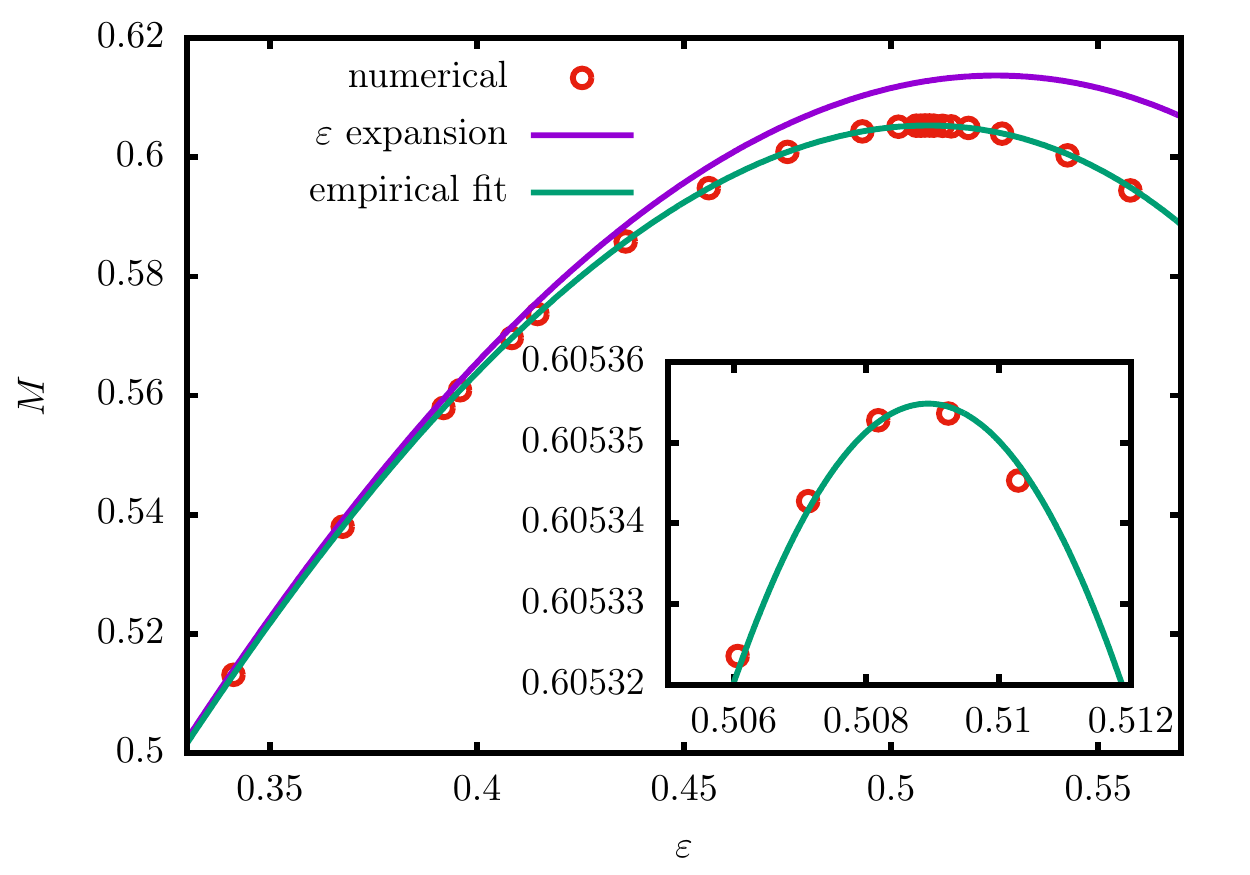}
\caption{\label{f:masseps4} Numerically and analytically computed oscillaton masses at various $\varepsilon$ values. The green curve is a precise empirical fit.}
\end{figure}

On Figure \ref{f:masseps4}
we give the mass of the oscillaton as a function of the parameter $\varepsilon$. The numerical data is the same as on Fig.~\ref{f:otnmeps}. The first two terms of the expansion given by \eqref{eq:totmass} gives an acceptable approximation, but the place and the value of the maximum is not particularly precise. This is not surprising, since the place of the maximum is $\varepsilon \approx 0.5$, which is not small at all. We can get a very good fitting curve on the numerical results in the following form:
\begin{equation}
 M^{\text{emp}}=\varepsilon M_1
 +\varepsilon^{3}M_2
 +\varepsilon^{5}M_3
 +\varepsilon^{7}M_4
 +\varepsilon^{9}M_5 \ , \label{eq:massfit}
\end{equation}
where for the value of $M_1$ and $M_2$ we keep the small-amplitude results $M_1=1.75266$ and $M_2=-2.11742$, while the fitted constants are $M_3=-0.24723$, $M_4=1.1749$ and $M_5=-4.1308$. The numerical value of the maximum's place is $\varepsilon_c=0.509$, which corresponds to the frequency $\omega_c=0.8608$, and the maximal mass is $M_c=0.60535$. The importance of these values is that for $\varepsilon>\varepsilon_c$, i.e.~for $\omega<\omega_c$, the oscillatons are unstable. In the earlier literature, in papers \cite{Urena02b,Alcubierre03} we can find the values $0.864$ for $\omega_c$, and $0.607$ for $M_c$. Restoring the scalar field mass $m$, and using SI units, we have written the maximal mass in \eqref{eqmcnatunit}.

\subsection{Mass loss rate at large amplitudes} \label{s:longevity}

The energy radiated away by the scalar field can be calculated in the same way as in Subsection \ref{secmasslosseps}. In contrast to the discussion there, we do not assume now that $\omega\approx 1$, in order to allow us to obtain precise results for large amplitude oscillatons. We have to calculate the energy carried out by the following spherical wave,
\begin{equation}
 \bar\phi=\frac{C_3}{r}\cos\left(\lambda_3 r-3\omega t+\alpha_3\right) \ , \label{e:outwave}
\end{equation}
where $\lambda_3=\sqrt{9\omega^2-1}$. Because of the rescaling \eqref{eq:rscpu}, we have $\phi=\bar\phi/\sqrt{8\pi}$. Equation \eqref{eqsaverage} giving the radiated energy current averaged for an oscillation period can also be applied now, where $\omega_f=3\omega$, $\lambda_f=\lambda_3$ and $\alpha=C_3$. As a result of this, the time averaged mass loss rate of oscillatons is
\begin{equation}
 \bar S=-\,\frac{\overline{\mathrm{d}M}}{\mathrm{d}t}=
 \frac{3}{4}C_3^2\omega\sqrt{9\omega^2-1} \ . \label{e:dmdt}
\end{equation}
Substituting the empirical expression \eqref{eq:c3emp} for $C_3$, we can obtain the result for the radiation rate, which according to our numerical calculations is valid for large amplitude states as well:
\begin{equation}
 \bar S=10.61\frac{\omega\sqrt{9\omega^2-1}}
 {\varepsilon^2}\left(1+\varepsilon^2\right)^{33.26}\,\exp
 \left[-\frac{22.4993}{\varepsilon}\left(1-0.2990
 \,\varepsilon^2\right)\right] \ , \label{e:dmdtemp}
\end{equation}
where $\omega=\sqrt{1-\varepsilon^2}$.

Using merely the small-amplitude expansion results, if we substitute into the formula \eqref{e:symradlaw} the constants for the $d=3$ spatial dimensional case from Table \ref{ctable}, we get the leading order in $\varepsilon$ result,
\begin{equation}
 \bar S=\frac{30.0}{\varepsilon^2}\exp\left(-\frac{22.4993}
 {\varepsilon}\right) \ . \label{e:dmdteps}
\end{equation}
For small $\varepsilon$ this is indeed the appropriate approximation of \eqref{e:dmdtemp}. However, for oscillatons that have a mass close to the maximal value, the empirical result based on the numerical calculations gives substantially stronger radiation.

Substituting into the empirical expression \eqref{e:dmdtemp} the value $\varepsilon_c=0.509$ belonging to the maximal mass $M_c=0.60535$, and dividing by $M_c$, we obtain the relative mass loss rate for the largest mass stable oscillaton,
\begin{equation}
 \left(\frac{1}{M}\frac{\overline{\mathrm{d}M}}{\mathrm{d}t}\right)_{M=M_c}
 =-5.917\times 10^{-13} m \ ,
\end{equation}
where $m$ is the scalar field mass in Planck units. This reliable numerical result is about $14000$ times larger than the leading $\varepsilon$ order estimation given in equation \eqref{eq:maxloss1}. The reason for this huge difference is that at the parameter value $\varepsilon=0.509$ the small-amplitude expansion underestimates the the amplitude of the radiating tail by about a factor of one hundred.

It is an interesting question that how long time it takes for an $M=M_c$ maximal mass oscillaton to loose a certain part of its mass. Since the elapsed time $t$ is inversely proportional to the scalar field mass $m$, we give the product $tm$ in Table \ref{ttable}.
\begin{table}[htbp]
\centering
\begin{tabular}{|l|l|l|l|l|l|l|}
\hline
& \multicolumn{3}{|c|}{$\varepsilon$ expansion}
& \multicolumn{3}{|c|}{numerical result}\\
\cline{2-7}
$\frac{M_c-M}{M_c}$ & $\varepsilon$
  & $t\,m$
  & $t[\text{year}]\,m\negthinspace\left[\frac{\mathrm{eV}}{c^2}\right]$  & $\varepsilon$
  & $t\,m$
  & $t[\text{year}]\,m\negthinspace\left[\frac{\mathrm{eV}}{c^2}\right]$\\
\hline
$0.01$ & $0.482$ & $3.45\cdot 10^{15}$ & $7.21\cdot 10^{-8}$ & $0.469$ & $7.03\cdot 10^{11}$ & $1.46\cdot 10^{-11}$ \\
$0.1$ & $0.383$ & $8.65\cdot 10^{20}$ & $0.0180$ & $0.376$ & $2.63\cdot 10^{18}$ & $5.48\cdot 10^{-5}$ \\
$0.2$ & $0.320$ & $6.69\cdot 10^{25}$ & $1400$ & $0.314$ & $1.16\cdot 10^{24}$ & $24.2$ \\
$0.3$ & $0.267$ & $2.55\cdot 10^{31}$ & $5.32\cdot 10^{8}$ & $0.264$ & $1.71\cdot 10^{30}$ & $3.56\cdot 10^{7}$ \\
$0.4$ & $0.224$ & $2.74\cdot 10^{38}$ & $5.72\cdot 10^{15}$ & $0.220$ & $5.87\cdot 10^{37}$ & $1.23\cdot 10^{15}$ \\
$0.5$ & $0.182$ & $9.54\cdot 10^{47}$ & $1.99\cdot 10^{25}$ & $0.180$ & $5.98\cdot 10^{47}$ & $1.25\cdot 10^{25}$ \\
$0.6$ & $0.144$ & $1.10\cdot 10^{62}$ & $2.29\cdot 10^{39}$ & $0.142$ & $2.03\cdot 10^{62}$ & $4.24\cdot 10^{39}$ \\
$0.7$ & $0.107$ & $1.78\cdot 10^{85}$ & $3.72\cdot 10^{62}$ & $0.105$ & $1.18\cdot 10^{86}$ & $2.45\cdot 10^{63}$ \\
\hline
\end{tabular}
\caption{\label{ttable}
The time $t$ necessary for an initially maximal mass oscillaton to lose a given part of its mass. 
We give the product of the scalar field mass $m$ and $t$ in Planck and also in ordinary units. For comparison we give both the small-amplitude formalism estimation and the reliable numerical results.}
\end{table}
We list the elapsed time based on the small-amplitude expansion approximation in the left-hand part of the table, and the results of the numerical calculations based on the Fourier expansion in the right-hand side of the table. In the first case we calculate the mass loss rate from \eqref{e:dmdteps}, and for the $\varepsilon$ dependence of the mass we apply the expression \eqref{eq:totmass}. For the numerical results we use the empirical expressions \eqref{e:dmdtemp} and \eqref{eq:massfit}.

\begin{figure}[!hbt]
\centering
\includegraphics[width=110mm]{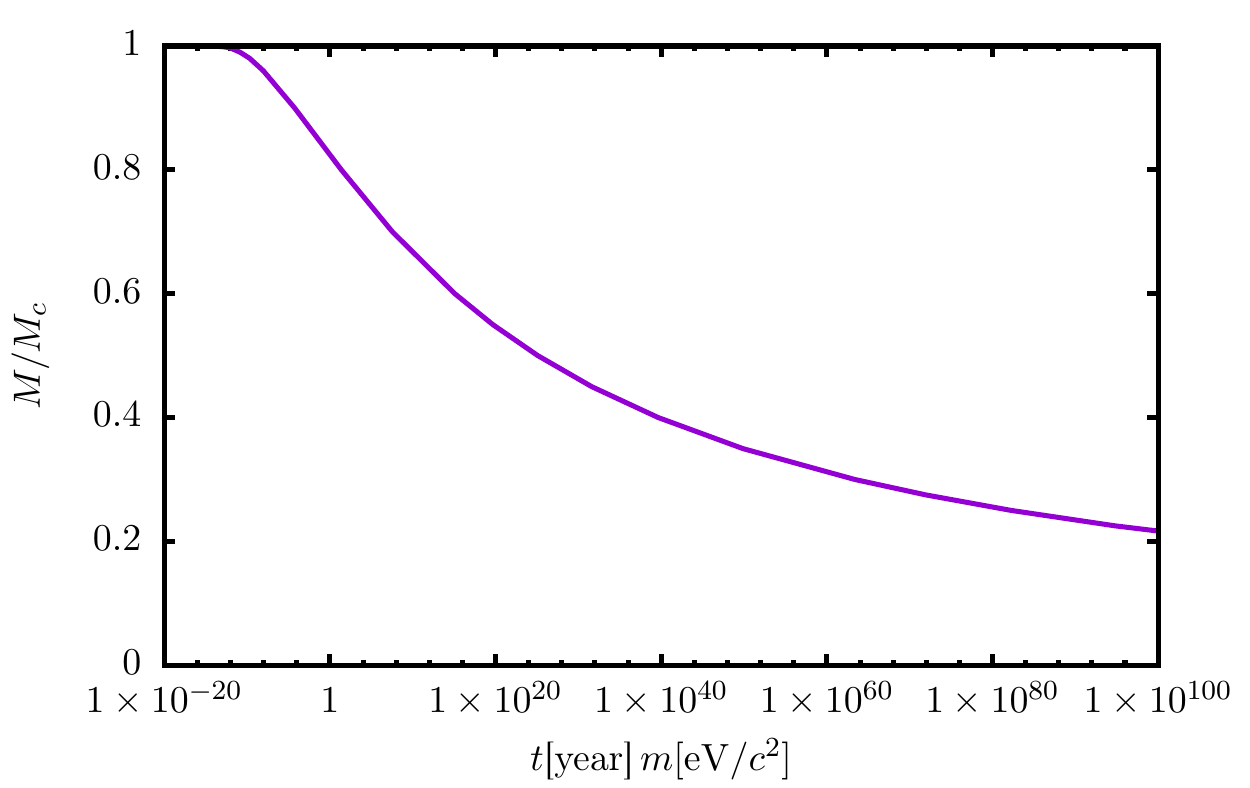}
\caption{\label{f:mpmtm} The mass change of the initially $M_c$ maximal mass oscillaton as a function of the product of the scalar mass $m$ and the elapsed time $t$.}
\end{figure}

On Figure \ref{f:mpmtm}
we show how the mass of an initially $M_c$ maximal mass oscillaton changes in time, giving the ratio $M/M_c$. We show the reliable numerical results. Because of the extremely slow decrease we take the product $tm$ logarithmically on the horizontal axis.

A further interesting problem is that an initially maximal mass oscillaton loses how big part of its mass during a time period corresponding to the age of the universe, which we take as $1.37\cdot 10^{10}$ year now. In Table \ref{mstable}
\begin{table}[htbp]
\begin{tabular}{|l|l|l|l|l|l|l|}
\hline
& \multicolumn{3}{|c|}{$\varepsilon$ expansion}
& \multicolumn{3}{|c|}{numerical result}\\
\cline{2-7}
$m\negthinspace\left[\frac{\mathrm{eV}}{c^2}\right]$ & $\varepsilon_c-\varepsilon$
  & $\frac{M}{M_\odot}$
  & $\frac{M_c-M}{M_c}$ & $\varepsilon_c-\varepsilon$
  & $\frac{M}{M_\odot}$
  & $\frac{M_c-M}{M_c}$ \\
\hline
$10^{-35}$ & $2.31\cdot 10^{-10}$
   & $8.20\cdot 10^{24}$ & $2.91\cdot 10^{-19}$ & $2.37\cdot 10^{-8}$
   & $8.09\cdot 10^{24}$ & $3.75\cdot 10^{-15}$ \\
$10^{-30}$ & $7.31\cdot 10^{-8}$
   & $8.20\cdot 10^{19}$ & $2.91\cdot 10^{-14}$ & $7.50\cdot 10^{-6}$
   & $8.09\cdot 10^{19}$ & $3.74\cdot 10^{-10}$ \\
$10^{-25}$ & $2.31\cdot 10^{-5}$
   & $8.20\cdot 10^{14}$ & $2.90\cdot 10^{-9}$ & $2.18\cdot 10^{-3}$
   & $8.09\cdot 10^{14}$ & $3.16\cdot 10^{-5}$ \\
$10^{-20}$ & $0.00621$
   & $8.20\cdot 10^{9}$ & $2.09\cdot 10^{-4}$ & $0.0544$
   & $7.94\cdot 10^{9}$ & $0.0183$ \\
$10^{-15}$ & $0.0883$
   & $7.87\cdot 10^{4}$ & $0.0400$ & $0.126$
   & $7.36\cdot 10^{4}$ & $0.0896$ \\
$10^{-10}$ & $0.169$
   & $7.06\cdot 10^{-1}$ & $0.139$ & $0.183$
   & $6.65\cdot 10^{-1}$ & $0.178$ \\
$10^{-5}$ & $0.226$
   & $6.24\cdot 10^{-6}$ & $0.238$ & $0.227$
   & $5.96\cdot 10^{-6}$ & $0.263$ \\
$1$ & $0.267$
   & $5.56\cdot 10^{-11}$ & $0.322$ & $0.262$
   & $5.36\cdot 10^{-11}$ & $0.337$ \\
$10^{5}$ & $0.298$
   & $4.98\cdot 10^{-16}$ & $0.392$ & $0.289$
   & $4.85\cdot 10^{-16}$ & $0.401$ \\
$10^{10}$ & $0.323$
   & $4.51\cdot 10^{-21}$ & $0.450$ & $0.311$
   & $4.41\cdot 10^{-21}$ & $0.454$ \\
$10^{15}$ & $0.342$
   & $4.11\cdot 10^{-26}$ & $0.499$ & $0.329$
   & $4.04\cdot 10^{-26}$ & $0.500$ \\
\hline
\end{tabular}
\caption{\label{mstable}
The mass $M$ of an originally maximal mass oscillaton, after a time period corresponding to the age of the universe, for various $m$ scalar field masses. We give the change of $\varepsilon$ from its original $\varepsilon_c$ value, and the relative mass change as well.
}
\end{table}
we list the remaining mass in $M_\odot$ solar mass units, and we also give the relative mass change. At the small-amplitude expansion we use $\varepsilon_c\approx\varepsilon_{\rm m}=0.525$, while at the numerical calculation we use the precise value $\varepsilon_c=0.509$.

\begin{figure}[!hbt]
\centering
\includegraphics[width=100mm]{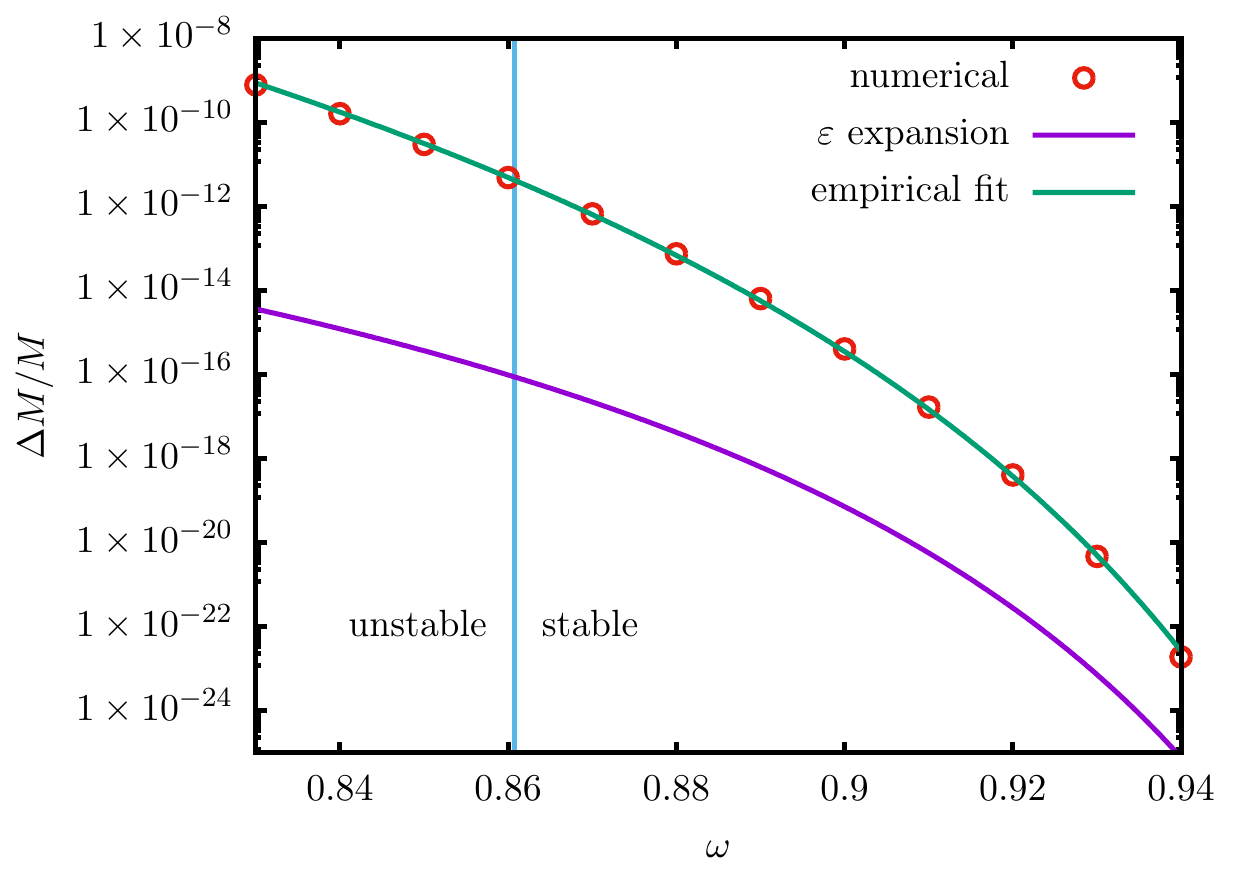}
\caption{\label{f:deltam2} Relative mass loss of oscillatons during one oscillation period.}
\end{figure}

On Figure \ref{f:deltam2}
we show that during one oscillation period an oscillaton loses how small part of its mass. This quantity is also interesting, because it is independent on the scalar field mass $m$,
\begin{equation}
 \frac{\Delta M}{M}=-\,\frac{2\pi}{\omega}
 \frac{1}{M}\frac{\overline{\mathrm{d}M}}{\mathrm{d}t} \ .
\end{equation}
From the figure it can be seen that the relative mass loss rate is so small that every stable oscillaton can be considered as time-periodic to a very good approximation.

\section{Oscillatons in case of a positive cosmological constant} \label{secosctnlambpos}

It is important to clarify how the expansion of the universe influences the formation, stability and lifetime of the localized objects formed by scalar fields. The dynamics of the universe is determined by the large scale properties of the matter that it contains. For the initial study of the problem it is natural to consider first the case when for large distances the matter is represented only by a nonzero cosmological constant. In this case at large distances from the localized oscillaton we have to solve the vacuum Einstein equations in the presence of a $\Lambda$ cosmological constant. From the viewpoint of direct physical applications the study of a positive cosmological constant is more relevant, since $\Lambda>0$ may generate the current accelerating expansion of the universe, and may also describe the inflational expansion of the early universe.

For a positive cosmological constant, at large distances from the oscillaton the spacetime must tend to the de Sitter metric, which in static Schwarzschild coordinates can be written as
\begin{equation}
 \mathrm{d}s^2=-\left(1-H^2\bar r^2\right)\mathrm{d}t^2
 +\frac{\mathrm{d}\bar r^2}{1-H^2\bar r^2}
 +\bar r^2\mathrm{d}\Omega^2_{d-1} \ .
\end{equation}
The relation between the Hubble constant $H$ and the cosmological constant $\Lambda$ in case of $d$ spatial dimensions is
\begin{equation}
 H^2=\frac{2\Lambda}{d(d-1)} \ . \label{caphlambda}
\end{equation}
As in the earlier sections, for our calculations we use spatially conformally flat coordinates, in which case the form of the de Sitter metric is
\begin{equation}
 \mathrm{d}s^2=-\frac{\left(1-\frac{1}{4}H^2r^2\right)^2}
 {\left(1+\frac{1}{4}H^2r^2\right)^2}\,\mathrm{d}t^2
 +\frac{1}
 {\left(1+\frac{1}{4}H^2r^2\right)^2}
 \left(\mathrm{d}r^2+r^2\mathrm{d}\Omega^2_{d-1}\right) \ .     \label{desittermet}
\end{equation}
The relation between the two kind of radial coordinates is $r=\bar r\left(1+\frac{1}{4}H^2r^2\right)$. The cosmological horizon, which in Schwarzschild coordinates is at the place $\bar r=1/H$, in spatially conformally flat coordinates is located at $r_h=2/H$.

The cosmological constant can also be considered as a perfect fluid with energy density and pressure given by
\begin{equation}
 \mu_\Lambda=\frac{\Lambda}{8\pi} \quad , \qquad
 p_\Lambda=-\frac{\Lambda}{8\pi} \ .   \label{mulambda}
\end{equation}
Since negative pressure implies an accelerated expansion in the universe, $\Lambda>0$ can be considered as an effective repulsing force, which makes the radiation of oscillatons larger compared to the asymptotically flat case. In the following subsections this extra radiation is described, based on the results in our paper \cite{fodor2010b}.

\subsection{Small-amplitude expansion}

For self-interacting scalar field oscillons on a fixed de Sitter background the magnitude of the radiation induced by a positive $\Lambda$ has been determined in article \cite{Farhi08}, by applying an expansion in terms of an amplitude parameter. In the following, we will present a similar method for the radiation of oscillatons formed by a scalar field interacting with gravitation. The method is based on a generalization for $\Lambda>0$ of the $\varepsilon$ expansion presented earlier in Section \ref{secosctnsmallampl}. It would be possible to set the mass of the scalar field to the value $m=1$ by using the scaling freedom \eqref{scaleprop} in this case as well. However, since the the relative magnitude of the scalar mass with respect to the cosmological constant is an important quantity for the interpretation of the results, we keep the value of $m$ arbitrary now.

We denote the small parameter describing the amplitude by $\varepsilon$. We can still assume that the size of the oscillaton increases with the decrease of the central amplitude, hence similarly to \eqref{eqrhoepsr} we introduce the rescaled radial coordinate $\rho$\,,
\begin{equation}
 \rho=\varepsilon m r \ . \label{eqrhoepmrlamb}
\end{equation}
The inclusion of the factor $m$ is motivated by the scaling freedom \eqref{scaleprop}.

Long-lived localized solutions are only expected to exist if their size is much smaller than the size of the cosmological horizon. Hence we introduce the rescaled Hubble constant $h$ by the relation
\begin{equation}
 H=\varepsilon^2 m h \ .  \label{caphlowh}
\end{equation}
We assume that the value of $h$ is appropriately small, i.e.~$h\ll 1$ even for the small $\varepsilon$ values used for the description of the configuration. This way we ensure that the typical size of the oscillatons, which is of order $1/(\varepsilon m)$ using the $r$ coordinate, is smaller than the radius of the cosmological horizon, $r_h=2/H=2/(\varepsilon^2 m h)$.

As we will see soon, the energy density $\mu$ of the scalar field in the central domain is proportional to $\varepsilon^4$. On the other hand, the energy density $\mu_\Lambda=\Lambda/(8\pi)$ associated to the cosmological constant, according to \eqref{caphlambda} and \eqref{caphlowh} can be written as
\begin{equation}
 \mu_\Lambda=\frac{d(d-1)}{16\pi}\varepsilon^4m^2h^2 \ . \label{mulambda2}
\end{equation}
The smallness of $h$ guaranties that $\mu_\Lambda$ remains smaller than the energy density $\mu$ of the scalar field, even in the small-amplitude limit.

We are looking for spatially localized solutions, for which if the scalar field $\phi$ becomes small then the spacetime approaches the de Sitter metric. Similarly to equations \eqref{eq:phiexp}-\eqref{eq:bexp}, we expand the scalar field and the metric functions in powers of $\varepsilon$,
\begin{align}
 \bar\phi&=\sum_{k=1}^\infty\epsilon^{2k}\phi_{2k} \ ,\label{eq:phiexp2}\\
 A&=\frac{\left(1-\frac{1}{4}\varepsilon^2h^2\rho^2\right)^2}
 {\left(1+\frac{1}{4}\varepsilon^2h^2\rho^2\right)^2}
 +\sum_{k=1}^\infty\epsilon^{2k}A_{2k} \ , \label{eq:aexp2}\\
 B&=\frac{1}{\left(1+\frac{1}{4}\varepsilon^2h^2\rho^2\right)^2}
 +\sum_{k=1}^\infty\epsilon^{2k}B_{2k} \ . \label{eq:bexp2}
\end{align}
The first terms of $A$ and $B$ correspond to the de Sitter spacetime according to \eqref{desittermet}. Since we intend to use asymptotically de Sitter coordinates, we look for such $\phi_{2k}$, $A_{2k}$ and $B_{2k}$ that tend to zero when $\rho$ goes to infinity.

Since the frequency depends on the amplitude, we introduce a rescaled time coordinate $\tau$ by
\begin{equation}
 \tau=\omega t \ ,
\end{equation}
and similarly to \eqref{eqom2expgr} we expand the square of the factor $\omega$ in powers of  $\varepsilon$,
\begin{equation}
 \omega^2=m^2\left(1+\sum_{k=1}^\infty\varepsilon^{2k}
 \omega_{2k} \right) \  .
\end{equation}

The field equations that we have to solve are still the Einstein equations \eqref{eq:eieq1}-\eqref{eq:eieq4} and the wave equation \eqref{eq:wave3}. The details of the calculations are so similar to those given in Subsection \ref{subsecvezeredm}, that we only give the results here. The $\varepsilon^2$ order expansion coefficients have the same form as in the $\Lambda=0$ case,
\begin{equation}
 \phi_2=p_2\cos\tau \quad , \qquad B_2=-\,\frac{A_2}{d-2} \ , \label{eqphi2p2b2a2}
\end{equation}
where $p_2$, $A_2$ and $B_2$ only depend on the radial coordinate $r$. In the same way as in  \eqref{eq:sands}, we write the solution in terms of the radial functions $s$ and $S$,
\begin{equation}
 A_2=\omega_2-s \quad , \qquad p_2=S\sqrt{\frac{d-1}{d-2}} \ . \label{eqa2om2p2s}
\end{equation}
The functions $S$ and $s$ satisfy the Schr\"odinger-Newton equations, which compared to \eqref{eqsn1}-\eqref{eqsn2} now also contain a cosmological term proportional to $h^2$,
\begin{align}
 &\frac{\mathrm{d}^2S}{\mathrm{d}\rho^2}
 +\frac{d-1}{\rho}\,\frac{\mathrm{d} S}{\mathrm{d}\rho}
 +(s+h^2\rho^2) S=0
 \ , \label{eqsn1cosm}\\
 &\frac{\mathrm{d}^2s}{\mathrm{d}\rho^2}
 +\frac{d-1}{\rho}\,\frac{\mathrm{d} s}{\mathrm{d}\rho}
 +S^2=0
 \ . \label{eqsn2cosm}
\end{align}
If $S(\rho)$ and $s(\rho)$ are solutions, then the transformed functions
\begin{equation}
\tilde S(\rho)=\lambda^2S(\lambda\rho) \ ,\quad
\tilde s(\rho)=\lambda^2s(\lambda\rho)  \label{eq:sntr}
\end{equation}
also solve the Schr\"odinger-Newton equations, with $\tilde h=\lambda^2h$.

If the scalar field, and consequently also $S$, tend to zero for $\rho\to\infty$, then based on \eqref{eqsn2cosm} the asymptotic behavior of $s$ is
\begin{equation}
 s=s_0+s_1\rho^{2-d} \ .  \label{sasympt}
\end{equation}
Since we are looking for solutions for which $A_2=\omega_2-s$ also tends to zero for large $\rho$, necessarily $\omega_2=s_0$ must hold. In order to make the solution of the Schr\"odinger-Newton equations \eqref{eqsn1cosm}-\eqref{eqsn2cosm} unique, we require that
\begin{equation}
 \omega_2=s_0=-1 \ . \label{eqom2s0m1}
\end{equation}
Setting $\omega_k=0$ for $k\geq3$, we fix the relation between the frequency $\omega$ and the parameter $\varepsilon$,
\begin{equation}
 \omega=m\sqrt{1-\varepsilon^2} \ . \label{omegaepsilon}
\end{equation}

Substituting the expression \eqref{sasympt} into \eqref{eqsn1cosm}, we obtain the equation determining the asymptotic behavior of $S$,
\begin{equation}
 \frac{\mathrm{d}^2S}{\mathrm{d}\rho^2}
 +\frac{d-1}{\rho}\,\frac{\mathrm{d} S}{\mathrm{d}\rho}
 +(h^2\rho^2-1+s_1\rho^{2-d}) S=0 \ . \label{eqswkb1}
\end{equation}

Substituting the small-amplitude expansion into the the energy density of the scalar field given in \eqref{eq:massen}, the leading order in $\varepsilon$ result is
\begin{equation}
 \mu=\frac{1}{16\pi}\varepsilon^4\frac{d-1}{d-2}m^2S^2 \ . \label{massenlead}
\end{equation}
Denoting the central value with an index $c$, and comparing with expression \eqref{mulambda2}, we obtain the ratio of the energy densities represented by the cosmological constant and the scalar field,
\begin{equation}
 \frac{\mu_\Lambda}{\mu_c}=\frac{d(d-2)}{S_c^2}h^2 \ . \label{mumulambda}
\end{equation}
This expression gives an important physical meaning to the rescaled Hubble constant $h$. As we will see, $h$ is the essential parameter that determines the radiation of oscillatons for $\Lambda>0$.

If $h$ is small, then in the core region the shape of the oscillatons agree very well with the profile already determined in the $\Lambda=0$ case. Substituting the central value of $S$ given in Table \ref{c1table} into the expression \eqref{massenlead} of the density, for $d=3$ spatial dimensions the central density in SI units can be written as
\begin{equation}
 \mu_c\left[\frac{\mathrm{kg}}{\mathrm{m}^3}\right]=\varepsilon^4\left(m\left[\frac{\mathrm{eV}}{c^2}\right]\right)^2\,
 1.436\times 10^{39} \ . \label{massendens}
\end{equation}
Even for very small $\varepsilon$ parameter values, this is still much larger than the the energy density corresponding to the present value of the cosmological constant, which is approximately $\mu_\Lambda\left[\mathrm{kg}/\mathrm{m}^3\right]=6.8\times 10^{-24}$ (see the Appendix D of  \cite{fodor2010b}).

\subsection{Magnitude of the radiation}

For the zero cosmological constant case in Subsection \ref{sec:SN} we have shown, that if the number of spatial dimensions is $3\leq d\leq 5$, then there exists a unique localized nodeless solution of the Schr\"odinger-Newton equations, for which the function $s$ tends to the value $s_0=-1$. For $\Lambda=0$ the function $S$ tends to zero exponentially, and its behavior for large $\rho$ is given in \eqref{eqssasympt}, where the numerical value of the constant $s_1$ can be found in Table \ref{c1table}, and $S_t$ in Table \ref{sttable}.

Even if $h$ is nonzero, we assume that it is small enough such that there exists a region in $\rho$ which is outside enough of the core domain, but where the influence of the cosmological constant is still negligible. In this region, to a good approximation, $S$ satisfies equation \eqref{eqswkb1} with the substitution of $h=0$. Consequently, in this intermediate region $S$ has the form given in \eqref{eqssasympt}, with the constants $s_1$ and $S_t$ given in the tables for the $\Lambda=0$ case.

If $h>0$, then for large enough distances $\rho\gg 1/h$, and then in equation \eqref{eqswkb1} the terms inside the brackets next to $h^2\rho^2$ are negligible,
\begin{equation}
 \frac{\mathrm{d}^2S}{\mathrm{d}\rho^2}
 +\frac{d-1}{\rho}\,\frac{\mathrm{d} S}{\mathrm{d}\rho}
 +h^2\rho^2 S=0 \ . \label{eqsheqstandw}
\end{equation}
The function $S$ has oscillating standing wave solutions in this domain, which for large $\rho$ can be written as
\begin{equation}
 S=\frac{\alpha}{\rho^{d/2}}
 \cos\left(\frac{h\rho^2}{2}+\beta\right) \ , \label{eqstail}
\end{equation}
where the constants $\alpha$ and $\beta$ give the amplitude and the phase. The amplitude $\alpha$ determines the mass loss rate of oscillatons resulting from $\Lambda>0$. In contrast to the zero cosmological constant case, for $\Lambda>0$ it is enough to calculate the leading $\varepsilon$ order of the small-amplitude expansion, since the tail already appears in the Schr\"odinger-Newton equations.

\begin{figure}[!ht]
    \begin{center}
    \includegraphics[width=10.0cm]{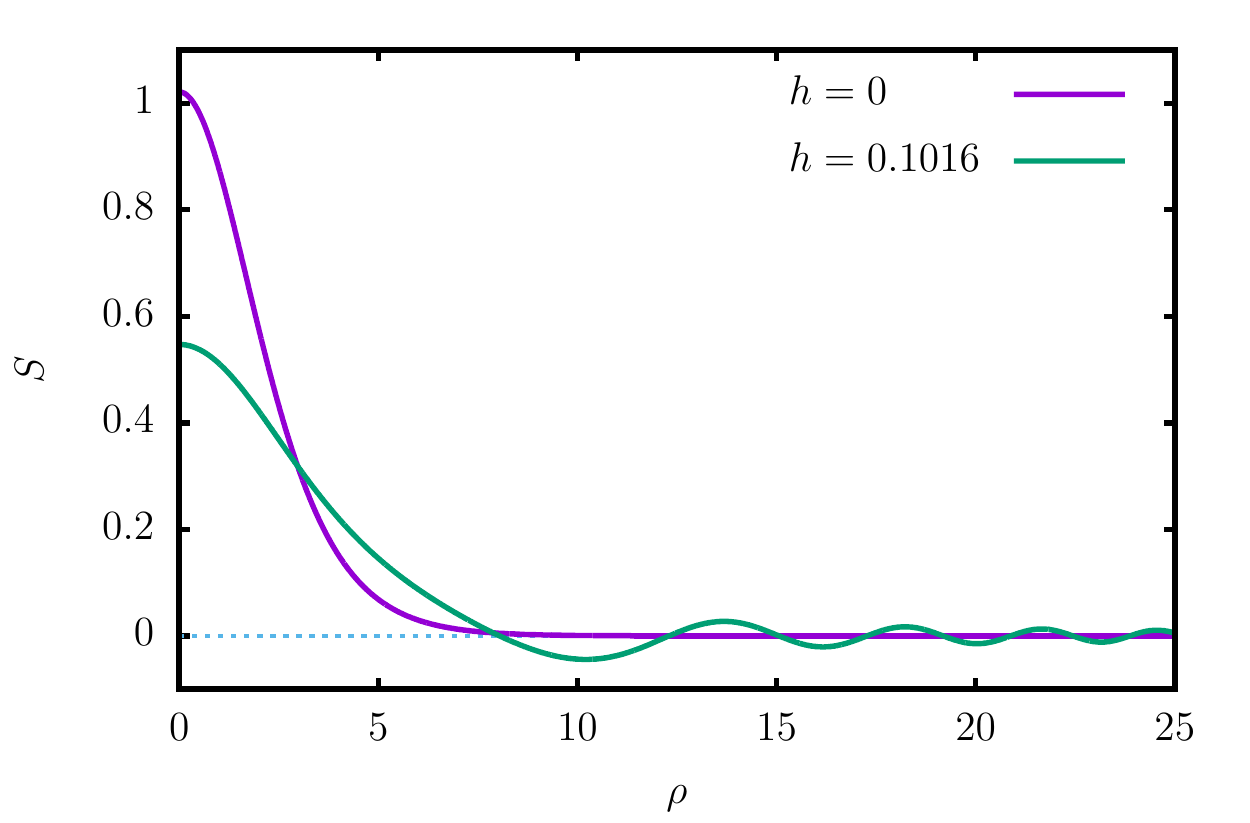}
    \end{center}
    \caption{Comparison of the radial dependence of the minimal tail $S$ function for $h=0.1016$ to the truly localized $S$ function in the $h=0$ case.
      \label{splot}}
\end{figure}

When solving the Schr\"odinger-Newton equations \eqref{eqsn1cosm}-\eqref{eqsn2cosm} numerically, the amplitude $\alpha$ of the standing wave tail depends on the values of the functions $s$ and $S$ in the center $\rho=0$. We intend to determine the minimal amplitude $\alpha=\alpha_{\mathrm{min}}$, with the requirement that for $\rho\to\infty$ the function $s$ tends to $-1$. We are interested in those solutions which have no nodes (zero crossings) in the core domain, where the influence of $h$ is negligible. If $h>0$, then necessarily $\alpha_{\mathrm{min}}$ is nonzero. In order to show the influence of the positive cosmological constant on the function $S$, on Figure \ref{splot} we give the function $S$ for a relatively large positive $h$ value, and also for the case $h=0$.

If $h$ is small, we can calculate the amplitude using WKB approximation. The details of the lengthy but elementary computation can be found in our paper \cite{fodor2010b}. The result of the calculation gives the dependence of the minimal amplitude on $h$, in terms of the constants $s_1$ and $S_t$,
\begin{equation}
 \alpha_{\mathrm{min}}=\left\{
 \begin{array}{ll}
 \displaystyle{\frac{S_t}{\sqrt{h}}
 \left(\frac{2}{h}\right)^{\displaystyle s_1/2}
 \exp\left(-\frac{\pi}{4h}\right)}
 &\text{ if    }\  d=3 \ , \\[4mm]
\displaystyle{\frac{S_t}{\sqrt{h}}\exp\left(-\frac{\pi}{4h}\right)}
 &\text{ if    }\  d>3 \ .
 \end{array}
 \right.  \label{alphamin}
\end{equation}

\begin{figure}[!ht]
    \begin{center}
    \includegraphics[width=10.0cm]{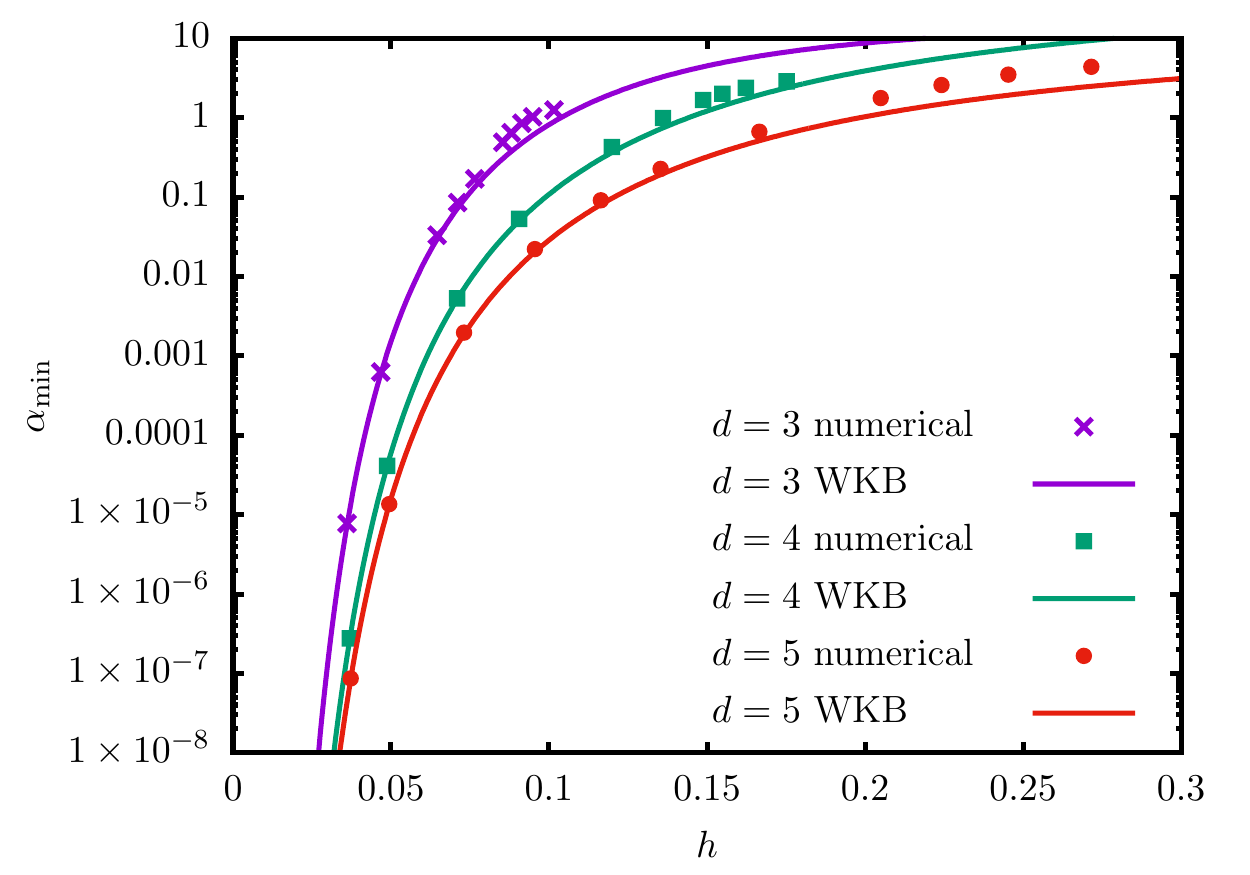}
    \end{center}
    \caption{The minimal tail amplitude of $S$ as a function of $h$.
      \label{figampl}}
\end{figure}

On Figure \ref{figampl} we show the values of the minimal amplitude $\alpha_{\mathrm{min}}$ obtained by the WKB approximation and also by a precise numerical method for $d=3,\,4,\,5$ spatial dimensions.

The coefficient $\alpha_{\mathrm{min}}$ gives the magnitude of the tail in the function $S$ of the solution of the Schr\"odinger-Newton equations. From this, we can calculate the tail of the scalar field using \eqref{eq:phiexp}, \eqref{eqphi2p2b2a2}, \eqref{eqa2om2p2s}, \eqref{eqstail} and \eqref{eqrhoepmrlamb}, obtaining
\begin{equation}
 \bar\phi=\varepsilon^2\phi_2=\varepsilon^2p_2\cos\tau
 =\varepsilon^2 S\sqrt{\frac{d-1}{d-2}}\cos\tau
 =\frac{\phi_A}{r^{d/2}}\cos(\omega t)
 \cos\left(\frac{h\varepsilon^2m^2r^2}{2}+\beta\right) \ , \label{phitosc}
\end{equation}
where $\omega=\sqrt{1-\varepsilon^2}$, and the amplitude is
\begin{equation}
 \phi_A=\frac{\varepsilon^{2-(d/2)}}{m^{d/2}}
 \sqrt{\frac{d-1}{d-2}}\,
 \alpha_{\mathrm{min}} \ . \label{phiaa}
\end{equation}

To this standing wave solution we can add a suppressed WKB term, which is proportional to $\sin(\omega t)\sin\left(h\varepsilon^2m^2r^2/2+\beta\right)$ at large distances, and decays exponentially when going inwards from the tail domain. Choosing the amplitude of the added term identical to the original one, we obtain the form of the outgoing wave responsible for the radiation of the oscillaton,
\begin{equation}
 \bar\phi=\frac{\phi_A}{r^{d/2}}
 \cos\left(\frac{h\varepsilon^2m^2r^2}{2}+\beta-\omega t\right) \ , \label{outwaosc}
\end{equation}
which is valid for $\rho\gg 1/h$, since we have obtained it by using \eqref{eqsheqstandw}.

For a general spherically symmetric spacetime the mass inside a given radius is given by the Misner-Sharp energy function $\hat m$ defined in \eqref{eq:massfunc}. For large distances $\hat m$ tends to the total mass $M$ of the oscillaton. In a general case, the time derivative of $\hat m$ can be written according to \eqref{eqhatmtgen}, in terms of the stress-energy tensor. The time derivative of $\hat m$ gives ${-1}$ times the mass loss rate $\mathrm{S}$. Even if $\rho\gg 1/h$ in the oscillating tail domain, if $\varepsilon$ is small enough then $\varepsilon h\rho\ll 1$, and according to \eqref{eq:aexp2}-\eqref{eq:bexp2} the functions $A$ and $B$ still remain very close to $1$. In this case, equation \eqref{eq:minkmt} presented for the flat background case can still be applied for the calculation of the time derivative of the mass. Substituting the outgoing wave form \eqref{outwaosc}, and averaging for one oscillation period, for the averaged mass loss rate we get the following result:
\begin{equation}
 \bar S=-\,\frac{\overline{\mathrm{d}M}}{\mathrm{d}t}=
 \frac{\pi^{d/2-1}h\varepsilon^2m^3\phi_A^2}
 {8\Gamma\left(\frac{d}{2}\right)} \ .
\end{equation}
Since we are looking for the leading order result in $\varepsilon$, in the calculation we have applied the substitution $\omega=m$.

Using equations \eqref{phiaa}, \eqref{alphamin}, and applying the substitution $h=H/(m\varepsilon^2)$, we can obtain that for $d>3$ spatial dimensions the mass loss rate is
\begin{equation}
 \bar S=\frac{\pi^{d/2-1}(d-1)S_t^2\varepsilon^{6-d}}
 {8m^{d-3}(d-2)\Gamma\left(\frac{d}{2}\right)}
 \exp\left(-\frac{\pi m\varepsilon^2}{2H}\right)
 \qquad \text{if} \qquad d>3 \ .
\end{equation}
If $d=3$ then there is a further factor containing the constant $s_1$ as an exponent,
\begin{equation}
 \bar S=\frac{1}{2}S_t^2\varepsilon^3
 \left(\frac{2m\varepsilon^2}{H}\right)^{s_1}
 \exp\left(-\frac{\pi m\varepsilon^2}{2H}\right)
 \qquad \text{if} \qquad d=3 \ .  \label{masslossd3}
\end{equation}

The mass of the oscillatons can be calculated in the same way as in Subsection \ref{secosctnmass} for a nonzero cosmological constant as well, and to leading order the result is still given by \eqref{eq:totmass}-\eqref{eqmm1totm},
\begin{equation}
 M=\varepsilon^{4-d}\frac{(d-1)\pi^{d/2}}
 {8\pi m^{d-2}\Gamma\left(\frac{d}{2}\right)}s_1 \ . \label{meps}
\end{equation}
For $d=3$ spatial dimensions,
\begin{equation}
 M=\varepsilon \frac{s_1 }{2m}  \ . \label{meps2}
\end{equation}
Substituting, we can obtain the time averaged mass loss rate as a function of the total mass $M$ of the oscillaton. The result for $d=3$ spatial dimensions is
\begin{equation}
 \bar S=
 -\frac{\overline{\mathrm{d}M}}{\mathrm{d}t}=4S_t^2\frac{m^3M^3}{s_1^3}
 \left(\frac{8m^3M^2}{Hs_1^2}\right)^{s_1}
 \exp\left(-\frac{2\pi m^3M^2}{Hs_1^2}\right) .
\end{equation}
For specific $m$ and $H$ the time dependence of the mass can be obtained by the numerical integration of this expression.

The Hubble time $T_H=1/H$ characterizes the time scale of the expansion of the universe. The relative mass loss rate extrapolated for a period corresponding to the Hubble time is
\begin{equation}
 \frac{T_H}{M}\frac{\overline{\mathrm{d}M}}{\mathrm{d}t}=\left\{
 \begin{array}{ll}
 \displaystyle{
 -\frac{S_t^2}{s_1h}
 \left(\frac{2}{h}\right)^{s_1}
 \exp\left(-\frac{\pi}{2h}\right)
 }
 &\text{ if    }\  d=3 \ , \\[4mm]
 \displaystyle{
 -\frac{S_t^2}{(d-2)s_1h}
 \exp\left(-\frac{\pi}{2h}\right)
 }
 &\text{ if    }\  d>3 \ .
 \end{array}
 \right.
\end{equation}
The result only depends on the rescaled cosmological constant $h=H/(m\varepsilon^2)$. 

\begin{figure}[!ht]
    \begin{center}
    \includegraphics[width=9.0cm]{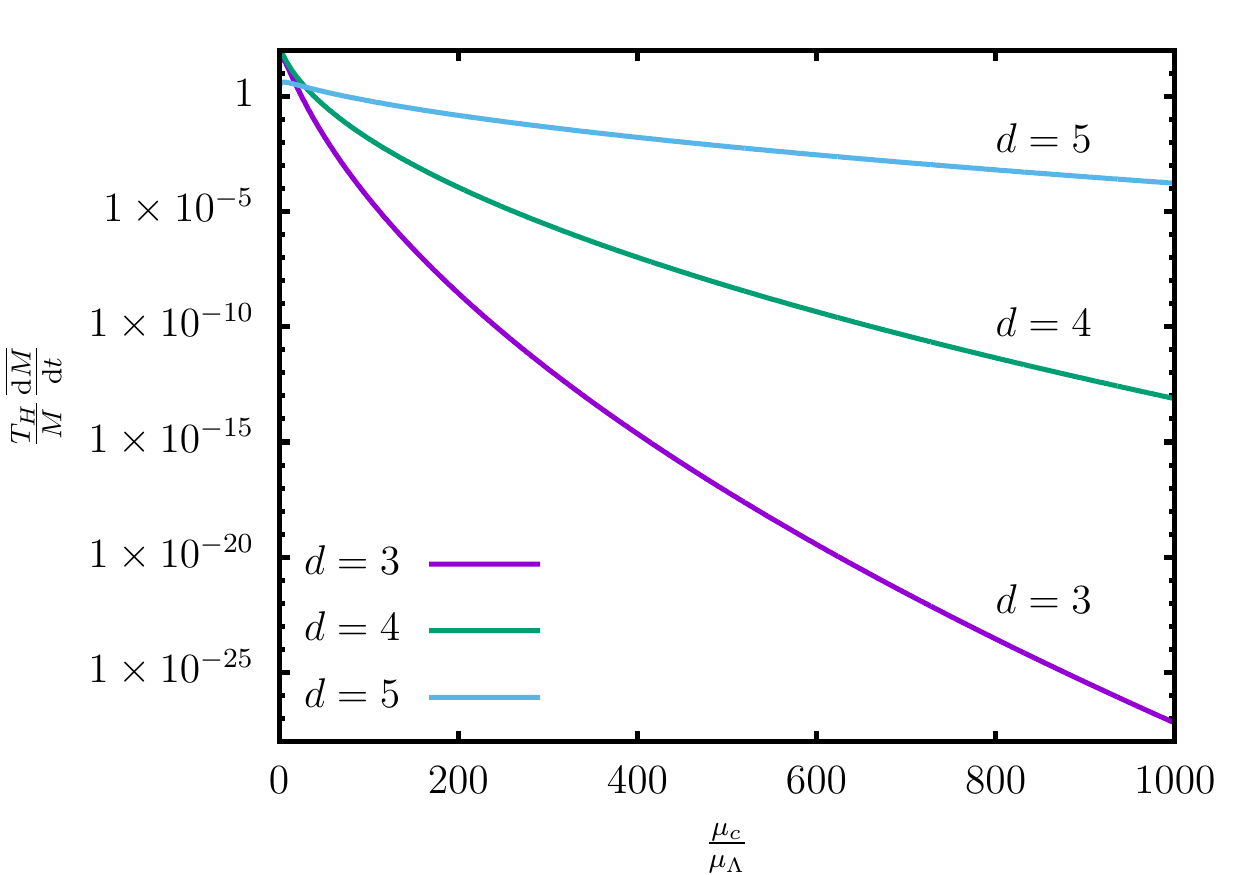}
    \end{center}
    \caption{The relative mass loss rate extrapolated for the Hubble time as a function of the ratio of the central density $\mu_c$ and the density $\mu_\Lambda$ belonging to the cosmological constant.
      \label{figmassl}}
\end{figure}

As we have seen in equation \eqref{mumulambda}, the quantity $h^2$ is proportional to the ratio of the central energy density $\mu_c$ of the scalar field and the energy density $\mu_\Lambda$ belonging to the cosmological constant. On Figure \ref{figmassl}
we show the mass loss rate extrapolated for the Hubble time as a function of the ratio $\frac{\mu_c}{\mu_\Lambda}$. As it can be seen from the figure as well, the radiation generated by the cosmological constant is only significant if the energy density of the oscillaton is not much larger than the energy density corresponding to the cosmological constant. This only happens for very small amplitude oscillatons, when the size of the oscillaton is already very large, approaching the order of the distance to the cosmological horizon. For close to maximal mass oscillatons, belonging to the amplitude parameter $\varepsilon\approx 0.5$, the radiation induced by $\Lambda>0$ is generally negligible. On the other hand, the radiation calculated in Sections \ref{secoscsugt} and \ref{seckvbrnum} for the $\Lambda=0$ case is the largest when the amplitude is close to maximal, and for smaller $\varepsilon$ values it decreases exponentially. If a large amplitude oscillaton is formed, then initially the radiation is described by equation \eqref{e:dmdtemp}, and much later, when because of the mass loss the size of the oscillaton becomes very large, the radiation given in \eqref{masslossd3} generated by $\Lambda>0$ becomes dominant. In the end, when the size approaches the size of the cosmological horizon, the oscillaton quickly decays.

\chapter{Summary and outlook}

Although static or exactly time-periodic localized regular solutions do not exist in real scalar theories, according to numerical simulations a large class of the possible initial data evolve into long-lived localized states. The frequency of these objects increases very slowly because of the weak energy loss resulting from the radiation of the scalar field. If the nonlinearity of the theory defined on a fixed background is provided by the self-interaction potential of the scalar field, then the formed object is generally called oscillon. The localized state formed from a self-gravitating scalar field in the framework of general relativity is known as oscillaton. In our papers on which this review is based, we have determined the shape and other properties of oscillons and oscillatons considerably more precisely than it was known earlier. Our numerical and analytical results consistently support each other. The extreme precision is mainly needed for the reliable and precise determination of the radiation loss of oscillons and oscillatons.

For the calculation of the outgoing radiation the necessary first step is the definition and study of the exactly time-periodic quasibreather state, which has a minimal amplitude standing wave tail. The periodicity in time makes it possible not only the application of precise spectral numerical codes, but also allows analytical methods by the extension of the equations into the complex plane. It can be shown relatively easily, that the radiation is exponentially small in terms of the parameter $\varepsilon$ determining the oscillation amplitude of the core region. However, the concrete determination of the factor before the exponential term is only possible by a rather complicated method consisting of several steps. In this review I tried to present the analytical method for the calculation of the radiation in a comprehensible and easily reproducible way. During our work it became apparent, that the analytical results based on the $\varepsilon$ expansion give precise results for small and intermediate amplitude states, while for the large amplitude states only the spectral numerical method is able to determine the radiation in a reliable way.

Because of the long lifetime of these configurations, various physical applications become feasible in cosmology and astrophysics. The radiation of $3+1$ dimensional oscillons is so small that several thousands of oscillation periods may happen before they move into the unstable domain and suddenly decay. The oscillatons coupled to gravity radiate so weakly, that even during a time period corresponding to the lifetime of the universe they do not lose more than half of their original mass. Oscillons and oscillatons can be expected to form in any theory containing several fields, if among them at least one real massive scalar field is included. But in case of appropriate nonlinearity, even the existence of the scalar is not necessarily needed, as it is shown by the existence of the oscillating localized states forming from self-gravitating Proca fields \cite{Garfinkle03,Brito16}.

It is important to clarify how a nonzero cosmological constant $\Lambda$ influences the formation, structure and radiation of localized states. A negative cosmological constant, acting as an effective attractive force, enhances the formation of oscillons and oscillatons, and decreases their radiation. The influence of a positive cosmological constant is just the opposite. The effect of a positive $\Lambda$ for oscillons has been investigated in papers \cite{Graham06,Farhi08,Gleiser10a}. For oscillatons in our paper \cite{fodor2010b}, and also in Section \ref{secosctnlambpos} of this review, we present in detail how a small $\Lambda>0$ induces a small mass loss rate. By a positive cosmological constant we can model the accelerated expansion of the current universe, or describe the early inflationary period.

The negative cosmological constant have indirect physical significance because of the AdS/CFT correspondence. Researchers began to study more extensively the behavior of self-gravitating scalar fields for the $\Lambda<0$ case after Bizo\'n and Rostworowski demonstrated by numerical methods in 2011 the instability of the anti-de Sitter spacetime \cite{Bizon11}. They have investigated the time evolution of spherically symmetric wave packets formed by a massless real Klein-Gordon field in general relativity, for $\Lambda<0$. According to their results, black holes can form from arbitrarily small amplitude initial data, by repeated bouncings of the wave packet through the center to infinity and back, becoming more and more concentrated in the meantime. However, not all type of initial data evolve into black holes.

If the cosmological constant is negative, there exist exactly time-periodic localized solutions, which do not radiate at all, and hence they can be considered as breathers \cite{Maliborski13}. For not too large perturbations, the deformed versions of these solutions remain near the time-periodic solution for all time, and no black hole formation occurs during their evolution. In this way, ``stability islands'' are formed around the breather solutions in the space of the possible initial data \cite{Dias12b,Choptuik18}. We have studied spherically symmetric scalar breather solutions on fixed AdS background in our paper \cite{Fodor14}, for various self-interaction potentials. Breather solutions with AdS asymptotics exist even for zero mass fields, in which case their typical size is determined by the cosmological constant. For negative $\Lambda$ we have studied spherically symmetric breather solutions formed by a self-gravitating real scalar field in our paper \cite{Fodor15}, by analytic and numerical methods, for $d$ spatial dimensions. We have used a method, which contrary to earlier work, can be applied for odd $d$ as well. For all $n\geq 0$ integers, a one-parameter family of breather solutions exists, where $n$ gives the number of nodes of the scalar field. For the small amplitude limit the oscillation frequency tends to the integer value $\omega=d+2n$.

The study of the scalar fields for $\Lambda<0$ in general relativity was mainly motivated by the fact that in this case, even if we are restricted to spherical symmetry, the system has a nontrivial dynamics, which makes it possible that energy is radiated to infinity. In this way, instead of the complicated theory of gravitational waves and their radiation, we have to deal with a technically much simpler but in many respects analogously behaving model. The formation of small  black holes in the scalar system makes it very likely that the system consisting of only vacuum gravitational waves is also unstable for the $\Lambda<0$ case. On the other hand, not all solutions evolve into black holes in this case either. When $\Lambda$ is negative, there exist exactly time-periodic, but not spherically symmetric, localized breather solutions of the vacuum Einstein equations \cite{Dias12a,Dias16,Rostworowski17,Horowitz15}. These solutions are not losing energy by radiation, and for large distances they approach the anti-de Sitter spacetime. The vacuum AdS breather solutions are generally known by the name AdS geons. The concept of geons was introduced by John Archibald Wheeler in 1955, for long-lived localized states formed by electromagnetic or gravitational waves in the asymptotically flat case \cite{Wheeler55,Brill64,Anderson97}. The structure of one-parameter families of AdS geons has been studied in our papers \cite{Martinon17} and \cite{Fodor17}, by spectral numeric and high order expansion methods. Among these solutions there are axially symmetric ones, rotating solutions with helical symmetry, and surprisingly, also non-symmetric ones with zero angular momentum as well.

\chapter*{Acknowledgements}
\addcontentsline{toc}{chapter}{Acknowledgements}

I would like to thank all my colleagues who have been co-authors in the papers \cite{Fodor2006,Fodor2008,Fodor2009a,Fodor2009b,fodor2010a,grandclement2011,fodor2010b,Fodor14,Fodor2009c,Fodor15} that we have published on the topic of this review.
I am especially grateful to Péter Forgács, who suggested me the topic of oscillons, oscillatons and similar geon-type configurations, and who has been working together with me enthusiastically for so many years since then. The only reason that this review has only one author is that it is the English translation of a sort of habilitation thesis wrote for the Hungarian Academy of Sciences.
I would like to offer my special thanks to Philippe Grandclément, who has made much effort to teach us how to apply spectral numerical methods, and who has calculated numerically all time-periodic quasibreather solutions presented in this review.
I am also grateful to István Rácz for encouraging me to learn numerical time-evolution methods. We have developed together a fourth-order method of lines code for general spherically symmetric systems, which employs spatial compactification to deal with the problem of the outer boundary.

\cleardoublepage
\phantomsection
\addcontentsline{toc}{chapter}{Bibliography}
\bibliographystyle{hhieeetr}
\bibliography{oscrev}

\begin{thebibliography}{100}

\bibitem{Derrick64}
G.~H. Derrick, \href {http://dx.doi.org/10.1063/1.1704233} {``Comments on
  nonlinear wave equations as models for elementary particles,''} {\em Journal
  of Mathematical Physics}, vol.~5, no.~9, pp.~1252--1254, 1964.

\bibitem{Kichenassamy91}
S.~Kichenassamy, \href {http://dx.doi.org/10.1002/cpa.3160440704} {``Breather
  solutions of the nonlinear wave equation,''} {\em Communications on Pure and
  Applied Mathematics}, vol.~44, no.~7, pp.~789--818, 1991.

\bibitem{Bogolubsky-77b}
I.~L. Bogolubsky, \href {http://dx.doi.org/10.1016/0375-9601(77)90138-4}
  {``Cascade evolution of spherically symmetric pulsons in multivacuum field
  theory models,''} {\em Physics Letters A}, vol.~61, no.~4, pp.~205 -- 206,
  1977.

\bibitem{bogolyubskii-77}
I.~L. Bogolyubskii and V.~G. Makhan'kov, \href
  {http://jetpletters.ac.ru/ps/1388/article_21071.shtml} {``Dynamics of
  spherically symmetrical pulsons of large amplitude,''} {\em JETP Letters},
  vol.~25, p.~107, 1977.

\bibitem{gleiser-94}
M.~Gleiser, \href {http://dx.doi.org/10.1103/PhysRevD.49.2978} {``Pseudostable
  bubbles,''} {\em Phys. Rev. D}, vol.~49, pp.~2978--2981, Mar 1994.

\bibitem{copeland-95}
E.~J. Copeland, M.~Gleiser, and H.-R. M\"uller, \href
  {http://dx.doi.org/10.1103/PhysRevD.52.1920} {``Oscillons: Resonant
  configurations during bubble collapse,''} {\em Phys. Rev. D}, vol.~52,
  pp.~1920--1933, Aug 1995.

\bibitem{kudryavtsev-75}
A.~E. Kudryavtsev, \href
  {http://www.jetpletters.ac.ru/ps/1522/article_23290.shtml} {``Solitonlike
  solutions for a {Higgs} scalar field,''} {\em JETP Letters}, vol.~22, p.~82,
  1975.

\bibitem{adib2002}
A.~B. Adib, M.~Gleiser, and C.~A.~S. Almeida, \href
  {http://dx.doi.org/10.1103/PhysRevD.66.085011} {``Long-lived oscillons from
  asymmetric bubbles: Existence and stability,''} {\em Phys. Rev. D}, vol.~66,
  p.~085011, Oct 2002.

\bibitem{Kolb94}
E.~W. Kolb and I.~I. Tkachev, \href
  {http://dx.doi.org/10.1103/PhysRevD.49.5040} {``Nonlinear axion dynamics and
  the formation of cosmological pseudosolitons,''} {\em Phys. Rev. D}, vol.~49,
  pp.~5040--5051, May 1994.

\bibitem{hindmarsh2006}
M.~Hindmarsh and P.~Salmi, \href {http://dx.doi.org/10.1103/PhysRevD.74.105005}
  {``Numerical investigations of oscillons in 2 dimensions,''} {\em Phys. Rev.
  D}, vol.~74, p.~105005, Nov 2006.

\bibitem{Amin10}
M.~A. Amin, R.~Easther, and H.~Finkel, \href
  {http://dx.doi.org/10.1088/1475-7516/2010/12/001} {``Inflaton fragmentation
  and oscillon formation in three dimensions,''} {\em Journal of Cosmology and
  Astroparticle Physics}, vol.~2010, pp.~001--001, dec 2010.

\bibitem{Zhou13}
S.-Y. Zhou, E.~J. Copeland, R.~Easther, H.~Finkel, Z.-G. Mou, and P.~M. Saffin,
  \href {http://dx.doi.org/10.1007/JHEP10(2013)026} {``Gravitational waves from
  oscillon preheating,''} {\em Journal of High Energy Physics}, vol.~2013,
  p.~26, Oct 2013.

\bibitem{Antusch17}
S.~Antusch, F.~Cefal\`a, and S.~Orani, \href
  {http://dx.doi.org/10.1103/PhysRevLett.118.011303} {``Gravitational waves
  from oscillons after inflation,''} {\em Phys. Rev. Lett.}, vol.~118,
  p.~011303, Jan 2017.

\bibitem{Antusch18a}
S.~Antusch, F.~Cefal{\`a}, S.~Krippendorf, F.~Muia, S.~Orani, and F.~Quevedo,
  \href {http://dx.doi.org/10.1007/JHEP01(2018)083} {``Oscillons from string
  moduli,''} {\em Journal of High Energy Physics}, vol.~2018, p.~83, Jan 2018.

\bibitem{LiuGuo18}
J.~Liu, Z.-K. Guo, R.-G. Cai, and G.~Shiu, \href
  {http://dx.doi.org/10.1103/PhysRevLett.120.031301} {``Gravitational waves
  from oscillons with cuspy potentials,''} {\em Phys. Rev. Lett.}, vol.~120,
  p.~031301, Jan 2018.

\bibitem{Antusch18b}
S.~Antusch, F.~Cefal{\`{a}}, and S.~Orani, \href
  {http://dx.doi.org/10.1088/1475-7516/2018/03/032} {``What can we learn from
  the stochastic gravitational wave background produced by oscillons?,''} {\em
  Journal of Cosmology and Astroparticle Physics}, vol.~2018, pp.~032--032, mar
  2018.

\bibitem{Amin18}
M.~A. Amin, J.~Braden, E.~J. Copeland, J.~T. Giblin, C.~Solorio, Z.~J. Weiner,
  and S.-Y. Zhou, \href {http://dx.doi.org/10.1103/PhysRevD.98.024040}
  {``Gravitational waves from asymmetric oscillon dynamics?,''} {\em Phys. Rev.
  D}, vol.~98, p.~024040, Jul 2018.

\bibitem{Kitajima18}
N.~Kitajima, J.~Soda, and Y.~Urakawa, \href
  {http://dx.doi.org/10.1088/1475-7516/2018/10/008} {``Gravitational wave
  forest from string axiverse,''} {\em Journal of Cosmology and Astroparticle
  Physics}, vol.~2018, pp.~008--008, oct 2018.

\bibitem{LiuGuo19}
J.~Liu, Z.-K. Guo, R.-G. Cai, and G.~Shiu, \href
  {http://dx.doi.org/10.1103/PhysRevD.99.103506} {``Gravitational wave
  production after inflation with cuspy potentials,''} {\em Phys. Rev. D},
  vol.~99, p.~103506, May 2019.

\bibitem{Sang19}
Y.~Sang and Q.-G. Huang, \href {http://dx.doi.org/10.1103/PhysRevD.100.063516}
  {``Stochastic gravitational-wave background from axion-monodromy oscillons in
  string theory during preheating,''} {\em Phys. Rev. D}, vol.~100, p.~063516,
  Sep 2019.

\bibitem{Lozanov19}
K.~D. Lozanov and M.~A. Amin, \href
  {http://dx.doi.org/10.1103/PhysRevD.99.123504} {``Gravitational perturbations
  from oscillons and transients after inflation,''} {\em Phys. Rev. D},
  vol.~99, p.~123504, Jun 2019.

\bibitem{Farhi05}
E.~Farhi, N.~Graham, V.~Khemani, R.~Markov, and R.~Rosales, \href
  {http://dx.doi.org/10.1103/PhysRevD.72.101701} {``An oscillon in the
  ${SU}(2)$ gauged {Higgs} model,''} {\em Phys. Rev. D}, vol.~72, p.~101701,
  Nov 2005.

\bibitem{Graham07a}
N.~Graham, \href {http://dx.doi.org/10.1103/PhysRevLett.98.101801} {``An
  electroweak oscillon,''} {\em Phys. Rev. Lett.}, vol.~98, p.~101801, Mar
  2007.

\bibitem{Graham07b}
N.~Graham, \href {http://dx.doi.org/10.1103/PhysRevD.76.085017} {``Numerical
  simulation of an electroweak oscillon,''} {\em Phys. Rev. D}, vol.~76,
  p.~085017, Oct 2007.

\bibitem{Arnold88}
P.~Arnold and L.~McLerran, \href {http://dx.doi.org/10.1103/PhysRevD.37.1020}
  {``The sphaleron strikes back: A response to objections to the sphaleron
  approximation,''} {\em Phys. Rev. D}, vol.~37, pp.~1020--1029, Feb 1988.

\bibitem{Rebbi96}
C.~Rebbi and R.~Singleton, \href {http://dx.doi.org/10.1103/PhysRevD.54.1020}
  {``Computational study of baryon number violation in high energy electroweak
  collisions,''} {\em Phys. Rev. D}, vol.~54, pp.~1020--1043, Jul 1996.

\bibitem{Gleiser07}
M.~Gleiser and J.~Thorarinson, \href
  {http://dx.doi.org/10.1103/PhysRevD.76.041701} {``Phase transition in {U}(1)
  configuration space: Oscillons as remnants of vortex-antivortex
  annihilation,''} {\em Phys. Rev. D}, vol.~76, p.~041701, Aug 2007.

\bibitem{Gleiser09}
M.~Gleiser and J.~Thorarinson, \href
  {http://dx.doi.org/10.1103/PhysRevD.79.025016} {``Class of nonperturbative
  configurations in {Abelian-Higgs} models: Complexity from dynamical symmetry
  breaking,''} {\em Phys. Rev. D}, vol.~79, p.~025016, Jan 2009.

\bibitem{Gleiser12}
M.~Gleiser and N.~Stamatopoulos, \href
  {http://dx.doi.org/10.1103/PhysRevD.86.045004} {``Information content of
  spontaneous symmetry breaking,''} {\em Phys. Rev. D}, vol.~86, p.~045004, Aug
  2012.

\bibitem{Achilleos13}
V.~Achilleos, F.~K. Diakonos, D.~J. Frantzeskakis, G.~C. Katsimiga, X.~N.
  Maintas, E.~Manousakis, C.~E. Tsagkarakis, and A.~Tsapalis, \href
  {http://dx.doi.org/10.1103/PhysRevD.88.045015} {``Oscillons and oscillating
  kinks in the {Abelian-Higgs} model,''} {\em Phys. Rev. D}, vol.~88,
  p.~045015, Aug 2013.

\bibitem{Diakonos15}
F.~K. Diakonos, G.~C. Katsimiga, X.~N. Maintas, and C.~E. Tsagkarakis, \href
  {http://dx.doi.org/10.1103/PhysRevE.91.023202} {``Symmetric solitonic
  excitations of the (1 + 1)-dimensional {Abelian-Higgs} classical vacuum,''}
  {\em Phys. Rev. E}, vol.~91, p.~023202, Feb 2015.

\bibitem{dashen-75}
R.~F. Dashen, B.~Hasslacher, and A.~Neveu, \href
  {http://dx.doi.org/10.1103/PhysRevD.11.3424} {``Particle spectrum in model
  field theories from semiclassical functional integral techniques,''} {\em
  Phys. Rev. D}, vol.~11, pp.~3424--3450, 1975.

\bibitem{Hertzberg10}
M.~P. Hertzberg, \href {http://dx.doi.org/10.1103/PhysRevD.82.045022}
  {``Quantum radiation of oscillons,''} {\em Phys. Rev. D}, vol.~82, p.~045022,
  Aug 2010.

\bibitem{Saffin14}
P.~M. Saffin, P.~Tognarelli, and A.~Tranberg, \href
  {http://dx.doi.org/10.1007/JHEP08(2014)125} {``Oscillon lifetime in the
  presence of quantum fluctuations,''} {\em Journal of High Energy Physics},
  vol.~2014, p.~125, Aug 2014.

\bibitem{Olle19}
J.~Oll\'e, O.~Pujol\`as, T.~Vachaspati, and G.~Zahariade, \href
  {http://dx.doi.org/10.1103/PhysRevD.100.045011} {``Quantum evaporation of
  classical breathers,''} {\em Phys. Rev. D}, vol.~100, p.~045011, Aug 2019.

\bibitem{Fodor2006}
G.~Fodor, P.~Forg\'acs, P.~Grandcl\'ement, and I.~R\'acz, \href
  {http://dx.doi.org/10.1103/PhysRevD.74.124003} {``Oscillons and
  quasibreathers in the ${\ensuremath{\phi}}^{4}$ {Klein-Gordon} model,''} {\em
  Phys. Rev. D}, vol.~74, p.~124003, Dec 2006.

\bibitem{Fodor2008}
G.~Fodor, P.~Forg\'acs, Z.~Horv\'ath, and A.~Luk\'acs, \href
  {http://dx.doi.org/10.1103/PhysRevD.78.025003} {``Small amplitude
  quasibreathers and oscillons,''} {\em Phys. Rev. D}, vol.~78, p.~025003, Jul
  2008.

\bibitem{Fodor2009a}
G.~Fodor, P.~Forg\'acs, Z.~Horv\'ath, and M.~Mezei, \href
  {http://dx.doi.org/10.1103/PhysRevD.79.065002} {``Computation of the
  radiation amplitude of oscillons,''} {\em Phys. Rev. D}, vol.~79, p.~065002,
  Mar 2009.

\bibitem{Fodor2009b}
G.~Fodor, P.~Forgács, Z.~Horváth, and M.~Mezei, \href
  {http://dx.doi.org/10.1016/j.physletb.2009.03.054} {``Radiation of scalar
  oscillons in 2 and 3 dimensions,''} {\em Physics Letters B}, vol.~674, no.~4,
  pp.~319 -- 324, 2009.

\bibitem{HondaChoptuik2002}
E.~P. Honda and M.~W. Choptuik, \href
  {http://dx.doi.org/10.1103/PhysRevD.65.084037} {``Fine structure of oscillons
  in the spherically symmetric ${\ensuremath{\varphi}}^{4}$ {Klein-Gordon}
  model,''} {\em Phys. Rev. D}, vol.~65, p.~084037, Apr 2002.

\bibitem{Kosevich75}
A.~M. Kosevich and A.~S. Kovalev, \href
  {http://www.jetp.ac.ru/cgi-bin/e/index/e/40/5/p891?a=list}
  {``Self-localization of vibrations in a one-dimensional anharmonic chain,''}
  {\em JETP}, vol.~40, p.~1973, 1975.

\bibitem{SegurKruskal87}
H.~Segur and M.~D. Kruskal, \href
  {http://dx.doi.org/10.1103/PhysRevLett.58.747} {``Nonexistence of
  small-amplitude breather solutions in $\phi^{4}$ theory,''} {\em Phys. Rev.
  Lett.}, vol.~58, pp.~747--750, Feb 1987.

\bibitem{Buslaev1977}
V.~S. Buslaev, \href {http://dx.doi.org/10.1007/BF01041234} {``Solutions of
  ``double soliton'' type for the multidimensional equation
  ${\ensuremath{\Box}} u={F}(u)$,''} {\em Theoretical and Mathematical
  Physics}, vol.~31, pp.~293--299, Apr 1977.

\bibitem{bogolyubskii-77c}
I.~L. Bogolyubskii, \href
  {http://www.jetpletters.ac.ru/ps/1817/article_27779.shtml} {``Oscillating
  particle-like solutions of the nonlinear {Klein-Gordon} equation,''} {\em
  JETP Letters}, vol.~24, p.~535, 1976.

\bibitem{seidel-91}
E.~Seidel and W.-M. Suen, \href {http://dx.doi.org/10.1103/PhysRevLett.66.1659}
  {``Oscillating soliton stars,''} {\em Phys. Rev. Lett.}, vol.~66,
  pp.~1659--1662, Apr 1991.

\bibitem{seidel-94}
E.~Seidel and W.-M. Suen, \href {http://dx.doi.org/10.1103/PhysRevLett.72.2516}
  {``Formation of solitonic stars through gravitational cooling,''} {\em Phys.
  Rev. Lett.}, vol.~72, pp.~2516--2519, Apr 1994.

\bibitem{PageDon04}
D.~N. Page, \href {http://dx.doi.org/10.1103/PhysRevD.70.023002} {``Classical
  and quantum decay of oscillations: Oscillating self-gravitating real scalar
  field solitons,''} {\em Phys. Rev. D}, vol.~70, p.~023002, Jul 2004.

\bibitem{fodor2010a}
G.~Fodor, P.~Forg\'acs, and M.~Mezei, \href
  {http://dx.doi.org/10.1103/PhysRevD.81.064029} {``Mass loss and longevity of
  gravitationally bound oscillating scalar lumps (oscillatons) in ${D}$
  dimensions,''} {\em Phys. Rev. D}, vol.~81, p.~064029, Mar 2010.

\bibitem{grandclement2011}
P.~Grandcl\'ement, G.~Fodor, and P.~Forg\'acs, \href
  {http://dx.doi.org/10.1103/PhysRevD.84.065037} {``Numerical simulation of
  oscillatons: Extracting the radiating tail,''} {\em Phys. Rev. D}, vol.~84,
  p.~065037, Sep 2011.

\bibitem{fodor2010b}
G.~Fodor, P.~Forg\'acs, and M.~Mezei, \href
  {http://dx.doi.org/10.1103/PhysRevD.82.044043} {``Boson stars and oscillatons
  in an inflationary universe,''} {\em Phys. Rev. D}, vol.~82, p.~044043, Aug
  2010.

\bibitem{Matos01}
T.~Matos and F.~S. Guzmán, \href
  {http://stacks.iop.org/0264-9381/18/i=23/a=303} {``On the spacetime of a
  galaxy,''} {\em Classical and Quantum Gravity}, vol.~18, no.~23, p.~5055,
  2001.

\bibitem{Alcubierre02}
M.~Alcubierre, F.~S. Guzmán, T.~Matos, D.~Núñez, L.~A. Ureña-López, and
  P.~Wiederhold, \href {http://stacks.iop.org/0264-9381/19/i=19/a=314}
  {``Galactic collapse of scalar field dark matter,''} {\em Classical and
  Quantum Gravity}, vol.~19, no.~19, p.~5017, 2002.

\bibitem{Susperregi03}
M.~Susperregi, \href {http://dx.doi.org/10.1103/PhysRevD.68.123509} {``Dark
  energy and dark matter from an inhomogeneous dilaton,''} {\em Phys. Rev. D},
  vol.~68, p.~123509, Dec 2003.

\bibitem{Malakolkalami16}
B.~Malakolkalami and A.~Mahmoodzadeh, \href
  {http://dx.doi.org/10.1103/PhysRevD.94.103505} {``Time-dependent scalar
  fields as candidates for dark matter,''} {\em Phys. Rev. D}, vol.~94,
  p.~103505, Nov 2016.

\bibitem{Tkachev86}
I.~I. Tkachev, \href {http://adsabs.harvard.edu/abs/1986SvAL...12..305T}
  {``Coherent scalar-field oscillations forming compact astrophysical
  object,''} {\em Soviet Astronomy Letters}, vol.~12, pp.~305--308, 1986.
\newblock Translation Pisma v Astronomicheskii Zhurnal, vol. 12, Sept. 1986, p.
  726-733.

\bibitem{Hogan88}
C.~Hogan and M.~Rees, \href {http://dx.doi.org/10.1016/0370-2693(88)91655-3}
  {``Axion miniclusters,''} {\em Physics Letters B}, vol.~205, no.~2, pp.~228
  -- 230, 1988.

\bibitem{Kolb93}
E.~W. Kolb and I.~I. Tkachev, \href
  {http://dx.doi.org/10.1103/PhysRevLett.71.3051} {``Axion miniclusters and
  {Bose} stars,''} {\em Phys. Rev. Lett.}, vol.~71, pp.~3051--3054, Nov 1993.

\bibitem{Arvanitaki10}
A.~Arvanitaki, S.~Dimopoulos, S.~Dubovsky, N.~Kaloper, and J.~March-Russell,
  \href {http://dx.doi.org/10.1103/PhysRevD.81.123530} {``String axiverse,''}
  {\em Phys. Rev. D}, vol.~81, p.~123530, Jun 2010.

\bibitem{Marsh16}
D.~J. Marsh, \href {http://dx.doi.org/10.1016/j.physrep.2016.06.005} {``Axion
  cosmology,''} {\em Physics Reports}, vol.~643, pp.~1 -- 79, 2016.

\bibitem{Marsh17}
D.~J.~E. Marsh, \href {http://arxiv.org/abs/1712.03018} {``Axions and {ALPs}: a
  very short introduction,''} 2017.
\newblock arXiv:1712.03018 [hep-ph].

\bibitem{Krippendorf18}
S.~Krippendorf, F.~Muia, and F.~Quevedo, \href
  {http://dx.doi.org/10.1007/JHEP08(2018)070} {``Moduli stars,''} {\em Journal
  of High Energy Physics}, vol.~2018, p.~70, Aug 2018.

\bibitem{Visinelli18}
L.~Visinelli, S.~Baum, J.~Redondo, K.~Freese, and F.~Wilczek, \href
  {http://dx.doi.org/10.1016/j.physletb.2017.12.010} {``Dilute and dense axion
  stars,''} {\em Physics Letters B}, vol.~777, pp.~64 -- 72, 2018.

\bibitem{Braaten18}
E.~Braaten and H.~Zhang, \href {http://arxiv.org/abs/1810.11473} {``Axion
  stars,''} 2018.
\newblock arXiv:1810.11473 [hep-ph].

\bibitem{Chavanis18}
P.-H. Chavanis, \href {http://dx.doi.org/10.1103/PhysRevD.98.023009} {``Phase
  transitions between dilute and dense axion stars,''} {\em Phys. Rev. D},
  vol.~98, p.~023009, Jul 2018.

\bibitem{Schive14}
H.-Y. Schive, T.~Chiueh, and T.~Broadhurst, \href
  {http://dx.doi.org/10.1038/NPHYS2996} {``Cosmic structure as the quantum
  interference of a coherent dark wave,''} {\em Nature Physics}, vol.~10,
  p.~496, 2014.
\newblock See supplementary infomation also.

\bibitem{Veltmaat16}
J.~Veltmaat and J.~C. Niemeyer, \href
  {http://dx.doi.org/10.1103/PhysRevD.94.123523} {``Cosmological
  particle-in-cell simulations with ultralight axion dark matter,''} {\em Phys.
  Rev. D}, vol.~94, p.~123523, Dec 2016.

\bibitem{HuAtal00}
W.~Hu, R.~Barkana, and A.~Gruzinov, \href
  {http://dx.doi.org/10.1103/PhysRevLett.85.1158} {``Fuzzy cold dark matter:
  The wave properties of ultralight particles,''} {\em Phys. Rev. Lett.},
  vol.~85, pp.~1158--1161, Aug 2000.

\bibitem{Marsh15}
D.~J.~E. Marsh and A.-R. Pop, \href {http://dx.doi.org/10.1093/mnras/stv1050}
  {``Axion dark matter, solitons and the cusp–core problem,''} {\em Monthly
  Notices of the Royal Astronomical Society}, vol.~451, no.~3, pp.~2479--2492,
  2015.

\bibitem{Schwabe16}
B.~Schwabe, J.~C. Niemeyer, and J.~F. Engels, \href
  {http://dx.doi.org/10.1103/PhysRevD.94.043513} {``Simulations of solitonic
  core mergers in ultralight axion dark matter cosmologies,''} {\em Phys. Rev.
  D}, vol.~94, p.~043513, Aug 2016.

\bibitem{HuiAtal17}
L.~Hui, J.~P. Ostriker, S.~Tremaine, and E.~Witten, \href
  {http://dx.doi.org/10.1103/PhysRevD.95.043541} {``Ultralight scalars as
  cosmological dark matter,''} {\em Phys. Rev. D}, vol.~95, p.~043541, Feb
  2017.

\bibitem{Brito15}
R.~Brito, V.~Cardoso, and H.~Okawa, \href
  {http://dx.doi.org/10.1103/PhysRevLett.115.111301} {``Accretion of dark
  matter by stars,''} {\em Phys. Rev. Lett.}, vol.~115, p.~111301, Sep 2015.

\bibitem{Brito16}
R.~Brito, V.~Cardoso, C.~F.~B. Macedo, H.~Okawa, and C.~Palenzuela, \href
  {http://dx.doi.org/10.1103/PhysRevD.93.044045} {``Interaction between bosonic
  dark matter and stars,''} {\em Phys. Rev. D}, vol.~93, p.~044045, Feb 2016.

\bibitem{Garfinkle03}
D.~Garfinkle, R.~Mann, and C.~Vuille, \href
  {http://dx.doi.org/10.1103/PhysRevD.68.064015} {``Critical collapse of a
  massive vector field,''} {\em Phys. Rev. D}, vol.~68, p.~064015, Sep 2003.

\bibitem{Fodor14}
G.~Fodor, P.~Forg\'acs, and P.~Grandcl\'ement, \href
  {http://dx.doi.org/10.1103/PhysRevD.89.065027} {``Scalar field breathers on
  anti--de {Sitter} background,''} {\em Phys. Rev. D}, vol.~89, p.~065027, Mar
  2014.

\bibitem{Avis78}
S.~J. Avis, C.~J. Isham, and D.~Storey, \href
  {http://dx.doi.org/10.1103/PhysRevD.18.3565} {``Quantum field theory in
  anti-de {Sitter} space-time,''} {\em Phys. Rev. D}, vol.~18, pp.~3565--3576,
  Nov 1978.

\bibitem{wald1984general}
R.~M. Wald, \href
  {http://press.uchicago.edu/ucp/books/book/chicago/G/bo5952261.html} {{\em
  General Relativity}}.
\newblock University of Chicago Press, 1984.

\bibitem{Eleonskii1984}
V.~M. Eleonskii, N.~E. Kulagin, N.~S. Novozhilova, and V.~P. Silin, \href
  {http://dx.doi.org/10.1007/BF01017891} {``Asymptotic expansions and
  qualitative analysis of finite-dimensional models in nonlinear field
  theory,''} {\em Theoretical and Mathematical Physics}, vol.~60, no.~3,
  pp.~896--902, 1984.

\bibitem{Eleonsky1991}
V.~Eleonsky, \href {http://dx.doi.org/10.1007/978-1-4757-0435-8_28} {{\em
  Problems of Existence of Nontopological Solitons (Breathers) for Nonlinear
  Klein-Gordon Equations}}, pp.~357--363.
\newblock Boston, MA: Springer US, 1991.

\bibitem{Vuillermot1987}
P.~A. Vuillermot, \href {http://dx.doi.org/10.1007/BF02564463} {``Nonexistence
  of spatially localized free vibrations for a class of nonlinear wave
  equations,''} {\em Commentarii Mathematici Helvetici}, vol.~62, no.~1,
  pp.~573--586, 1987.

\bibitem{Hormuzdiar99b}
J.~N. Hormuzdiar and S.~D.~H. Hsu, \href {http://arxiv.org/abs/hep-th/9906058}
  {``On spherically symmetric breathers in scalar theories,''} 1999.
\newblock arXiv:hep-th/9906058.

\bibitem{Arodz08}
H.~Arod\ifmmode~\acute{z}\else \'{z}\fi{}, P.~Klimas, and T.~Tyranowski, \href
  {http://dx.doi.org/10.1103/PhysRevD.77.047701} {``Compact oscillons in the
  signum-{G}ordon model,''} {\em Phys. Rev. D}, vol.~77, p.~047701, Feb 2008.

\bibitem{Arodz11}
H.~Arod\ifmmode~\acute{z}\else \'{z}\fi{} and Z.~\ifmmode \acute{S}\else
  \'{S}\fi{}wierczy\ifmmode~\acute{n}\else \'{n}\fi{}ski, \href
  {http://dx.doi.org/10.1103/PhysRevD.84.067701} {``Swaying oscillons in the
  signum-{G}ordon model,''} {\em Phys. Rev. D}, vol.~84, p.~067701, Sep 2011.

\bibitem{Swierczynski17}
Z.~Świerczyński, \href {http://dx.doi.org/10.1080/14029251.2016.1274112}
  {``On the oscillons in the signum-{G}ordon model,''} {\em Journal of
  Nonlinear Mathematical Physics}, vol.~24, no.~1, pp.~20--28, 2017.

\bibitem{Klimas18}
P.~Klimas, J.~S. Streibel, A.~Wereszczynski, and W.~J. Zakrzewski, \href
  {http://dx.doi.org/10.1007/JHEP04(2018)102} {``Oscillons in a perturbed
  signum-{G}ordon model,''} {\em Journal of High Energy Physics}, vol.~2018,
  p.~102, Apr 2018.

\bibitem{Maslov90}
E.~M. Maslov, \href {http://dx.doi.org/10.1016/0375-9601(90)90845-F}
  {``Pulsons, bubbles, and the corresponding nonlinear wave equations in n+1
  dimensions,''} {\em Physics Letters A}, vol.~151, no.~1, pp.~47 -- 51, 1990.

\bibitem{Koutvitsky05}
V.~Koutvitsky and E.~Maslov, \href
  {http://dx.doi.org/10.1016/j.physleta.2004.12.083} {``Parametric instability
  of the real scalar pulsons,''} {\em Physics Letters A}, vol.~336, no.~1,
  pp.~31 -- 36, 2005.

\bibitem{Koutvitsky06}
V.~A. Koutvitsky and E.~M. Maslov, \href {http://dx.doi.org/10.1063/1.2167918}
  {``Instability of coherent states of a real scalar field,''} {\em Journal of
  Mathematical Physics}, vol.~47, no.~2, p.~022302, 2006.

\bibitem{Bialynicki79}
I.~Bialynicki-Birula and J.~Mycielski, \href
  {http://dx.doi.org/10.1088/0031-8949/20/3-4/033} {``Gaussons: Solitons of the
  logarithmic {S}chrödinger equation,''} {\em Physica Scripta}, vol.~20,
  pp.~539--544, sep 1979.

\bibitem{Koutvitsky11}
V.~A. Koutvitsky and E.~M. Maslov, \href
  {http://dx.doi.org/10.1103/PhysRevD.83.124028} {``Gravipulsons,''} {\em Phys.
  Rev. D}, vol.~83, p.~124028, Jun 2011.

\bibitem{Makhankov-78}
V.~G. Makhankov, \href {http://dx.doi.org/10.1016/0370-1573(78)90074-1}
  {``Dynamics of classical solitons (in non-integrable systems),''} {\em
  Physics Reports}, vol.~35, no.~1, pp.~1 -- 128, 1978.

\bibitem{Makhankovbook}
V.~G. Makhankov, \href {http://dx.doi.org/10.1007/978-94-009-2217-4} {{\em
  Soliton Phenomenology}}.
\newblock Springer Netherlands, 1990.

\bibitem{geicke-83}
J.~Geicke, \href {http://dx.doi.org/10.1016/0370-2693(83)90158-2} {``How stable
  are pulsons in the $(\phi^4_-)_2$ field theory?,''} {\em Physics Letters B},
  vol.~133, no.~5, pp.~337 -- 340, 1983.

\bibitem{geicke-84}
J.~Geicke, \href {http://dx.doi.org/10.1088/0031-8949/29/5/003} {``Cylindrical
  pulsons in nonlinear relativistic wave equations,''} {\em Physica Scripta},
  vol.~29, no.~5, p.~431, 1984.

\bibitem{GleiserSornborger}
M.~Gleiser and A.~Sornborger, \href
  {http://dx.doi.org/10.1103/PhysRevE.62.1368} {``Long-lived localized field
  configurations in small lattices: Application to oscillons,''} {\em Phys.
  Rev. E}, vol.~62, pp.~1368--1374, Jul 2000.

\bibitem{christiansen97}
P.~L. Christiansen, N.~Grønbech-Jensen, P.~S. Lomdahl, and B.~A. Malomed,
  \href {http://stacks.iop.org/1402-4896/55/i=2/a=002} {``Oscillations of
  eccentric pulsons,''} {\em Physica Scripta}, vol.~55, no.~2, p.~131, 1997.

\bibitem{Minzoni2004167}
A.~A. Minzoni, N.~F. Smyth, and A.~L. Worthy, \href
  {http://dx.doi.org/10.1016/j.physd.2003.09.047} {``Evolution of
  two-dimensional standing and travelling breather solutions for the
  {Sine–Gordon} equation,''} {\em Physica D: Nonlinear Phenomena}, vol.~189,
  no.~3–4, pp.~167 -- 187, 2004.

\bibitem{christiansen81}
P.~L. Christiansen and P.~S. Lomdahl, \href
  {http://dx.doi.org/10.1016/0167-2789(81)90023-3} {``Numerical study of 2+1
  dimensional sine-{Gordon} solitons,''} {\em Physica D: Nonlinear Phenomena},
  vol.~2, no.~3, pp.~482 -- 494, 1981.

\bibitem{piette98}
B.~Piette and W.~J. Zakrzewski, \href
  {http://stacks.iop.org/0951-7715/11/i=4/a=020} {``Metastable stationary
  solutions of the radial d -dimensional sine-{Gordon} model,''} {\em
  Nonlinearity}, vol.~11, no.~4, p.~1103, 1998.

\bibitem{alfimov2000}
G.~L. Alfimov, W.~A.~B. Evans, and L.~Vázquez, \href
  {http://stacks.iop.org/0951-7715/13/i=5/a=313} {``On radial sine-{Gordon}
  breathers,''} {\em Nonlinearity}, vol.~13, no.~5, p.~1657, 2000.

\bibitem{Bratsos2007251}
A.~Bratsos, \href {http://dx.doi.org/10.1016/j.cam.2006.07.002} {``The solution
  of the two-dimensional sine-{Gordon} equation using the method of lines,''}
  {\em Journal of Computational and Applied Mathematics}, vol.~206, no.~1,
  pp.~251 -- 277, 2007.

\bibitem{Schiappacasse18}
E.~D. Schiappacasse and M.~P. Hertzberg, \href
  {http://dx.doi.org/10.1088/1475-7516/2018/01/037} {``Analysis of dark matter
  axion clumps with spherical symmetry,''} {\em Journal of Cosmology and
  Astroparticle Physics}, vol.~2018, pp.~037--037, jan 2018.

\bibitem{Vaquero19}
A.~Vaquero, J.~Redondo, and J.~Stadler, \href
  {http://dx.doi.org/10.1088/1475-7516/2019/04/012} {``Early seeds of axion
  miniclusters,''} {\em Journal of Cosmology and Astroparticle Physics},
  vol.~2019, pp.~012--012, apr 2019.

\bibitem{Eggemeier19}
B.~Eggemeier and J.~C. Niemeyer, \href
  {http://dx.doi.org/10.1103/PhysRevD.100.063528} {``Formation and mass growth
  of axion stars in axion miniclusters,''} {\em Phys. Rev. D}, vol.~100,
  p.~063528, Sep 2019.

\bibitem{Namjoo18}
M.~H. Namjoo, A.~H. Guth, and D.~I. Kaiser, \href
  {http://dx.doi.org/10.1103/PhysRevD.98.016011} {``Relativistic corrections to
  nonrelativistic effective field theories,''} {\em Phys. Rev. D}, vol.~98,
  p.~016011, Jul 2018.

\bibitem{Braaten18b}
E.~Braaten, A.~Mohapatra, and H.~Zhang, \href
  {http://dx.doi.org/10.1103/PhysRevD.98.096012} {``Classical nonrelativistic
  effective field theories for a real scalar field,''} {\em Phys. Rev. D},
  vol.~98, p.~096012, Nov 2018.

\bibitem{Rosen68}
G.~Rosen, \href {http://dx.doi.org/10.1063/1.1664693} {``Particlelike solutions
  to nonlinear complex scalar field theories with positive‐definite energy
  densities,''} {\em Journal of Mathematical Physics}, vol.~9, no.~7,
  pp.~996--998, 1968.

\bibitem{Coleman85}
S.~Coleman, \href
  {http://dx.doi.org/https://doi.org/10.1016/0550-3213(85)90286-X}
  {``Q-balls,''} {\em Nuclear Physics B}, vol.~262, no.~2, pp.~263 -- 283,
  1985.

\bibitem{Copeland14}
E.~J. Copeland, P.~M. Saffin, and S.-Y. Zhou, \href
  {http://dx.doi.org/10.1103/PhysRevLett.113.231603} {``Charge-swapping {Q}
  balls,''} {\em Phys. Rev. Lett.}, vol.~113, p.~231603, Dec 2014.

\bibitem{Shnir18}
Y.~M.~. Shnir, \href {http://dx.doi.org/10.1017/9781108555623} {{\em
  Topological and Non-Topological Solitons in Scalar Field Theories}}.
\newblock Cambridge University Press, 2018.

\bibitem{Nugaev19}
E.~Nugaev and A.~Shkerin, \href {http://arxiv.org/abs/arXiv:1905.05146}
  {``Review of non-topological solitons in theories with {U(1)}-symmetry,''}
  2019.
\newblock arXiv:1905.05146 [hep-th].

\bibitem{Kasuya03}
S.~Kasuya, M.~Kawasaki, and F.~Takahashi, \href
  {http://dx.doi.org/10.1016/S0370-2693(03)00344-7} {``I-balls,''} {\em Physics
  Letters B}, vol.~559, no.~3, pp.~99 -- 106, 2003.

\bibitem{Mukaida14}
K.~Mukaida and M.~Takimoto, \href
  {http://dx.doi.org/10.1088/1475-7516/2014/08/051} {``Correspondence of {I}-
  and {Q}-balls as non-relativistic condensates,''} {\em Journal of Cosmology
  and Astroparticle Physics}, vol.~2014, pp.~051--051, aug 2014.

\bibitem{Kawasaki15}
M.~Kawasaki, F.~Takahashi, and N.~Takeda, \href
  {http://dx.doi.org/10.1103/PhysRevD.92.105024} {``Adiabatic invariance of
  oscillons/{I}-balls,''} {\em Phys. Rev. D}, vol.~92, p.~105024, Nov 2015.

\bibitem{Kawasaki14}
M.~Kawasaki and M.~Yamada, \href
  {http://dx.doi.org/10.1088/1475-7516/2014/02/001} {``Decay rates of
  gaussian-type {I}-balls and bose-enhancement effects in 3+1 dimensions,''}
  {\em Journal of Cosmology and Astroparticle Physics}, vol.~2014,
  pp.~001--001, feb 2014.

\bibitem{Mukaida17}
K.~Mukaida, M.~Takimoto, and M.~Yamada, \href
  {http://dx.doi.org/10.1007/JHEP03(2017)122} {``On longevity of
  {I}-ball/oscillon,''} {\em Journal of High Energy Physics}, vol.~2017,
  p.~122, Mar 2017.

\bibitem{Ibe2019}
M.~Ibe, M.~Kawasaki, W.~Nakano, and E.~Sonomoto, \href
  {http://dx.doi.org/10.1007/JHEP04(2019)030} {``Decay of {I}-ball/oscillon in
  classical field theory,''} {\em Journal of High Energy Physics}, vol.~2019,
  p.~30, Apr 2019.

\bibitem{Eby19}
J.~Eby, K.~Mukaida, M.~Takimoto, L.~C.~R. Wijewardhana, and M.~Yamada, \href
  {http://dx.doi.org/10.1103/PhysRevD.99.123503} {``Classical nonrelativistic
  effective field theory and the role of gravitational interactions,''} {\em
  Phys. Rev. D}, vol.~99, p.~123503, Jun 2019.

\bibitem{Copeland08}
E.~J. Copeland, A.~Padilla, and P.~M. Saffin, \href
  {http://dx.doi.org/10.1088/1126-6708/2008/01/066} {``No resonant tunneling in
  standard scalar quantum field theory,''} {\em Journal of High Energy
  Physics}, vol.~2008, pp.~066--066, jan 2008.

\bibitem{Saffin08}
P.~M. Saffin, A.~Padilla, and E.~J. Copeland, \href
  {http://dx.doi.org/10.1088/1126-6708/2008/09/055} {``Transmission of an
  inhomogeneous state via resonant tunnelling,''} {\em Journal of High Energy
  Physics}, vol.~2008, pp.~055--055, sep 2008.

\bibitem{SalmiThesis2008}
P.~E. Salmi, ``Oscillons.''
  \url{https://openaccess.leidenuniv.nl/handle/1887/13117}, 2008.
\newblock PhD thesis, Lorenz-Institute, Faculty of Science, Leiden University,
  ISBN: 978-90-9023437-3.

\bibitem{SiciliaThesis2011}
D.~P. Sicilia, ``Analytical characterization of scalar-field oscillons in
  quartic potentials.'' \url{http://www.dartmouth.edu/~dpsicilia/Thesis.pdf},
  2011.
\newblock PhD thesis, Darthmouth College, Hanover, New Hampshire.

\bibitem{Sugiyama79}
T.~Sugiyama, \href {http://dx.doi.org/10.1143/PTP.61.1550} {``Kink-antikink
  collisions in the two-dimensional $\phi^4$ model,''} {\em Progress of
  Theoretical Physics}, vol.~61, no.~5, pp.~1550--1563, 1979.

\bibitem{Moshir81}
M.~Moshir, \href {http://dx.doi.org/10.1016/0550-3213(81)90320-5}
  {``Soliton-antisoliton scattering and capture in $\lambda\phi^4$ theory,''}
  {\em Nuclear Physics B}, vol.~185, no.~2, pp.~318 -- 332, 1981.

\bibitem{Campbell83}
D.~K. Campbell, J.~F. Schonfeld, and C.~A. Wingate, \href
  {http://dx.doi.org/10.1016/0167-2789(83)90289-0} {``Resonance structure in
  kink-antikink interactions in $\phi^4$ theory,''} {\em Physica D: Nonlinear
  Phenomena}, vol.~9, no.~1, pp.~1 -- 32, 1983.

\bibitem{Wingate83}
C.~A. Wingate, \href {http://dx.doi.org/10.1137/0143010} {``Numerical search
  for a $\phi^4$ breather mode,''} {\em SIAM Journal on Applied Mathematics},
  vol.~43, no.~1, pp.~120--140, 1983.

\bibitem{Romanczukiewicz10}
T.~Romanczukiewicz and Y.~Shnir, \href
  {http://dx.doi.org/10.1103/PhysRevLett.105.081601} {``Oscillon resonances and
  creation of kinks in particle collisions,''} {\em Phys. Rev. Lett.},
  vol.~105, p.~081601, Aug 2010.

\bibitem{Gleiser06}
M.~Gleiser and R.~C. Howell, \href
  {http://dx.doi.org/10.1103/PhysRevLett.94.151601} {``Resonant nucleation,''}
  {\em Phys. Rev. Lett.}, vol.~94, p.~151601, Apr 2005.

\bibitem{GleiserRogers08}
M.~Gleiser, B.~Rogers, and J.~Thorarinson, \href
  {http://dx.doi.org/10.1103/PhysRevD.77.023513} {``Bubbling the false vacuum
  away,''} {\em Phys. Rev. D}, vol.~77, p.~023513, Jan 2008.

\bibitem{Broadhead05}
M.~Broadhead and J.~McDonald, \href
  {http://dx.doi.org/10.1103/PhysRevD.72.043519} {``Simulations of the end of
  supersymmetric hybrid inflation and nontopological soliton formation,''} {\em
  Phys. Rev. D}, vol.~72, p.~043519, Aug 2005.

\bibitem{Gleiser07b}
M.~Gleiser, \href {http://dx.doi.org/10.1142/S0218271807009954} {``Oscillons in
  scalar field theories: Applications in higher dimensions and inflation,''}
  {\em International Journal of Modern Physics D}, vol.~16, no.~02n03,
  pp.~219--229, 2007.

\bibitem{Gleiser11}
M.~Gleiser, N.~Graham, and N.~Stamatopoulos, \href
  {http://dx.doi.org/10.1103/PhysRevD.83.096010} {``Generation of coherent
  structures after cosmic inflation,''} {\em Phys. Rev. D}, vol.~83, p.~096010,
  May 2011.

\bibitem{Amin12a}
M.~A. Amin, R.~Easther, H.~Finkel, R.~Flauger, and M.~P. Hertzberg, \href
  {http://dx.doi.org/10.1103/PhysRevLett.108.241302} {``Oscillons after
  inflation,''} {\em Phys. Rev. Lett.}, vol.~108, p.~241302, Jun 2012.

\bibitem{Amin12b}
M.~A. Amin, P.~Zukin, and E.~Bertschinger, \href
  {http://dx.doi.org/10.1103/PhysRevD.85.103510} {``Scale-dependent growth from
  a transition in dark energy dynamics,''} {\em Phys. Rev. D}, vol.~85,
  p.~103510, May 2012.

\bibitem{Sainio12}
J.~Sainio, \href {http://dx.doi.org/10.1088/1475-7516/2012/04/038} {``{PyCOOL}
  {\textemdash} a cosmological object-oriented lattice code written in
  python,''} {\em Journal of Cosmology and Astroparticle Physics}, vol.~2012,
  pp.~038--038, apr 2012.

\bibitem{Gleiser14}
M.~Gleiser and N.~Graham, \href {http://dx.doi.org/10.1103/PhysRevD.89.083502}
  {``Transition to order after hilltop inflation,''} {\em Phys. Rev. D},
  vol.~89, p.~083502, Mar 2014.

\bibitem{Adshead15}
P.~Adshead, J.~T.~G. Jr., T.~R. Scully, and E.~I. Sfakianakis, \href
  {http://dx.doi.org/10.1088/1475-7516/2015/12/034} {``Gauge-preheating and the
  end of axion inflation,''} {\em Journal of Cosmology and Astroparticle
  Physics}, vol.~2015, pp.~034--034, dec 2015.

\bibitem{Antusch16}
S.~Antusch and S.~Orani, \href
  {http://dx.doi.org/10.1088/1475-7516/2016/03/026} {``Impact of other scalar
  fields on oscillons after hilltop inflation,''} {\em Journal of Cosmology and
  Astroparticle Physics}, vol.~2016, pp.~026--026, mar 2016.

\bibitem{Lozanov14}
K.~D. Lozanov and M.~A. Amin, \href
  {http://dx.doi.org/10.1103/PhysRevD.90.083528} {``End of inflation,
  oscillons, and matter-antimatter asymmetry,''} {\em Phys. Rev. D}, vol.~90,
  p.~083528, Oct 2014.

\bibitem{Lozanov18}
K.~D. Lozanov and M.~A. Amin, \href
  {http://dx.doi.org/10.1103/PhysRevD.97.023533} {``Self-resonance after
  inflation: Oscillons, transients, and radiation domination,''} {\em Phys.
  Rev. D}, vol.~97, p.~023533, Jan 2018.

\bibitem{Hasegawa18}
F.~Hasegawa and J.-P. Hong, \href
  {http://dx.doi.org/10.1103/PhysRevD.97.083514} {``Inflaton fragmentation in
  {E} models of cosmological $\ensuremath{\alpha}$-attractors,''} {\em Phys.
  Rev. D}, vol.~97, p.~083514, Apr 2018.

\bibitem{Hong18}
J.-P. Hong, M.~Kawasaki, and M.~Yamazaki, \href
  {http://dx.doi.org/10.1103/PhysRevD.98.043531} {``Oscillons from pure natural
  inflation,''} {\em Phys. Rev. D}, vol.~98, p.~043531, Aug 2018.

\bibitem{Hindmarsh08}
M.~Hindmarsh and P.~Salmi, \href {http://dx.doi.org/10.1103/PhysRevD.77.105025}
  {``Oscillons and domain walls,''} {\em Phys. Rev. D}, vol.~77, p.~105025, May
  2008.

\bibitem{Braden15a}
J.~Braden, J.~R. Bond, and L.~Mersini-Houghton, \href
  {http://dx.doi.org/10.1088/1475-7516/2015/03/007} {``Cosmic bubble and domain
  wall instabilities {I}: parametric amplification of linear fluctuations,''}
  {\em Journal of Cosmology and Astroparticle Physics}, vol.~2015,
  pp.~007--007, mar 2015.

\bibitem{Braden15b}
J.~Braden, J.~R. Bond, and L.~Mersini-Houghton, \href
  {http://dx.doi.org/10.1088/1475-7516/2015/08/048} {``Cosmic bubble and domain
  wall instabilities {II}: fracturing of colliding walls,''} {\em Journal of
  Cosmology and Astroparticle Physics}, vol.~2015, pp.~048--048, aug 2015.

\bibitem{Bond15}
J.~R. Bond, J.~Braden, and L.~Mersini-Houghton, \href
  {http://dx.doi.org/10.1088/1475-7516/2015/09/004} {``Cosmic bubble and domain
  wall instabilities {III}: the role of oscillons in three-dimensional bubble
  collisions,''} {\em Journal of Cosmology and Astroparticle Physics},
  vol.~2015, pp.~004--004, sep 2015.

\bibitem{Dymnikova00}
I.~Dymnikova, L.~Koziel, M.~Khlopov, and S.~Rubin, \href
  {http://arxiv.org/abs/hep-th/0010120} {``Quasilumps from first order phase
  transitions,''} 2000.
\newblock arXiv:hep-th/0010120, Grav.Cosmol. 6 (2000) 311-318.

\bibitem{Johnson12}
M.~C. Johnson, H.~V. Peiris, and L.~Lehner, \href
  {http://dx.doi.org/10.1103/PhysRevD.85.083516} {``Determining the outcome of
  cosmic bubble collisions in full general relativity,''} {\em Phys. Rev. D},
  vol.~85, p.~083516, Apr 2012.

\bibitem{Mersini14}
L.~Mersini-Houghton, \href {http://dx.doi.org/10.1142/S0218271814500552}
  {``Antioscillons from bubble collisions at finite temperature,''} {\em
  International Journal of Modern Physics D}, vol.~23, no.~06, p.~1450055,
  2014.

\bibitem{Gleiser96}
M.~Gleiser and R.~M. Haas, \href {http://dx.doi.org/10.1103/PhysRevD.54.1626}
  {``Oscillons in a hot heat bath,''} {\em Phys. Rev. D}, vol.~54,
  pp.~1626--1632, Jul 1996.

\bibitem{Gleiser03}
M.~Gleiser and R.~C. Howell, \href
  {http://dx.doi.org/10.1103/PhysRevE.68.065203} {``Resonant emergence of
  global and local spatiotemporal order in a nonlinear field model,''} {\em
  Phys. Rev. E}, vol.~68, p.~065203, Dec 2003.

\bibitem{Amin14}
M.~A. Amin, I.~Banik, C.~Negreanu, and I.-S. Yang, \href
  {http://dx.doi.org/10.1103/PhysRevD.90.085024} {``Ultrarelativistic oscillon
  collisions,''} {\em Phys. Rev. D}, vol.~90, p.~085024, Oct 2014.

\bibitem{Hormuzdiar99a}
J.~N. Hormuzdiar and S.~D.~H. Hsu, \href
  {http://dx.doi.org/10.1103/PhysRevC.59.889} {``Pion breather states in
  {QCD},''} {\em Phys. Rev. C}, vol.~59, pp.~889--893, Feb 1999.

\bibitem{Sfakianakis12}
E.~I. Sfakianakis, \href {http://arxiv.org/abs/arXiv:1210.7568} {``Analysis of
  oscillons in the {SU(2)} gauged {Higgs} model,''} 2012.
\newblock arXiv:1210.7568 [hep-ph].

\bibitem{Correa18}
R.~Correa, L.~Ospedal, W.~de~Paula, and J.~Helayël-Neto, \href
  {http://dx.doi.org/10.1016/j.physletb.2018.02.067} {``Supersymmetry and
  fermionic modes in an oscillon background,''} {\em Physics Letters B},
  vol.~780, pp.~159 -- 165, 2018.

\bibitem{Charukhchyan14}
M.~V. Charukhchyan, E.~S. Sedov, S.~M. Arakelian, and A.~P. Alodjants, \href
  {http://dx.doi.org/10.1103/PhysRevA.89.063624} {``Spatially localized
  structures and oscillons in atomic {Bose-Einstein} condensates confined in
  optical lattices,''} {\em Phys. Rev. A}, vol.~89, p.~063624, Jun 2014.

\bibitem{SuAtal15}
S.-W. Su, S.-C. Gou, I.-K. Liu, A.~S. Bradley, O.~Fialko, and J.~Brand, \href
  {http://dx.doi.org/10.1103/PhysRevA.91.023631} {``Oscillons in coupled
  {Bose-Einstein} condensates,''} {\em Phys. Rev. A}, vol.~91, p.~023631, Feb
  2015.

\bibitem{Borsanyi09}
S.~Bors\'anyi and M.~Hindmarsh, \href
  {http://dx.doi.org/10.1103/PhysRevD.79.065010} {``Low-cost fermions in
  classical field simulations,''} {\em Phys. Rev. D}, vol.~79, p.~065010, Mar
  2009.

\bibitem{Saffin17}
P.~M. Saffin, \href {http://dx.doi.org/10.1007/JHEP07(2017)126}
  {``Recrudescence of massive fermion production by oscillons,''} {\em Journal
  of High Energy Physics}, vol.~2017, p.~126, Jul 2017.

\bibitem{Cotner18}
E.~Cotner, A.~Kusenko, and V.~Takhistov, \href
  {http://dx.doi.org/10.1103/PhysRevD.98.083513} {``Primordial black holes from
  inflaton fragmentation into oscillons,''} {\em Phys. Rev. D}, vol.~98,
  p.~083513, Oct 2018.

\bibitem{Graham06}
N.~Graham and N.~Stamatopoulos, \href
  {http://dx.doi.org/10.1016/j.physletb.2006.06.070} {``Unnatural oscillon
  lifetimes in an expanding background,''} {\em Physics Letters B}, vol.~639,
  no.~5, pp.~541 -- 545, 2006.

\bibitem{Farhi08}
E.~Farhi, N.~Graham, A.~H. Guth, N.~Iqbal, R.~R. Rosales, and N.~Stamatopoulos,
  \href {http://dx.doi.org/10.1103/PhysRevD.77.085019} {``Emergence of
  oscillons in an expanding background,''} {\em Phys. Rev. D}, vol.~77,
  p.~085019, Apr 2008.

\bibitem{Gleiser10a}
M.~Gleiser, N.~Graham, and N.~Stamatopoulos, \href
  {http://dx.doi.org/10.1103/PhysRevD.82.043517} {``Long-lived time-dependent
  remnants during cosmological symmetry breaking: From inflation to the
  electroweak scale,''} {\em Phys. Rev. D}, vol.~82, p.~043517, Aug 2010.

\bibitem{GleiserSicilia08}
M.~Gleiser and D.~Sicilia, \href
  {http://dx.doi.org/10.1103/PhysRevLett.101.011602} {``Analytical
  characterization of oscillon energy and lifetime,''} {\em Phys. Rev. Lett.},
  vol.~101, p.~011602, Jul 2008.

\bibitem{GleiserSicilia09}
M.~Gleiser and D.~Sicilia, \href {http://dx.doi.org/10.1103/PhysRevD.80.125037}
  {``General theory of oscillon dynamics,''} {\em Phys. Rev. D}, vol.~80,
  p.~125037, Dec 2009.

\bibitem{Andersen2012}
E.~A. Andersen and A.~Tranberg, \href
  {http://dx.doi.org/10.1007/JHEP12(2012)016} {``Four results on $\phi^4$
  oscillons in ${D}+1$ dimensions,''} {\em Journal of High Energy Physics},
  vol.~2012, p.~16, Dec 2012.

\bibitem{HondaThesis2000}
E.~Honda, \href {http://arxiv.org/abs/hep-ph/0009104} {{\em Resonant dynamics
  within the nonlinear Klein-Gordon Equation: Much ado about Oscillons}}.
\newblock PhD thesis, The University of Texas at Austin, 2000.
\newblock arXiv:hep-ph/0009104.

\bibitem{geicke-94}
J.~Geicke, \href {http://dx.doi.org/10.1103/PhysRevE.49.3539} {``Logarithmic
  decay of ${\mathrm{\ensuremath{\varphi}}}^{4}$ breathers of energy
  ${E}\ensuremath{\lesssim}1$,''} {\em Phys. Rev. E}, vol.~49, pp.~3539--3542,
  Apr 1994.

\bibitem{Gleiser18}
M.~Gleiser, M.~Stephens, and D.~Sowinski, \href
  {http://dx.doi.org/10.1103/PhysRevD.97.096007} {``Configurational entropy as
  a lifetime predictor and pattern discriminator for oscillons,''} {\em Phys.
  Rev. D}, vol.~97, p.~096007, May 2018.

\bibitem{Israeli81}
M.~Israeli and S.~A. Orszag, \href
  {http://dx.doi.org/10.1016/0021-9991(81)90082-6} {``Approximation of
  radiation boundary conditions,''} {\em Journal of Computational Physics},
  vol.~41, no.~1, pp.~115 -- 135, 1981.

\bibitem{ChoptuikThesis1986}
M.~W. Choptuik, {\em A Study of Numerical Techniques for Radiative Problems in
  General Relativity}.
\newblock PhD thesis, The Univeristy of British Columbia, 1986.
\newblock \url{https://open.library.ubc.ca/media/download/pdf/831/1.0085044/1}.

\bibitem{Marsa96}
R.~L. Marsa and M.~W. Choptuik, \href
  {http://dx.doi.org/10.1103/PhysRevD.54.4929} {``Black-hole--scalar-field
  interactions in spherical symmetry,''} {\em Phys. Rev. D}, vol.~54,
  pp.~4929--4943, Oct 1996.

\bibitem{Balakrishna98}
J.~Balakrishna, E.~Seidel, and W.-M. Suen, \href
  {http://dx.doi.org/10.1103/PhysRevD.58.104004} {``{Dynamical evolution of
  boson stars. II. Excited states and self-interacting fields},''} {\em Phys.
  Rev. D}, vol.~58, p.~104004, Sep 1998.

\bibitem{Honda10}
E.~Honda, \href {http://dx.doi.org/10.1103/PhysRevD.82.024038} {``Fractal
  boundary basins in spherically symmetric ${\ensuremath{\phi}}^{4}$ theory,''}
  {\em Phys. Rev. D}, vol.~82, p.~024038, Jul 2010.

\bibitem{FodorRacz08}
G.~Fodor and I.~R\'acz, \href {http://dx.doi.org/10.1103/PhysRevD.77.025019}
  {``{Numerical investigation of highly excited magnetic monopoles in $SU(2)$
  Yang-Mills-Higgs theory},''} {\em Phys. Rev. D}, vol.~77, p.~025019, Jan
  2008.

\bibitem{FodorRacz04}
G.~Fodor and I.~R\'acz, \href {http://dx.doi.org/10.1103/PhysRevLett.92.151801}
  {``{What Does a Strongly Excited 't Hooft-Polyakov Magnetic Monopole Do?},''}
  {\em Phys. Rev. Lett.}, vol.~92, p.~151801, Apr 2004.

\bibitem{Gustafsson13}
B.~Gustafsson, H.-O. Kreiss, and J.~Oliger, \href
  {http://dx.doi.org/10.1002/9781118548448} {{\em Time-Dependent Problems and
  Difference Methods}}.
\newblock John Wiley Sons, Inc., 2013.

\bibitem{FodorRacz03}
G.~Fodor and I.~R\'acz, \href {http://dx.doi.org/10.1103/PhysRevD.68.044022}
  {``Massive fields tend to form highly oscillating self-similarly expanding
  shells,''} {\em Phys. Rev. D}, vol.~68, p.~044022, Aug 2003.

\bibitem{Saffin2007}
P.~M. Saffin and A.~Tranberg, \href
  {http://stacks.iop.org/1126-6708/2007/i=01/a=030} {``Oscillons and
  quasi-breathers in {D}+1 dimensions,''} {\em Journal of High Energy Physics},
  vol.~2007, no.~01, p.~030, 2007.

\bibitem{Salmi12}
P.~Salmi and M.~Hindmarsh, \href {http://dx.doi.org/10.1103/PhysRevD.85.085033}
  {``Radiation and relaxation of oscillons,''} {\em Phys. Rev. D}, vol.~85,
  p.~085033, Apr 2012.

\bibitem{libquadmath}
``{GCC libquadmath}.'' \url{https://gcc.gnu.org/onlinedocs/libquadmath/}.

\bibitem{boostmultiprec}
``{boost C++ libraries, multiprecision}.''
  \url{https://www.boost.org/doc/libs/1_69_0/libs/multiprecision/doc/html/index.html}.

\bibitem{Bailey15}
D.~H. Bailey and J.~M. Borwein, \href {http://dx.doi.org/10.3390/math3020337}
  {``High-precision arithmetic in mathematical physics,''} {\em Mathematics},
  vol.~3, no.~2, pp.~337--367, 2015.

\bibitem{Ikeda17}
T.~Ikeda, C.-M. Yoo, and V.~Cardoso, \href
  {http://dx.doi.org/10.1103/PhysRevD.96.064047} {``Self-gravitating oscillons
  and new critical behavior,''} {\em Phys. Rev. D}, vol.~96, p.~064047, Sep
  2017.

\bibitem{Boyd1989a}
J.~P. Boyd, \href {http://dx.doi.org/10.1016/S0422-9894(08)70180-6} {``Weakly
  non-local solitary waves,''} in {\em Mesoscale/Synoptic Coherent structures
  in Geophysical Turbulence} (J.~Nihoul and B.~Jamart, eds.), vol.~50 of {\em
  Elsevier Oceanography Series}, pp.~103 -- 112, Elsevier, 1989.

\bibitem{Boyd-book1998}
J.~P. Boyd, \href {http://dx.doi.org/10.1007/978-1-4615-5825-5} {{\em Weakly
  Nonlocal Solitary Waves and Beyond-All-Orders Asymptotics}}.
\newblock Springer US, 1998.

\bibitem{Boyd1989}
J.~P. Boyd, \href {http://dx.doi.org/10.1016/S0065-2156(08)70194-7} {``New
  directions in solitons and nonlinear periodic waves: Polycnoidal waves,
  imbricated solitons, weakly nonlocal solitary waves, and numerical boundary
  value algorithms,''} vol.~27 of {\em Advances in Applied Mechanics}, pp.~1 --
  82, Elsevier, 1989.

\bibitem{Boyd1990}
J.~P. Boyd, \href {http://stacks.iop.org/0951-7715/3/i=1/a=010} {``A numerical
  calculation of a weakly non-local solitary wave: the $\phi^4$ breather,''}
  {\em Nonlinearity}, vol.~3, no.~1, p.~177, 1990.

\bibitem{Boyd1995}
J.~P. Boyd, \href {http://dx.doi.org/10.1016/0165-2125(95)00005-4} {``Weakly
  nonlocal envelope solitary waves: numerical calculations for the
  {Klein-Gordon} ($\phi^4$) equation,''} {\em Wave Motion}, vol.~21, no.~4,
  pp.~311 -- 330, 1995.

\bibitem{Watkins96}
R.~Watkins, ``Theory of oscillons.'' DART-HEP-96 preprint, unpublished, 1996.

\bibitem{Lorene-home}
``{LORENE, Langage Objet pour la RElativité NumériquE}.''
  \url{http://www.lorene.obspm.fr/}.

\bibitem{Lorene-school}
``{School on spectral methods: Application to General Relativity and Field
  Theory, 2005, Meudon Observatory, France}.''
  \url{http://www.lorene.obspm.fr/school/index.html}.

\bibitem{Grandclement-01}
P.~Grandclément, S.~Bonazzola, E.~Gourgoulhon, and J.-A. Marck, \href
  {http://dx.doi.org/10.1006/jcph.2001.6734} {``A multidomain spectral method
  for scalar and vectorial {Poisson} equations with noncompact sources,''} {\em
  Journal of Computational Physics}, vol.~170, no.~1, pp.~231 -- 260, 2001.

\bibitem{gsl-minim}
``{GSL - GNU Scientific Library, Multidimensional Minimization}.''
  \url{https://www.gnu.org/software/gsl/doc/html/multimin.html}.

\bibitem{Gleiser04}
M.~Gleiser, \href {http://dx.doi.org/10.1016/j.physletb.2004.08.064}
  {``d-{Dimensional} oscillating scalar field lumps and the dimensionality of
  space,''} {\em Physics Letters B}, vol.~600, no.~1, pp.~126 -- 132, 2004.

\bibitem{Pokrovskii61}
V.~L. Pokrovskii and I.~M. Khalatnikov, \href
  {http://www.jetp.ac.ru/cgi-bin/e/index/e/13/6/p1207?a=list} {``On the problem
  of above-barrier reflection of high-energy particles,''} {\em JETP}, vol.~13,
  p.~1207, 1961.

\bibitem{Pomeau1988}
Y.~Pomeau, A.~Ramani, and B.~Grammaticos, \href
  {http://dx.doi.org/10.1016/0167-2789(88)90018-8} {``Structural stability of
  the {Korteweg-de Vries} solitons under a singular perturbation,''} {\em
  Physica D: Nonlinear Phenomena}, vol.~31, no.~1, pp.~127 -- 134, 1988.

\bibitem{dlmflibrary}
``{\it NIST Digital Library of Mathematical Functions}.''
  \url{http://dlmf.nist.gov/}.
\newblock F.~W.~J. Olver, A.~B. {Olde Daalhuis}, D.~W. Lozier, B.~I. Schneider,
  R.~F. Boisvert, C.~W. Clark, B.~R. Miller and B.~V. Saunders, eds.

\bibitem{Kaup68}
D.~J. Kaup, \href {http://dx.doi.org/10.1103/PhysRev.172.1331}
  {``{Klein-Gordon} geon,''} {\em Phys. Rev.}, vol.~172, pp.~1331--1342, Aug
  1968.

\bibitem{Ruffini69}
R.~Ruffini and S.~Bonazzola, \href {http://dx.doi.org/10.1103/PhysRev.187.1767}
  {``Systems of self-gravitating particles in general relativity and the
  concept of an equation of state,''} {\em Phys. Rev.}, vol.~187,
  pp.~1767--1783, Nov 1969.

\bibitem{Jetzer92}
P.~Jetzer, \href {http://dx.doi.org/10.1016/0370-1573(92)90123-H} {``Boson
  stars,''} {\em Physics Reports}, vol.~220, no.~4, pp.~163 -- 227, 1992.

\bibitem{Schunck03}
F.~E. Schunck and E.~W. Mielke, \href
  {http://stacks.iop.org/0264-9381/20/i=20/a=201} {``General relativistic boson
  stars,''} {\em Classical and Quantum Gravity}, vol.~20, no.~20, p.~R301,
  2003.

\bibitem{Liebling17}
S.~L. Liebling and C.~Palenzuela, \href
  {http://dx.doi.org/10.1007/s41114-017-0007-y} {``Dynamical boson stars,''}
  {\em Living Reviews in Relativity}, vol.~20, p.~5, Nov 2017.

\bibitem{Urena02a}
L.~A. Ureña-López, \href {http://stacks.iop.org/0264-9381/19/i=10/a=307}
  {``Oscillatons revisited,''} {\em Classical and Quantum Gravity}, vol.~19,
  no.~10, p.~2617, 2002.

\bibitem{Urena02b}
L.~A. Ureña-López, T.~Matos, and R.~Becerril, \href
  {http://stacks.iop.org/0264-9381/19/i=23/a=320} {``Inside oscillatons,''}
  {\em Classical and Quantum Gravity}, vol.~19, no.~23, p.~6259, 2002.

\bibitem{Guzman03}
F.~S. Guzm\'an and L.~A. Ure\~na L\'opez, \href
  {http://dx.doi.org/10.1103/PhysRevD.68.024023} {``Newtonian collapse of
  scalar field dark matter,''} {\em Phys. Rev. D}, vol.~68, p.~024023, Jul
  2003.

\bibitem{Guzman04}
F.~S. Guzm\'an and L.~A. Ure\~na L\'opez, \href
  {http://dx.doi.org/10.1103/PhysRevD.69.124033} {``Evolution of the
  {Schr\"odinger-Newton} system for a self-gravitating scalar field,''} {\em
  Phys. Rev. D}, vol.~69, p.~124033, Jun 2004.

\bibitem{Guzman06}
F.~S. Guzmán and L.~A. Ureña-López, \href
  {http://stacks.iop.org/0004-637X/645/i=2/a=814} {``Gravitational cooling of
  self-gravitating {Bose} condensates,''} {\em The Astrophysical Journal},
  vol.~645, no.~2, p.~814, 2006.

\bibitem{Alcubierre03}
M.~Alcubierre, R.~Becerril, F.~S. Guzmán, T.~Matos, D.~Núñez, and L.~A.
  Ureña-López, \href {http://stacks.iop.org/0264-9381/20/i=13/a=332}
  {``Numerical studies of $\phi^2$ -oscillatons,''} {\em Classical and Quantum
  Gravity}, vol.~20, no.~13, p.~2883, 2003.

\bibitem{Brady97}
P.~R. Brady, C.~M. Chambers, and S.~M. C.~V. Gon\c{c}alves, \href
  {http://dx.doi.org/10.1103/PhysRevD.56.R6057} {``Phases of massive scalar
  field collapse,''} {\em Phys. Rev. D}, vol.~56, pp.~R6057--R6061, Nov 1997.

\bibitem{Okawa14}
H.~Okawa, V.~Cardoso, and P.~Pani, \href
  {http://dx.doi.org/10.1103/PhysRevD.89.041502} {``Collapse of
  self-interacting fields in asymptotically flat spacetimes: Do
  self-interactions render {Minkowski} spacetime unstable?,''} {\em Phys. Rev.
  D}, vol.~89, p.~041502, Feb 2014.

\bibitem{Rozali18}
M.~Rozali and B.~Way, \href {http://dx.doi.org/10.1007/JHEP11(2018)106}
  {``Gravitating scalar stars in the large {D} limit,''} {\em Journal of High
  Energy Physics}, vol.~2018, p.~106, Nov 2018.

\bibitem{Obregon05}
O.~Obreg\'on, L.~A. Ure\~na L\'opez, and F.~E. Schunck, \href
  {http://dx.doi.org/10.1103/PhysRevD.72.024004} {``Oscillatons formed by
  nonlinear gravity,''} {\em Phys. Rev. D}, vol.~72, p.~024004, Jul 2005.

\bibitem{Urena12}
L.~A. Ureña-López, S.~Valdez-Alvarado, and R.~Becerril, \href
  {http://stacks.iop.org/0264-9381/29/i=6/a=065021} {``Evolution and stability
  $\phi^4$ oscillatons,''} {\em Classical and Quantum Gravity}, vol.~29, no.~6,
  p.~065021, 2012.

\bibitem{Mahmoodzadeh18b}
A.~Mahmoodzadeh and B.~Malekolkalami, \href
  {http://dx.doi.org/10.1016/j.dark.2017.11.001} {``Oscillatons described by
  self-interacting quartic scalar fields,''} {\em Physics of the Dark
  Universe}, vol.~19, pp.~21 -- 26, 2018.

\bibitem{Balakrishna12}
J.~Balakrishna, R.~Bondarescu, G.~Daues, and M.~Bondarescu, \href
  {http://dx.doi.org/10.1103/PhysRevD.77.024028} {``Numerical simulations of
  oscillating soliton stars: Excited states in spherical symmetry and ground
  state evolutions in {3D},''} {\em Phys. Rev. D}, vol.~77, p.~024028, Jan
  2008.

\bibitem{Muia19}
F.~Muia, M.~Cicoli, K.~Clough, F.~Pedro, F.~Quevedo, and G.~P. Vacca, \href
  {http://dx.doi.org/10.1088/1475-7516/2019/07/044} {``The fate of dense scalar
  stars,''} {\em Journal of Cosmology and Astroparticle Physics}, vol.~2019,
  pp.~044--044, jul 2019.

\bibitem{Okawa15}
H.~Okawa, \href {http://stacks.iop.org/0264-9381/32/i=21/a=214003} {``Nonlinear
  evolutions of bosonic clouds around black holes,''} {\em Classical and
  Quantum Gravity}, vol.~32, no.~21, p.~214003, 2015.

\bibitem{Helfer17}
T.~Helfer, D.~J. Marsh, K.~Clough, M.~Fairbairn, E.~A. Lim, and R.~Becerril,
  \href {http://stacks.iop.org/1475-7516/2017/i=03/a=055} {``Black hole
  formation from axion stars,''} {\em Journal of Cosmology and Astroparticle
  Physics}, vol.~2017, no.~03, p.~055, 2017.

\bibitem{Michel18}
F.~Michel and I.~G. Moss, \href
  {http://dx.doi.org/10.1016/j.physletb.2018.07.063} {``Relativistic collapse
  of axion stars,''} {\em Physics Letters B}, vol.~785, pp.~9 -- 13, 2018.

\bibitem{Widdicombe18}
J.~Y. Widdicombe, T.~Helfer, D.~J. Marsh, and E.~A. Lim, \href
  {http://stacks.iop.org/1475-7516/2018/i=10/a=005} {``Formation of
  relativistic axion stars,''} {\em Journal of Cosmology and Astroparticle
  Physics}, vol.~2018, no.~10, p.~005, 2018.

\bibitem{Iwazaki15}
A.~Iwazaki, \href {http://dx.doi.org/10.1103/PhysRevD.91.023008} {``Axion stars
  and fast radio bursts,''} {\em Phys. Rev. D}, vol.~91, p.~023008, Jan 2015.

\bibitem{Raby16}
S.~Raby, \href {http://dx.doi.org/10.1103/PhysRevD.94.103004} {``Axion star
  collisions with neutron stars and fast radio bursts,''} {\em Phys. Rev. D},
  vol.~94, p.~103004, Nov 2016.

\bibitem{Clough18}
K.~Clough, T.~Dietrich, and J.~C. Niemeyer, \href
  {http://dx.doi.org/10.1103/PhysRevD.98.083020} {``Axion star collisions with
  black holes and neutron stars in full {3D} numerical relativity,''} {\em
  Phys. Rev. D}, vol.~98, p.~083020, Oct 2018.

\bibitem{Edwards18}
F.~Edwards, E.~Kendall, S.~Hotchkiss, and R.~Easther, \href
  {http://stacks.iop.org/1475-7516/2018/i=10/a=027} {``Pyultralight: a
  pseudo-spectral solver for ultralight dark matter dynamics,''} {\em Journal
  of Cosmology and Astroparticle Physics}, vol.~2018, no.~10, p.~027, 2018.

\bibitem{Madarassy15}
E.~J.~M. Madarassy and V.~T. Toth, \href
  {http://dx.doi.org/10.1103/PhysRevD.91.044041} {``Evolution and dynamical
  properties of {Bose-Einstein} condensate dark matter stars,''} {\em Phys.
  Rev. D}, vol.~91, p.~044041, Feb 2015.

\bibitem{Amin19}
M.~A. Amin and P.~Mocz, \href {http://dx.doi.org/10.1103/PhysRevD.100.063507}
  {``Formation, gravitational clustering, and interactions of nonrelativistic
  solitons in an expanding universe,''} {\em Phys. Rev. D}, vol.~100,
  p.~063507, Sep 2019.

\bibitem{Becerril2006}
R.~Becerril, T.~Matos, and L.~Ure{\~{n}}a-L{\'o}pez, \href
  {http://dx.doi.org/10.1007/s10714-006-0253-x} {``Geodesics around
  oscillatons,''} {\em General Relativity and Gravitation}, vol.~38,
  pp.~633--641, Apr 2006.

\bibitem{Boskovic18}
M.~Bo\v{s}kovi\'{c}, F.~Duque, M.~C. Ferreira, F.~S. Miguel, and V.~Cardoso,
  \href {http://dx.doi.org/10.1103/PhysRevD.98.024037} {``Motion in
  time-periodic backgrounds with applications to ultralight dark matter halos
  at galactic centers,''} {\em Phys. Rev. D}, vol.~98, p.~024037, Jul 2018.

\bibitem{Mahmoodzadeh18}
A.~Mahmoodzadeh and B.~Malekolkalami, \href
  {http://dx.doi.org/10.1007/s10509-018-3389-8} {``Geodesics around oscillatons
  made of exponential scalar field potential,''} {\em Astrophysics and Space
  Science}, vol.~363, p.~173, Jul 2018.

\bibitem{Brito16b}
R.~Brito, V.~Cardoso, C.~A. Herdeiro, and E.~Radu, \href
  {http://dx.doi.org/10.1016/j.physletb.2015.11.051} {``Proca stars:
  Gravitating {Bose–Einstein} condensates of massive spin 1 particles,''}
  {\em Physics Letters B}, vol.~752, pp.~291 -- 295, 2016.

\bibitem{Helfer19}
T.~Helfer, E.~A. Lim, M.~A.~G. Garcia, and M.~A. Amin, \href
  {http://dx.doi.org/10.1103/PhysRevD.99.044046} {``Gravitational wave emission
  from collisions of compact scalar solitons,''} {\em Phys. Rev. D}, vol.~99,
  p.~044046, Feb 2019.

\bibitem{Hawley00Thesis}
S.~H. Hawley, ``Scalar analogues of compact astrophysical systems.''
  \url{http://hedges.belmont.edu/~shawley/diss.ps}, 2000.
\newblock PhD thesis, University of Texas at Austin.

\bibitem{Hawley03}
S.~H. Hawley and M.~W. Choptuik, \href
  {http://dx.doi.org/10.1103/PhysRevD.67.024010} {``Numerical evidence for
  ``multiscalar stars'',''} {\em Phys. Rev. D}, vol.~67, p.~024010, Jan 2003.

\bibitem{Giovanni18}
F.~Di~Giovanni, N.~Sanchis-Gual, C.~A.~R. Herdeiro, and J.~A. Font, \href
  {http://dx.doi.org/10.1103/PhysRevD.98.064044} {``Dynamical formation of
  proca stars and quasistationary solitonic objects,''} {\em Phys. Rev. D},
  vol.~98, p.~064044, Sep 2018.

\bibitem{Barranco11}
J.~Barranco and A.~Bernal, \href {http://dx.doi.org/10.1103/PhysRevD.83.043525}
  {``Self-gravitating system made of axions,''} {\em Phys. Rev. D}, vol.~83,
  p.~043525, Feb 2011.

\bibitem{Eby15}
J.~Eby, P.~Suranyi, C.~Vaz, and L.~C.~R. Wijewardhana, \href
  {http://dx.doi.org/10.1007/JHEP03(2015)080} {``Axion stars in the infrared
  limit,''} {\em Journal of High Energy Physics}, vol.~2015, p.~80, Mar 2015.

\bibitem{Eby16}
J.~Eby, P.~Suranyi, and L.~C.~R. Wijewardhana, \href
  {http://dx.doi.org/10.1142/S0217732316500905} {``The lifetime of axion
  stars,''} {\em Modern Physics Letters A}, vol.~31, no.~15, p.~1650090, 2016.

\bibitem{Braaten17}
E.~Braaten, A.~Mohapatra, and H.~Zhang, \href
  {http://dx.doi.org/10.1103/PhysRevD.96.031901} {``Emission of photons and
  relativistic axions from axion stars,''} {\em Phys. Rev. D}, vol.~96,
  p.~031901, Aug 2017.

\bibitem{Eby18}
J.~Eby, M.~Ma, P.~Suranyi, and L.~C.~R. Wijewardhana, \href
  {http://dx.doi.org/10.1007/JHEP01(2018)066} {``Decay of ultralight axion
  condensates,''} {\em Journal of High Energy Physics}, vol.~2018, p.~66, Jan
  2018.

\bibitem{Misner64}
C.~W. Misner and D.~H. Sharp, \href
  {http://dx.doi.org/10.1103/PhysRev.136.B571} {``Relativistic equations for
  adiabatic, spherically symmetric gravitational collapse,''} {\em Phys. Rev.},
  vol.~136, pp.~B571--B576, Oct 1964.

\bibitem{Nakao95}
K.~ichi Nakao, \href {http://arxiv.org/abs/gr-qc/9507022} {``On a quasi-local
  energy outside the cosmological horizon,''} 1995.
\newblock arXiv:gr-qc/9507022.

\bibitem{Kodama80}
H.~Kodama, \href {http://dx.doi.org/10.1143/PTP.63.1217} {``Conserved energy
  flux for the spherically symmetric system and the backreaction problem in the
  black hole evaporation,''} {\em Progress of Theoretical Physics}, vol.~63,
  no.~4, pp.~1217--1228, 1980.

\bibitem{Hayward96}
S.~A. Hayward, \href {http://dx.doi.org/10.1103/PhysRevD.53.1938}
  {``Gravitational energy in spherical symmetry,''} {\em Phys. Rev. D},
  vol.~53, pp.~1938--1949, Feb 1996.

\bibitem{Tangherlini63}
F.~R. Tangherlini, \href {http://dx.doi.org/10.1007/BF02784569}
  {``Schwarzschild field in n dimensions and the dimensionality of space
  problem,''} {\em Il Nuovo Cimento (1955-1965)}, vol.~27, pp.~636--651, Feb
  1963.

\bibitem{Friedberg87}
R.~Friedberg, T.~D. Lee, and Y.~Pang, \href
  {http://dx.doi.org/10.1103/PhysRevD.35.3640} {``Mini-soliton stars,''} {\em
  Phys. Rev. D}, vol.~35, pp.~3640--3657, Jun 1987.

\bibitem{Kichenassamy08}
S.~Kichenassamy, \href {http://stacks.iop.org/0264-9381/25/i=24/a=245004}
  {``Soliton stars in the breather limit,''} {\em Classical and Quantum
  Gravity}, vol.~25, no.~24, p.~245004, 2008.

\bibitem{Fodor2009c}
G.~Fodor, P.~Forgács, Z.~Horváth, and M.~Mezei, \href
  {http://stacks.iop.org/1126-6708/2009/i=08/a=106} {``Oscillons in
  dilaton-scalar theories,''} {\em Journal of High Energy Physics}, vol.~2009,
  no.~08, p.~106, 2009.

\bibitem{Moroz98}
I.~M. Moroz, R.~Penrose, and P.~Tod, \href
  {http://stacks.iop.org/0264-9381/15/i=9/a=019} {``Spherically-symmetric
  solutions of the {Schrödinger-Newton} equations,''} {\em Classical and
  Quantum Gravity}, vol.~15, no.~9, p.~2733, 1998.

\bibitem{TodMoroz99}
P.~Tod and I.~M. Moroz, \href {http://stacks.iop.org/0951-7715/12/i=2/a=002}
  {``An analytical approach to the {Schrödinger-Newton} equations,''} {\em
  Nonlinearity}, vol.~12, no.~2, p.~201, 1999.

\bibitem{Harrison03}
R.~Harrison, I.~Moroz, and K.~P. Tod, \href
  {http://stacks.iop.org/0951-7715/16/i=1/a=307} {``A numerical study of the
  {Schrödinger–Newton} equations,''} {\em Nonlinearity}, vol.~16, no.~1,
  p.~101, 2003.

\bibitem{Choquard08}
P.~Choquard, J.~Stubbe, and M.~Vuffray, \href {http://arxiv.org/abs/0712.3103}
  {``Stationary solutions of the {Schrödinger-Newton} model---an {ODE}
  approach,''} {\em Differential Integral Equations}, vol.~21, no.~7-8,
  pp.~665--679, 2008.

\bibitem{Diosi84}
L.~Diósi, \href {http://dx.doi.org/10.1016/0375-9601(84)90397-9}
  {``Gravitation and quantum-mechanical localization of macro-objects,''} {\em
  Physics Letters A}, vol.~105, no.~4, pp.~199 -- 202, 1984.

\bibitem{Penrose98}
R.~Penrose, \href {http://dx.doi.org/10.1098/rsta.1998.0256} {``Quantum
  computation, entanglement and state reduction,''} {\em Philosophical
  Transactions of the Royal Society of London. Series A: Mathematical, Physical
  and Engineering Sciences}, vol.~356, p.~1927, 1998.

\bibitem{Ferrell89}
R.~Ferrell and M.~Gleiser, \href {http://dx.doi.org/10.1103/PhysRevD.40.2524}
  {``Gravitational atoms: Gravitational radiation from excited boson stars,''}
  {\em Phys. Rev. D}, vol.~40, pp.~2524--2531, Oct 1989.

\bibitem{Moroz17}
V.~Moroz and J.~Van~Schaftingen, \href
  {http://dx.doi.org/10.1007/s11784-016-0373-1} {``A guide to the {Choquard}
  equation,''} {\em Journal of Fixed Point Theory and Applications}, vol.~19,
  pp.~773--813, Mar 2017.

\bibitem{kadathwebp}
``{KADATH Spectral Solver}.''
  \url{https://luth.obspm.fr/~luthier/grandclement/kadath.html}.
\newblock author: Philippe Grandclément.

\bibitem{grandclement2010}
P.~Grandclément, \href {http://dx.doi.org/10.1016/j.jcp.2010.01.005}
  {``Kadath: A spectral solver for theoretical physics,''} {\em Journal of
  Computational Physics}, vol.~229, no.~9, pp.~3334 -- 3357, 2010.

\bibitem{Bizon11}
P.~Bizo\'{n} and A.~Rostworowski, \href
  {http://dx.doi.org/10.1103/PhysRevLett.107.031102} {``Weakly turbulent
  instability of anti--de {Sitter} spacetime,''} {\em Phys. Rev. Lett.},
  vol.~107, p.~031102, Jul 2011.

\bibitem{Maliborski13}
M.~Maliborski and A.~Rostworowski, \href
  {http://dx.doi.org/10.1103/PhysRevLett.111.051102} {``Time-periodic solutions
  in an {Einstein AdS}--massless-scalar-field system,''} {\em Phys. Rev.
  Lett.}, vol.~111, p.~051102, Aug 2013.

\bibitem{Dias12b}
{\'{O}}.~J.~C. Dias, G.~T. Horowitz, D.~Marolf, and J.~E. Santos, \href
  {http://dx.doi.org/10.1088/0264-9381/29/23/235019} {``On the nonlinear
  stability of asymptotically anti-de {Sitter} solutions,''} {\em Classical and
  Quantum Gravity}, vol.~29, p.~235019, nov 2012.

\bibitem{Choptuik18}
M.~W. Choptuik, J.~E. Santos, and B.~Way, \href
  {http://dx.doi.org/10.1103/PhysRevLett.121.021103} {``Charting islands of
  stability with multioscillators in anti--de {Sitter} space,''} {\em Phys.
  Rev. Lett.}, vol.~121, p.~021103, Jul 2018.

\bibitem{Fodor15}
G.~Fodor, P.~Forg\'acs, and P.~Grandcl\'ement, \href
  {http://dx.doi.org/10.1103/PhysRevD.92.025036} {``Self-gravitating scalar
  breathers with a negative cosmological constant,''} {\em Phys. Rev. D},
  vol.~92, p.~025036, Jul 2015.

\bibitem{Dias12a}
{\'{O}}.~J.~C. Dias, G.~T. Horowitz, and J.~E. Santos, \href
  {http://dx.doi.org/10.1088/0264-9381/29/19/194002} {``Gravitational turbulent
  instability of anti-de {Sitter} space,''} {\em Classical and Quantum
  Gravity}, vol.~29, p.~194002, aug 2012.

\bibitem{Dias16}
{\'{O}}.~J.~C. Dias and J.~E. Santos, \href
  {http://dx.doi.org/10.1088/0264-9381/33/23/23lt01} {``{AdS} nonlinear
  instability: moving beyond spherical symmetry,''} {\em Classical and Quantum
  Gravity}, vol.~33, p.~23LT01, nov 2016.

\bibitem{Rostworowski17}
A.~Rostworowski, \href {http://dx.doi.org/10.1088/1361-6382/aa71cc} {``Comment
  on `{AdS} nonlinear instability: moving beyond spherical symmetry' (2016
  class. quantum grav. 33 23lt01),''} {\em Classical and Quantum Gravity},
  vol.~34, p.~128001, may 2017.

\bibitem{Horowitz15}
G.~T. Horowitz and J.~E. Santos, \href
  {http://dx.doi.org/10.4310/SDG.2015.v20.n1.a13} {``Geons and the instability
  of anti-de {Sitter} spacetime,''} in {\em One hundred years of general
  relativity} (Y.~Bieri, ed.), vol.~20 of {\em Surveys in Differential
  Geometry}, pp.~321 -- 335, International Press of Boston, 2015.

\bibitem{Wheeler55}
J.~A. Wheeler, \href {http://dx.doi.org/10.1103/PhysRev.97.511} {``Geons,''}
  {\em Phys. Rev.}, vol.~97, pp.~511--536, Jan 1955.

\bibitem{Brill64}
D.~R. Brill and J.~B. Hartle, \href
  {http://dx.doi.org/10.1103/PhysRev.135.B271} {``Method of the self-consistent
  field in general relativity and its application to the gravitational geon,''}
  {\em Phys. Rev.}, vol.~135, pp.~B271--B278, Jul 1964.

\bibitem{Anderson97}
P.~R. Anderson and D.~R. Brill, \href
  {http://dx.doi.org/10.1103/PhysRevD.56.4824} {``Gravitational geons
  revisited,''} {\em Phys. Rev. D}, vol.~56, pp.~4824--4833, Oct 1997.

\bibitem{Martinon17}
G.~Martinon, G.~Fodor, P.~Grandcl{\'{e}}ment, and P.~Forg{\'{a}}cs, \href
  {http://dx.doi.org/10.1088/1361-6382/aa6f48} {``Gravitational geons in
  asymptotically anti-de {Sitter} spacetimes,''} {\em Classical and Quantum
  Gravity}, vol.~34, p.~125012, may 2017.

\bibitem{Fodor17}
G.~Fodor and P.~Forg\'acs, \href {http://dx.doi.org/10.1103/PhysRevD.96.084027}
  {``Anti--de {Sitter} geon families,''} {\em Phys. Rev. D}, vol.~96,
  p.~084027, Oct 2017.

\end{thebibliography}


\end{document}